\pdfoutput=1

\documentclass[twoside,12pt]{article}
\usepackage{graphicx}

\def\lsim{\mathrel{\rlap{\lower3pt\hbox{\hskip1pt$\sim$}}
     \raise1pt\hbox{$<$}}} 
\def\gsim{\mathrel{\rlap{\lower3pt\hbox{\hskip1pt$\sim$}}
     \raise1pt\hbox{$>$}}} 
      \newcommand{\CD}{\mathcal{D}}

\def\barc{$\bar c c\,\,\,$}
\def\Jp{$J/\psi$ }
\def\bz{b^{(0)}}

\def\({\left(}
\def\){\right)}
\def\[{\left[}
\def\]{\right]}

\newcommand{\be}{\begin{equation}}
\newcommand{\ee}{\end{equation}}
\newcommand{\ba}{\begin{eqnarray}}
\newcommand{\ea}{\end{eqnarray}}

\begin{document}





\title{Physics of Strongly coupled Quark-Gluon Plasma}

\author{Edward Shuryak\\
Department of Physics and Astronomy\\
University at Stony Brook,NY 11794 USA \\
shuryak@tonic.physics.sunysb.edu}
\maketitle
\begin{abstract}
 This review cover our current understanding of strongly coupled Quark-Gluon Plasma
 (sQGP), especially theoretical progress in (i) explaining the RHIC
 data by hydrodynamics, (ii) describing lattice data using electric-magnetic
 duality; (iii) understanding of gauge-string duality known as AdS/CFT
 and its application for ``conformal" plasma. In view of interdisciplinary
 nature of the subject, we include brief introduction
 into several topics ``for pedestrians". 
Some fundamental questions addressed are:
 Why is sQGP such a
good liquid? What is the nature of (de)confinement and 
what do we know about ``magnetic'' objects creating it? Do they play any important
role in sQGP physics?
Can we understand
the AdS/CFT predictions, from the gauge theory side?
Can they be tested  experimentally?
Can AdS/CFT duality help us understand  rapid
equilibration/entropy production? Can we work out
a complete dynamical ``gravity dual'' to heavy ion collisions?
\end{abstract}

\section{ Introduction}
Soon after discovery of QCD, people used asymptotic freedom to argue
that very hot/dense matter must be weakly coupled \cite{Collins:1974ky} and thus deconfined.  
My own entrance to this field started with the question: while $vacuum$ fluctuations
of gauge field lead to {\em anti-screening}, the famous negative
coefficient leading to asymptotic freedom, what  their $thermal$ fluctuations would
do? Explicit calculation \cite{Shuryak:1977ut}, using the Coulomb gauge, have produced
$positive$ sign of the Debye mass, opposite to that of virtual gluons
and the same as in QED!
Thus I called this phase of matter ``plasma", putting it even in the paper title\footnote{A small anecdote is related to that: in those days plasma research in Russia was semi-classified and thus the paper returned to me with a note saying that it lacks proper
permissions. I found a quick fix to this problem: a letter from
renown plasma physicist saying this plasma has nothing to do with ``real plasma physics". And perhaps it was true: I got an invitation to speak about it at some plasma physics meeting only 30 years later. }. 
One more important finding of that calculation was that static magnetic 
screening is absent, also like in QED: we will see that this conclusion would be
seriously modified non-perturbatively.

The next 25 years theory of QGP was based on pQCD.
The program to do so, with a number of ``signals" was proposed in my other
paper \cite{Shuryak:1978ij} : those were based on high energy hadronic/nuclear collisions
and included several perturbative process, from gluonic production of new quark
flavors to charmonium dissociation by gluonic ``photoeffect".
A lot of efforts have been made to derive perturbative series
for finite-$T$ thermodynamics, and eventually all
calculable terms have been calculated, see e.g. \cite{Andersen:2004fp}.
Convergence was bad, but we thought that some clever re-summation can still
make it work.

 But whether QGP can or cannot be created experimentally was not at all clear.
 In fact all the way to the year 2000, when RHIC started, most theorists
 argued  that nothing else but a ``firework of mini-jets" can possibly
 be seen at RHIC.

  And yet, in my talk at QM99 I predicted based on hydro
calculations, that 
 elliptic flow at RHIC would be twice
  $larger$ than at SPS. 
 It did not take long after the start of RHIC operation to see that
this is indeed what happens: in fact 
a rather perfect case for hydrodynamical explosion
was made both from radial and elliptic flows.
 After lots of
debates, this period culminated with the ``discovery"
workshop of 2004 and subsequent ``white papers" from 4 experimental
collaborations  which documented it.

Theory of QGP is still profoundly affected by
this ``paradigm shift"  to 
the so called strong-coupling regime.
 We are still in so-to-say non-equilibrium transition, as 
 huge amount of physics issues required  to be
learned.
Some came from other fields, including physics of {\em 
 strongly coupled QED plasmas} and
 trapped {\em  ultracold gases} with large scattering length.
{\em String theory} provided  a remarkable tool -- the AdS/CFT
correspondence -- which related heavy ions to the
the fascinating physics of strong gravity and {\em black
  holes}. Another important trend is that
{\em transport properties} of QGP and {\em non-equilibrium dynamics}
 came to the forefront: and for those the Euclidean approaches 
(lattice, instantons) we used before is much less suited
than for thermodynamics. All of it made the last 5 years
the time of unprecedented challenges.

Because the issues we discuss
incorporate several fields of physics, some introductory parts of this review are marked
``for pedestrians''. Indeed, heavy ion physics did not have much in
common with string theory and black holes, or dilute quantum gases,
so some basic definitions and main physics statements (made at an ``intuitive'' 
level) may be  helpful to some readers. We start with two such
introductory subsections, about classical strongly coupled plasmas and
quantum ultracold gases, which we will not discuss in this review in depth.

\subsection{Strongly coupled plasmas for pedestrians}

 By definition, plasmas are states of matter in which
particles are ``charged'' and thus interact via {\em
long range} (massless) gauge\footnote{
Why only gauge and not scalar fields? Indeed, supersymmetric models
have massless scalars which in many cases create the so called BPS situation,
in which gauge repulsion is canceled by scalar attraction: and we will
call them plasmas as well. However this is as far as it goes: a generic
massless scalar, attractive in all channels and not restricted by supersymmetry,
is just a recipe for instability and should not be considered at all.
} fields. This separate it from ``neutral''
gases, liquids or solids in which the interparticle
interaction is short range.
Sometimes plasmas were called ``the 4-th state of matter'',
but this does not comply with standard terminology: in fact
plasmas can themselves be gases, liquids or solids.

Classical plasmas are of course those which does not involve
quantum mechanics or $\hbar$.
Let me start with counting the parameters of the problem. There are
4 variables
\footnote{Recall that the problem is not
only classical but it is also nonrelativistic: thus no
$\hbar$ or $c$.} -- the particle mass and density, the temperature and the Coulomb charge. Three of them can be used as units of mass, length and time: thus only
one combination remains. The standard choice is the
so called {\em plasma parameter}, which can be loosely defined as
the ratio of interaction energy to kinetic energy, and is
more technically defined as
 \be \Gamma= (Ze)^2/(a_{WS}T) \ee
where $Ze,a_{WS},T$ are respectively the ion charge, the Wigner-Seitz radius
$a_{WT}=(3/4\pi n)^{1/3}$  and the temperature: this form 
is convenient 
to use because it only involves the {\it input} parameters, 
such as the 
temperature and density, while average potential energy
is not so easily available. The meaning of it is the same,
and the values
of all observables - like transport coefficients we will discuss
below -- are usually expressed as a function
of $\Gamma$.

Depending on magnitude of this parameter $\Gamma$
classical plasmas have the following regimes:\\
{\bf i.} a weakly coupled or gas regime, for $\Gamma<1$;\\
 {\bf ii.} a
liquid regime for $\Gamma\approx 1- 10$;\\
  {\bf iii.} a glassy liquid regime
for $\Gamma\approx 10-100$;\\
 {\bf iv.} a solid regime for $\Gamma > 300$.
 
 Existence of permanent correlation between the particles is seen
 in the simplest way via
density-density correlation functions 
\be
G(r,t)={1\over N} \left<\sum^{N}_{i=1} \sum^{N}_{j=1} \,
\delta \left (\vec{x}+\vec{x}_{i}(0)-\vec{x}_{j}(t) \right) \right> \,,
\label{Grt}
\ee
with $N$ is the number of particles, $\vec{x}_{i}(t)$ is the position
of the i-th-particle at time $t$. $G(r, t)$ characterizes the 
likelihood to find 2 particles a distance $r$ away from each other at 
time $t$. Here are some examples,
 from our own (non-Abelian) MD simulations \cite{Gelman:2006xw},
which show that liquid regime demonstrate nearest-neighbor peaks,
and crystals have peaks corresponding to longer range order.  They also show ``healing'' of correlations with time in gases, but much less so in liquids and solids.

\begin{figure}[ht]
\begin{center}
\includegraphics*[height=8.cm,width=6.cm]{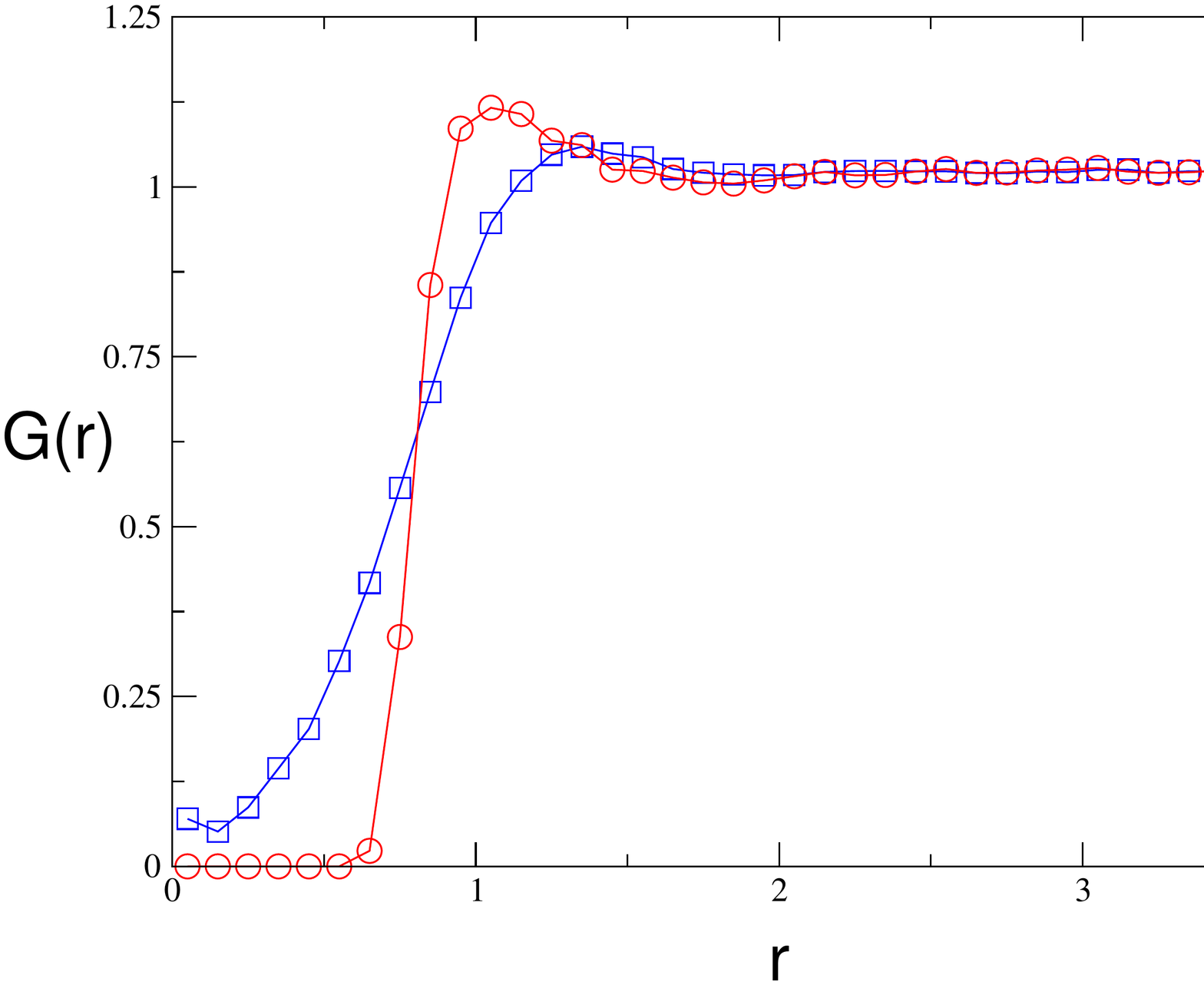}
\includegraphics*[height=8.cm,width=6.cm]{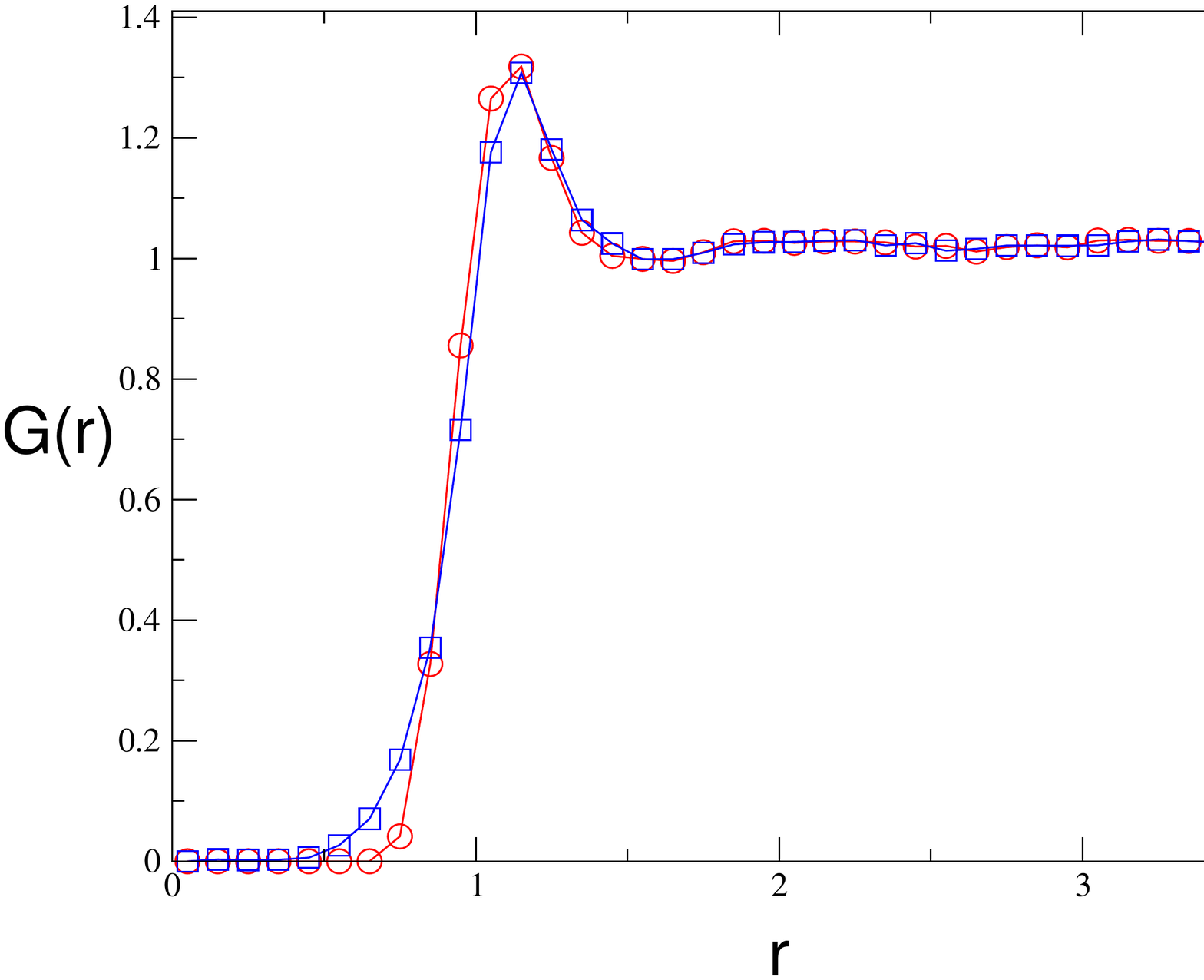}
\includegraphics*[height=8.cm,width=6.cm]{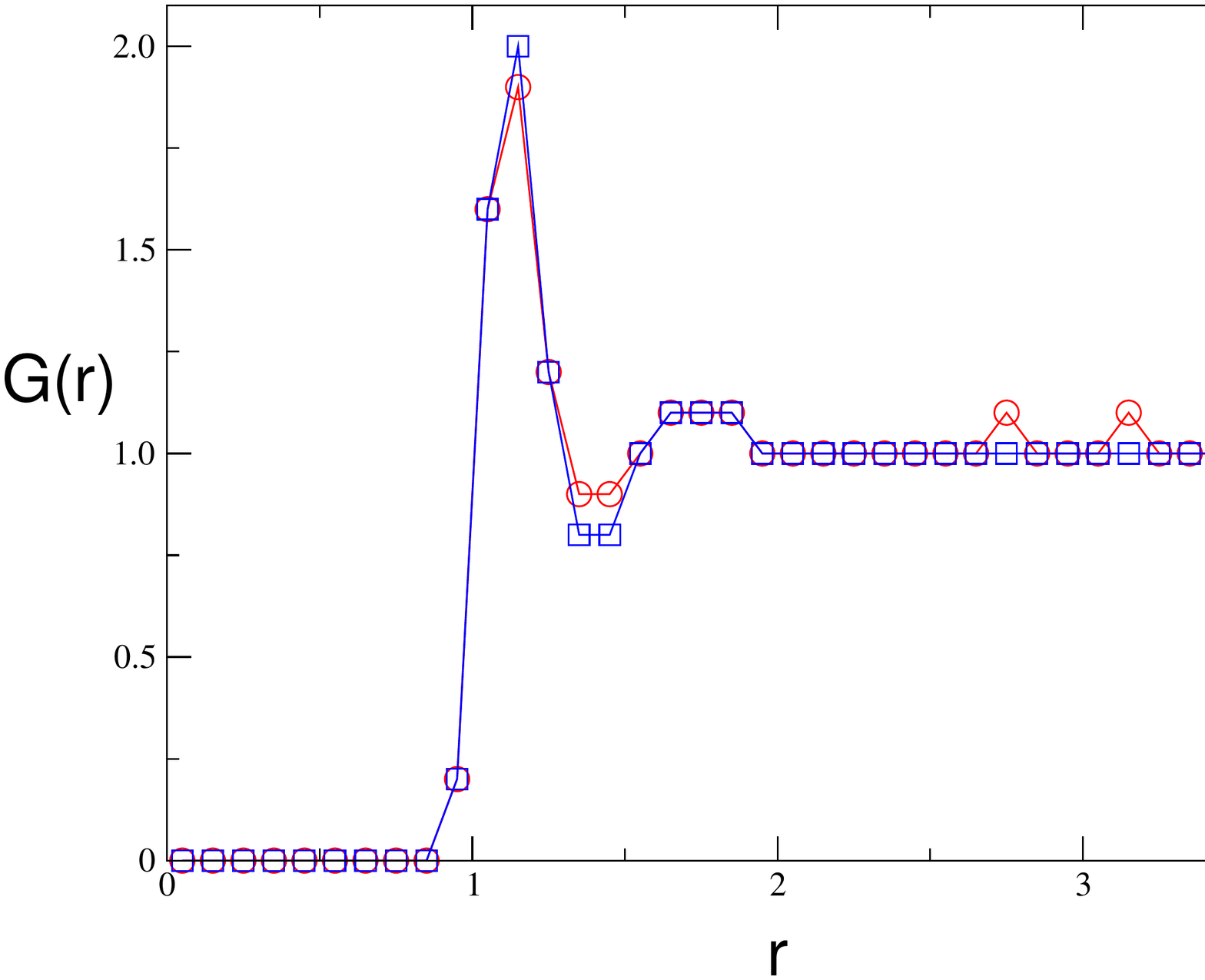}
\caption{(Color online) $G_{d}$ correlation function for $\Gamma=0.83,31.3,131$,
respectively. Red circles correspond
to $t^{*}=0$, and blue squares correspond to $t^{*}=6$.
}
\label{Gd1}
\end{center}
\end{figure}

The case of small $\Gamma$ is widely discussed in statistical
mechanics courses: let me just remind the reader that it is in this
case when one can use Boltzmann eqn, cascades and other simple tools
appropriate for a gas. Unlike gases and solids, the interplay of local order and
randomness at large distances makes
 liquids difficult to treat theoretically\footnote{I heard an opinion,
ascribed to a lecture of V.Weisskopf, that if theorists would invent
the Universe from scratch, without any experiment, they would never
think about liquids.}.
Thus, in spite of their crucial importance for a lot of chemistry
in general and our life
in particular, most  physics and statistical mechanics courses
tend to either omit them completely or 
tell as little as possible about them.
It is possibly worth reminding heavy ion practitioners, that
for liquids neither Boltzmann equation nor cascades  can be used
 because particle are strongly correlated with several neighbors
at all times. The very idea of ``scattering'' and cross section
involves particles coming from and going to infinity: it is
appropriate
for dilute gases but not condensed matter where interparticle
distances do not exceed the range of the forces at any time. 

 Strongly coupled classical electromagnetic
 plasmas can be studied experimentally: they are not at all
exotic objects. For example, table solt $NaCl$ can be considered
a crystalline plasma made of permanently charged ions $Na^+$ and
 $Cl^-$.  At $T\sim 10^3K$ (still too small to ionize non-valence
electrons) one gets
a molten solt, 
which is  liquid plasma with $\Gamma\sim 60$.
A more famous object of recent experimentation is a charged
  {\em dust}: in space (e.g. at International Space Station)  
it has been put 
  into a nice crystal and studied in depth. For example,
one can get a particle piercing it and creating Mach
cones, several if there are excitations other than the sound.  
One can find more introductory information and references
on the subject in Mrowczynski and Thoma review 
\cite{Mrowczynski:2007hb}.
  
We will not discuss any theory of it, but just note that
starting from about 1980's availability of computational resources
get sufficient to use ``Molecular Dynamics" (MD). This means
that one can write equations of motion (EOM) for the
interparticle forces and directly solve them for say
$10^3$ particles. It is this simplest but powerful
 theoretical tool we will use to access
properties of strongly coupled QGP by classical simulations.

\subsection{Strongly coupled ultracold gases }
 
Back in 1999 G.Bertsch formulated a ``many-body challenge problem,'' asking:
what are the ground state properties of a two-species fermion system 
in the limit that the scattering length $(a_s)$ of its interaction
approaches infinity? (This limit is usually referred to as the ``unitary
limit'' because the scattering cross section reaches its unitarity limit
per s-wave.) Such problem
was originally set up as a parameter-free model for a fictitious dilute neutron matter: recall that nn scattering length is indeed huge because of near-zero-energy
isoscalar virtual state.

The answer was provided experimentally by atomic experimentalists,
who found a way to modify the interaction strength in
ultracold trapped systems  by the magnetic field, which can shift level
positions till they cross and form  the so called 
``Feshbach resonances". As a result, the
interaction measure  -- the scattering length in units of
interparticle spacing $a*n^{1/3}$ -- can be changed in a wide
interval, practically from $-\infty$ (very strong attraction) to $+\infty$ (very strong 
repulsion).  
Systems of ultracold atoms  became a
very hot topic a decade earlier, when laser cooling techniques were developed so that
temperatures got  low
 enough (so that atomic thermal de Brogle wavelength get
 comparable to interparticle distances) to observe Bose-Einstein
 condensation (BEC)  for bosonic atoms. Last years
 was the time of strongly coupled quantum gases made of {\em fermions},
possessing  superfluidity and huge pairing gaps.
Never before one had manybody quantum systems with
widely tunable interparticle interaction amenable to experimentation, and clearly it created a kind
of revolution in quantum manybody physics. 

For some technical reasons, BEC in strongly  systems (of interest to us
in respect to monopole condensation/confinement) is not yet studied, 
while strongly coupled fermions have been investigated quite extensively. 
Let me briefly review the questions which were of the main interest of the atomic community. One of them is whether
 two
known weakly coupled phases -- BCS superconductor for small and negative $a$ and
BEC of bound atom pairs at small positive $a$ -- join smoothly or there
exist a discontinuity at the Feshbach resonance. The answer seems to be
the former, namely two phases do join smoothly.
Another issue is whether low-$T$ strongly coupled fermions is in a
 gaped  superfluid state: it has been answered positively,
 and phenomena as complex as Abrikosov's lattice of vortexes
 has been observed.
 
  Clearly, there are many fascinating phenomena in this field,
 but let us focus on just two issues most relevant for this review and sQGP:\\
 (i)  whether transition from weakly to strongly coupling regime is reflected in an
 unusually small viscosity, seen in onset of hydrodynamical behavior in
 unexpectedly small systems;\\
 (ii) possible universal limits in the infinite interaction limit $a*n^{1/3}\rightarrow \infty$  for pairing, in connection to color superconductivity of quark matter.

\begin{figure}[t] 
\begin{minipage}[h]{3.5cm}
\vskip -3cm   \includegraphics*[clip,height=10cm]{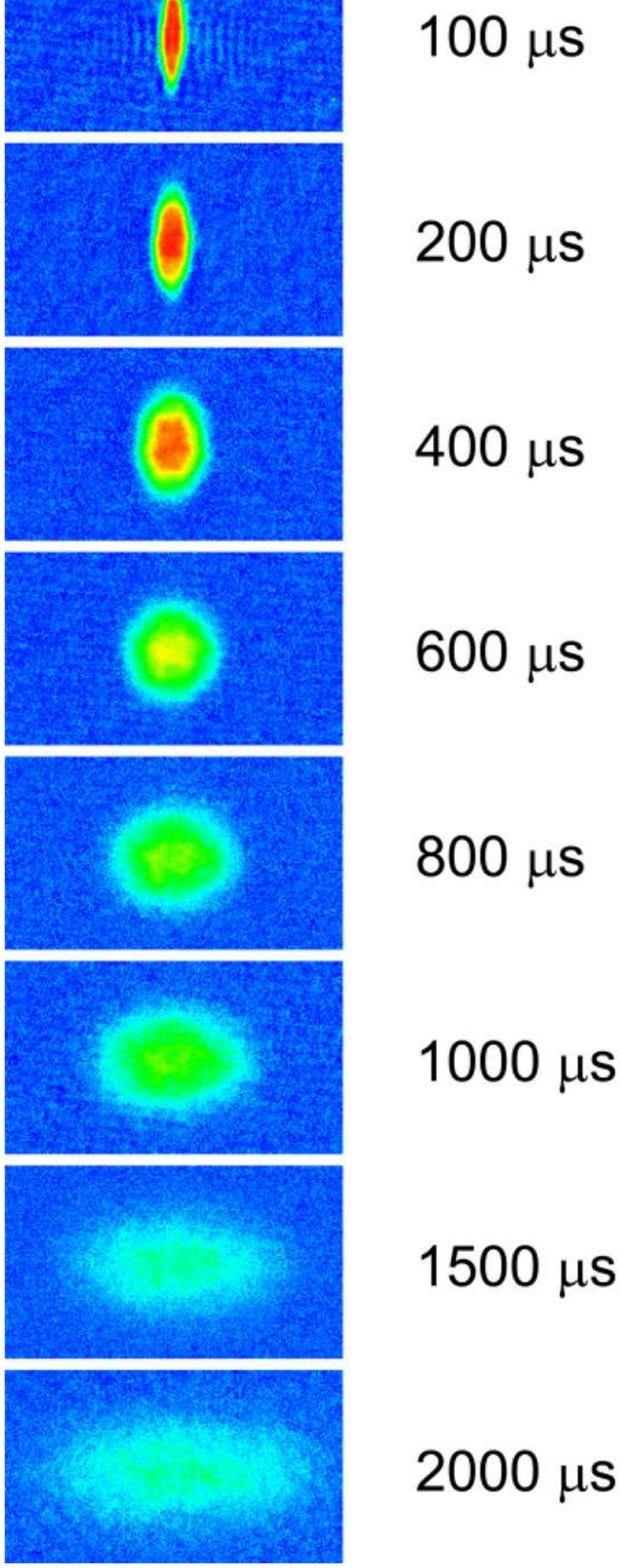} 
\end{minipage} \hspace{2cm}
\begin{minipage}[h]{10cm}
\vskip -5cm
 \includegraphics*[width=8cm,clip]{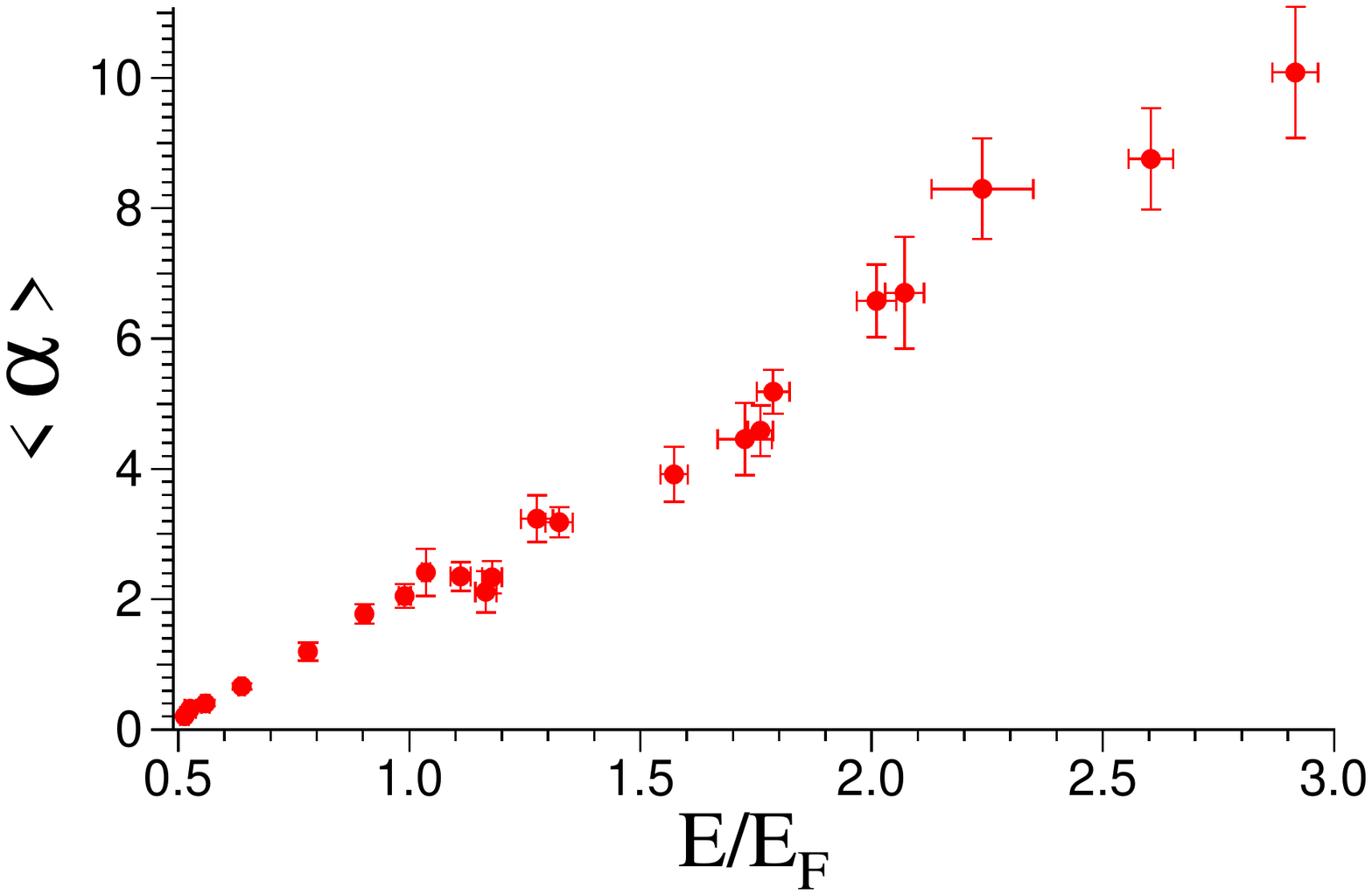}\\
\vskip -7cm 
\includegraphics*[width=10cm,clip]{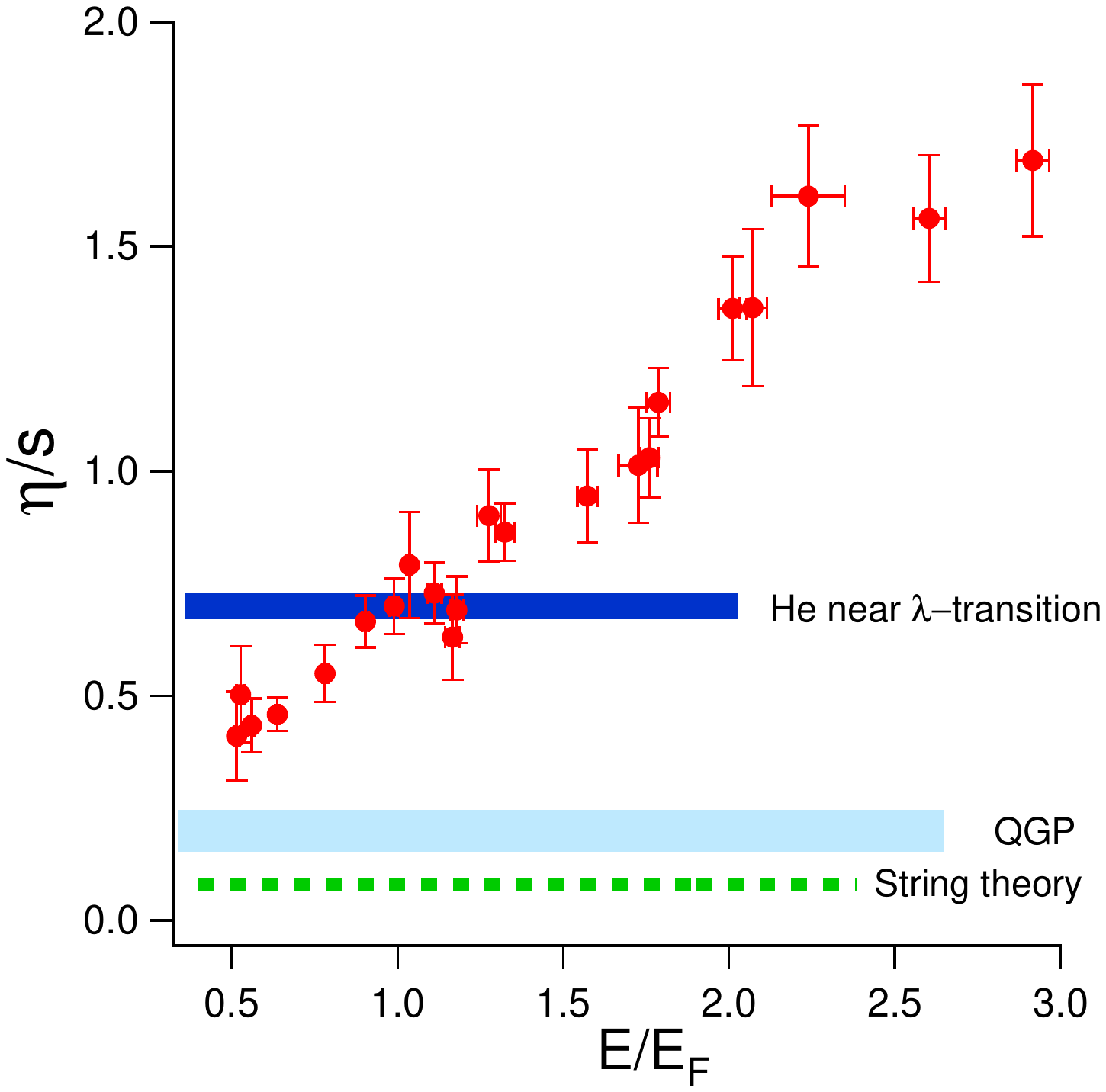}
\end{minipage} 
\caption{\protect
(a) False color absorption images of a strongly interacting
degenerate Fermi gas of ultracold $^6$Li atoms as a function of time after
release from a laser trap.  From O'Hara 
{\it et al.}\protect\cite{OHGGT02}
(b)``Quantum viscosity" in strongly-interacting Fermi gas
 $\alpha=\eta/\hbar\,n$ (trap-averaged).
(c) Same data for the shear viscosity as $\eta/\hbar s$ in units of  the entropy
density $s$  as a function of
energy $E$. The lower green dotted line shows
the string theory prediction $1/(4\pi)$. The light blue bar shows
the estimate for a quark-gluon plasma (QGP)
while the blue solid bar shows the estimate for  $^4$He,
near the $\lambda$- point.
\label{fig:atomtrap}
}
\vspace*{-3mm}
\end{figure}

Rather early it has been observed 
 that the answer to the former question
is clearly affirmative. When the trap is switched off, the atoms in a weakly coupled gas
simply fly away with their thermal/quantum velocities, displaying $isotropic$
angular distribution of velocities $irrespective$ of the trap shape. However 
in strongly coupled regime, with the particle mean free path smaller than
the system size,  hydrodynamical flow develops. Deformed traps thus develop
``elliptic flows"  in the direction of maximal pressure gradient, in a way
analogous to what is happening in heavy ion collisions: see Fig.\ref{fig:atomtrap}(a).
This is by no means trivial: the interparticle distances are about 1000 times
the atomic size, and the total number of atoms is only $\sim 10^4$ (only few times
more than in central heavy ion collisions at RHIC): similar number of water
molecules would $not$ show any hydro!

%

The next questions  was whether one can use hydrodynamics quantitatively,
find out its accuracy and quantify the viscosity.
From ``released traps" the experiments  switched to quadrupole vibrational modes:
as the trapped system has a cigar-shape with much weaker
focusing along $z$ axes compared to axial ones, there is softer $z$-vibrations
and higher frequency ``axial" mode.
I will not go into vast literature and simply say that the value of the
vibrational frequencies are indeed given by hydro, reaching near-percent
accuracy at the Feshbach resonance point (where $|a|=\infty$).

 The first study of viscosity has been made by  Gelman, myself and Zahed 
 \cite{Gelman:2004fj}: from available 
data on $two$ different vibrational modes -- z-mode and axially
symmetric radial mode of a cigar-shaped atomic cloud -- we tried
to deduce viscosity and found that the values are
roughly consistent with each other.  Instead of going into details of that work, let us discuss what
one would expect based on ``universality" arguments.
The main point is that if certain observable
are finite in the $a*n^{1/3}\rightarrow \infty$ limit, the value can only depend
on few parameters -- e.g. the particle mass $m$, the density $n$ and
Plank constant $\hbar$ when the temperature is zero.
 By dimensional analysis, at $T=0$,
the energy density of infinitely strongly coupled gas can only be a number
times the energy density of the ideal Fermi gas with the same density,
\be \epsilon(n)=\beta{\hbar^2 n^{2/3} \over m }\ee
with some $universal$ constant $\beta$.

Assuming that there can be nonzero ``quantum viscosity" at zero $T$
for large amplitude oscillations, we
 \cite{Gelman:2004fj}  proposed to measure viscosity in units of
 $\alpha= \eta / \hbar n $, the only combination with the right dimension.
 Experimental data  by Turlapov et al.
\protect\cite{Turlapov_etal} for viscosity in such units are shown 
 in Fig.\ref{fig:atomtrap}(b). It is plotted as a function
of energy per particle in Fermi energy units, $E/E_F$, which  basically
 characterizes the excitation temperature:  $E/E_F\approx .5$
 (corresponding to the left side of the figure) is close to the ground state $T=0$.
 From the measured points it seems that  $\alpha= \eta / \hbar n $
is actually vanishing in the ground state, about linearly in $T$. 
Another way to express viscosity -- familiar from Black Hole physics--
is to express it as  $\eta/ \hbar s$ with
 $s$ being the entropy density. Such ratio is
 shown in Fig.\ref{fig:atomtrap}(c): and it looks like this one reaches
 nonzero value as $T,s\rightarrow 0$. Its magnitude makes
 strongly coupled fermionic atoms to
 be the ``second best liquid"
known, in between the sQGP (light blue band below)
 and the ``bronze winner" (former champion) , liquid $He4$, shown by dark blue.

  Few people (myself included) may find it fascinating,
but most of atomic quantum gases community do not care and
even unaware that some qualitatively new regime happens close to
the Feshbach resonance:  global parameters (like $\beta$
in total energy which was subject of many theoretical and numerical
works) are smooth there.
Small viscosity of quantum gases does not yet have $any$ microscopic
explanations: certainly not by ``unitary" cross section and kinetic theory.
 Linear behavior of $\eta$ and $s$ in $T$ reminds that of electrons in
solids: but strong coupling actually destroys the Fermi surface (as experiments
measuring momentum distribution show quite clearly) and makes a very strong
superconductor. The critical point 
in units  is at $E/E_F\approx.85$: a look at the
discussed figure reveals no  changes in viscosity (nor in
oscillation frequencies themselves) visible by an eye. 

\begin{figure}[h]
\begin{minipage}[c]{8.cm}
 \centering 
\includegraphics[width=7.5cm]{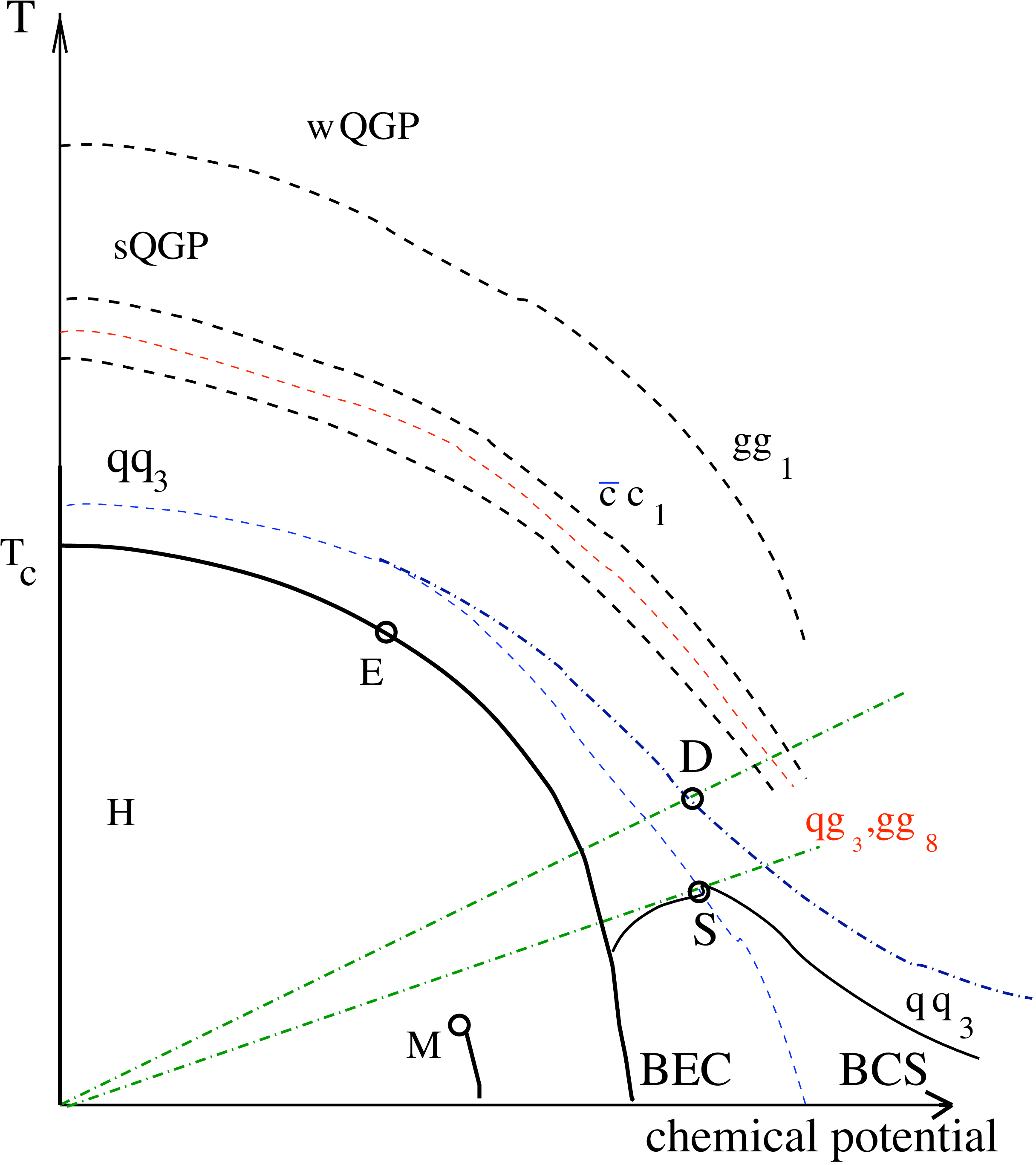}
\end{minipage}
\hspace{2cm} 
\begin{minipage}[c]{8.cm}
\vskip -.4cm
\includegraphics[width=8cm]{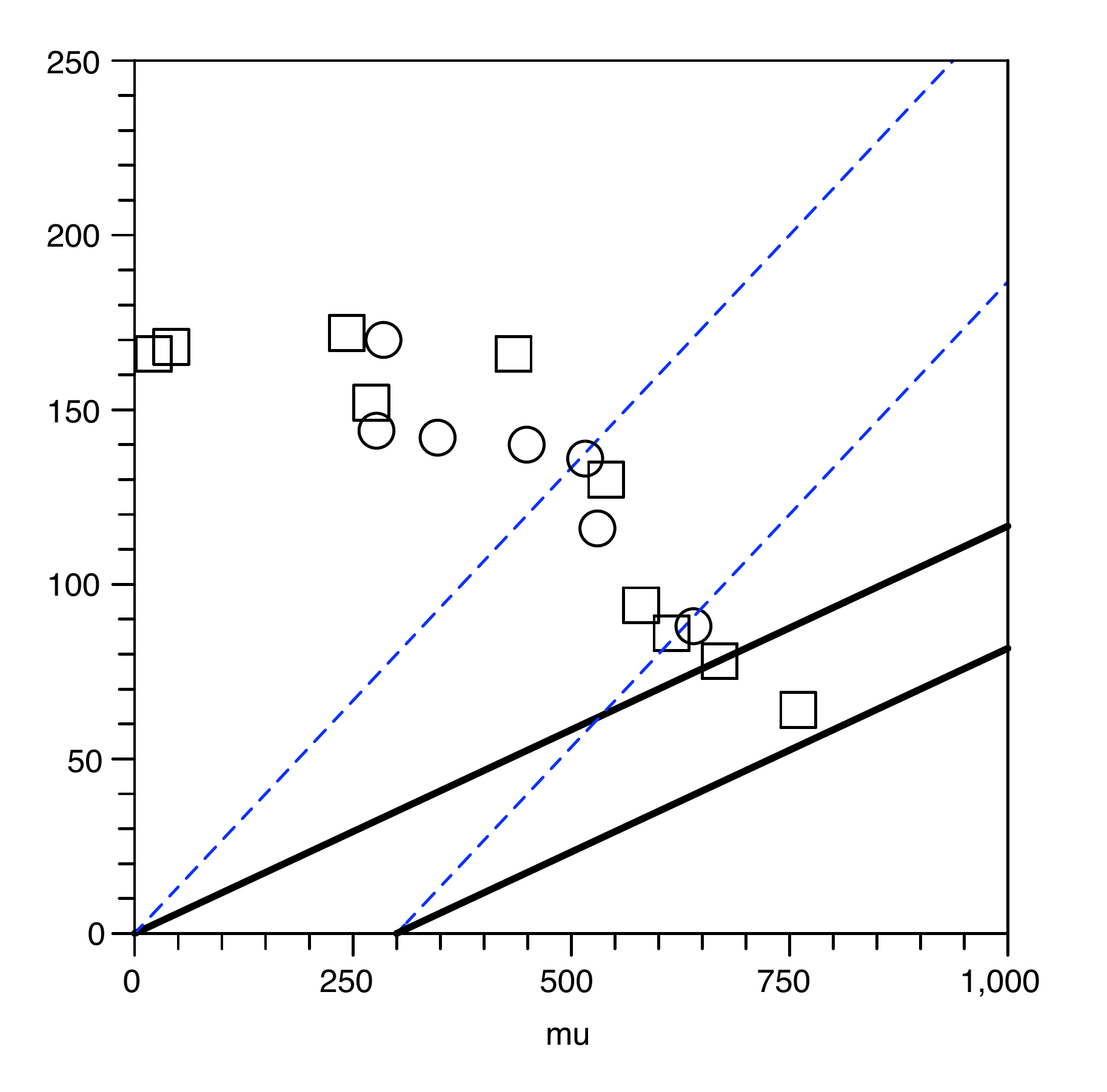}
\end{minipage}
\caption{(a) Schematic phase diagram for QCD, in the plane baryon
  chemical potential - temperature.
M (multifragmentation) point is the endpoint of nuclear gas-liquid transition.
E is a similar endpoint separating the first order transition
to the right from a crossover to the left of it.
(Black) solid lines show
phase boundaries, dashed  lines are curves
of marginal stability of indicated states. 
Two dash-dotted straight lines are related with bounds from atomic
experiments we discuss in the text, they intersect with
unbinding of diquark Cooper pairs (D) and most strongly coupled point (S),
which is at the maximum of the transition line and is
also a divider between BCS-like and BEC-like color superconductor.
Other lines are zero binding for singlet $gg$, singlet $\bar c c$,
octet $qg,gg$ and finally triplet $qq$ states.
(right) Compilation of experimental data on the 
chemical freezeout
from different experiments:
 squares (circles) are for fits at mid-rapidity (all
particles), respectively. Two solid lines are the phase transition lines
with the quark effective mass $M_1=0$ and $100\, MeV$, two dashed
  lines show pair unbinding lines for
 the same masses.
}
\label{fig_phase_diag_CS}
\end{figure}

Apart of being in general related to the issue of sQGP as ``the most perfect liquid",
the strongly coupled atomic superconductor provides some
valuable information on how large the {\em pairing gaps } can possibly
become in cold quark matter with color superconductivity. Assuming
that near deconfinement  cold quark matter is also strongly coupled,
with weakly bound diquarks playing the role of Feshbach resonance,
in \cite{Shuryak:2006ap}
I have used $universality$ of the ratio $T_c/\epsilon_F$ (the pairing critical temperature to 
the Fermi
energy) and data on strongly coupled fermionic atoms
to get {\em upper limit} on the corresponding transition
to the color superconductivity. Fig.\ref{fig_phase_diag_CS} explains the argument. 
``Universality" tells that the critical temperature 
must be simply proportional to the Fermi energy 
\be \label{eqn_alpha_def}
T_c=\alpha_{T_c} E_F\ee
with the universal constant $\alpha_{T_c}$. In fact two values of $T$
show some change: 
 the phase transition to superfluidity at
$ \alpha_{T_c}={T_c\over E_F}=.35$ and  $\alpha_2={T_2\over T_F} \approx 0.7-0.8 $
which experimentalists (Kinast et al)
 interpret as a transition to a regime where
not only there is no $condensate$ of atomic pairs, but
even the  pairs themselves are melted.
Using these two values as slopes of solid and dashed lines
in Fig.\ref{fig_phase_diag_CS}(b), one can see that 
all chemical freezeouts  in heavy ion collisions (points) are above
possible domain of pairing, even at infinite coupling.  At finite coupling the gaps and $T_c$ is even smaller.

Let me finish with one recent development, providing quite fascinating
prospects if true. The idea is that infinitely strongly interacting atoms
may have ``gravity dual" description similar to AdS/CFT. Son
\cite{Son:2008ye} and also Balasubramanian and McGreevy \cite{Balasubramanian:2008dm}
have argued that since at $a*n^{1/3}\rightarrow \infty$ the interaction
can be treated as a kind of boundary condition when two atoms meet, and
this problem has what he called ``Schreodinger symmetry", a subgroup
of conformal symmetry. They
 found examples of the metric possessing this
very symmetry and suggested it $may$ be the desired ``gravity dual" .
(The reader should however be reminded  that there is no proof
of its existence or in fact any empirical
confirmation of this idea so far.)

\section{Heavy Ion Collisions}

\subsection{Heavy ion collisions and flows for pedestrians}
Let me start with brief history. At one hand, high energy physics
was for all of its history been interested in ``high energy
asymptotic'' $s\rightarrow \infty$. In practice this
mostly was reduced to global features like total and elastic
cross sections, as it was thought that multibody
final state produced is too complicated to get some
sufficiently simple theoretical treatment.  
 Nuclear physicists started from nuclear collisions
at nonrelativistic domain: their cross section was obvious
and the main objective was understanding of ``excited
matter''. Experimental program at Relativistic Heavy Ion Collider
(RHIC) at Brookhaven and Large Hadronic Collider (LHC) at CERN
is so to say a brainchild of both communities.

 What do we mean saying that collisions produced ``excited
matter''? Should one expect simplification if it is the case? The answer is that
we are not  interested in $any$ excited system produced
(e.g. in ``elementary'' pp collisions), but mostly in a
{\em macroscopically large fireball} whose size (macro scale) $R$
greatly exceed the {\em micro scale $l$} of correlations inside it.
\be R>> l\ee
In weakly coupled systems (gases) $l$ is the mean free path
length, or relaxation time.
 In strongly coupled setting such as AdS/CFT  
 the temperature $T$ would be the only dimensional parameter
 describing the microscopic physics:
 most dissipative phenomena have the scale $1/T$ as a relaxation
 scale.

{\em Have we reached this regime?} Perhaps good illustration that the
 question is nontrivial are 
my own two $unsuccessful$ attempts to apply macroscopic physics
for high energy processes. 
In 1971 I proposed
 ``spherical explosion" in $e^+e^-$ annihilation
into many pions \cite{Shuryak:1971hn}, only to be killed by
discovery of asymptotic freedom in 1973 and jets in  $e^+e^-$
evens in 1976. In 1979 
Zhirov and myself \cite{Shuryak:1979ds} looked at
 fresh results from pp collisions
at then-new ISR collider at CERN\footnote{
L. van Hove and few others, including myself, 
put forward a proposal to
accelerate heavy ions in ISR, to see what happens. CERN leaders
 basically rejected it; the  experiments 
never went beyond the alpha-particles, and then
 this first hadronic collider was physically
destroyed. As we know now, QGP could have been discovered and studied
at ISR 20 years prior to RHIC.
}.
 The general idea 
of the paper was to look for collective transverse flows.
We argued that since secondaries
have different masses, the kinematic effect
of a collective motion (flow velocity) can be
separated from their thermal spectra. 
More specifically, if the ``matter flow" is 
only longitudinal, along the beam direction, the transverse momenta
 spectra of different secondaries would be just thermal
\be  {dN\over dp_t^2}\sim exp(-{M_t\over T}) \ee
where the mass and transverse momentum combine into a ``transverse mass" $M_t^2=M^2+p_t^2$. The ISR data indeed showed this  ``$M_t$ scaling":
but there was no sign of the deviations from it due
to transverse flow. (Lacking the effect, we argued that it is the
``vacuum pressure'' which stops it.)

Heavy ion collisions are used to make the system larger. The
  macro scale -- nuclear sizes -- for Au  (or Pb) used
in Brookhaven (or CERN) have typical $A\sim 200$ and radius $R\sim 6 fm$.
At RHIC the temperature
 T changes from the initial value of about 
$T_i\sim 400 MeV\sim 1/(.5 fm)$, thus $RT\sim 10>>1$ can be viewed
as a large parameter. In fact experiments with $Cu$ beam, $A=64$,
also showed good macroscopic behavior, as are rather peripheral
collisions. The boundary between macroscopic
and microscopic systems should be somewhere, as $pp$ collisions
are not hydro-like: but the
exact location of it  is not yet known: small systems of several nucleons
are subject of large fluctuations, and it is not easy\footnote{On top of that,
there is experimental background, as very peripheral collisions of two beams
is hard to tell from beam-residual gas interaction.}
to study them. 

Once produced, the fireball is expanding hydrodynamically
up to the so called freezeout conditions.
People familiar with Big Bang can recognize existence of multiple
freezeout, for each reaction at its own time: and in the Little
Bang it is similar. The so called {\em chemical} freezeout (at which
particle composition gets frozen) is at $T\sim 170\, MeV$ at RHIC,
while re-scattering continue to $T_f\sim 90\, MeV$ (for central
collisions). In units of QCD critical temperature, it is
 $T/T_c$ variation from about $2$ to  $1/2$, conveniently bracketing the QCD
 deconfinement/chiral restoration transition.
 
Flow velocity is decomposed into longitudinal
(along the beam) and transverse components. The latter
is further split into {\em radial flow} (present even for axially
symmetric central collisions)  and  {\em elliptic flow} which
exists only for non-central collisions. The reason  elliptic flow to be 
very important  was pointed out 
by H.Sorge:  it is developed $earlier$ 
than the radial one, and thus most sensitive to the
 QGP era than other flows.

The radial flow has Hubble-like profile, with transverse
rapidity growing roughly linear till the edge of the fireball.
The maximal radial flow  velocity 
 at RHIC is about $.7\,c$.
This radial flow is firmly established from a
combined
analysis of particle spectra, HBT correlations, a
deuteron coalescence and other observables: so we will not show 

At non-zero impact parameter the
original excited system has a deformed almond-like
shape: thus its expansion in the transverse plane can be
 described by the (Fourier) series in azimuthal
angle $\phi$
\be v_n(s,p_t,M_i,y,b,A) =<cos(n\phi)> \ee
where average is over all particles\footnote{This is written as if the direction
of impact parameter of the event is known. I would not go into how it is determined
in detail but just say that it usually comes from counters sitting at
rapidity $y\approx \pm 3$ far from the region where most observations are
done, at mid-rapidity $y\approx 0$.
}. At mid-rapidity $y=0$ only even harmonics
are allowed, the second one
characterizes the so called {\em elliptic flow}.
Multiple arguments of the parameter $v_2$
 stand for
the collision energy $s$, transverse momentum $p_t$, 
particle mass/type $M_i$, rapidity $y$, centrality\footnote{In real
experiments the measure of impact parameter is the so called
 the number of participants  $N_{p}$, which changes from zero
at most peripheral collisions to $2A$ at central collisions.
This number is the total number of nucleons $2A$ minus the so called
spectators, which fly forward and are directly observed
by forward-backward calorimeters.
} $b$ 
and the atomic number $A$
characterizing the colliding system.

   If in  high
 energy collisions of hadrons and/or nuclei a
 macroscopically large  excited system is produced, its expansion
and decay
can be described by relativistic hydrodynamics.
Its history starts with the pioneer
paper by Fermi
of 1951 \cite{fermi} who proposed a statistical model
to Lorentz-contracted initial state.
Pomeranchuk \cite{Pomeranchuk:1951ey} then pointed out that 
initial Fermi stage  cannot
be the final stage of the collisions since strong interaction
in the system  persists: he proposed a
freezeout temperature $T_f\sim m_\pi$.
  L.D.Landau
 \cite{Landau:1953gs} then explain that one should use
relativistic hydrodynamics in between those two stages,
saving Fermi's prediction of the multiplicity by entropy
conservation.

 (Before we go into details, a comment:  often  hydrodynamics
 is considered as some consequence of kinetic  equations, but  in
 fact applicability conditions of both approaches are far from being coincident.
In particular, for the former approach the stronger the interaction in
 the  system is,  the  better.  Kinetic  approach,  on   the
 contrary, was never formulated but  for  weakly  interacting
 systems: and as we repeatedly emphasize in this review, it is 
$ not$ so for QGP.)

The  
     conceptual basis of the ideal hydrodynamics  is  very  simple: it is just
 a set of local conservation laws for the
 stress tensor ($T^{\mu\nu}$) 
and for the conserved currents ($J_{i}^{\mu}$), 
\begin{eqnarray}
 \partial_{\mu}T^{\mu \nu}&=& 0   \\ 
 \partial_{\mu}J_{i}^{\mu}&=& 0  \nonumber
\end{eqnarray}
The $assumption$ is in $local$ form for  $T^{\mu\nu}$ and $J_{i}^{\mu}$
 related to the bulk properties of the fluid and its 4-velocity $u^\mu$ by
the relations,
\begin{eqnarray} \label{eqn_tmunu}
T^{\mu \nu} &=& (\epsilon + p) u^{\mu} u^{\nu} - p g^{\mu \nu}   \\
J_{i}^{\mu} &=& n_{i} u^{\mu} \nonumber
\end{eqnarray}
Here $\epsilon$ is the energy density, $p$ is the pressure,
$n_i$ is the number density of the corresponding current, and 
$u^{\mu}=\gamma(1, \bf{v})$ is the proper velocity of the fluid. 
In strong interactions, the conserved currents are isospin ($J_{I}^{\mu}$),
strangeness ($J_{S}^{\mu}$), and baryon number ($J_{B}^{\mu}$). 
For the hydrodynamic evolution, 
isospin symmetry is assumed and  the net strangeness is set to zero;
 therefore only the baryon current $J_{B}$ is considered below.

Let me add a simple heuristic argument why the first term in the stress
tensor has $(\epsilon + p)$ and not any other combination. The point
is $\epsilon$ and $ p$ themselves are defined up to a constant
$\pm B$ (which depending on context we call the bag constant, or
cosmological
term or dark energy). The combination $(\epsilon + p)$ does not have
it, and it is also proportional to entropy which is defined uniquely.
How this argument complies with the last term in 
(\ref{eqn_tmunu})? Well it has B but without velocity, so
in hydro eqns there is only pressure gradient
and this B term disappears as well.

 In order to close
 up this set of equations, one needs  also  the  equation  of
 state (EoS) $ p(\epsilon )$. One should also be aware of two
 thermodynamical
differentials
   \be d\epsilon=Tds \qquad dp=sdT \ee
and the definition of the sound velocity
\be c_s^2={\partial p \over\partial \epsilon}=
{s\over T}{\partial T \over \partial s}\ee
and that $\epsilon+p=Ts$.
     Using these equations and the thermodynamical relations in the form
 \be  
     {\partial_\mu \epsilon \over \epsilon+p}= {\partial_\mu s \over s}
  \ee
  one may show that  these  equations imply another  nontrivial
 conservation law, namely, the conservation of the entropy
  \be \partial_\mu (s u_\mu)=0\ee
  Therefore in the ideal hydro  all  the  entropy
 is  produced  only  in  the discontinuities --  shock  waves -- which are
 not actually there is application we discuss. Thus the ``initial entropy"
is simply passed to the solution as an initial parameter, determined
in the earlier (violent) stage of the collision: (this is similar to Big Bang cosmology,
in which ``entropy production" stage is also very different from stages
of cosmological evolution we can observe by e.g. Nucleosynthesis.)

Next order effects in micro-to-macro expansion is the domain of ``viscous hydrodynamics": we will discuss their applications to data description
as well as their derivation from AdS/CFT settings.

\begin{figure}[t]
   \begin{center}
      \includegraphics[width=8cm]{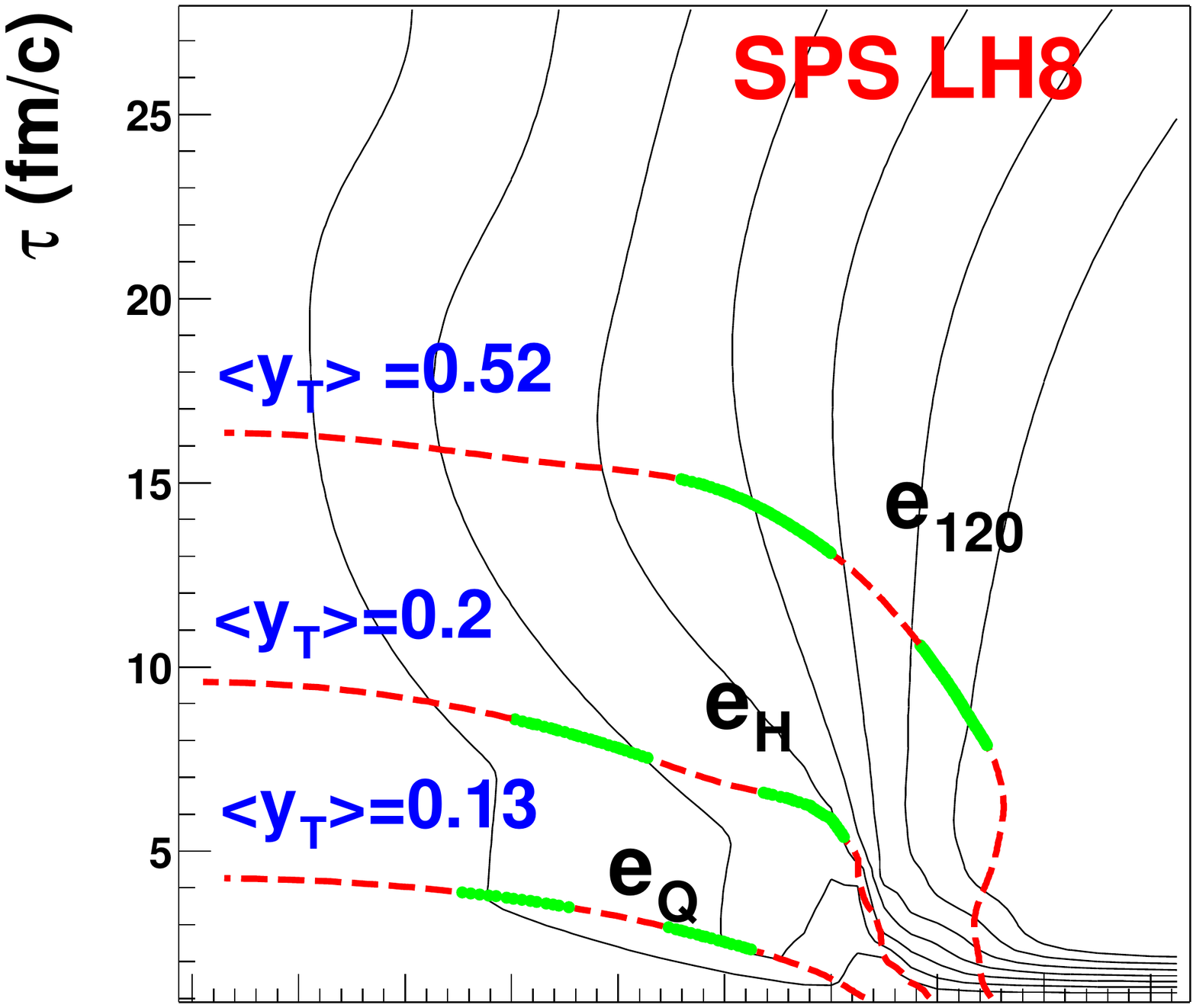}
      \includegraphics[width=8cm]{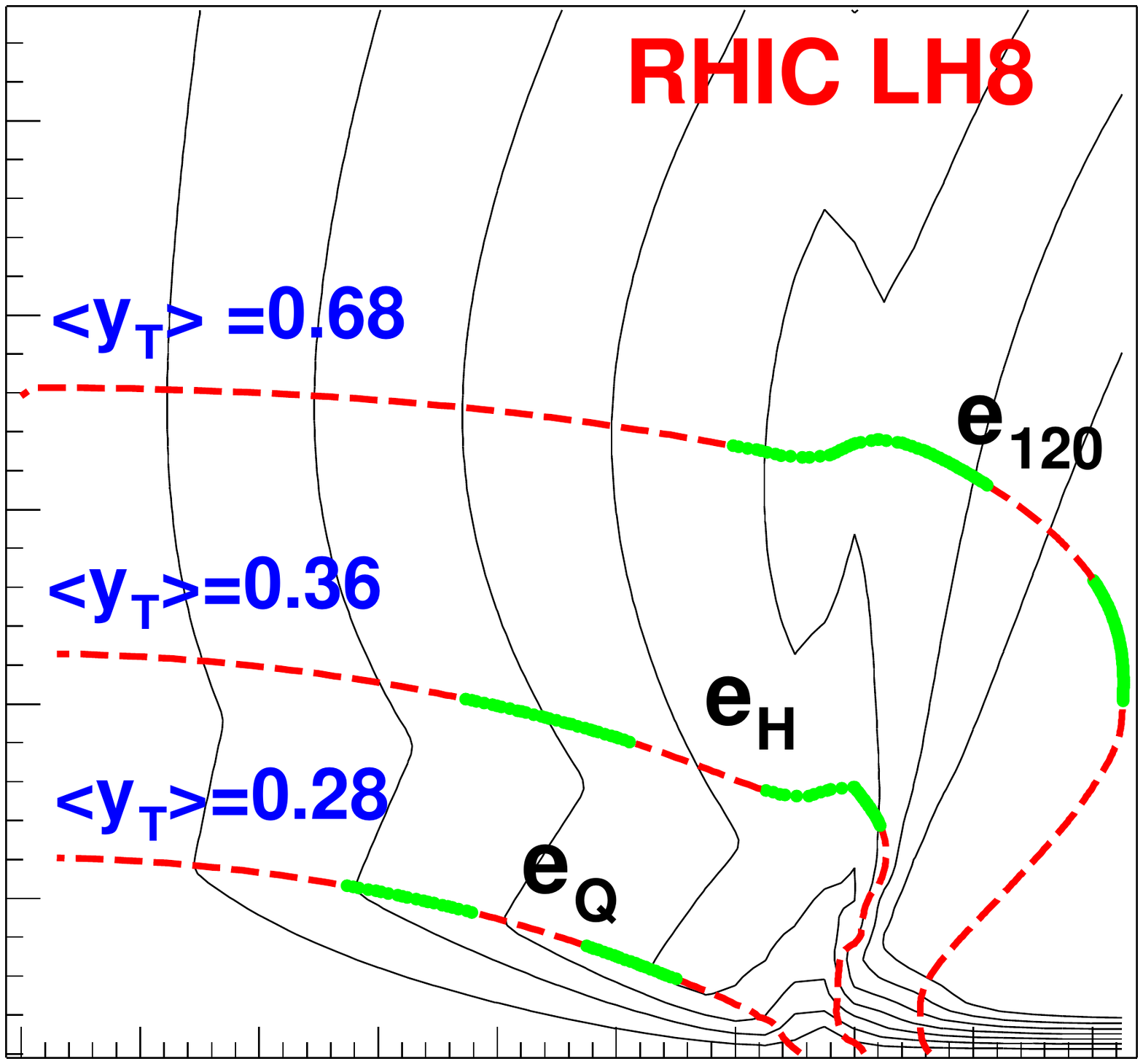}
   \end{center}
   \vspace{-0.1in}
   \caption[An illustration of the hydrodynamic solution
   at the SPS and RHIC for three different EOS.]{
   \label{Fig_Surface} 
      The left and right sides show the hydrodynamic solution at 
      the SPS and RHIC.
      The thin lines show
      contours of constant transverse fluid rapidity ($v_{T} = \tanh(y_{T})$) 
      with values 0.1,0.2,...,0.7\,.  The thick 
      lines show contours of constant energy density. $e_{120}$
      denotes the energy density where $T=120\,\mbox{MeV}$. $e_{H}$ 
      and $e_{Q}$
      denote the energy density  where the matter shifts from
      hadronic to mixed and mixed to a QGP, respectively. 
      The shift to hadronic cascade is made at $e_{H}$. $\langle y_{T} \rangle$ 
                denotes the mean transverse rapidity weighted with 
                the total entropy flowing through the energy density contours.
      Walking along these contours, the line
      is broken into segments by dashed and then solid lines.
      20\% of the total entropy 
      passing through the entire arc passes through each segment.
   }
\end{figure}

 The simplest Bjorken 1+1 dim solution is a good example ``for
 pedestrians", reminding how to write hydro in arbitrary coordinates.
 \be
   {T^{mn}}_{;m} = 0\,, \qquad 
   {j^m}_{;m} = 0\,,
 \ee
where the semicolon indicates a covariant derivative. For tensors of 
rank 1 and 2 it reads explicitly
 \be
   j^i_{;p} =j^i_{,p} + \Gamma^i_{pk}\, j^k \,,
 \\
 \label{A3}
   {T^{ik}}_{;p} = {T^{ik}}_{,p} + \Gamma^i_{pm} T^{mk} 
                                   + \Gamma^k_{pm} T^{im}\,,
 \ee
where the comma denotes a simple partial derivative and the Christoffel 
symbols $\Gamma^s_{ij}$ are given by derivatives of the metric tensor
$g^{ab}(x)$:
 \be
    \Gamma^s_{ij} =
    (1/2) g^{ks} \bigl( g_{ik,j} + g_{jk,i} - g_{ij,k}\bigr)\,.
 \ee
As an example,
let us do the following transformation from Cartesian to light 
cone coordinates:
 $$
   x^\mu=(t,x,y,z)  \longrightarrow   \bar{x}^m=(\tau,x,y,\eta)
 $$
  $$ t = \tau \cosh\eta \hspace{1cm} \tau = \sqrt{t^2-z^2}
 $$
  $$ z = \tau \sinh\eta \hspace{1cm}   \eta = (1/2) \ln{t{+}z\over t{-}z} \,.
 $$
In the new coordinate system the velocity field (after inserting $v_z=z/t$)
is given by
 \be
   \bar{u}^m = \bar{\gamma}(1,\bar{v}_x,\bar{v}_y,0)
 \ee
with $\bar{v}_i \equiv v_i\cosh\eta$, $i=x,y$, and
$\bar{\gamma} \equiv 1/{\textstyle{\sqrt{1{-}\bar{v}_x^2{-}\bar{v}_y^2}}}$.

Now we turn to the metric of the new system. We have 
 \ba
   ds^2 = g_{\mu \nu} dx^\mu dx^\nu 
       &=& dt^2 - dx^2 - dy^2 - dz^2
  \nonumber \\
       &=& d\tau ^2 - dx^2 - dy^2 - \tau^2 d\eta^2
 \ea
 and therefore
 \ba
   g_{mn}=\left( \begin{array}{*{4}{c}} 
        1 & 0 & 0 & 0 \\ 
        0 & -1 & 0 & 0 \\
        0 & 0 & -1 & 0 \\ 
        0 & 0 & 0 & -\tau^2 \\
        \end{array} \right)\,,
 \ea
The only non-vanishing Christoffel symbols are
 \be
   \Gamma^{\eta}_{\eta \tau} = 
   \Gamma^{\eta}_{\tau \eta} = {1\over\tau}\,,\qquad
   \Gamma^\tau_{\eta \eta} = \tau\,.
 \ee
  The 1+1d equations for boost-invariant solution can be written
in the following way
\be \label{eq_entropy_1d} 
{\partial \over \partial t}(s \, \cosh y)+ 
{\partial \over \partial z}(s \,\sinh y)=0\ee
\be  \label{eq_pressure_1d}
{\partial \over \partial t}(T \,\sinh y)+ 
{\partial \over \partial z}(T \, \cosh y)=0\ee

The so called
 Bjorken \cite{Bjorken} solution\footnote{ We call it following
established tradition, although the existence  of
 such simple solution was first noticed by Landau and it  was
 included in his classic paper as some intermediate step.  The
  space-time picture connected with such  scaling  regime  was
 discussed in refs  \cite{preB1,preB2} before Bjorken,
 and some estimates for the    energy  above  which  the
 transition to the  scaling regime were  expected to happen
 were also discussed in my paper \cite{Shuryak:1978ij} as well. }
is obtained if the velocity is given by the velocity
$ u_\mu=(t,0,0,z)/\tau $
where $\tau^2=t^2-z^2$ is the proper time. In
     this 1-d-Hubble regime 
  there   is   no   longitudinal
 acceleration at all: all  volume  elements  are
 expanded linearly with time  and  move  along   straight
 lines from the collision point. The spatial $\tanh^{-1}(z/t)$
 and the energy-momentum rapidity
$y\tanh^{-1} v$ are just equal to each other.
 Exactly as in the Big Bang, for each "observer" ( the volume
 element ) the picture is just the same,  with  the  pressure
 from the left  compensated  by  that  from  the  right.  The
 history is also the same for all volume
elements, if  it  is  expressed  in  its  own
 proper time $\tau$.

Thus the
 entropy conservation  
becomes the following (ordinary) differential equation in proper time $\tau$
\be {ds(\tau) \over d\tau}+{s\over \tau}=0 \ee
which has the obvious solution 
\be s={const \over \tau}\ee
    Let us compare three simple cases:
                  (i) hadronic matter, (ii) quark-gluon plasma  and 
(iii) the  mixed
 phase (existing if there are first order transitions in  the
 system). In the first case we adopt the equation  of  state
 suggested in \cite{Shu_res_gas}
$c^2=\partial p/\partial \epsilon=const(\tau)\approx .2$. If so, 
   the   decrease of  the  energy
 density with time is given by 
  \be \epsilon(\tau)=\epsilon(0)({\tau_0\over \tau})^{1+c^2} \ee
 In the QGP case the same law holds, but with $c^2=1/3$   
  
    In the mixed phase the pressure remains  constant  $p=p_c$  ,
 therefore
  \be \epsilon(\tau)=(\epsilon(0)+p_c)({\tau_{mix}\over \tau})-p_c \ee
    So far all dissipative phenomena were ignored. Including
   first dissipative terms into our equations one has
  \be 
{1 \over\epsilon+p} {d\epsilon \over d\tau} ={1\over s}{ds\over d\tau}=-{1 \over
   \tau}\left(1-{(4/3)\eta+\zeta \over (\epsilon+p)  \tau}\right)
      \ee
Note that ignoring $\zeta$ one finds in the r.h.s. exactly
the combination which also appears in the sound attenuation, so the correction
to ideal case is  $(1-\Gamma_s/\tau)$. Thus the length $\Gamma_s$
   directly tells us the magnitude of the dissipative corrections. At time
$\tau\sim\Gamma_s$ one 
has to abandon the hydrodynamics altogether, as the dissipative
   corrections
cannot be ignored.
Since the correction is negative,
 it reduces the rate of the entropy decrease with time.
Another way to say that, is that the total positive sign shows that
some amount of entropy is
generated by the dissipative term. We will discuss
``gravity dual'' to this solution in the last chapter.

\subsection{Collective flows and hydrodynamics}

Any treatment
  of the  explosion by hydro  
in general includes (i) the initiation; (ii) hydro evolution and (iii)  freezeout.
Geometrically, (i) and (iii) form two 3-hypersurfaces in 4
dimensions, together constituting a boundary of 4-volume region
in which hydro eqns are solved.
Only the stage (ii) deals with  the Equation of State and transport properties of new form of matter, the QGP -- while (i) and (iii)
have to deal with stages at which hydrodynamics is not applicable. 
The stage (i) is least understood theoretically: yet the uncertainties
 it produces are rather minor.  Correct treatment of the freezeout 
 stage (iii) is much more important for
calculations of the final spectra which are compared to data: and that is where
most intensive debates were.  
Fortunately the latest  hadronic stage
is not that mysterious, as some theorists think: it happens
in a dilute gas-like  medium made of mostly pions and their resonances. We understand their low energy 
scattering cross sections quite well,  the corresponding cascade
codes have been extensively tested using the low energy
heavy ion collisions
 at AGS (BNL) and SPS(CERN). It just needs a bit of extra work.

  Solving hydro eqns is not simple -- they are nonlinear and prone to
  instabilities -- but solutions are easily controllable by the energy, momentum and
  entropy (for ideal hydro) conservation: thus I presume it was done
  correctly by all groups. At years 2000-2004 most groups used an
  approximation in which one out of 3+1 (coordinates+time) 
   -- the longitudinal spatial rapidity -- was taken as irrelevant,
  thus switching to 2+1.
Example of output from such hydro calculations \cite{Teaney:2001av}, with properly chosen EoS\footnote{LH8 mentioned in the figure means  the``latent heat" of
magnitude $.8 GeV/fm^3$ at the QCD transition temperature.}
is shown in Fig.\ref{Fig_Surface}, for two collision energies, 
$s^{1/2}=
15,200\, GeV$ per nucleon, marked SPS and RHIC and shown at left and right 
figure. The only thing I would like to mention is that the fraction of time
spent in the QGP phase (below the $e_Q$ curve) is not that large:
but its existence crucially change flow pattern at RHIC. As one can see
from the right plot, all lines of constant rapidity are nearly vertical and nearly
equidistant: it means after QGP the flow gets a simple Hubble-like flow 
with $v_t=H r_t$ and time-{\em independent} constant $H$.

 Location of the freezeout surface is also a nontrivial task. 
In our first application of
 hydrodynamics for radial transverse flow at SPS \cite{Hung:1997du}
we developed 
rather tedious ``differential freezeout", for each geometry and
each secondary particle was applied. We calculated individual reaction rates
for different secondaries and matched them with
 hydro expansion rates locally. In spite of
 relative success of the work, there were no followers to this approach
 at RHIC.
Selecting hydrodynamics
as a Ph.D. topic for my graduate student, Derek Teaney 
\cite{Teaney:2001av}, we had in mind a different procedure
for freezout, namely switching at the onset of hadronic phase
($e_H$ in the figure above) to a hadronic cascade (RQMD)
which  automatically leads to earlier freezeout for smaller
systems or particles with smaller hadronic cross sections\footnote{Note that e.g. $\phi$ meson and the nucleon have about the same mass of 1 GeV, but cross section
of re-scattering on pions ranging from few mb for the former to 200 mb for the latter,
at the $\Delta$ resonance peak dominating re-scattering.}.
The same approach was later used by Hirano et al, who generalized our
 rapidity-independent hydro to the full 3+1 dimensional case, successfully
reproducing also the $v_2$ dependence on rapidity $y$ 
\cite{Hirano:2004ta}. They confirmed that  $v_2$ linearly decreases from its maximum
at mid-rapidity linearly, for the same reason as it decreases toward smaller energies:
there is less matter there, shorter QGP phase and also shorter hadronic phase
due to earlier freezeout.
Independently developed code by Nonaka and Bass
\cite{Nonaka:2006yn}, with a different cascade code UrQMD, later also 
 confirmed these calculations and well reproduce data for all
dependences.  These three groups have basically covered  all
outstanding issues related to  applications of the ideal hydrodynamics
to RHIC data.

One might think that, after a couple of groups checked the
calculations
themselves,
 the rest would be history if all heavy ion community would 
 recognize/accept it. Unfortunately, it is not the case. 
Many hydro groups 
 have not implemented hadronic freezeout,
ending  hydro 
at some (arbitrarily selected) {\em isoterms } $T=T_f$.
There is absolutely no 
 reason to think it is the right choice.
 Even thinking about freezout as a local concept,
one finds that it is determined not by local density but by 
  local expansion rate $\dot n=\partial_\mu u_\mu$ of matter.
It is well known for more than a decade \cite{Hung:1997du} that
  the isoterms do  $not$  resemble
 the lines of {\em fixed expansion rate}. Furthermore, 
 while hydro solutions simply scale with the size/time of the system\footnote{Hydro eqns have only first derivative over coordinates, which can thus be rescaled.}, the freezout 
 conditions (involving the reaction rates) do not.  
 It is thus not at all surprising, that many
 results based on unrealistic freezeout  show qualitatively wrong
dependencies.
 
\begin{figure}[t!]
	\includegraphics[width=8.cm]{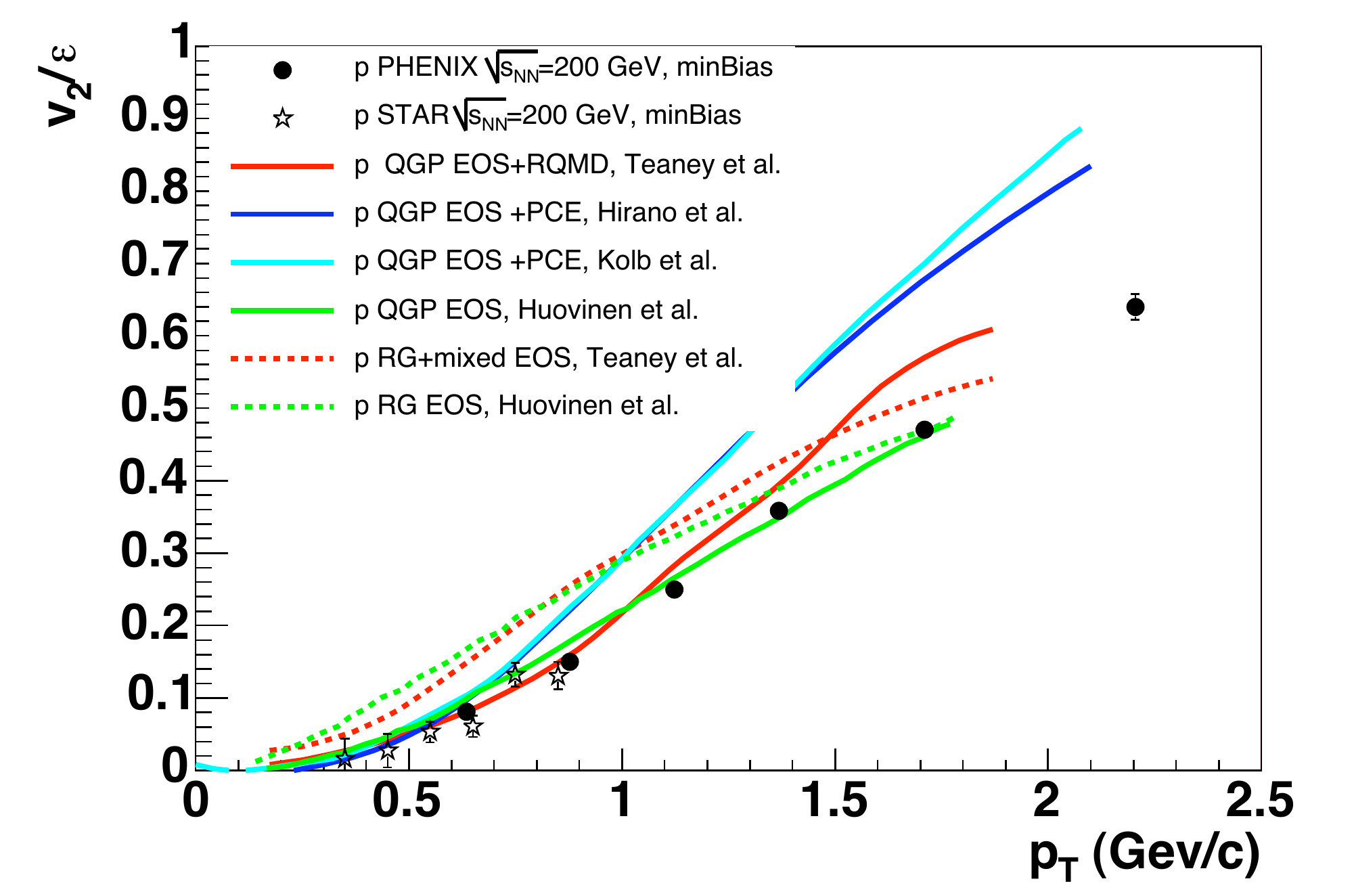} 
	\includegraphics[width=8.cm]{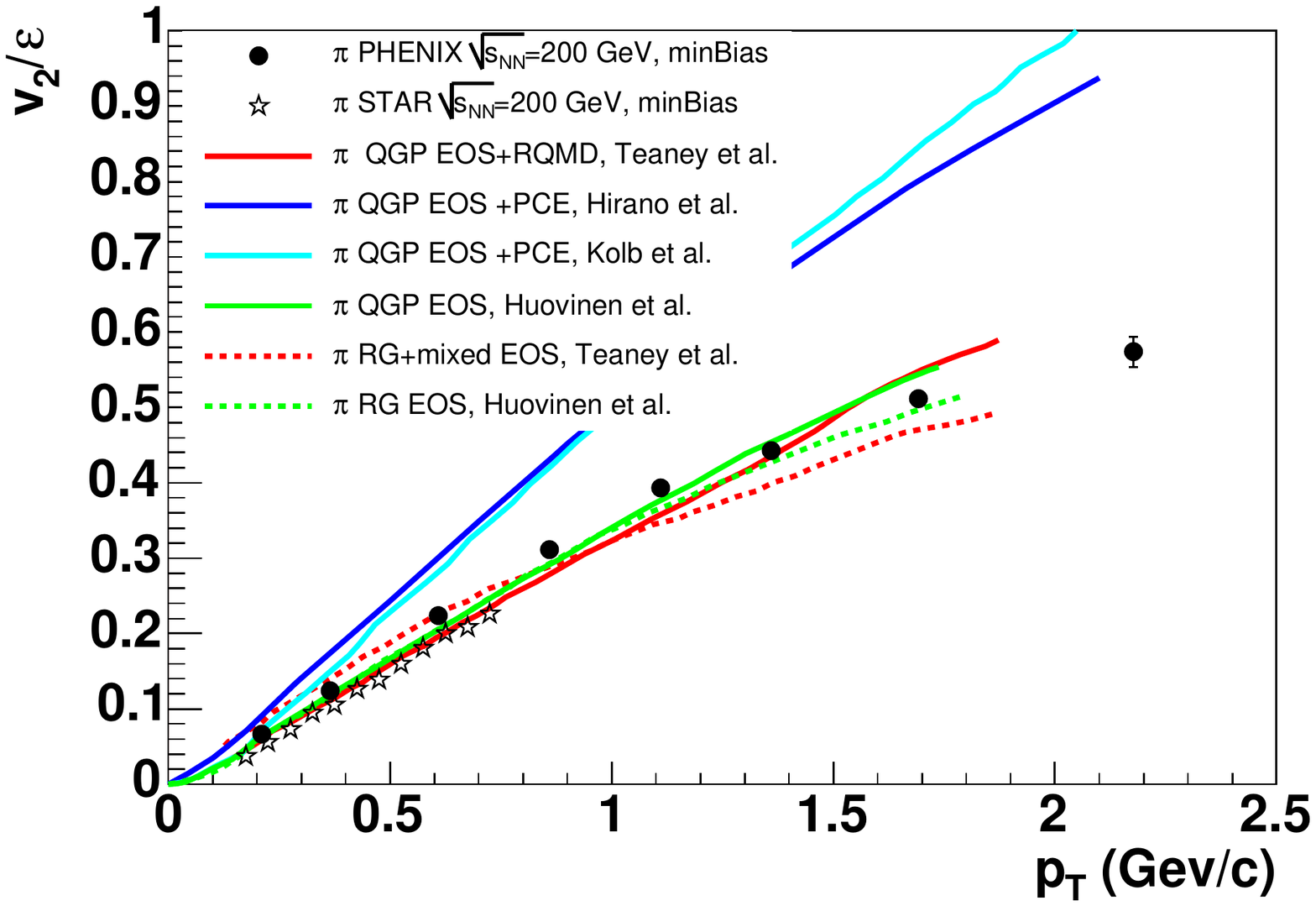} \\
	\includegraphics[width=8.cm]{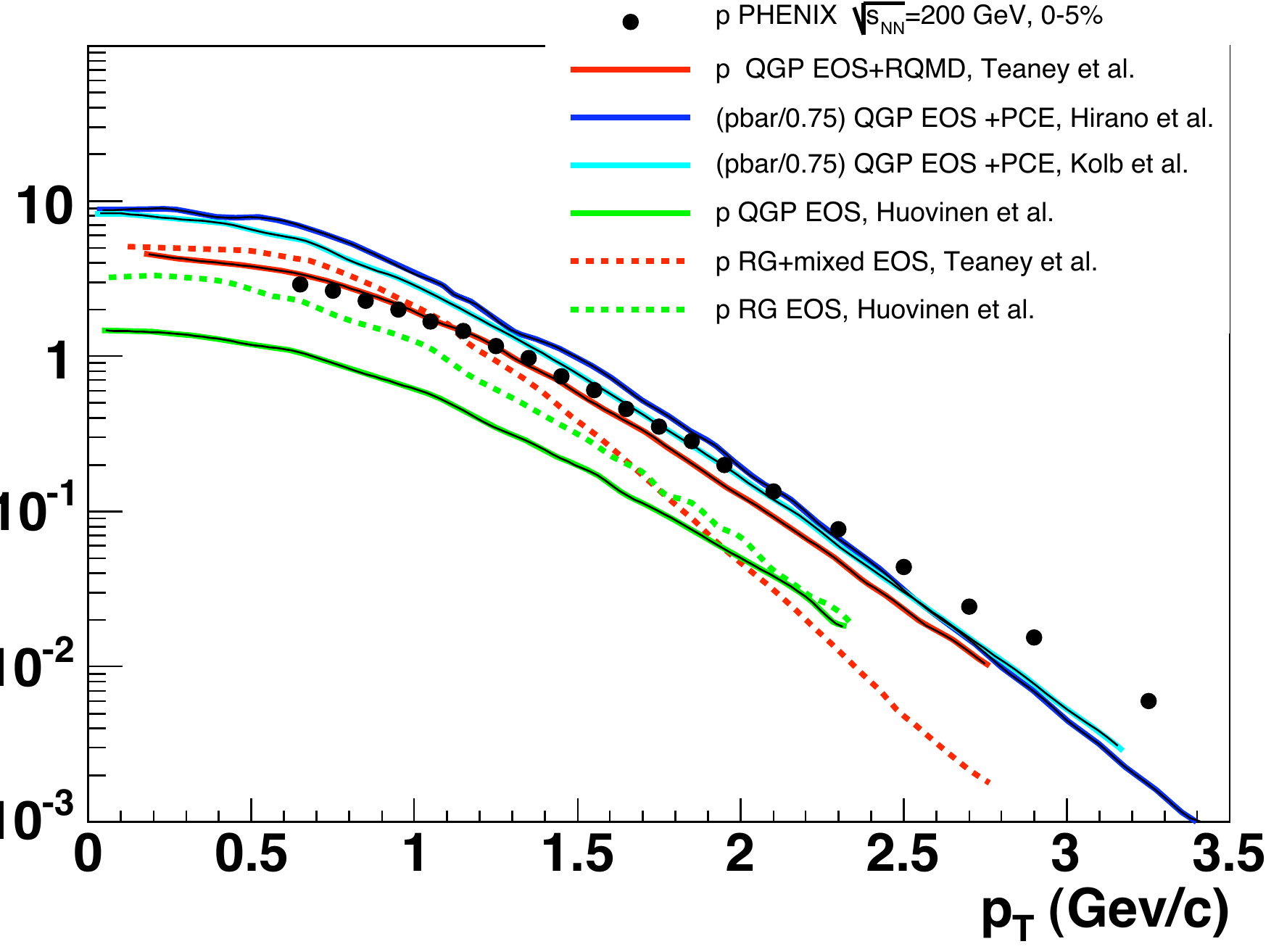} 
	\includegraphics[width=8.cm]{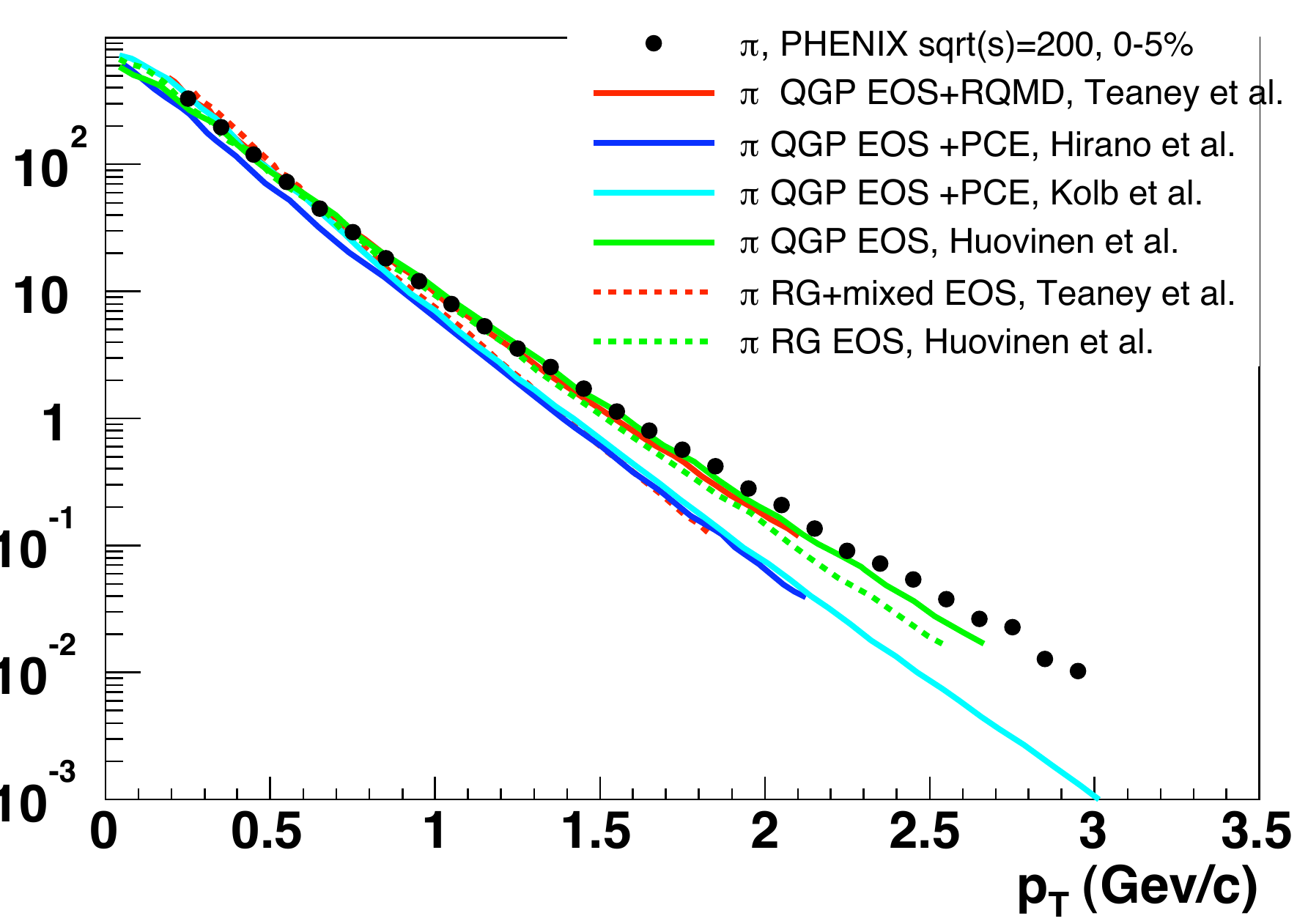} 
\caption{Top two panels: On the left, proton 
$v_2(p_T)/\epsilon$ vs. $p_T$ for 
minimum-bias collisions at RHIC 
\protect\cite{Adler:2003kt,Adler:2001nb} 
are compared with hydro 
calculations \protect\cite{Teaney:2001av,Hirano:2002ds,Kolb:2003dz,Huovinen:2001cy}, 
and on the right is the same comparison for pions. 
Bottom two panels: $(1/2\pi)d^2N/p_{t}dp_tdy$. On the left, 
for protons, for 0--5\% 
centrality bin collisions at RHIC \protect\cite{Adler:2003cb} are compared 
with the same hydro calculations. On the the right,
 the same comparison for pions.}
\label{fig:hydro_v2_spectra_RHIC}
\end{figure}

  Let me start with few plots from PHENIX ``white paper" \cite{Adcox:2004mh} in Fig.\ref{fig:hydro_v2_spectra_RHIC}
  for protons and pions, to illustrate some important physics points explaining
  different level of success of different authors. 
The lower part of Fig.\ref{fig:hydro_v2_spectra_RHIC}
shows measured $p_t$ spectra of protons and pions, in comparison with
different hydro calculations. The shape of those is very different mostly because
heavy protons and light pions have different thermal motion at
the time of freezout, in spite of the {\em same} collective radial flow. Note that
nearly every group has a correct shape of the spectra (and thus correct
radial flow velocity), but they don't aways have
 the normalization (for the nucleons) correctly: it is because of ``chemical freezout"
 not implemented by some groups. There is no problem for hydro+cascade
 model \cite{Teaney:2001av} (the red curve).   



\begin{figure}[h] 
        \includegraphics[width=7.cm]{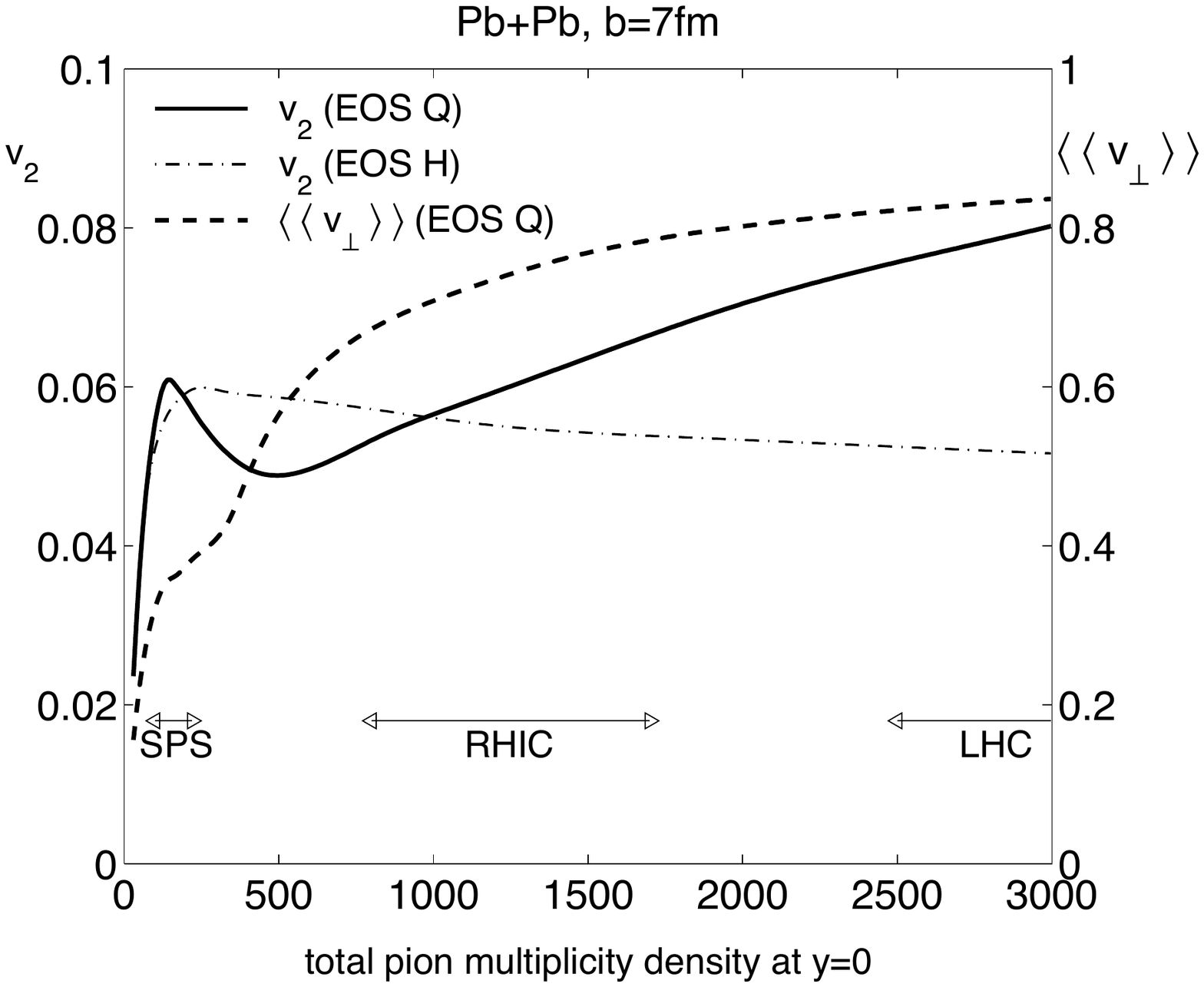}
            \includegraphics[width=7.cm]{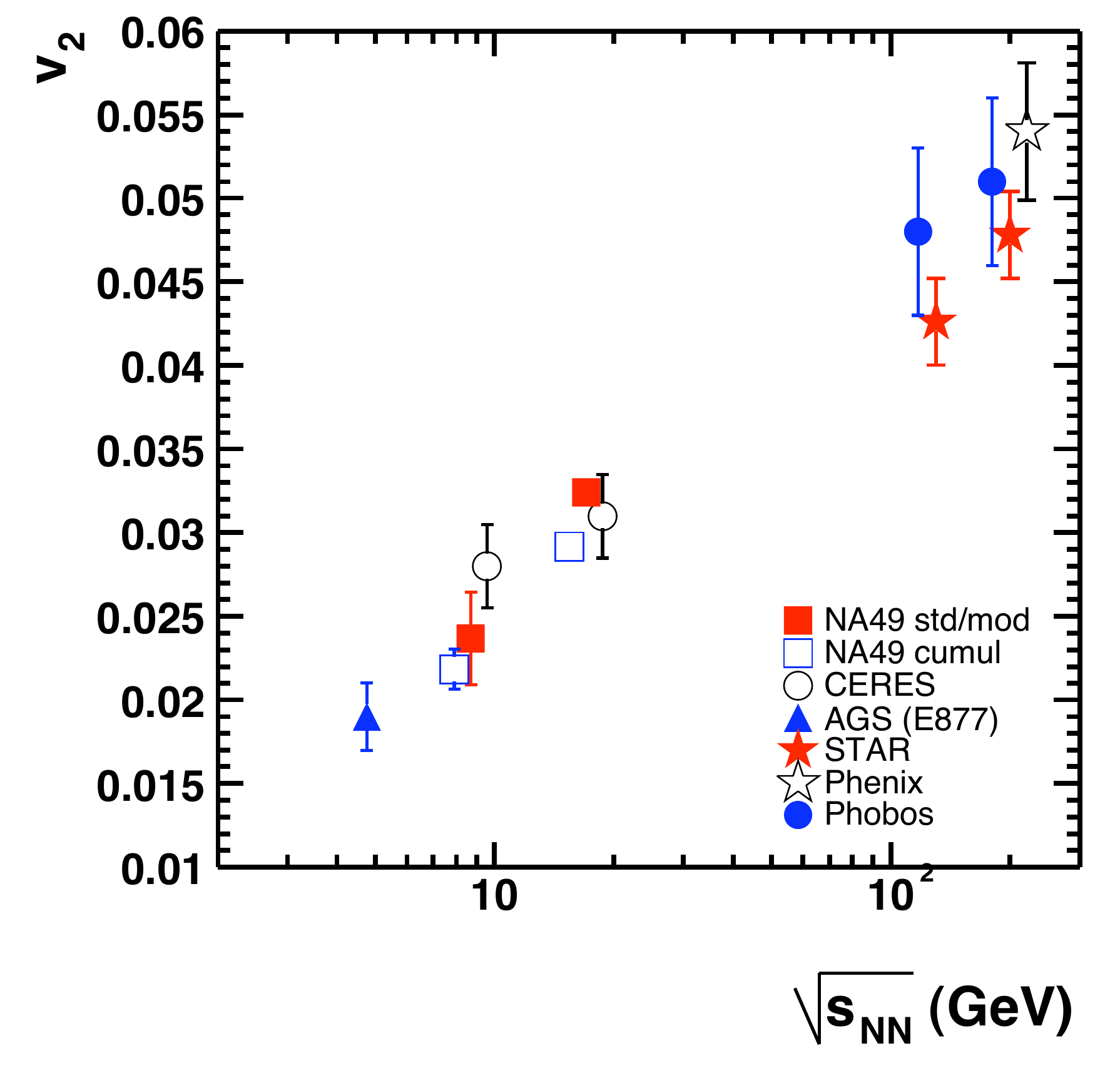}
  \\
             \includegraphics[height=6.cm]{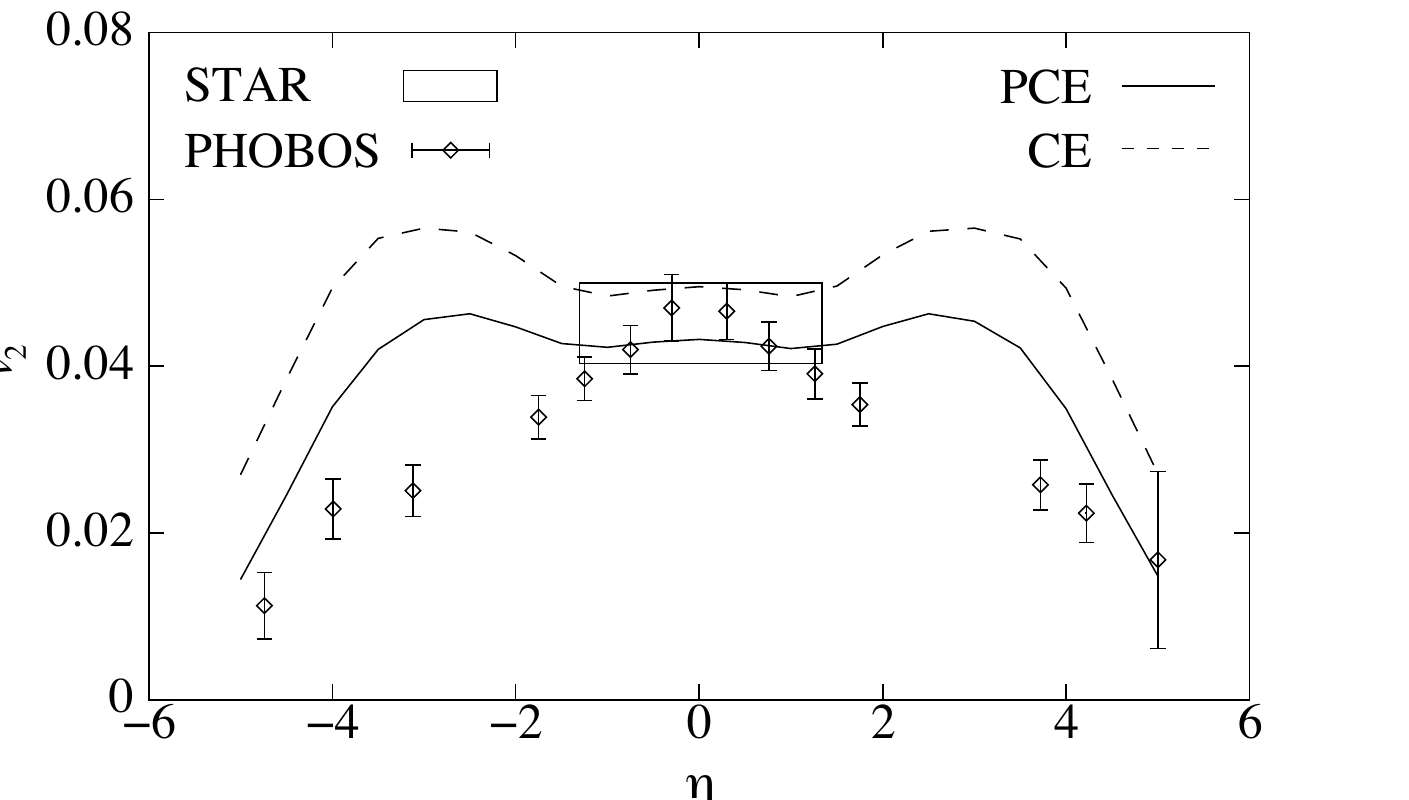} 
             \includegraphics[height=6.cm]{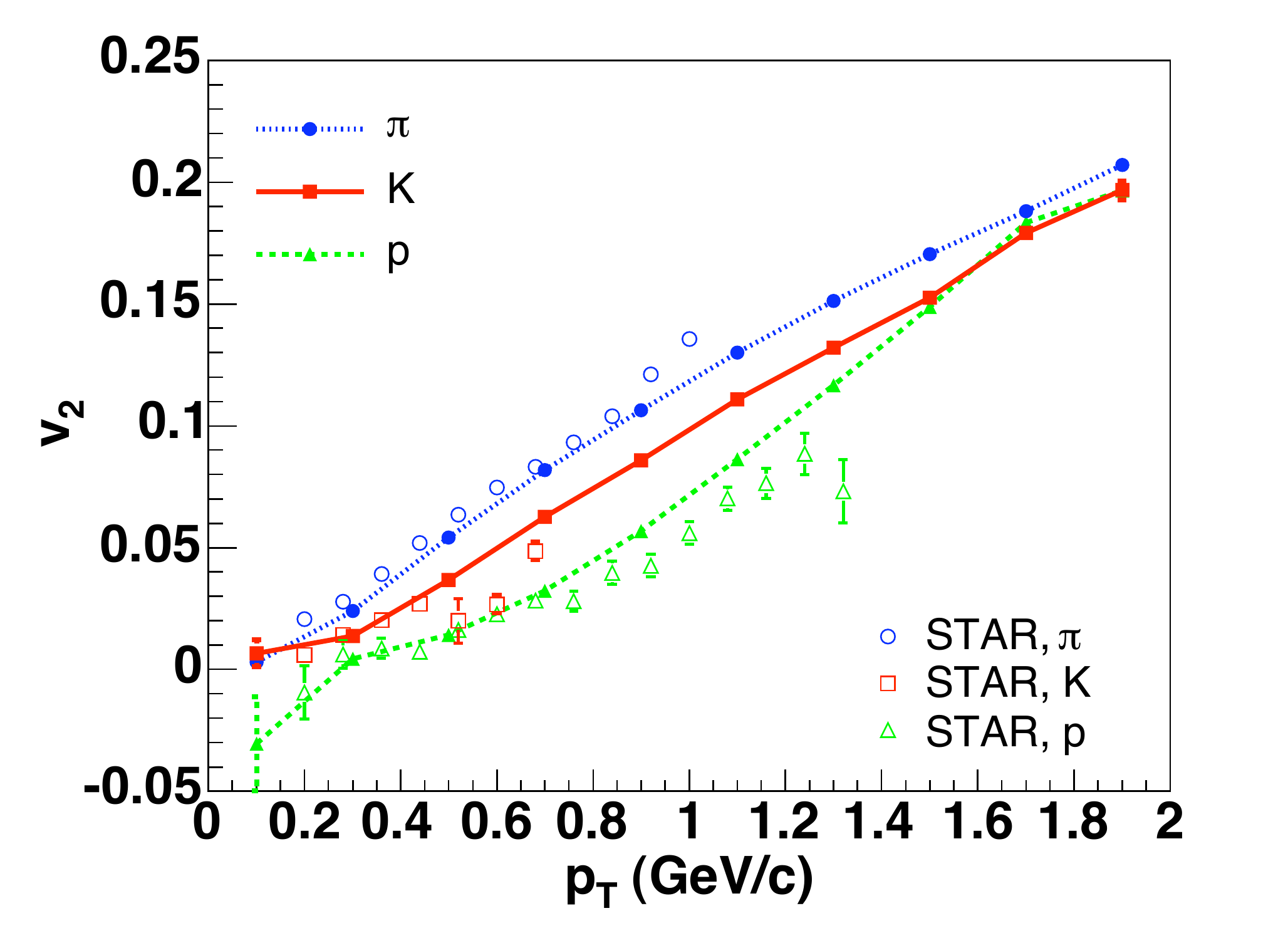}
\caption{(a) Excitation function of the elliptic (solid) and radial 
         (dashed) flow for Pb+Pb or Au+Au collisions at $b=7$~fm from 
         a hydrodynamic calculation.\protect\cite{Kolb:2000sd} The collision 
         energy is parametrized on the horizontal axis in terms of 
         total particle multiplicity density $dN/dy$ at this impact 
         parameter. (b) A compilation of $v_2$ data vs. collision 
         energy from mid-central (12--34\% of the total cross section)
         Pb+Pb and Au+Au collisions.
         (c) $p_t$-integrated elliptic flow for minimum bias Au+Au collisions
         at $s^{1/2}=130\,A$\,GeV as a function of 
         pseudorapidity, compared with  
         data from PHOBOS and STAR. (d)Transverse momentum dependence of $v_2$ for pions, kaons and protons.
Filled plots are the results from the hybrid model.
The impact parameter in the model simulation is 7.2 fm
which corresponds to 20-30\% centrality.
\label{fig:v2excitation} 
} 
\vspace*{-3mm}
\end{figure}

Now we switch to elliptic flow, shown as the ratio of observed momentum anisotropy
$v_2(p_t)$ divided by calculated spatial elliptic anisotropy $\epsilon$, shown in the 
upper part of Fig.\ref{fig:hydro_v2_spectra_RHIC}. Although some calculations
are not too close to the data, the overall magnitude of the effect and its
$p_t$ dependence is clearly reproduced.
However, that is only true for the ``good" dependence $v_2(p_t)$.
(Actually there is another one, the dependence of $v_2(m)$
on the particle mass, which everybody get right. See  one 
nice example of that from Hirano
 in Fig.~\ref{fig:v2excitation}(d).)

Unfortunately
other dependences of $v_2$ are not so forgiving as $v_2(p_t)$ and they show 
qualitative differences between models which do and do not
include hadronic freezeout properly.
Those include the
dependences of the elliptic flow on (i) collision energy $s^{1/2}$, 
(ii) centrality $b$ or number of participants $N_p$, (iii) and rapidity $y$.
Let me start with the collision energy: Heinz and Kolb  
in their large hydro review \cite{Kolb:2003dz} 
give
 their excitation curve for  elliptic flows  shown in Fig.\ref{fig:v2excitation}(a).
All variants of their prediction for the elliptic flow has rapid rise
on the left side of the plot (at low collision energies) with about constant 
saturating values at higher energies 
(one variant even reaches a peak  followed by a $decrease$ ). 
As one sees from see Fig.\ref{fig:v2excitation}(b) from the same review,
this is $not$ the trend observed in the RHIC domain:
the data show a slow rise without peaks or saturation. Kolb and Heinz thus
concluded that hydro is $not$ supported by the data, at all
collision energies
below  RHIC.
This lead to a {\em myth} about a ``hydro limit" which was ``never reached before RHIC" which
was (and still is) repeated from one conference to another. 
Finally Fig.\ref{fig:v2excitation}(c) from that review display
rapidity dependence of $v_2(y)$ from a calculation by Hirano (before
 he switched to hydro+cascade): the conclusion was that hydro only works
at mid-rapidity.


%

\begin{figure}[tbh] 
\centering
{\includegraphics[height=6.cm]{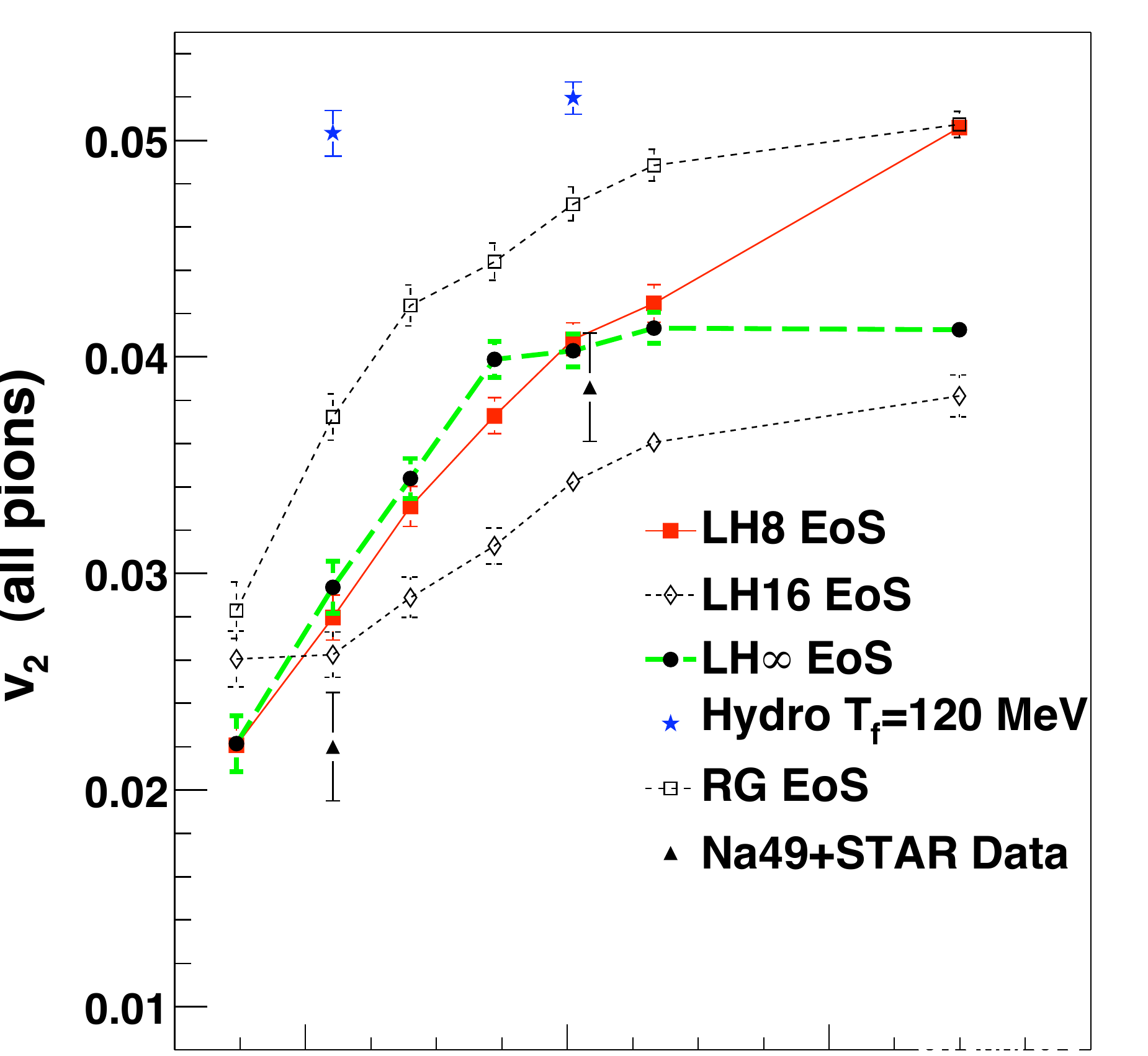}}
\includegraphics[height=6.cm]{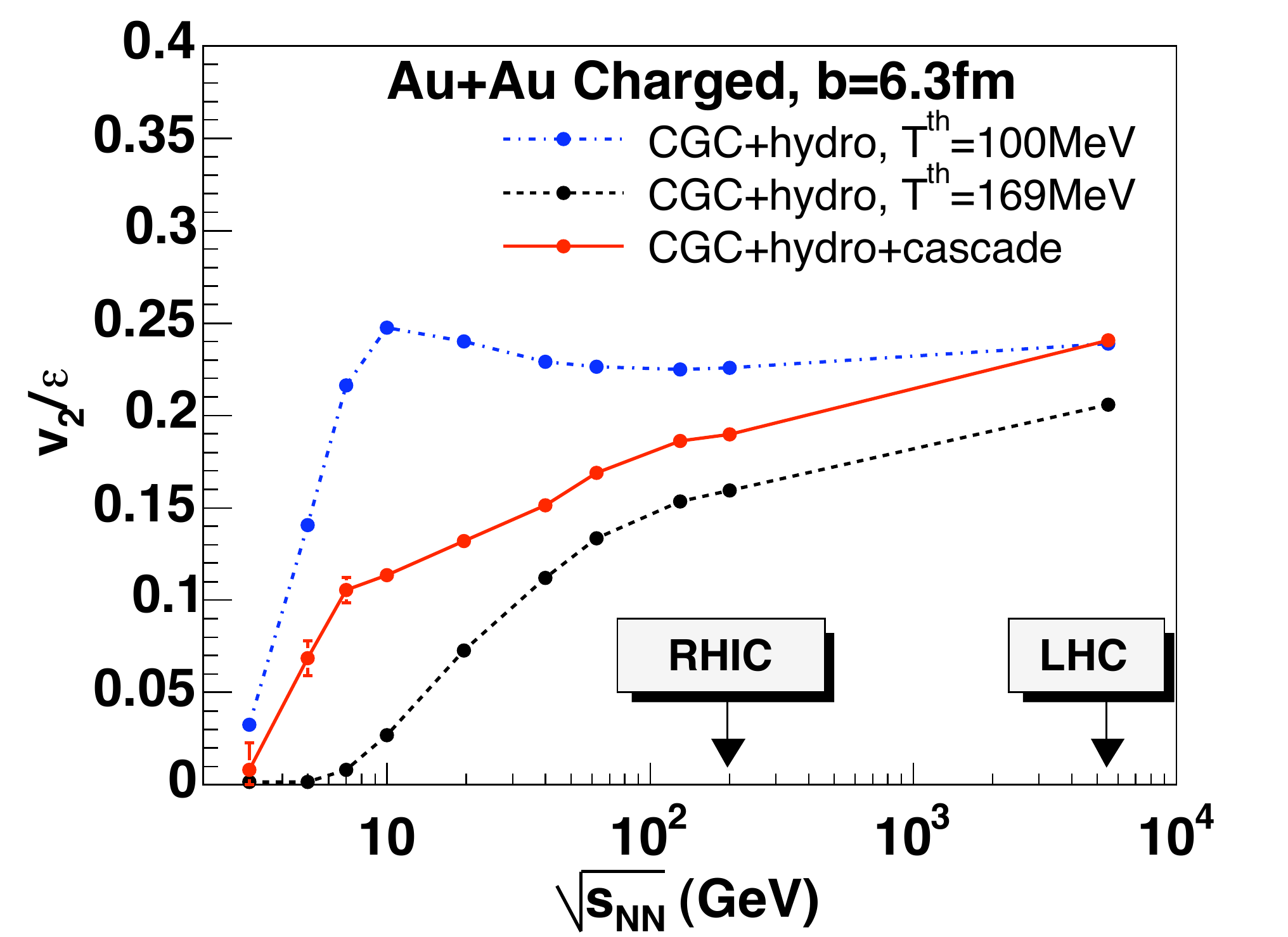} 
\caption{( Two examples of the 
excitation function of $v_2$ in Au+Au 
collisions:(a) from \protect\cite{Teaney:2001av} and
(b) from \protect\cite{Hirano:2004ta}.
 Points obtained with fixed-$T_f$ (120 Mev in (a) and 100 MeV in (b)) freezeout
strongly overpredict the data at low energies, but with cascade freezeout (red curves in both)
those describe data quite well. (Two triangular points in (a) correspond to SPS and RHIC
collision energies.)
} 
\label{fig:v2epssnn} 
\end{figure}

All those results are for fixed-T freezeout, which is not based on anything and thus is simply wrong. 
Here  what hydro+cascades approach finds for all of these
observables. The energy excitation curves of $v_2$ from 
\cite{Teaney:2001av} and   \cite{Hirano:2004ta} 
are shown in Fig. \ref{fig:v2epssnn}, left and right.
When correct freezout is implemented, the elliptic flow is rising steadily
all the way from SPS to RHIC, as the data do.
The reason Heinz et al (as well as curved marked 120 Mev and 100 MeV
in left and right plots) strongly overshoot the data is simply because
the freezout does $not$ occur at the same $T$ at different collision energy.
In fact, it is independently measured (from radial flow for central collisions) that while the 
freezeout temperature at SPS is about 140 MeV,  it is as low as  
 90 MeV  at RHIC.
 The trend is well understood: {\em the larger} is the system, the $hotter$ it is at the beginning, and the $cooler$ it 
 gets at the end of the explosion! 

\begin{figure}[tbh]
\centering
\includegraphics[width=8cm]{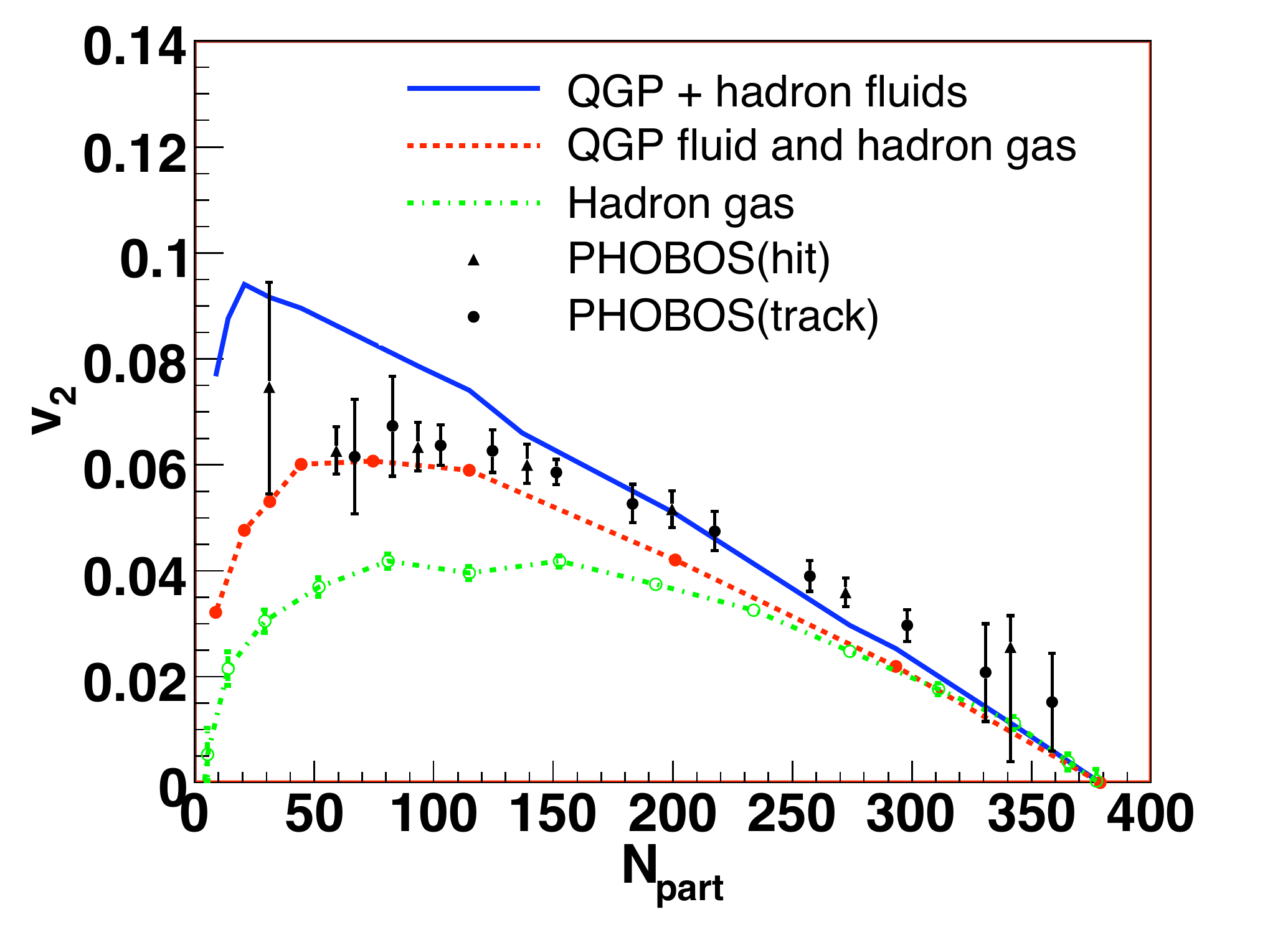}
\includegraphics[width=8cm]{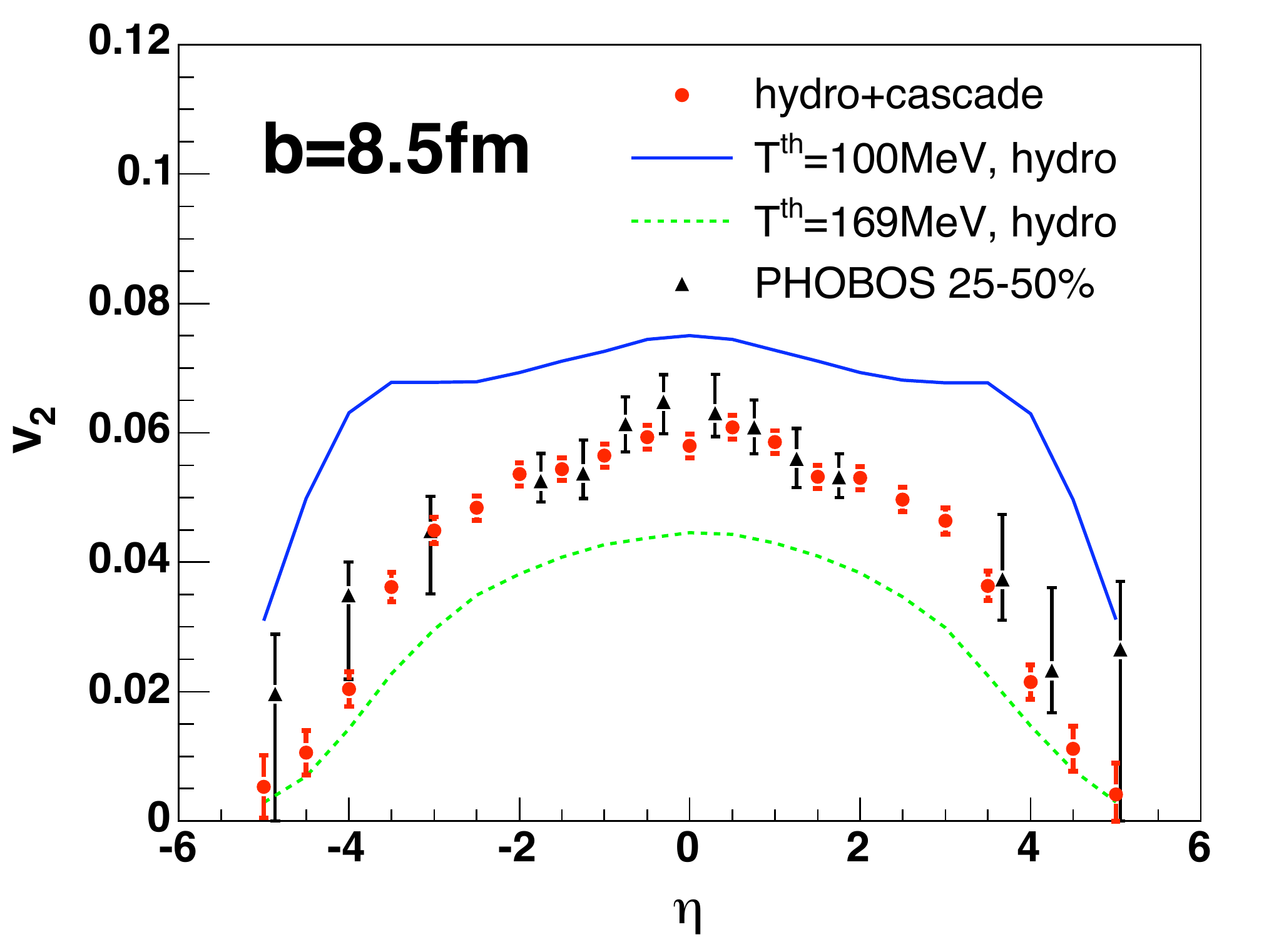}
\caption{(From \protect\cite{Hirano:2004ta}) 
(Left) Centrality dependence of $v_2$. The solid (dashed) line 
results from a full ideal fluid dynamic approach (a hybrid model).
For reference, a result from a hadronic cascade model
is also shown (dash-dotted line).
(Right) Pseudorapidity dependence of $v_2$. The solid (dashed) line
is the result from a full ideal hydrodynamic approach with $T^{\mathrm{th}} = 100$ MeV
($T^{\mathrm{th}} = 169$ MeV).
Filled circles are the result from the hybrid model.
All data are from the PHOBOS Collaboration. \cite{Back:2004mh}
}
\label{fig:v2_rapidity}
\end{figure}

Another ``hydro problem" discussed in Kolb-Heinz review, the
rapidity dependence of $v_2$, also went away as soon as correct
freezeout was used 
by Hirano et al.
The results (circles in  Fig.\ref{fig:v2_rapidity}  ) are right on top of the data
(triangles), without any change of any parameters. The reason
is exactly the same as for the energy
dependence: in fact one can check that $v_2(y)$ 
show good ``limiting fragmentation properties, depending
basically on $y-Y$, the distance to beam rapidity $Y$.  
One can see the difference in centrality dependence as well, in the left side of
Fig.\ref{fig:v2_rapidity} .

Intermediate summary:
  these ``problems'' (and associated myths) were caused
 by wrong freezeout. Matter at this time is a dilute pion gas,
which is not a good liquid,  neither at SPS,
not at RHIC and will not be at LHC, and cascade
is the best\footnote{
This does not imply that we have complete confidence in
many details of those cascades. To name one outstanding
issue: the precise in-matter modification of hadronic resonances
like $\rho,\Delta$ etc,
dominating the cross sections, is being addressed but
still far from been solved.  
}
 approach we have to describe it. As far as we can now test,
two other evolution eras -- sQGP and ``mixed'' or near-$T_c$ one --
can be surprisingly well described by the $ideal$ hydrodynamics.

\begin{figure}[tbh]
\centering
\includegraphics[width=8cm]{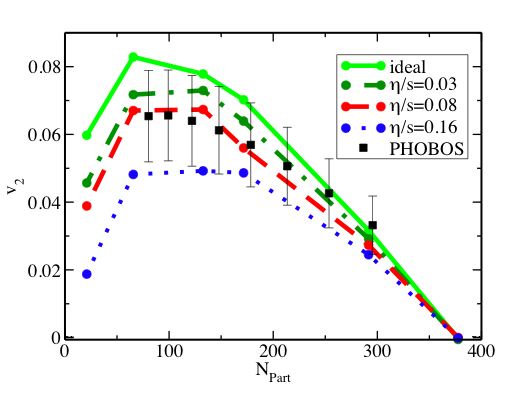}
\includegraphics[width=8cm]{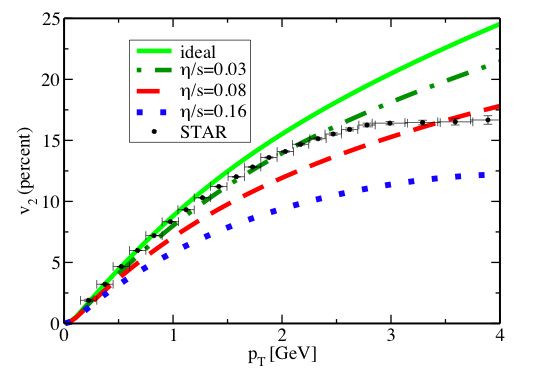}
\caption{(From \protect\cite{Romatschke:2007mq}) 
Elliptic flow  $v_2$ dependence on (Left) centrality  and (Right)
$p_t$,  compared to viscous hydro with variable viscosity.
}
\label{fig:v2_Roma}
\end{figure}

The last statement should not be understood that
the agreement with all details of the data
 is  perfect.  Theoretically, one should always ask 
about accuracy and
applicability limits of this hydrodynamical description.  
As emphasized  by Teaney \cite{Teaney:2003kp} the answer
 should be obtained by calculation the role of viscous
effects.
 Since we will have extensive derivation of ``derivative expansion"
in AdS/CFT language at the end of this paper, I will not
go into details here, going directly into new developments. 

In order to get more accurate account of viscosity effect on flow, a new round
of studies has been performed during the last year.
Relativistic Navier-Stokes has some problem with causality, thus
`higher order" methods has been used. Apart from viscosity  those methods have another
parameter, the relaxation time $\tau_r$, which
 is can either be used as a regulator  -- and its value put to zero at the end --
 or as a real representation of two-gradient terms. There are 4 groups who have reported solving 2+1 dim higher order
hydrodynamics. P. and U. Romatschke
\cite{Romatschke:2007mq}, Dusling and Teaney \cite{Dusling:2007gi} ,
Heinz and Song\footnote{This group originally found viscosity effects about
twice larger than others: but it was found in their later work that
this happened because of different account for some higher order term. In the $\tau_r\rightarrow 0$
limit, in which Navier-Stokes
limits is supposed to be recovered, all results are now consistent with others.} \cite{Heinz:2008qm}. 
Molnar \cite{Molnar:2007an} have compared viscous hydro with
some version of his parton cascade and found good agreement when the
parameters are tuned appropriately: but cascade describe $v_2(p_t)$
even at larger momenta. 
the solution is supposed to converge to that of the   Navier-Stokes eqn,
avoiding the causality problems.

 In Fig.\ref{fig:v2_Roma} we show Romatschke's results: literally taken they
 favor very small viscosity, even less than the famous lower bound.
 Now, is the accuracy level really
 allows us to extract  $\eta/s$? 
The uncertainties in the initial state deformation \cite{Lappi:2006xc,Gelis:2006dv}
 are
 at  the 10\% level, comparable to the viscosity effect itself.
EoS  can probably be constrained better, but I think
uncertainties related to freezeout  -- not yet discussed
at all -- are also at 10 percent level, although they
can also be reduced down to few percent 
 level provided more efforts to understand  hadronic resonances/interactions
at the hadronic stage will be made. All of it leaves us with a  statement 
  that  while literally fits require $\eta/s\sim 0.1$ or less, we can only conclude that it is definitely
 $\eta/s< .2$.
\footnote{Unfortunately
I am skeptical about magnitude of systematic 
errors of any lattice results for  
$\eta/s$
(such as \cite{Meyer:2007fc}): while the Euclidean correlation functions
themselves are quite accurate, the spectral density
is obtained by rather arbitrary choice
between many excellent possible fits.
}.
Even so,  sQGP is still {\it the most
perfect liquid known}.

In summary: hydrodynamics+hadronic cascades reproduces all RHIC data on radial and elliptic flows
of various secondaries, as a function of centrality,rapidity
or energy 
are reproduced till $p_t\sim 2 GeV$, which is 99\% of particles.
Contrary to predictions of some,
CuCu data match AuAu well, so  Cu is large
enough to be treated hydrodynamically. New round of studies last year included
viscosity and relaxation time parameters on top of ideal hydro:
viscosity values is limited to very small value.

\subsection{Jets quenching and correlations}
 
Pairs of  partons can collide at small impact parameter:
in pp collisions this produces a pair of large $p_t$ hadronic jets, which are
(nearly) back-to-back in transverse plane (because total transverse
momentum due to ``intrinsic' parton $p_t$ is small).    
We can use therefore those high-$p_t$ partons as a kind of x-rays,
penetrating through the medium on its way outward and 
in principle providing its ``tomography". 

\begin{figure}[h]
\begin{center}
\includegraphics[height=6.cm]{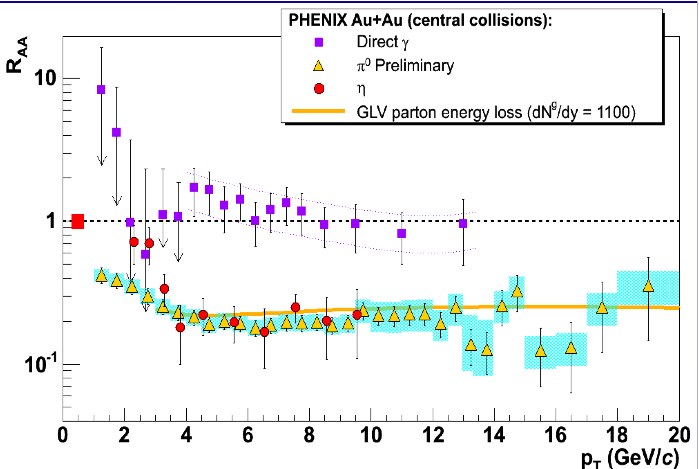}
\includegraphics[height=6.cm]{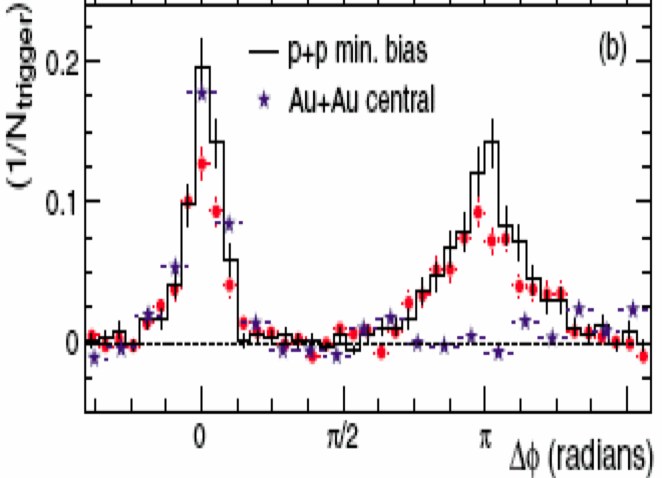}\\
\hspace{.5cm}\includegraphics[height=6.cm]{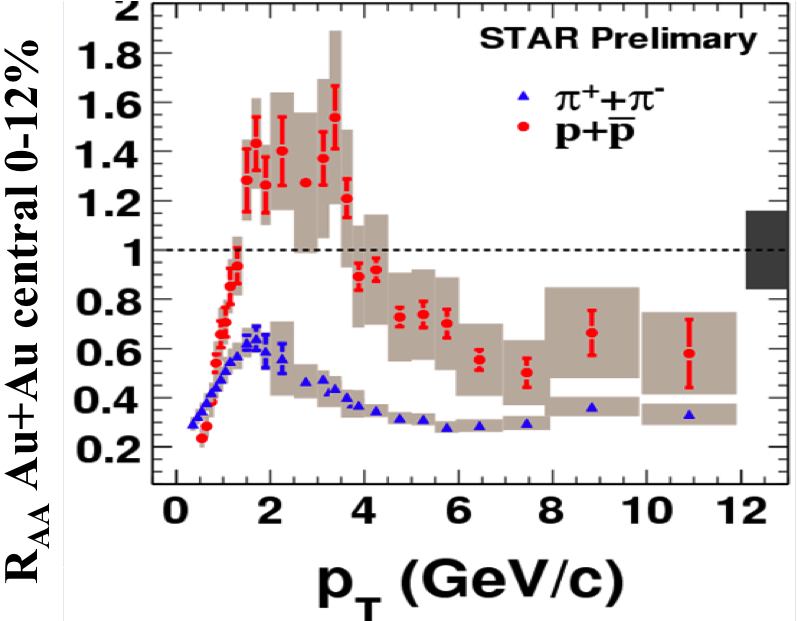}
\hspace{1cm}\includegraphics[height=6.cm]{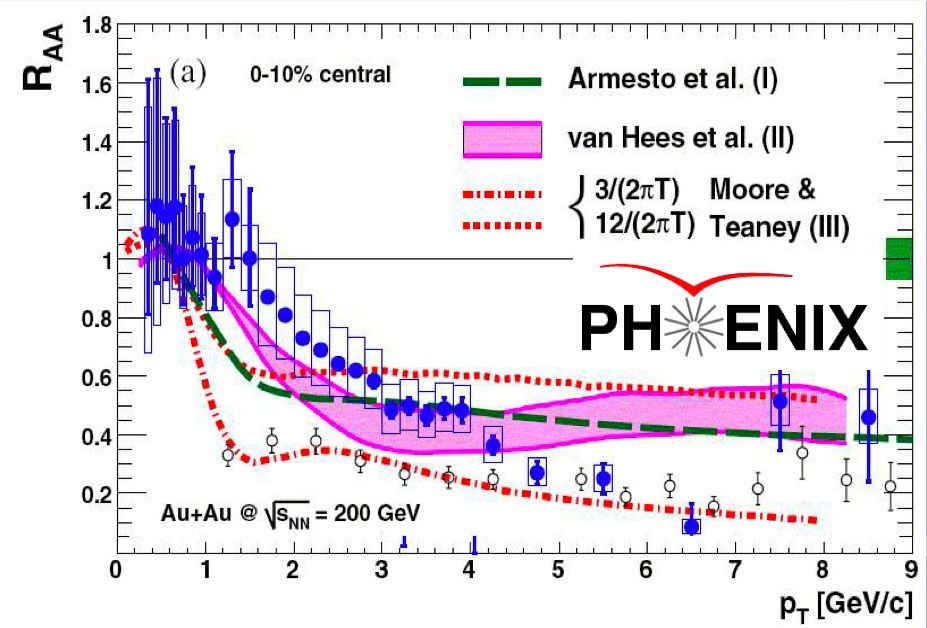}
\caption{ 
(left) $R_{AA}$ from PHENIX for $\gamma,\pi^0,\eta$ and STAR
for protons and charged pions. 
(right top) Two particle correlation function from STAR;
(right bottom) $R_{AA}$ from PHENIX for single leptons (closed points) and $\pi^0$
(open points) compared to theoretical calculations
\label{fig:raa}
}
\end{center}
\end{figure}

We will not go into this subject in depth (see e.g.
PHENIX ``white paper'' \cite{Adcox:2004mh} but just note that
accurate calibration of structure functions have been made in pp and 
dAu collisions, as well as with hard photon measurements (which are
not interacting with the QGP). Thus we know quite well how
many jets are being produced, for any impact parameter. The
number of hard hadrons $observed$ at transverse momentum $p_t$
relative to those {\em expected to be produced} as calculated
from the parton model is called $R_{AA}(p_t)$.
If this quantity is 1, it means the jets are all accounted for and none is lost
in the medium.
This is what indeed is observed with direct photons, not interacting
with the matter, see Fig.\ref{fig:raa}(a). It was quite unexpected that
for mesons this ratio $R_{AA}(p_t)$ was found to be
rapidly decreasing 
and then
 in a wide
range of momenta $p_t>4\, GeV$ its value is only $R_{AA}(p_t)\sim .2$
for central AuAu collisions, which means that $80\%$
of jets are absorbed. 

Theory of quenching mechanisms  included 
gluon radiation on uncorrelated centers (with Landau-Pomeranchuk-Migdal
effect) \cite{Baier:1996sk}, synchrotron-like radiation on coherent fields
\cite{Shuryak:2002ai,Kharzeev:2008qr}, as well as losses due to elastic scattering.
Comparing two radiation mechanisms in general let me just remind that
 the synchrotron-like radiation gives 
the energy loss  $dE/dt\sim E^2$ growing quadratically with energy, it 
is stronger at high energies than radiation from uncorrelated kicks  
which gives only the first power $dE/dt\sim E$:
but then correlation length in ``GLASMA" fields and their lifetime is limited.
 As for elastic scattering losses, it depends on what are the couplings and 
 especially masses of quasiparticles (quarks, gluons or maybe monopoles
 near $T_c$) on which scattering occurs.   The rate of energy
loss itself is  an order of magnitude larger than
pQCD predictions.  Multiple phenomenological fits for the $R_{AA}(p_t)$ 
were made: but they depend on models.

 Furthermore, as noted in
my paper \cite{Shuryak:2001me}, any such model has predictions for
ellipticity $v_2$ in the range below its ``geometric limit" for infinite quenching,
while the data showed  $v_2$ exceeding such limits for all models used.
In simpler words, those models could fit $R_{AA}(p_t)$ but not a double plot
$R_{AA}(p_x,p_y)$. The reason of such large $v_2$ is not yet found,
to my knowledge.

Further crucial test of this theory came from experimental observation of ``single lepton"
quenching and $v_2$: those leptons 
come from semileptonic decays of $c,b$ quarks. At the same $p_t$
heavy quarks have smaller velocity, and if the main quenching
mechanism be radiative, it should reduce quenching accordingly.
The data however do $not$ show any serious reduction, with the same 
 $R_{AA}(p_t)\approx .2$ value for single leptons as for pions (coming
 mostly from gluon jets).  This fact cast doubts at any perturbative
 mechanism of energy loss, since re-scattering of a gluon should be
 larger than that of a quark by the Casimir (color charge) ratio 9/4.

\begin{figure}
\begin{center}
\includegraphics[height=3.0in,width=3.0in, angle=-90]{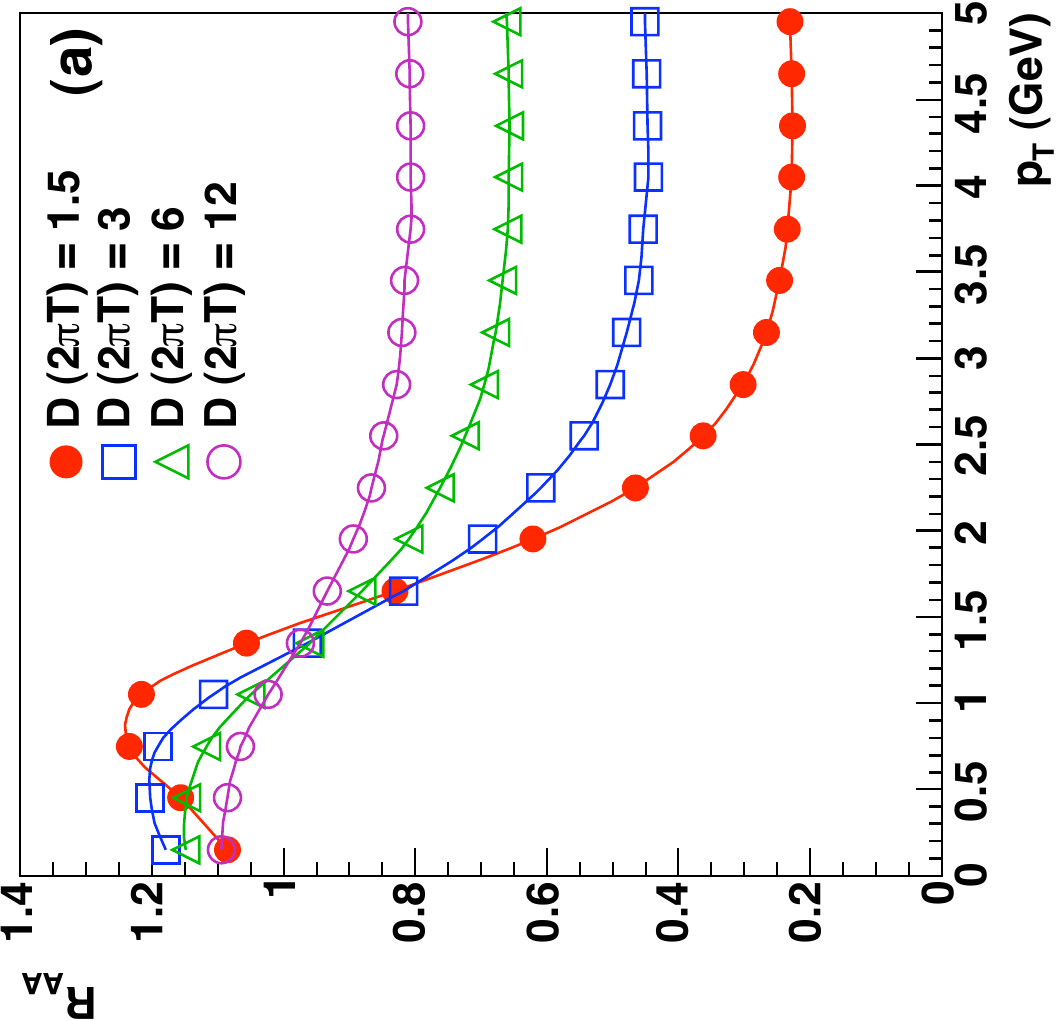}
\hspace{0.2in}
\includegraphics[height=3.0in,width=3.0in, angle=-90]{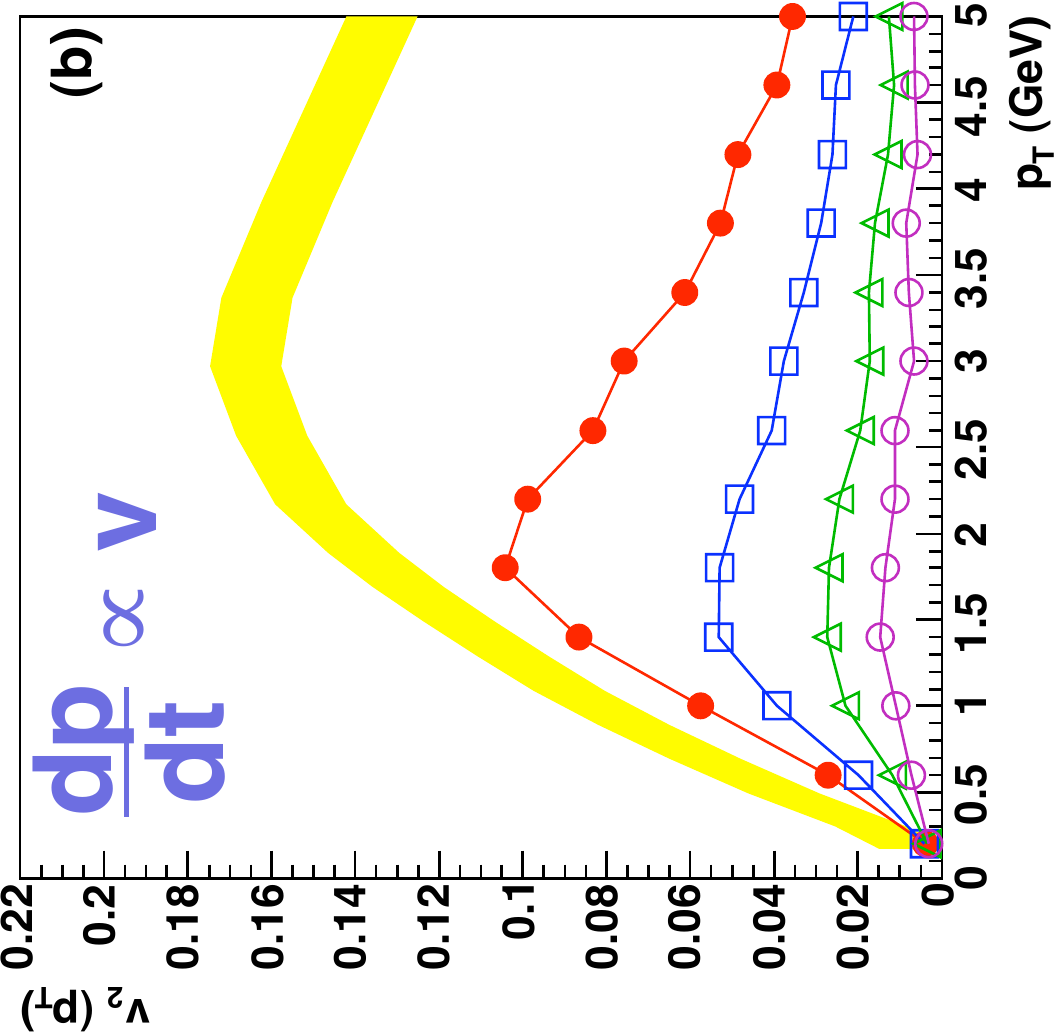}
\caption{ (from Moore and Teaney)
(a) The charm quark nuclear modification factor $R_{AA}$  and  (b) elliptic flow for representative values of the diffusion coefficient given in the
legend.
In this model, the drag is proportional to the velocity,
$\frac{dp}{dt}\propto v$.  
\label{fig:moore_teaney}
}
\end{center}
\end{figure}

 Moore and Teaney \cite{Moore:2004tg} developed a general framework
 of dealing with heavy quark dynamics in QGP, by invoking Focker-Plank
 or Langevin eqn. They have provided a general argument that if
 quark mass relative to temperature $M/T$ is large, relaxation of heavy
 quark is happening slowly and thus justify the Langevin's uncorrelated
kicks assumption. 
\begin{equation}
\label{newton}
\frac{dp_i}{dt} = \xi_i(t) - \eta_D p_i \; , \qquad
\langle \xi_i(t) \xi_j(t') \rangle = \kappa \delta_{ij}
	\delta(t-t') \; .
\end{equation}
Here $\eta_D$ is a momentum drag coefficient and $\xi_i(t)$ delivers random momentum kicks which are uncorrelated in time. $3\kappa$ is the mean
squared momentum transfer per unit time.  The
usual diffusion constant $D$ in space is related to those parameters by
\begin{equation}
D = \frac{T}{M \eta_D} = \frac{2 T^2}{\kappa} \; .
\label{eq:D}
\end{equation}
In Fig.\ref{fig:moore_teaney} we show the calculated dependence of
quenching $R_{AA}$ and elliptic flow $v_2$ for leptons, resulting
from the Langevin process (calculated on top of hydro evolution).
As one can see stronger coupling leads to smaller $R_{AA}$ and larger 
 $v_2$: comparison with data (value about .2 for $R_{AA}$  and
 yellow band for $v_2$) 
 clearly favor the smallest diffusion constant, about  
$D p\pi T\sim 1$. Further work on heavy quark diffusion by Rapp
and collaborators \cite{vanHees:2007me} have tried to specify
the diffusion constant from data better, and also suggested
its explanation using heavy-light resonances.

\begin{figure} 
\begin{minipage}[c]{6.cm}
\includegraphics[width=6.5cm]{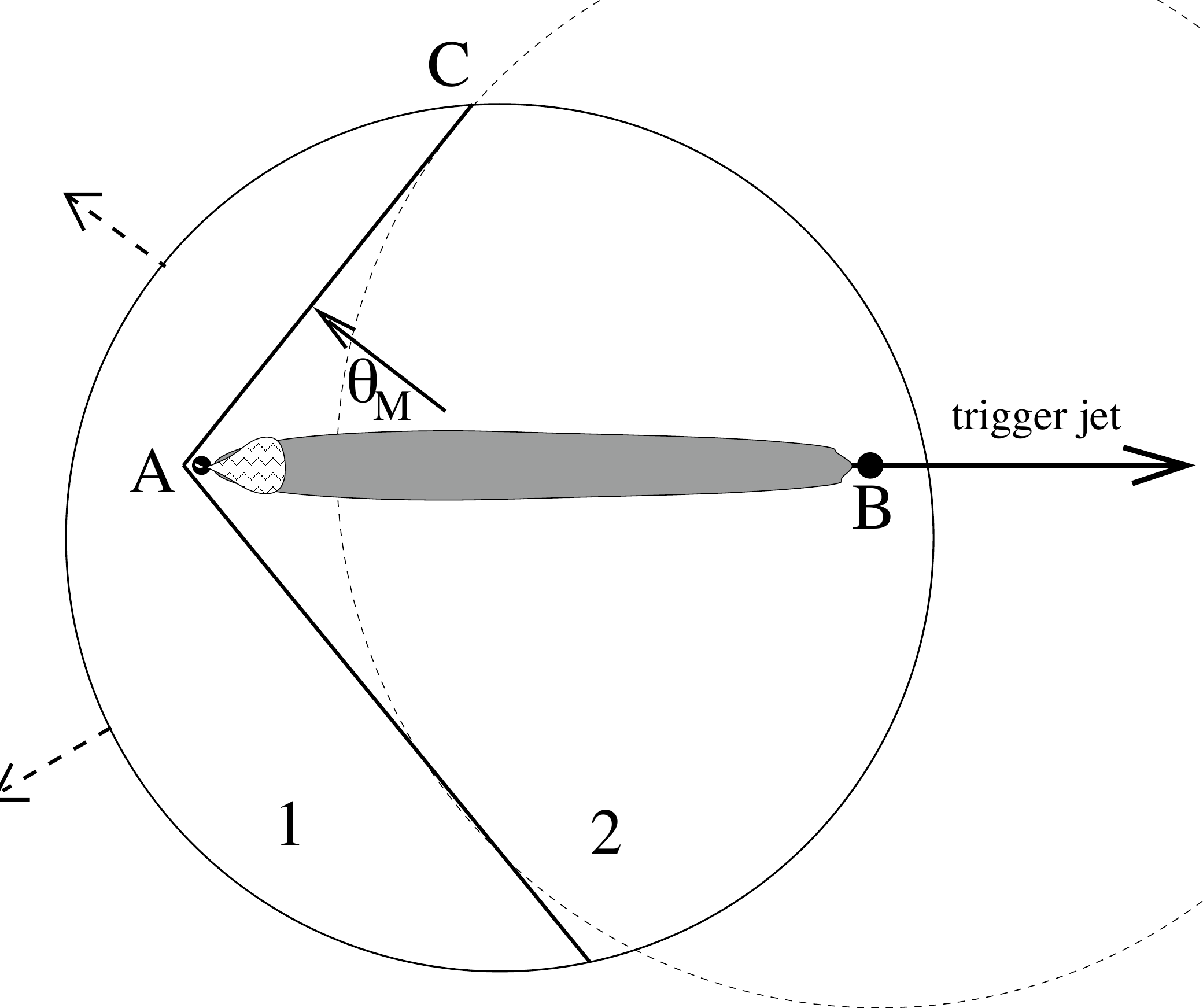}
\end{minipage}
\hfill
\begin{minipage}[c]{10.cm}
\vskip -.6cm
\includegraphics[width=10cm]{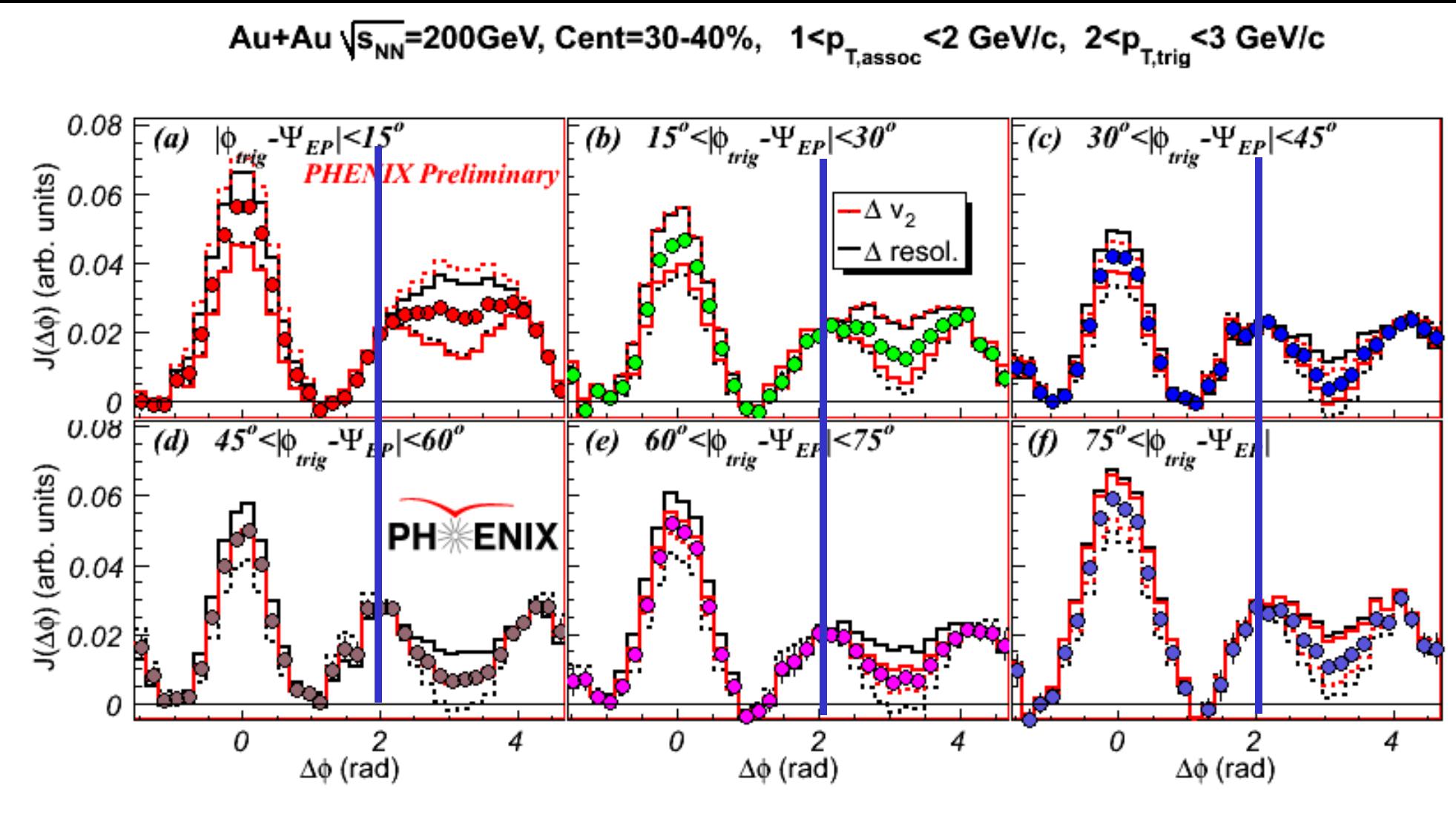}
 \includegraphics[width=10cm]{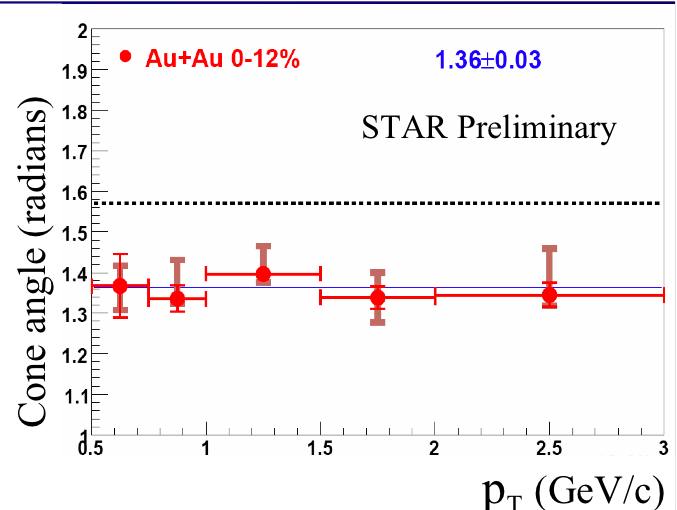}
\end{minipage}
 \caption[a]{
(left) A schematic picture of flow created by a jet going through
the fireball. The trigger jet is going to the right
from the origination point  B. The
  companion quenched jet is moving to the left, heating the matter
(in shadowed  area) and producing a shock cone
with a flow normal to it, at
the Mach angle $cos\theta_M=c_s/v$, where $v,c_s$ are jet and sound velocities.\\
(right top)The background subtracted correlation functions from PHENIX
experiments, a distribution in azimuthal
angle $\Delta\phi$ between the trigger jet and associated particle.
Unlike in pp and $dAu$ collisions where the decay of the companion jet
create a peak at  $\Delta\phi=\pi$ (STAR plot), central $AuAu$
collisions show a minimum at that angle and a maximum corresponding
to the Mach angle }
\label{eos-shuryak-fig_shocks}
 \end{figure}

The next RHIC discovery was associated with ``jet correlations",
which means that in events triggered by one hard particle with large $p_t$
one look for a second ``companion particle" correlated\footnote{It means that
thousands of particles $not$ correlated with the trigger are statistically
subtracted.} with it. While in $pp$ and peripheral collisions one sees
``back-to-back jets", with two peaks and relative azimuthal angle
$\Delta \phi$ values near zero or $\pi$, nuclear collisions typically
show $no$ (or strongly reduced) peak at $\pi$. Further subtraction of flow
in the correlation functions revealed new peaks shifted from the direction
of the companion jet by a
large angle $\Delta \phi\sim 1.2 rad\sim 80^o$, see Fig.\ref{eos-shuryak-fig_shocks}(b).

  Where the energy of the quenched jets go? Thinking about
  this question at the time we came
  with the answer: provided energy is
  deposited locally, hydrodynamics should provide a detailed prediction.
Thus new hydrodynamical phenomenon\footnote{In the field of heavy ion
   collisions Mach cone emission was actively discussed in 1970's by Greiner et al:
but it turned out not to work because nuclear matter -- unlike
   sQGP --
is not a particularly good liquid. The nucleon m.f.p. in nuclear matter is about 1.5 $fm$ is not much smaller than the nuclear size.
} suggested in \cite{Stoecker:2004qu,Casalderrey-Solana:2004qm},
-- the so called {\em conical} flow -- is induced by
jets quenched in sQGP.
The kinematics is explained in Fig.\ref{eos-shuryak-fig_shocks}(a)
 which show
 a plane transverse to the beam.  Two oppositely
moving jets originate from   the hard collision point B.
 Due to strong quenching, the survival of the trigger
jet biases it to be produced close to the surface and to
 move outward. This  forces its companion to 
move inward through matter and to be maximally quenched.
The energy deposition starts at point B, thus a spherical sound wave
appears (the dashed circle in Fig.\ref{eos-shuryak-fig_shocks}left ). Further 
 energy deposition is along the jet line, and is propagating with a speed of
light, 
till the leading parton is found at point A
at the moment of the snapshot.
The expected Mach cone angle is given by
\be cos(\theta_M)={<c(sound)>\over c}\ee
Here angular bracket means not only the ensemble average but also the $time$
average over the time from appearance of the wave to its observation\footnote{Recall
that the speed of sound changes significantly during the evolution, becoming small
 near $T_c$}.

\begin{figure}[t]
\vskip -3.cm
\begin{minipage}{8.cm}
\vskip -5.cm
\includegraphics*[clip,width=12.cm]{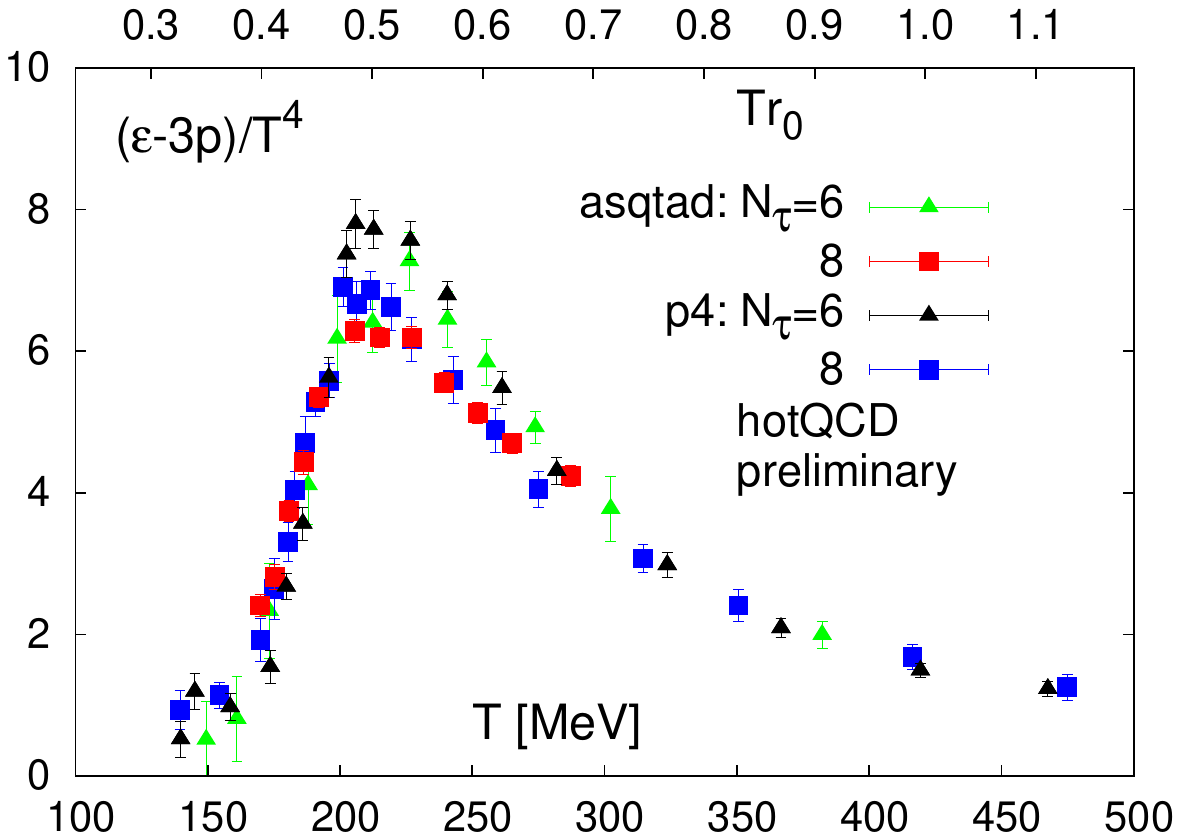}
\end{minipage}  
\begin{minipage}{8.cm} \vskip -5.cm
\includegraphics*[clip,width=12.cm]{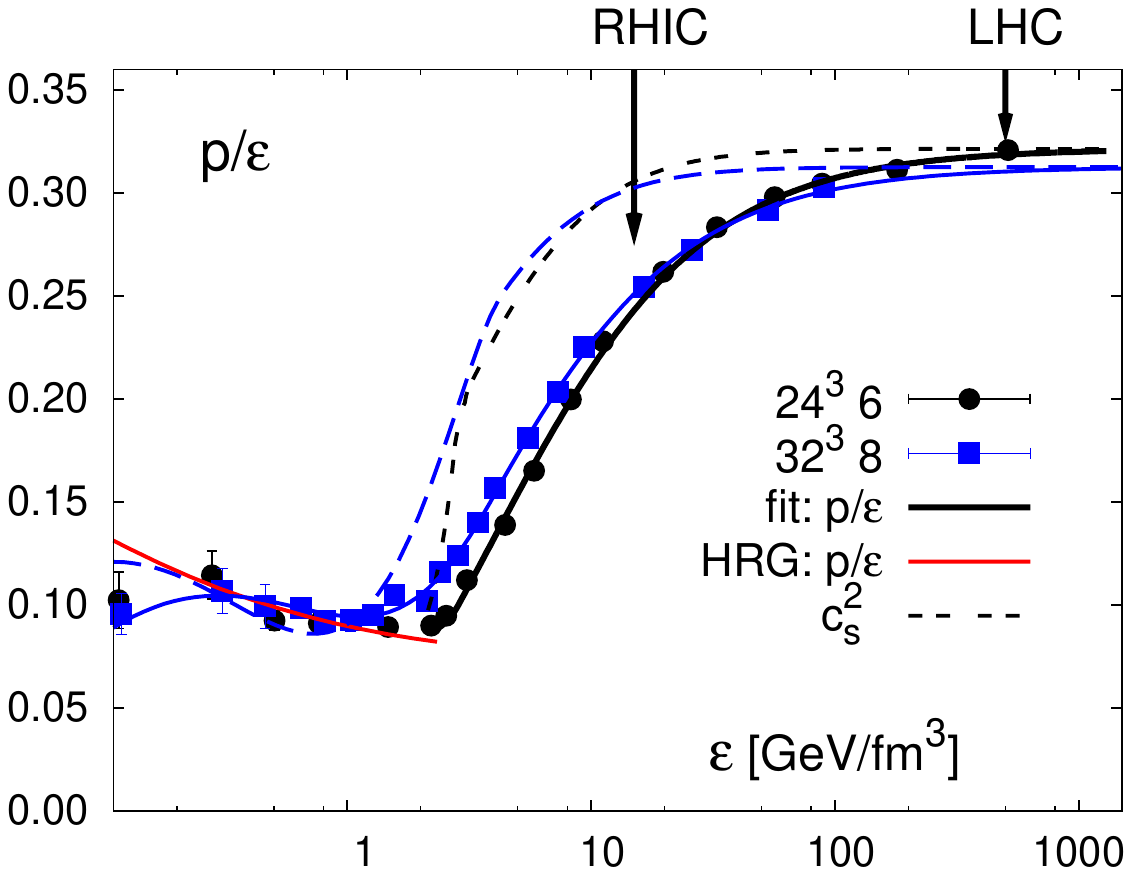}
\end{minipage} 
\caption{\label{fig:eos} The trace anomaly, $(\epsilon -3p)/T^4$ (left) 
calculated on lattices with temporal extent $N_\tau =6,~8$ and the 
ratio of pressure and energy density as well as the velocity of sound 
obtained in calculations with the p4fat3 action on $N_\tau =6$ 
(short dashes)  
and $8$ (long dashes).
\vspace{-0.3cm}
}
\end{figure}

Experimental correlation functions include the usual elliptic flow
and the serious experimental issue was whether the peaks I just described
are not the artifact of elliptic flow subtraction. By Quark Matter 05
this was shown $not$ to be the case, see Fig.\ref{eos-shuryak-fig_shocks}( top right),
which shows PHENIX data selected in bins with specific angle between trigger
jet and the reaction plane: the shape and position of the
maximum (shown by blue lines) are the same while elliptic flow has a
very different phase at all these bins.
The position of the cone is $independent$ on angle relative to
reaction plane Fig.\ref{eos-shuryak-fig_shocks}(right top), centrality (not shown here)
 and $p_t$ Fig.\ref{eos-shuryak-fig_shocks}(right down): so one may
 think it is an universal property of the medium.
The angle values themselves are a bit different, with
 1.2 rad preferred by Phenix and 1.36 rad by Star data:
 those correspond to amazingly
small velocity of the sound wave\footnote{Note, it is the velocity
not squared. The speed of sound $squared$ can be seen as a
dash curve in Fig.\ref{fig:eos}(right).} $<c_s>=0.36-0.2$,
indicating perhaps that what we see was related to the 
$near-T_c$ region. Another evidence for that is observation
of conical structure at low collision (SPS) energies, reported at QM08 by CERES collaboration: at such energies near-$T_c$ region dominates.

 At the last  Quark Matter 08 large set of 3-particle correlation
 data ( a hard trigger plus $two$ companions) have been presented
 both by STAR and PHENIX collaborations. 
 Although those are too technical to be shown here, the overall conclusion
 is that Mach cone structure is more likely explanation of the data
 than other possibilities such as ``deflected jets".

A number of authors have by now reproduced the very
existence of conical flow in
hydro, see e.g. Baeuchle et al\cite{Baeuchle:2007qw}: 
but really quantitative study of
its excitation 
 is still to be done. Casalderrey and myself\cite{CasalderreySolana:2005rf}
 have shown, using 
conservation of adiabatic invariants, that fireball expansion 
should in fact  greatly enhance the sonic boom: the 
reason is similar to
enhancement
of a sea wave (such as tsunami) as it goes onshore. They also showed
that data exclude 1-st order phase transition, because in this case
conical flow would stop and split into two, which is not observed.

Antinori and myself\cite{Antinori:2005tu} suggested 
a decisive test
by  b-quark jets. Those
 can be tagged experimentally even when semi-relativistic:
  the Mach
cone should then shrink, till it goes to zero at the critical velocity
$v=c_s=1/\sqrt{3}$. (Gluon radiation behaves oppositely, expanding
 with $decreasing$
$v$.) 

The experiments with tagged
$b$-jets seem to be even more important in view of recent studies by
Guylassy et al \cite{Gyulassy:2008fa,Betz:2008wy} who found
(using AdS/CFT results from  \cite{Gubser:2008vz} and specific model relating
it to Cooper-Fry formula)
that when the 
velocity of the jet $v$ approach $c_s$ the angle of the peak 
does not accurately follow
the Mach angle but remains always larger. 
They have further found
that the main part of the peak comes not from cones but from the 
non-hydrodynamical near-jet zone 
(they call the ``neck"): what is the nature of those large angle
emission remains unknown.
This group have further studied weak-coupling (pQCD-based) version
of the near zone   \cite{Betz:2008wy}, finding that in this case most
of the flow remains at small angles, 
with very small but visible peaks at Mach angle, but no trace
of large-angle emission predicted by AdS/CFT.

 Let me finally mention the main open question,
which is {\em the absolute and relative amplitudes} of excitation of two
hydrodynamical modes, the $sound$ (responsible for the Mach cone
structure) and the $``diffuson"$ mode, which show matter co-moving
forward behind the jet. As emphasized in our works, this question cannot
be answered from hydro itself, as close to the jet it looses its applicability.
As we will see later, this ratio was recently found from AdS/CFT:
 but we don't know yet of it does or does not agree with the data.
  
\subsection{Charmonium suppression}

Charmonium suppression is one of the classic probes: since charm quark pairs originate during early hard 
processes, they go through all stages of the evolution of the system. A 
small fraction of such pairs $\sim O(10^{-2})$ 
produce bound $\bar c c$ states. By comparing
the yield of these states in heavy ion collisions to that in pp collisions 
(where matter is absent) one can observe their survival probability, giving
us important information about the properties of the medium.

Many mechanisms of $J/\psi$  suppression
in matter were proposed over the years.
The first was suggested by myself in the original ``QGP paper"
 \cite{Shuryak:1978ij}, it is
 a  gluonic analog to ``photo-effect''
 $g J/\psi\rightarrow \bar c c$. Perturbative calculations
of its rate  (see e.g. Kharzeev et al
\cite{Kharzeev:1995id}) leads to a  large 
excitation rates. Indeed, since charmonia
are surrounded by many gluons in QGP, and nearly each has energy
sufficient for excitation, one may think $J/\psi$ would have hard
time surviving. That was  
 the first preliminary conclusion:    nearly all charmonium
states at RHIC should be rapidly destroyed.  
If so, the observed   $J/\psi$ may only come from
$recombined$ charm quarks at chemical freezout, as advocated
e.g. by Andronic et al \cite{Andronic:2003zv}.

However the argument given above is  valid  only if QGP is a weakly coupled
gas,
so that charm quarks would fly away from each other as soon as enough energy 
is available. As was recently shown by Young and myself \cite{Young:2008he},
in strongly coupled QGP the fate charmonium
is very different. Multiple momentum exchanges with matter will lead to
$rapid$ equilibration in momentum space , while equilibration in position space 
is {\em very slow} and diffusive in nature. Persistent attractions between 
$\bar c$ and $c$   
makes  the possibility of returning back to the ground state for  the
$J/\psi$ quite substantial, leading to a substantially higher survival 
probability. For the sake of argument, imagine the matter  so dense
that any diffusion of $\bar c$ and $c$ is completely stopped: then, after this
situation changes by hadronization, one would still find them close to each other
and thus $J/\psi$ -- with its by far the largest density at the origin -- will be
obtained again. Thus, strongly coupled -- sticky - plasma may actually
$preserve$ the $J/\psi$.
 
  Matsui and Satz  \cite{Matsui:1986dk} have proposed 
another idea and asked a different question: up to which $T$
does   charmonium  survive {\em as a bound state}?
They argued that  
because of the deconfinement and the
Debye screening, the effective \barc  attraction in QGP 
is simply too small to hold them together. Satz
and others in 1980's have used the {\em free energy} potential,
 obtained from the
lattice, as an effective potential in Schreodinger eqn.
\be F(T,r)\approx -{4 \alpha(s)\over 3 r }\exp(-M_D(T)r)+F(T,\infty)
\label{eqn_F} \ee
They have shown that as the Debye screening radius $M_D^{-1}$
decreases with $T$ and becomes smaller than the r.m.s. radii
of corresponding states $\chi,\psi',J/\psi,\Upsilon'',\Upsilon',\Upsilon...$, 
those states should subsequently melt. Furthermore, it was found that
for \Jp the melting point is nearly exactly $T_c$, making it
a famous ``QGP signal''.

\begin{figure}[th]
\includegraphics[width=8.5cm]{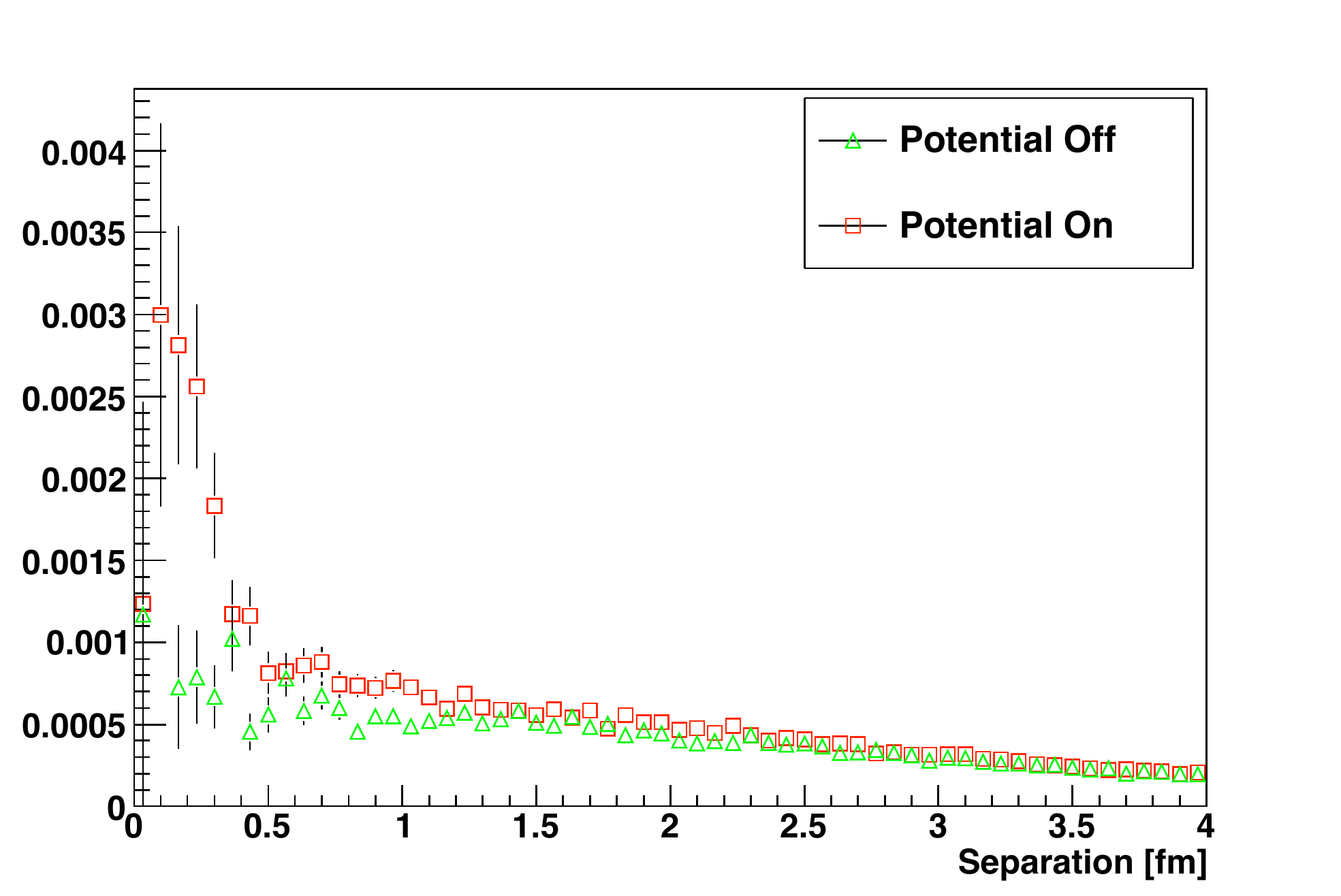}
\includegraphics[width=8cm]{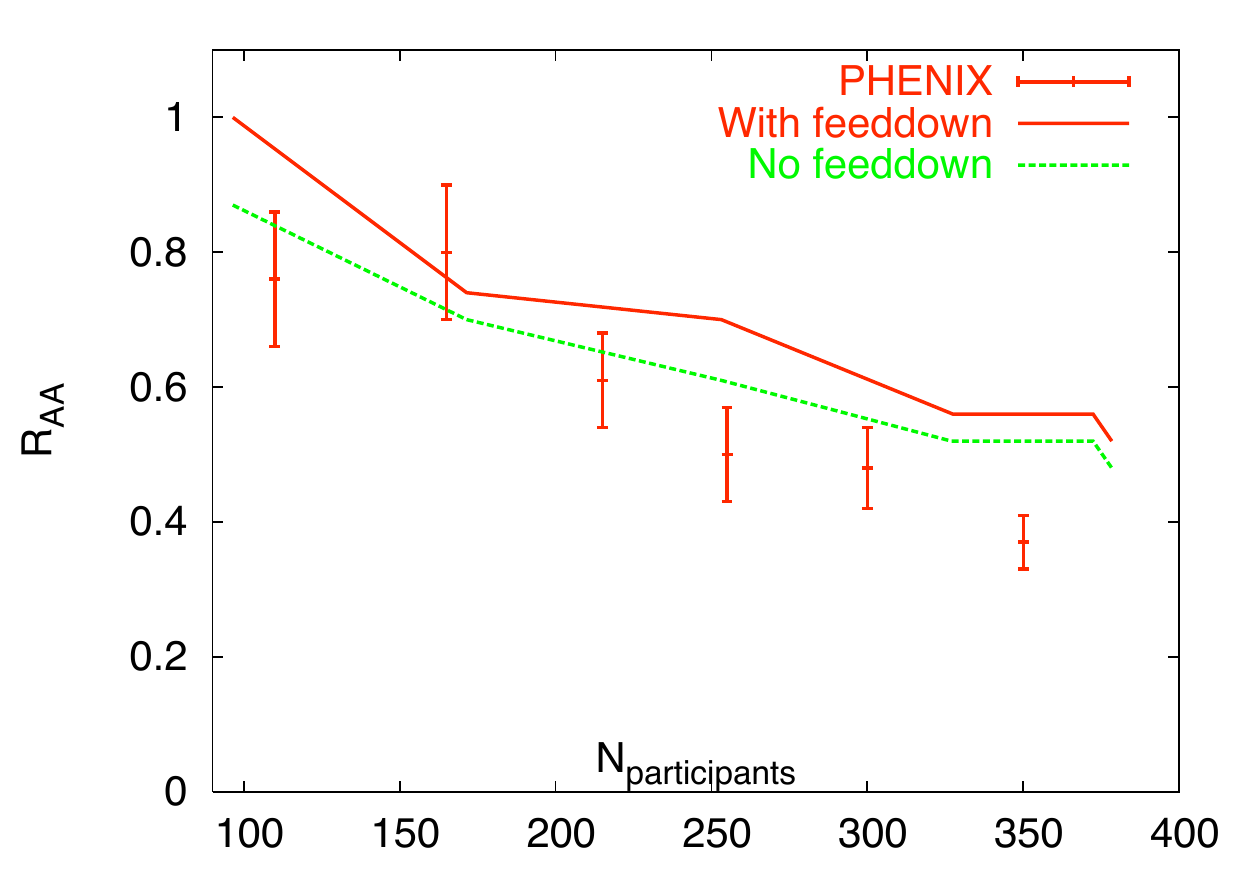}
\caption{ (From \protect\cite{Young:2008he}). 
(a)Distribution over quark pair separation at 
fixed $T=1.5T_c$ after 9 fm/c of Langevin evolution,
 $with$ (red squares) and $without$ (green triangles)
the $\bar c c$ potential. Strong enhancement at small distances due to
potential is revealed.
(b)The points are PHENIX data for anomalous suppression
of $J/\psi$ in AuAu min bias collisions $ R^{anomalous}_{AA}(y=0)$.
Two curves are Langevin model, with (solid, upper) and without (dashed, lower)
 feed-down.}
\label{plot_feeddown}
\end{figure}

Dedicated lattice studies \cite{Asakawa:2003re,Datta:2002ck} extracted quarkonia spectral
densities using the so called maximal entropy method (MEM) to 
analyze the temporal (Euclidean time) correlators.
Contrary to the above-mentioned predictions, the peaks
corresponding to $\eta_c,J/\psi$ states remains basically
unchanged with $T$ in this region,  indicating the
dissolution temperature is as high as $T_\psi\approx (2.5-3)T_c$.
Mocsy et al  \cite{Mocsy:2006qz} have used the Schr\"odinger equation for 
the Green function in order to find an effective potential which would 
describe best
not only the lowest s-wave states, but the whole spectral density.
Recently \cite{Mocsy:2007bk} they have argued that a near-threshold
enhancement is hard to distinguish from a true bound state: 
 according to these authors, the  above mentioned
 MEM dissolution temperature is perhaps too high. My view is that
 the collisional width of states in sQGP is probably large and thus
 the discussion of MEM data is rather academic: in any case we dont
 observe $J/\psi$ in plasma but after it, and a survival is a real-time
issue which cannot be answered by the lattice anyway.

Let us now briefly review the experimental situation. 
For a long time it was dominated by the  SPS experiments
NA38/50/60, who have observed both ``normal'' nuclear absorption
and an ``anomalous'' suppression, maximal in central  PbPb
collisions. Since at RHIC QGP has a longer lifetime
and reaches a higher energy density, 
straightforward extrapolations of the naive \Jp melting scenarios 
predicted near-total suppression. And yet, the first RHIC data
apparently indicate a survival probability very similar to that 
observed at the SPS.

One possible explanation \cite{Thews:2006ia,Grandchamp:2004tn}
is that the \Jp suppression 
is (nearly exactly) $canceled$ by a recombination 
process from unrelated
(or non-diagonal)  $\bar c c$ pairs floating in the medium.
However this scenario needs quite
accurate fine-tuning of two mechanisms.
It also would require rapidity and momentum distributions  of 
the $J/\psi$  at RHIC  be
completely different from those in a single hard production.

 Another logical possibility advocated by Karsch, Kharzeev and Satz \cite{Karsch:2005nk} 
is that $J/\psi$ actually does $survive$  both 
at SPS and RHIC: all  the (so called anomalous, or nonnuclear) 
suppression observed is simply due to suppression of feed-down from 
 higher charmonium states, $\psi'$ and
 $\chi$. (Those are  feeding down about 40\% of  $J/\psi$
 in pp collisions.) These authors however have not
explained $why$  $J/\psi$ survival probability
can be close to one.

 Young and myself \cite{Young:2008he} did exactly that, followed Langevin
dynamics of charm quark pairs, propagating on top of (hydro) expanding
fireball. The treatment basically is the same as that discussed
in the preceding section, where heavy quark diffusion constant
has been derived. One new important element though
is the $\bar c c$ effective potential, which we found is slowing down
dissolution of the pair quite substantially, l
see Fig.\ref{plot_feeddown}(a)  leading to ``quasiequilibrium"
situation in which ratios of different charmonium states are close
to equilibrium ones at corresponding $T$, while the probability
is continue to leak into unbound pairs which occupy slowly growing
volume. 
The main finding of this work is that the lifetime of sQGP 
is not sufficient to reach the
equilibrium  distribution of the pairs in space,
allowing for a significant fraction of \Jp $\sim 1/2$ to survive through $\sim 5\, fm/c$
of the sQGP era.
This probability 
for charmonium dissociation in sQGP is much large than in perturbative estimates, or for Langevin
diffusion which would not include strong mutual interaction.
We have not yet answered many other questions: e.g.
what happens during  $\sim 5\, fm/c$ of the ``mixed phase".
(In view of it seem to be a magnetic plasma, as we will argue below,
the mutual attraction of charmed quarks gets only stronger there,)

\section{From lattice QCD to sQGP}
This section has been more difficult to write than others, because
a connection between lattice results and the ``strongly coupled" regime
of QGP at $T=(1-2)T_c$ 
remains indirect. Perhaps by itself it is rather unconvincing
for a critical reader, as it was for many lattice practitioners,
who are still quite reluctant to
accept the ``paradigm shift" of 2004.
There are quite serious reasons for that.
One (which we will discuss in the AdS/CFT section) is that
the difference between weakly and strongly coupled
regimes is deceivingly small in thermodynamical quantities. 
The second reason is that by performing Euclidean
rotation of the formalism and correlators, one indeed gets rid of
the unwanted phase factors, but  a heavy price for that
is extremely limited ability to understand real-time transport properties --
diffusion constants, viscosity and so on -- which turned out to be at the heart 
of this debate. As usual, we start with introduction for pedestrians which
experts should jump over.

\subsection{The QCD phase diagram for pedestrians}

QCD phase diagram is quite multidimensional:
apart from the temperature $T$ one can introduce chemical potentials $\mu_f$ for each quark flavors $u,d,s$. One however only consider 2 combinations of
those,the  $baryonic$ $\mu_b=(1/N_f)\sum_f \mu_f$ and
$isospin$ $\mu_I=\mu_u-\mu_d$ chemical potentials. 
Then one can vary parameters of the theory itself, such as
the number of colors $N_c$ or quark masses $m_f$:
in many cases however we will only discuss certain limits, for example
in our discussion below a shorthand notation ``2 flavors" $N_f=2$ would imply
massless $u,d$ quarks and infinitely heavy (or just absent) $s,c,b,t$ quarks.
 
Three main phenomena will be under discussion: (i) confinement,  (ii) chiral symmetry breaking
and (iii) color superconductivity. The minimalistic phase diagram may have
only three main phases: (a) hadronic, at low $T$ and $\mu_b$,
 which is both confined and has broken chiral symmetry; (b) Quark-Gluon
 plasma  (QGP) at high  $T$ and $\mu_f$, where all kind of condensates
 are absent; and (c) color superconductor (CS) at high $\mu_b$ and low $T$.
There can of course be many more phases, as these features
are not really exclusive-- e.g. there can be coexisting chiral and CS
condensates.

Note that two last phenomena are due to different kinds of
$pairing$, chiral breaking (ii) is due to  quark-antiquark 
pairing, its nonzero order parameter
is the ``quark condensate" $<\bar q q>$; while color superconductivity
 (iii) is due to quark-quark pairing and its nonzero order parameters
are a set of  diquark condensates $<q_i^aq_j^b>$, with color indices a,b and the flavor
ones i,j arranged in various ways. (The 
 spinor indices will be always suppressed because it is believed that
whatever color-flavor structure of the condensates can be, the
diquark spin remains $zero$, in most cases except very exotic ones.)

By confinement we mean more specifically $electric$ confinement,
which means that electric color field is expelled into {\em flux tubes}, making
quark-antiquark potentials linear. $Magnetic$ component of the 
gauge field is $not$ expelled from the low temperature $T<Tc$ phase: we will
in fact see that it actually dominates
it\footnote{We will not discuss AdS/QCD in this review: let me still
remind its practitioners that popular  models of confinement 
-- hard or soft ``walls" ending
 the holographic space and forcing strings to generate linear potential --
 are over-simplistic, because the same thing happens
 with electric and  magnetic strings.}.
Linear potential is synonymous to the ``Wilson area law" of a large Wilson loop $C$
\be <W_C> \sim exp[-\sigma Area(C)]\ee
with the nonzero string tension. From pioneering lattice calculation
by Creutz in 1980's we know that pure gauge theories indeed have this feature,
although mathematical ``proof" of that remains famously elusive.
Polyakov introduced in 1978 his famous loop and argued that
it provides a $disorder$ parameter, with nonzero value at $T>T_c$.
We will discuss issues related to this below in some detail.

 The area law and Polyakov loop average
 are of course   true order parameters only
in pure gauge theory without quarks. Since quark pair can be produced,
the lattice static potentials show linear behavior up to certain distances. 
Many people thus think that in QCD with quarks the confining phase
cannot be strictly defined and thus there should be  no real
phase transition separating it from the deconfined QGP.  
 
This is however still disputed. Another possible order parameter (albeit
nonlocal) is based on the idea of the ``dual 
superconductor" by tHooft and Mandelstamm \cite{'tHooft-Mandelstamm} which suggested long ago an
existence of  a {\em nonzero magnetic condensate}
of some  bosonic objects. Di Giacomo and collaborators  
have implemented the corresponding observable -- called ``Pisa order parameter" -  which is an operator adding one explicit
monopole to the vacuum.  If so, the deconfinement should be a true phase transition:
 but numerical lattice data for generic nonzero quark masses
 show only a rapid crossover. This contradiction is not yet resolved.

\begin{figure}
\includegraphics[height=5.cm]{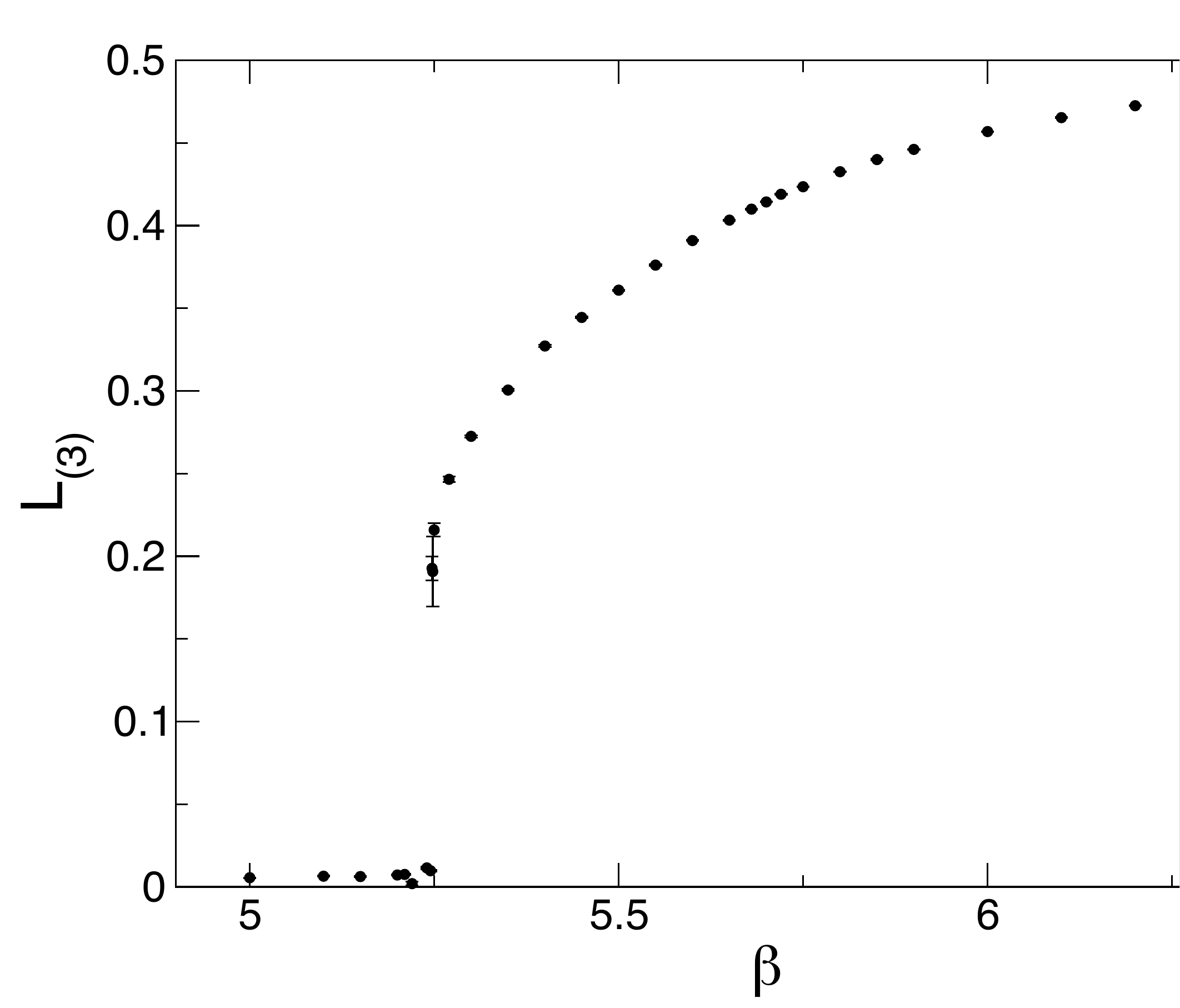}
\includegraphics[height=5.cm]{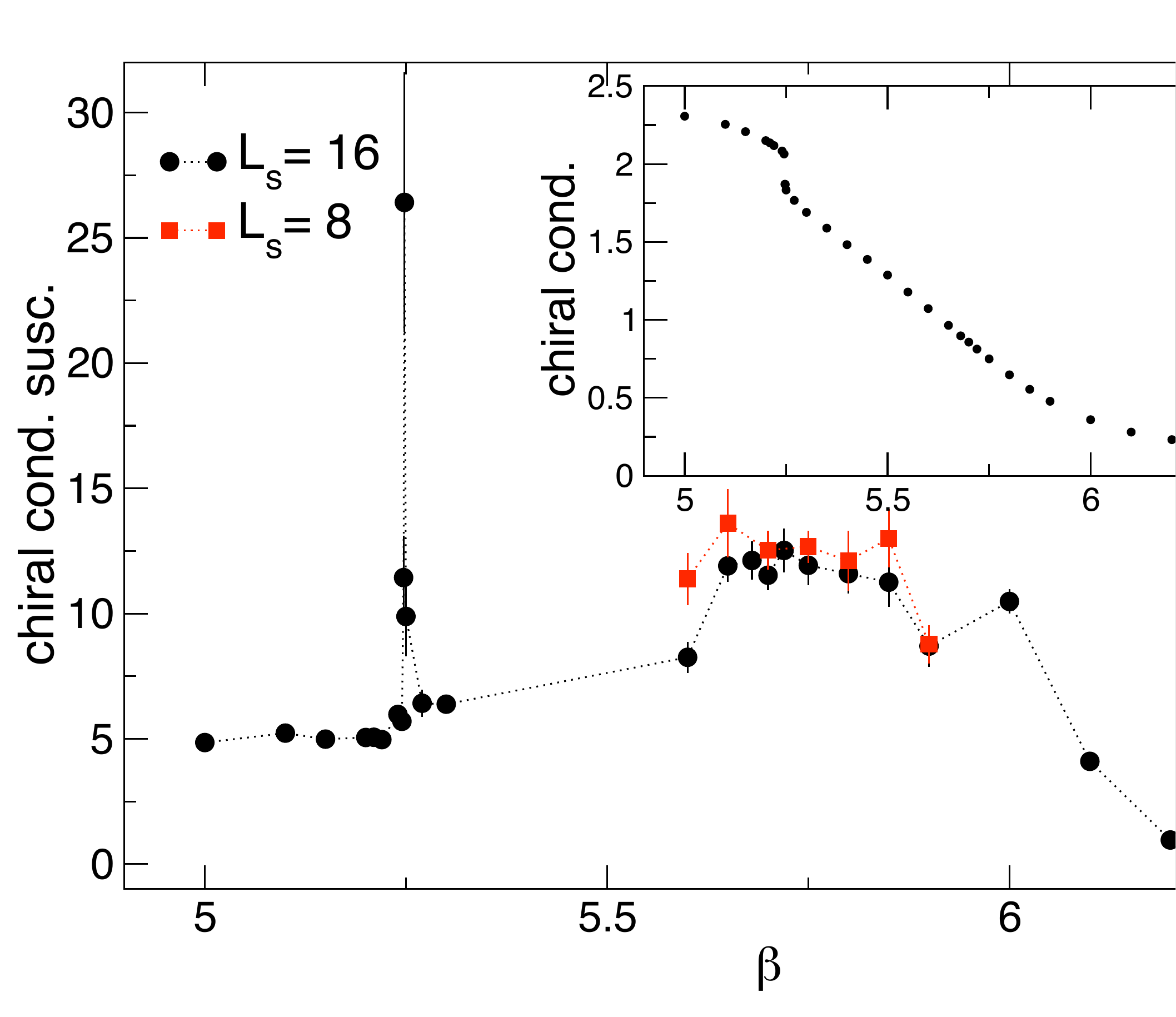}
 \includegraphics[height=5.cm]{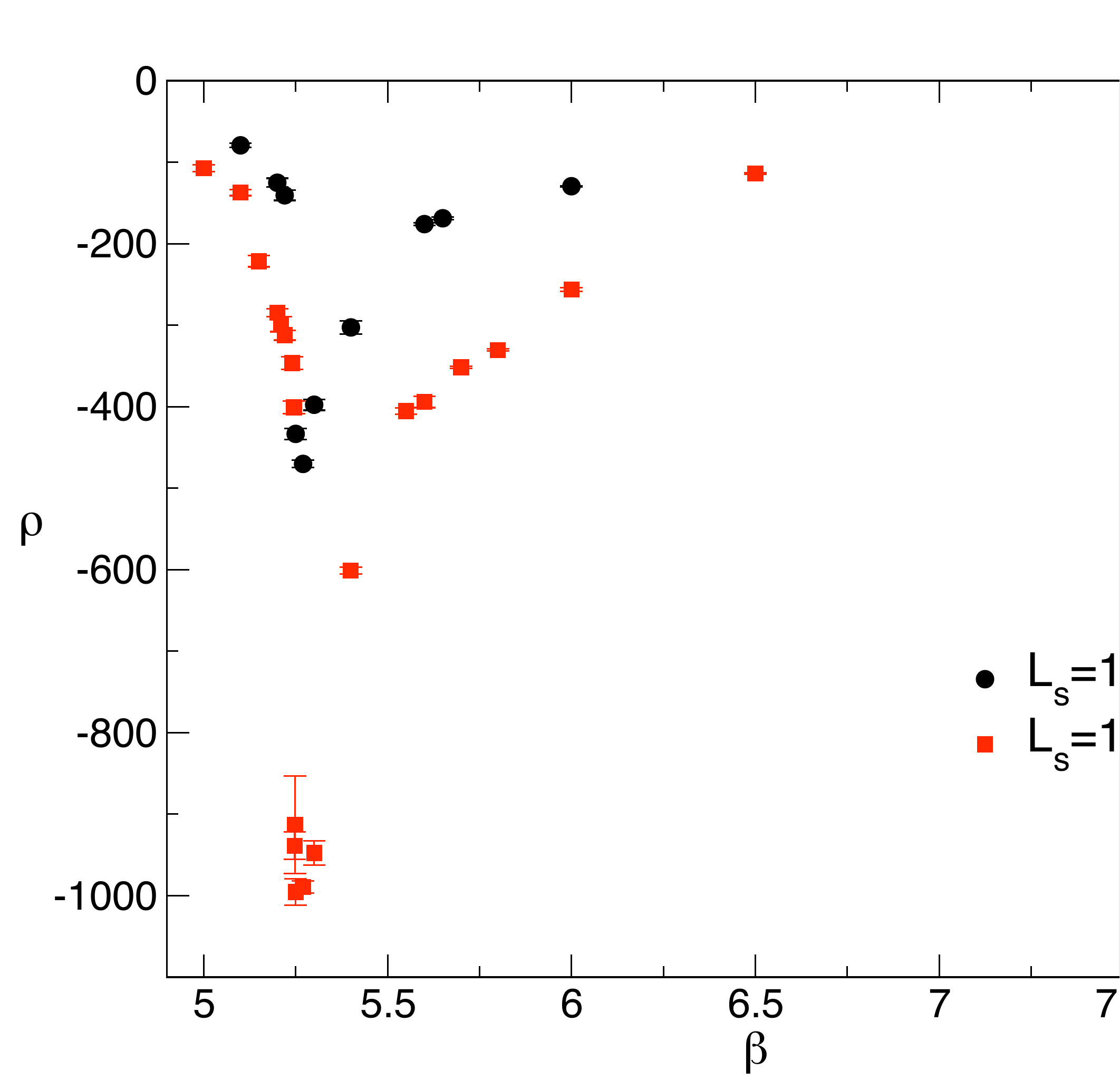}
\caption{(a) The Polyakov loop, with $am=0.01$ and $16^3\times 4$
lattice. (b)
Susceptibility of the chiral condensate, with $am=0.01$ and two different lattices: $8^3\times 4$, $16^3\times 4$. The data are compatible with a crossover at the chiral transition. In the inset, the chiral condensate within the same range of $\beta$ values. A clear jump is visible at the deconfinement phase transition.
(c) The $\rho$ parameter, with $am=0.01$, $L_t=4$, for two different
  spatial volumes.}
\label{fig_lattice-adj}
\end{figure}

To show one example in which all used order parameters
are studied together, let me discuss the work
by Pisa group on $adjoint$ QCD \cite{Cossu:2008wh}
 shown in Fig.\ref{fig_lattice-adj}. In this theory a quark has the same color
as gauge particles, and -- unlike the usual quarks -- it makes
chiral restoration temperature distinctly different\footnote{Most  lattice works agree 
on coinciding deconfinement and chiral symmetry in fundamental -- real-world -- QCD, but it is still debated and I refrained from taking sides and show one but not the other plot. 
} from
deconfinement.  (Larger color representation (charge) of quarks is believed
to lead to  stronger $\bar q q$ pairing and thus higher melting temperature for
  the chiral condensate.)
The Polyakov loop shown in figure  (a) behaves as disorder parameter at
deconfinement indeed, the same point as indicated by Pisa parameter (c),
while the chiral condensate and its susceptibility (b) indicate much smoother 
 higher-$T$ chiral restoration transition. Thus in adjoint QCD we see a presence
  of one more phase (on top of the minimal list of three given above),
a  deconfined but chirally broken phase which is usually called {\em ``a plasma
  of constituent quarks"}. It has all the requisites of a chirally broken phase,
  such as massless Goldstone bosons -- pions, etc. 

In general it is believed that with growing $\mu$ all transition become
sharper. 
Arguments about likely shift in real-world QCD from
 near-crossover at zero $\mu$ and $T_c\approx 170\, MeV$ to
real second-order point  (the so called QCD critical point) and then first order
line has been put forward by Rajagopal,Stephanov and  myself 
\cite{Stephanov:1998dy},
as well as few proposals how one can experimentally search 
 for it. RHIC specialized run with greatly reduced beam energy
 is planned for 2009: perhaps it will shed light on this issue.

 What happens if one changes another famous parameter, the number of colors  by increasing $N_c$? Because
  gluons are adjoint and their effects are $\sim N_c^2$, they dominate
all  quark-induced  $\sim N_c^1$ effects.  It has been recently argued
  by McLerran and Pisarski \cite{McLerran:2007qj} that   
this implies that at large enough $N_c$ $T_{deconfinement}>>T_{chiral}$.
The phase  in between --chirally restored and confined -- they called a ``quarkionic" phase, thinking about
quark-filled Fermi sphere but with baryonic excitations at the surface. Glozman
\cite{Glozman:2008kn}  further suggested even more exotic possibility: particle excitations
in form of (chirally symmetric) baryons, $without$ baryonic holes.
The issue is not yet settled and such phases
have not been seen on the lattice\footnote{Chirally symmetric baryons
can certainly exist: in fact Liao and myself \cite{Liao:2005pa}  have argued that
those are needed to explain lattice susceptibilities in the usual $N_c=3$ QGP
near deconfinement at $T\sim \mu$.}. My view is that  both these exotic ideas
are perhaps excluded in a
confined phase, so it should be just baryonic.

In the discussion above we have ignored short 1-st order transition 
line between the ``vacuum-like" and ``nuclear matter-like" regions, 
at $\mu\approx M_N/3,T< T_c\sim 8\, MeV$.
  Since the usual $N_c=3$ nuclear matter is Fermi-liquid, it is not
qualitatively different from the bulk of hadronic phase: thus one can go around
this phase transition. This is not the case for larger $N_c$, as in this case
baryons are becoming heavier $M_B\sim N_c$ while the nuclear forces
are believed to have a smooth  limit 
\be V(r,N_c)\rightarrow V(r,N_c=\infty)\ee 
The obvious consequence of this is crystallization, as kinetic energy
gets subleading to the potential one. Examples of specific calculations
with the skyrmions are well known\footnote{These works also found
another phase at higher density, in which skyrmions ``fragment" into 
objects of fractional topology: they interpreted those as a
"chirally restored" baryonic phase. I however don't know if one can
trust the model all the way to this phase.
} \cite{Castillejo:1989hq}. 
Is there a possibility that solidification happens even for physical $N_c=3$
 QCD? Old Migdal's pion condensation was of this nature, and
 for more recent study of crystalline $quark$ matter see 
 Rapp, myself and Zahed \cite{Rapp:2000zd}. There can also be
a crystalline color superconductor, known also as LOFF phase,
see more in the review \cite{Alford:2007xm}. Instead of going any further
into the zoo of possibilities, let me stop
with a  joke: perhaps QCD would not have less phases than
water does, and this is quite a lot.

Finally, for completeness, let me mention the opposite direction, increasing
the number of light quarks $N_f$. At some point asymptotic freedom will
disappear, but before that there should be a critical line (or more)
at which Banks-Zaks infrared fixed point will make the theory conformal in
IR. As the corresponding fixed point coupling
gets less and less, chiral symmetry breaking and confinement must
go: again either together or separately. All this territory
is amenable to lattice analysis but remains largely unexplored.

\subsection{Main QGP properties from the lattice}
The thermodynamical observables -- pressure and energy density $p(T),
\epsilon(T)$ -- from the lattice is the simplest global observable,
thus they were calculated more than a decade ago (and
used in hydro calculations). Let me show two recent plots from Karsch
\cite{Karsch:2008fe} which depict them in a combinations which reveal
somewhat more than standard plots of $p(T),
\epsilon(T)$.

The first combination $\epsilon-3p=T_{\mu\mu}$ is related to the
famous scale anomaly: the fact that its value relative to 
$p,\epsilon$ themselves goes to zero means that QGP gets
more and more conformal. However the way it goes down is
not $1/T^4$ --  as simple bag model predicts: this phenomenon
is not yet explained.

As shown in \cite{Karsch:2008fe} and earlier papers, 
quark and gluon quasiparticles dominate at $T>1.5T_c$.
But what happens below that? Gluon/quark  masses are too high
$M\sim (3-4)T$ to explain the peak of  $\epsilon-3p$. 
Chernodub et al \cite{Chernodub:2007sy} have shown that
lattice magnetic objects -- monopoles and vortices -- 
reproduce the shape\footnote{Unfortunately not
the absolute normalization: this issue deserves further studies
as the lattice used in this work is rather small.}
 of this curve well.

The second combination\footnote{Its special
 role in hydrodynamics was emphasized
\cite{Hung:1997du} and its minimum got a spatial name - ``the softest
 point''. Indeed, the gradient of pressure provides a force
and energy density a mass to be moved,
so $p/\epsilon$ it is  proportional to hydrodynamical 
acceleration of matter.} is $p/\epsilon$ - now conformity is seen
as the place where this ratio reaches 1/3. One can clearly see
that it is not yet reached at RHIC, but it is the case at LHC. 
This is one of 
the reasons why LHC experiments will be decisive
in proving (or disproving) whether the
AdS/CFT duality can (or cannot) be used 
in the {\em conformal window} of finite-$T$ QCD.

Let me now turn to the screening lengths. As I already mentioned,
QGP got its name after it was found  \cite{Shuryak:1977ut} that
thermal gluons -- unlike virtual ones -- lead to {\em electric screening}
of the charge in weakly coupled regime (high $T$). The 
corresponding electric (or a Debye) mass
is $M(electric)\sim gT$. 
Static $magnetic$ screening does not appear
via perturbative diagrams; but it has been soon
conjectured by Polyakov \cite{Polyakov:1978vu} that magnetic
screening should appear
non-perturbatively, at the smaller ``magnetic scale'' 
 $ M(magnetic)\sim g^2 T$.

\begin{figure}[h]
\begin{minipage}[h]{8cm}\hspace{-.5cm} 
\includegraphics*[clip,width=7cm]{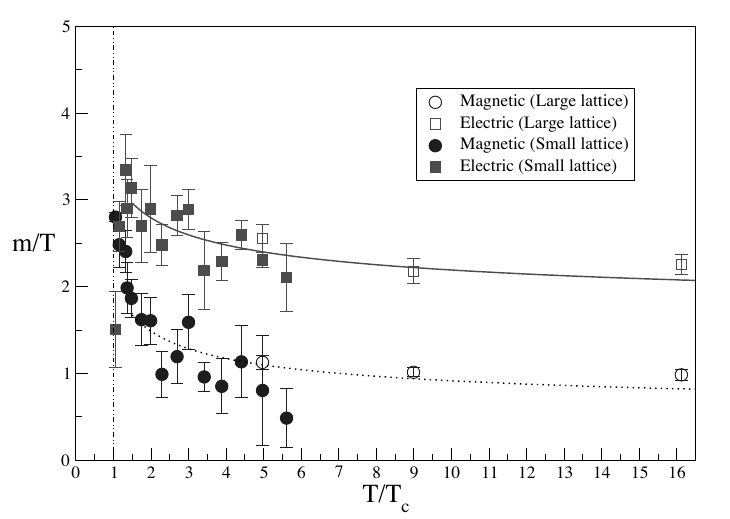}
\end{minipage}
\begin{minipage}[h]{8cm} \hspace{-.8cm} 
\includegraphics*[clip,width=7cm]{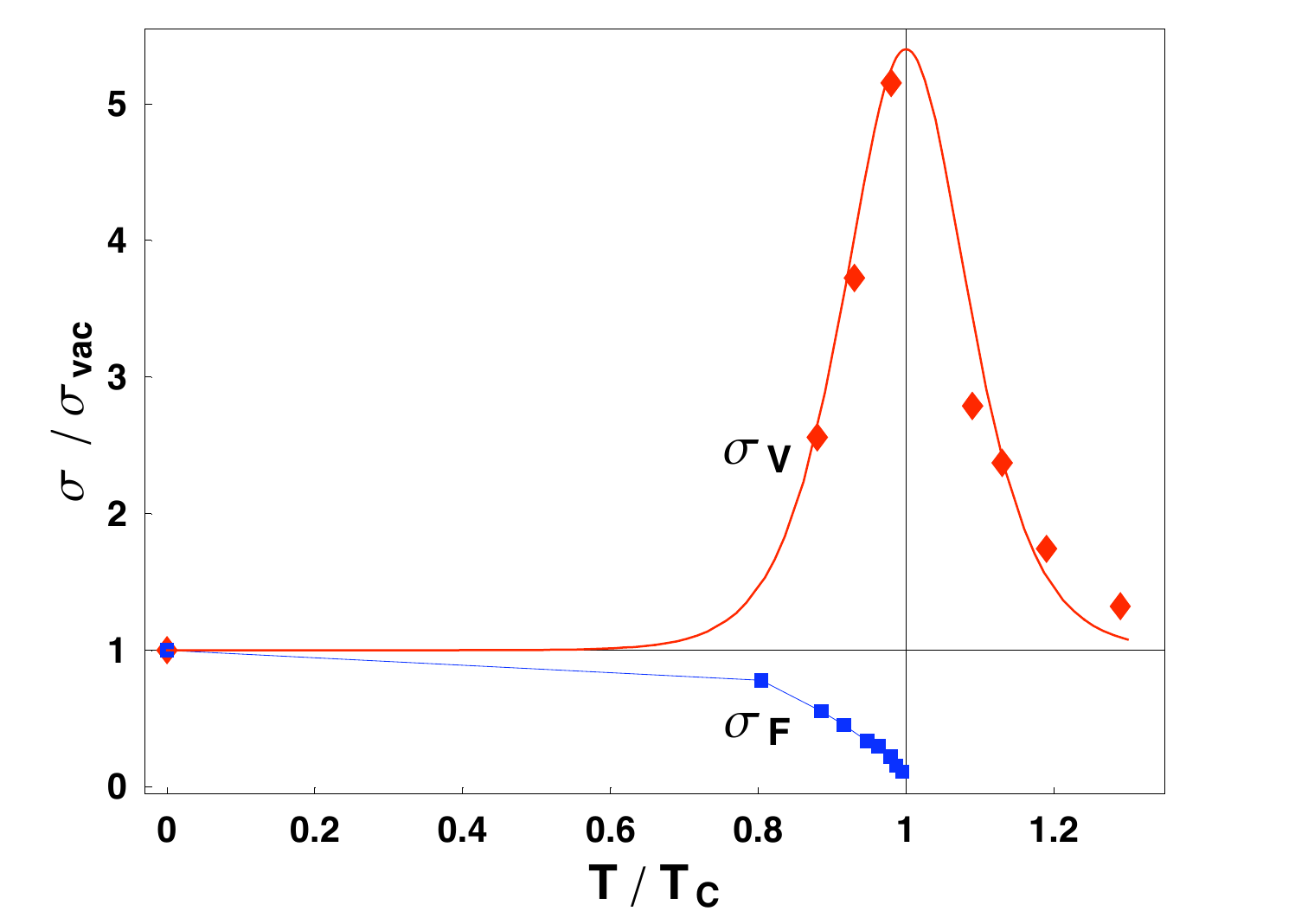}
\end{minipage}\\
\begin{minipage}[h]{8cm} 
\includegraphics*[clip,width=7cm]{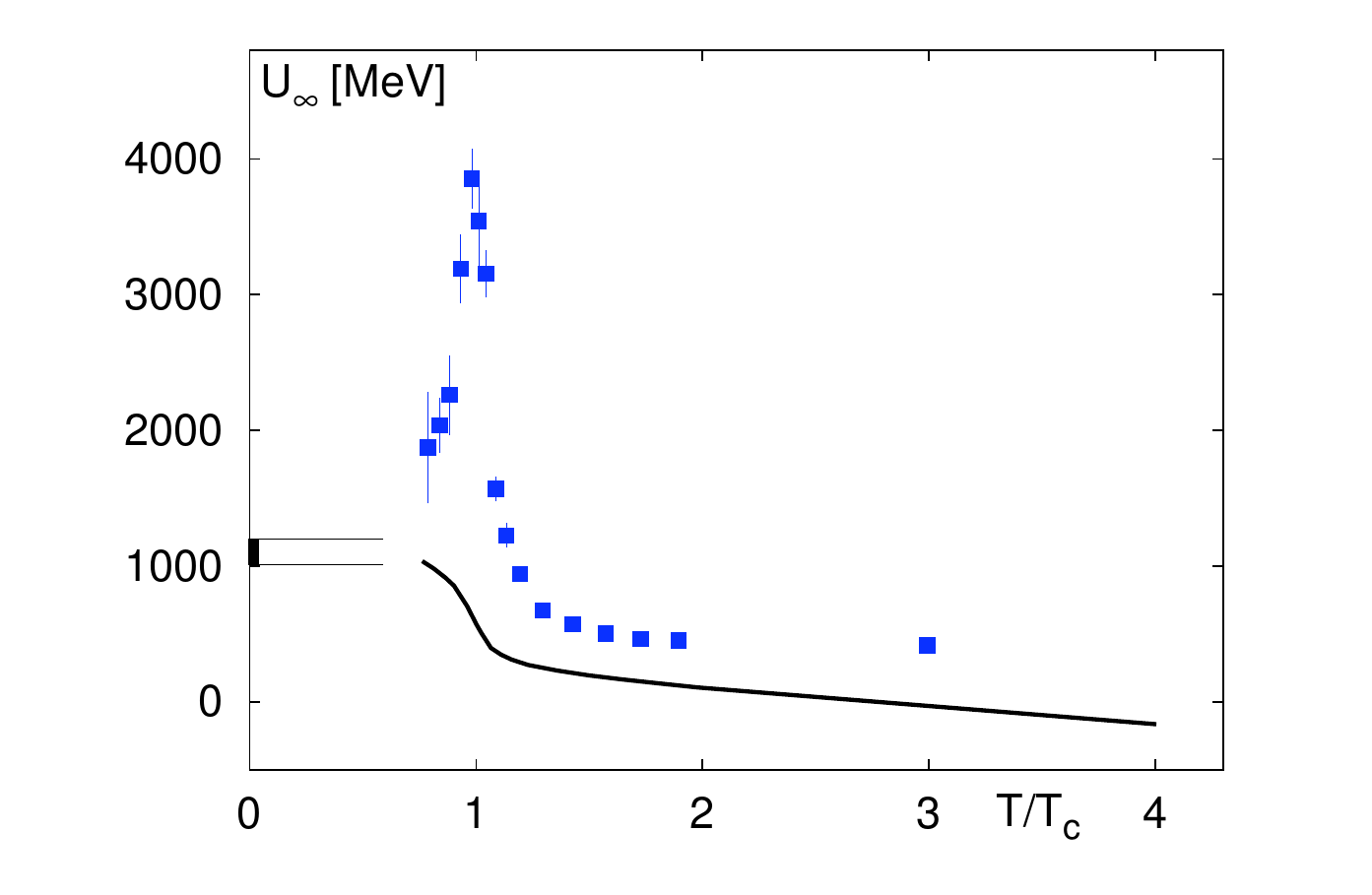}
\end{minipage}
\begin{minipage}[h]{8cm} \hspace{-.8cm} 
\includegraphics*[clip,width=7cm]{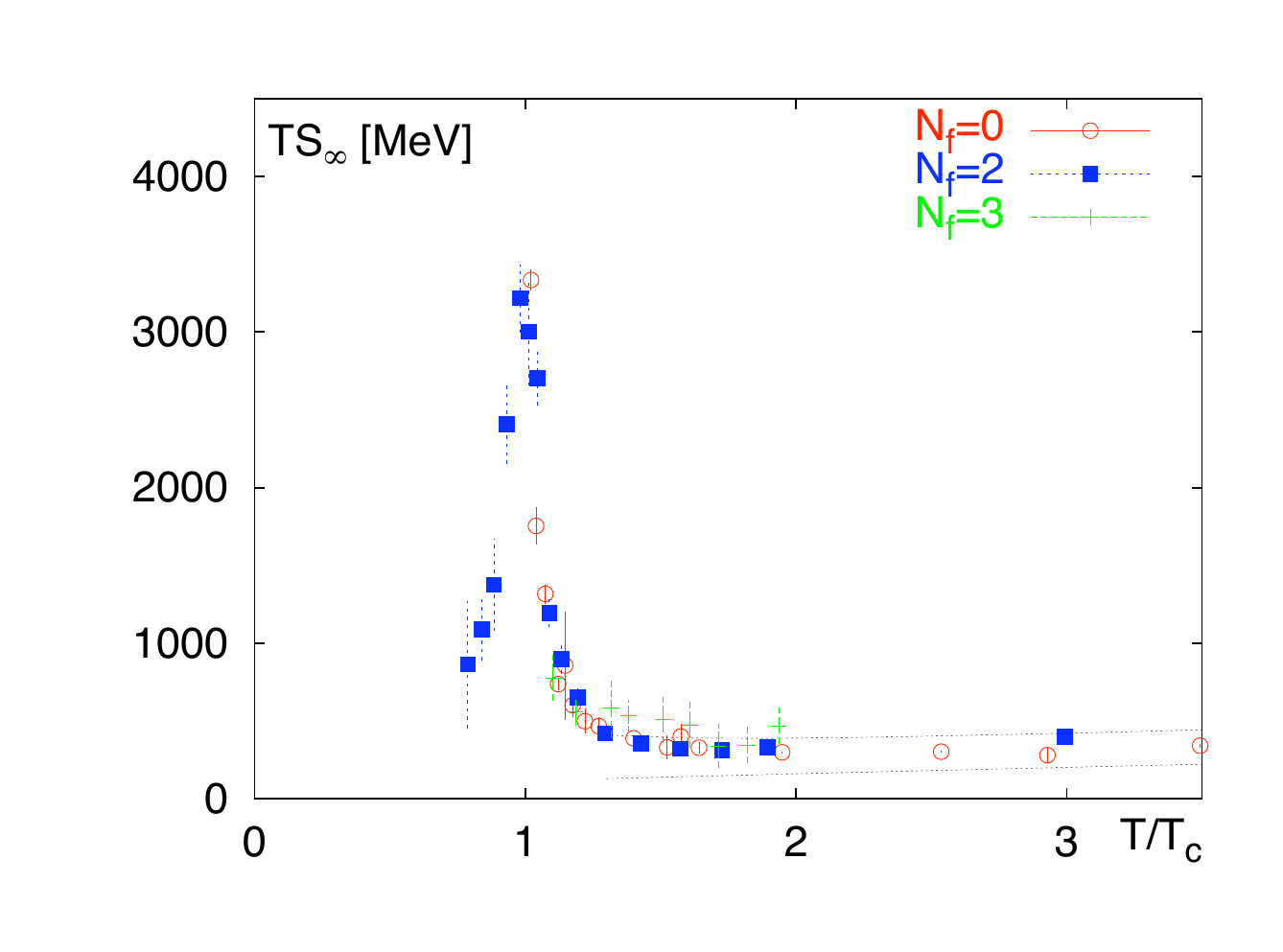}
\end{minipage}
\caption{ 
(upper left) Temperature
dependence of electric and magnetic screening
  masses according to Nakamura et al \protect\cite{Nakamura:2003pu}.
The dotted line is fitted by the assumption, $m_g \sim g^2T$. For
the electric mass, the dashed and solid lines represent LOP and
HTL re-summation results, respectively. 
(upper right)  Temperature dependence of the effective string tensions
of the free and potential energies $\sigma_F,\sigma_V$.
(down) The energy and entropy 
(as $TS_\infty(T)$) 
of two static quarks separated by large distance,
in $2$-flavor QCD according to
 \protect\cite{Kaczmarek:2005zp}. 
}
 \label{fig_tension}
\end{figure}

  To illustrate whether lattice results on the screening masses
  are or are not in agreement with that, we
show their $T$-dependence 
 calculated by Nakamura et al \cite{Nakamura:2003pu}, see
Fig.\ref{fig_tension}(a). Note that at high $T$
the electric mass is indeed significantly larger than the magnetic
one, but it vanishes at $T_c$  -- here electric
objects gets too heavy and ``electric part'' of QGP
effectively disappears.
However magnetic
screening mass  grows continuously toward $T_c$: thus the two cross
each other, around $T=1.5T_c$.

These observation were in fact the starting point
for Liao and myself in thinking about ``magnetic scenario'' for the near-$T_c$
region.
We had used the screening masses to get an idea about density of
electric and magnetic objects, one should conclude
that QGP switched from electric to magnetic plasma
somewhere around
of \be T_{E=M}\approx 1.5T_c \sim 300\, MeV\ee.

The static potential between quark and anti-quark is 
another traditional observable, by means of which
quark confinement in Non-Abelian gauge theories was
established. It was originally inferred from  heavy
meson spectrum and Regge trajectories, and then studied in great detail
numerically, through lattice gauge theories, for review see e.g.
\cite{Bali:1998de}. It is usually represented
as
 a sum of a Coulomb part $V\sim 1/r$, dominant at small distances,
and a linear part $V=\sigma r$ dominant at large distances.
The latter, related with existence of electric flux tubes,
  is a manifestation of
quark confinement. The string tension  in the vacuum ($T=0$) has been consistently
determined by different methods to be about \be \sigma_{vac.}
\approx (426MeV)^2 \approx 0.92\,GeV/fm \ee

 Studies of the static $\bar Q Q$
potential have been extended to finite $T$. In particular,
deconfinement temperature $T_c$ is defined as
a disappearance of the linear behavior as a signal of
deconfinement at $T>T_c$ in the corresponding free energy
$F(T,r)$. Bielefeld-BNL group has  published lattice results
for static $\bar Q Q$ free energy, as well as internal energy and
entropy
\be V(T,r)=F-TdF/dT=F+TS \ee
 at  $T$ both below and above $T_c$,
see \cite{Kaczmarek:2005ui,Petreczky:2004pz}.

 Remarkable features of these results 
 include:\\
1. The linear (in r) part of the potentials. Their effective tensions  are
shown in Fig.\ref{fig_tension}(top right). While that for free energy
vanishes at $T_c$ (by definition), that for   
potential energy  extends till at least about 1.3$T_c$,  with
a peak values 
 about 5 times (!) the $\sigma_{vac.}$. Similar behavior is seen in
 entropy,while canceling in free energy. The widths of these peaks
 provide a natural definition of ``near-$T_c$" region as
 $T/T_c=0.8-1.2$ \\
2.Although potentials at large distances $r\rightarrow\infty$ are  finite
 $V(T,\infty)$, near $T_c$ their values
reach very large magnitudes, see Fig.\ref{fig_tension}(down).
 The corresponding large entropy $S(T_c,\infty)\approx 20$
means that really huge $\sim exp(20)$ number of states is involved ; \\
 The origin of this large energy and entropy
associated with static $\bar Q Q$ pairs near $T_c$, remains mysterious:
many attempts (e.g. 
\cite{Antonov:2006wz}) failed to explain it. Below we will return
to this phenomenon in connection with ``magnetic plasma" scenario.

 Before looking for explanations, however, let us focus on physical
  difference between F and U, based on papers by Zahed, Liao and myself
 \cite{Shuryak:2003ty,Shuryak:2004tx}, in which they are related to
 what happens for $slow$ and $fast$ motion of the charges.
 To be specific, let's consider a pair of static
charges
 held by external hands. Suppose
they are close initially  $L_{ini.}\to 0$,
and then are separated to some finite $L$. This can be done
in two possible
ways, adiabatically slowly or very fast. 
The difference between them in thermodynamical and quantum-mechanical
contexts are  known in many fields of physics. Perhaps the oldest is
the so called
Landau-Zener problem \cite{Landau_Zener} of electron motion, following
the motion of 
 two nucleus in a diatomic molecule. While nuclei change their
relative distance $L$
 with velocity $v_{12}=dL/dt$,
the electrons are in a specific
 quantum state with the energy depending in $L$.
 The issue is probability of the level crossings, which appear
when there are two quantum
levels $E_1(L)\approx \sigma_1 L+C_1$ $E_2(L) \approx \sigma_2
L+C_2$ crossing each other, at some  separation $L_0$. 
 When the two
nucleus approach the crossing at $L_0$ very slowly, then the
electrons may jump from one energy state to another, always selecting
the  lowest energy state. However if the two nucleus
move fast, there is large probability for the electron to remain in
the original state. More quantitatively,
Landau-Zener showed that this probability
is given by
\begin{equation}
P_{fast}=exp{\bigg [}  -\frac{2\pi |H_{12}|^2}{ v_{12} |\sigma_1-\sigma_2|}
{\bigg ]}
\end{equation}
where $H_{12}$ is the non-diagonal transition matrix element of the
  Hamiltonian; but we will only need the limits of large and small
$ v_{12}$. The adiabatic limit
obviously corresponds to free energy $F(T,L)$
measured on the lattice. The ``potential energy'' $V(T,L)$
means that no entropy is generated: this implies that
there was no transition from the original pure state at T=0
into multiple states as level crossing occur: thus it corresponds
to fast motion limit. The positivity of entropy
means that $V>F$ always.

   This discussion is very relevant for the problem of effective potential
   to be used in Schreodinger eqn for the bound states, e.g. in charmonium
   problem.
  Zahed  and myself \cite{Shuryak:2003ty}
  argued that  in this case one should use the $internal$ energy:  
 provide much more stable
bound states, delaying $J/\psi$ melting to higher $T\sim 3T_c$.
Several authors (e.g. \cite{Alberico:2006vw}) have used  effective
potentials in between those two limiting cases. 
With such potentials not only charmonium but also light
quark mesons get bound, as also are baryons and ``gluonic chains"
\cite{Liao:2005hj} and also colored binary states. However,
in a liquid with the parameters we expect from such interaction
the number of nearest neighbors associated with one charge
is expected to be $\sim 4$, and thus it is not clear what
is the role of the binary states. Quantum manybody studies
of these systems
are not attempted yet, and  we don't know
if there is any sense to identify them, and if so how wide those states
may be.

\subsection{Polyakov loop, ``Higgsing" and deformations of QCD}
  Physics of monopoles, to which we will turn shortly below, has been
  originally developed in the Georgi-Glashow model or $\cal N$=2 SUSY which
  has adjoint scalars. Those may have some nonzero expectation values --
  this phenomenon would be colloquially referred to as
  ``Higgsing"  below. If so,  
  the color group is broken, generically to diagonal $U(1)^N_c$.  QCD-like theories are much more difficult
  precisely because they do $not$ have elementary scalars, making Higgsing  much
  more subtle, with its role at finite $T$ presumably
  played by the zeroth component of the gauge field $A_0$.
    
The definition of the Polyakov loop \cite{Polyakov:1978vu} 
is a holonomy of the gauge field across the periodic direction
\be  U= Pexp(i \int_0^\beta A_0 d\tau)\ee
where $\beta=1/T$ is the Matsubara time. If it has VEV one can think of
$<U>$ as a diagonal color matrix, with some eigenvalues $v_1...v_N$: Polyakov's
view on confinement is that the eigenvalues widely fluctuate
and $<Tr(U)>=0$: in the deconfined (plasma phase) these eigenvalues
fluctuate little near minima of the effective action, at $v_k=e^{2\pi i k/N}$.
At high $T$ -- that is in weak coupling -- one can calculate this effective
potential perturbatively. In zeroth order, $A_0=const$ costs no energy,
but its coupling to gluons in the heat bath leads to shifting the gluon states\footnote{Here we mean gluodynamics only: we turn to quarks a bit later.}
and as a result one gets the following effective action 
\be  V_{eff}= -{2 T^4 \over \pi^2}\sum_{n=1}{ |Tr U^n | \over n^4} \ee 
This pushes $Tr U$ away from zero, to the minima mentioned.  
One set of lattice data on $<P(T)>$ 
has been already shown in Fig.\ref{fig_lattice-adj}(a): as one can see
the breaking indeed happens and thus it is indeed a disorder parameter.

Opinions on the role of the breaking of the $Z_N$ symmetry
vastly duffer: while some think it the very essence of confinement
others think it has no dynamical meaning at all.
  To exemplify this polemics the reader may e.g. see  
 Ref. \cite{Smilga:1993vb} in which Smilga pointed out that it is highly suspicious that 
the pure gauge theory  -- which has no such symmetry and can be formulated as
 $SO(3)$ gauge theory rather than $SU(2)$, with $Z_2$ explicitly eliminated.
Smilga further argued using simpler examples that effective potential is
gauge-dependent concept and should be treated with care, he emphasized that 
although at finite lattices/coupling one may apparently see $Z_N$ domains,
only one minimum is physical and the so called
 $Z(N)$ bubbles with nonzero surface tension are not really there.
For comparison of these two lattice formulation see deForcrand
 \cite{deForcrand:2002vx}. 
The question is no longer debatable when there are fundamental 
fermions in the theory, as they see different $Z_N$ phases and
thus no longer respect the symmetry
and one vacuum with real Polyakov line VEV is preferred.  

\begin{figure}[t]
\begin{minipage}[h]{7cm}
\hspace{4cm} 
 \includegraphics[height=8cm]{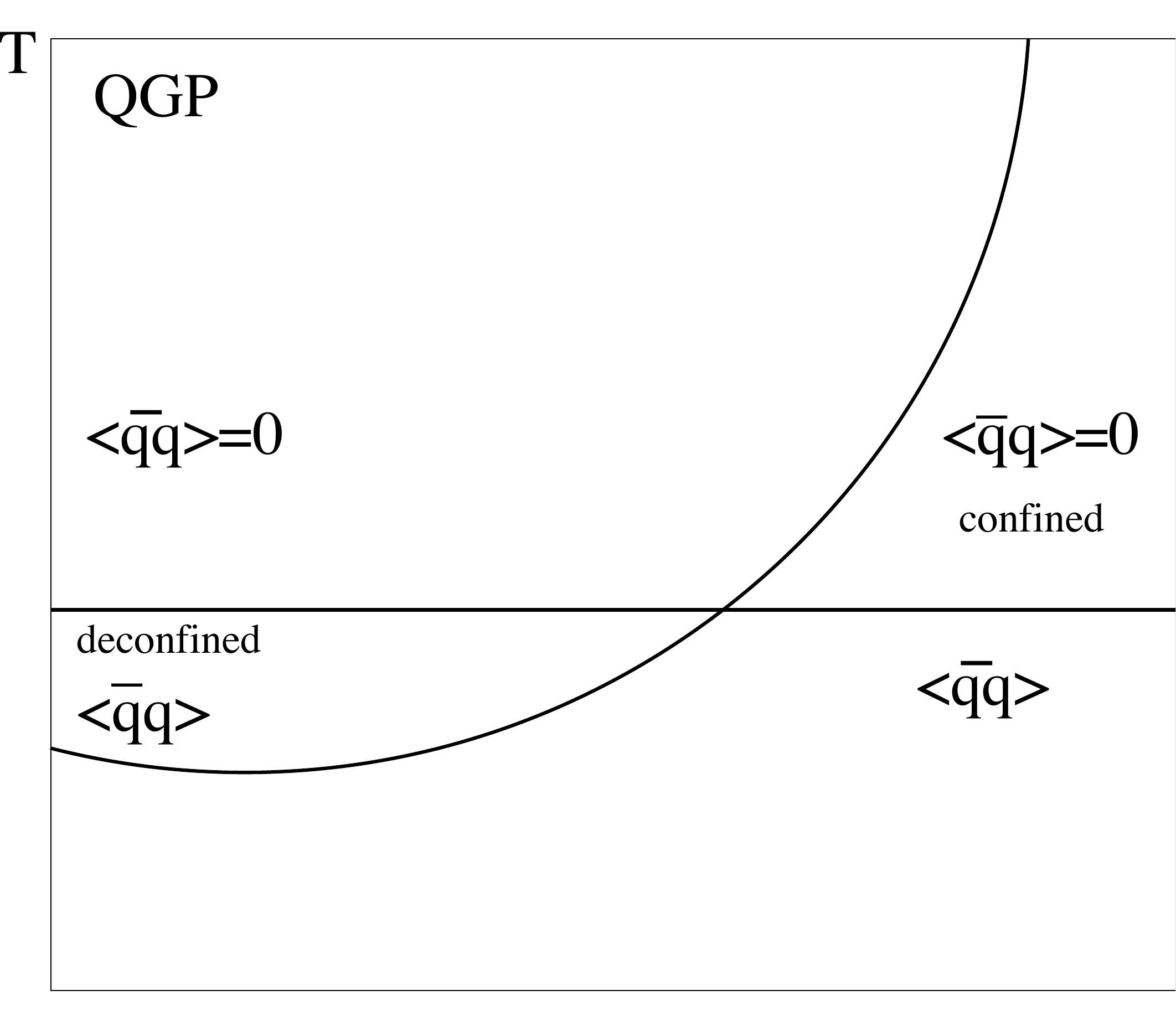} 
\end{minipage} 
\caption{ Schematic phase diagram for adjoint QCD with Shifman-Unsal deformations,
on a plane deformation parameter $a$ - temperature $T$.
At zero deformation $a=0$ (left side of the plot) one has the
usual adjoint QCD in which deconfinement happens below chiral symmetry restoration.
However, as $a$ grows and we deform into a theory without deconfinement,
the order must be reversed, giving place to a new phase
in the right upper corner, which is confined but chirally symmetric.
\label{fig_Unsal_plot}
}\end{figure}

Before we discuss phenomenology of the Polyakov line and its effective
potential in strong coupling (on the lattice) in real-world QCD, let us turn to interesting
``deformations" of QCD recently discussed by Unsal \cite{Unsal:2008ch} and
Shifman and Unsal \cite{Shifman:2008ja}.
The central idea is to deform some QCD-like theory $continuously$ into something else
to which the answer is either known or can be obtained perturbatively. 
 Although they consider several different fermion representations,
for simplicity we will only consider in this section the case
 of $adjoint$ fermions, for which the interplay of gluons and fermions is simpler.
Before going into details, here is the list of deformations:\\
(i) rotating fermionic boundary conditions along the time from anti-periodic
to periodic\\
(ii) putting the theory into a spatial box with variable size\\
(iii) introducing additional potential for the Polyakov loop. 

The {\em first  deformation} is obtained via the introduction of the phase into
fermion boundary condition,
allowing to interpolate smoothly between the periodic and anti-periodic
  boundary condition 
\be (-) =>  exp(i\alpha+i\pi) \ee  into fermionic contribution into a line, with $\alpha$
 changing from $0$ (anti-periodic)  to $\pi$ (periodic) continuously. The effective potential becomes then
 \be V_{eff}\sim\sum_n (N_f e^{(i\alpha+i\pi) n}-1) {|Tr U^n|^2 \over n^4} 
 \ee
  When $\alpha=\pi$
 ($periodic$ boundary conditions) and
  the number of flavors $N_f=1$ two terms in the effective action simply cancel:  this happens because 
this theory is the $\cal N$=1 SUSY gluodynamics with
 the supersymmetry remaining unbroken\footnote{We remind the reader that normal thermal boundary conditions obviously break
 the symmetry between fermions and bosons.}. The famous argument based
 on Witten index applies in this case, telling that the number of vacua cannot be changed
 with any deformation.
 When  $N_f>1$ the fermionic terms dominate and the sign of the potential
 is reversed. It means in this case one has the theory in which $U$ is not pushed to
 large values and there is no $Z_N$ breaking and thus no deconfinement even at weak coupling.
 
 Let us now think about deconfinement and chiral symmetry of the adjoint QCD
  for $2\leq N_f \leq 5$ (the upper limit from asymptotic freedom)
in the $\alpha - T$ plane, shown in  Fig.\ref{fig_Unsal_plot}.
 When  $\alpha=0$ we have the usual (undeformed) adjoint QCD
in which (see Fig.\ref{fig_lattice-adj}) $T_{\chi}>T_{conf}$. But when 
 $\alpha=\pi$ there should be no deconfinement phase at all, which means
that  $T_{conf}(\alpha)$ grows indefinitely before crossing the $\alpha=\pi$ vertical
line.  There should however still be chiral symmetry breaking
at any $\alpha$: in fact the $\alpha=\pi$ theory with periodic fermions
has well known dyons with $2N_f$ fermionic zero modes, which generate
NJL-type interaction and chiral symmetry breaking provided $T$ is 
small enough to get the coupling large enough\footnote{The
physics of chiral transition is the same transition between
``instanton liquid" and ``instanton-antiinstanton molecules
described by Ilgenfritz and myself \cite{Ilgenfritz:1988dh}, see also references for later studies in the
review \cite{Schafer:1996wv}.}.
As a result, there should be an intersection between the deconfinement
line $T_{conf}(\alpha)$ 
and the chiral restoration line $T_{\chi}(\alpha)$    at some $\alpha_{crossing}$,
as
 shown in Fig.\ref{fig_Unsal_plot}:  after the deconfinement and chiral symmetry
 restoration lines have crossed one  finds  a qualitatively new phase, $with$
confinement but with $unbroken$ chiral symmetry .

 Now let us turn to the  {\em second Shifman-Unsal deformation} of 
 the QCD-like theories: if one formulates the theory on $R_3\times S_1$ space  
 with  (periodic for fermions) compact direction of the variable length $L$,
 one can gradually interpolate between 4-d and 3-d gauge theories. The major
 difference between those, as explained by Polyakov \cite{Polyakov:1976fu} many years ago,
 is that while 4-d instanton-antiinstanton interaction is short range $1/r^4$, the
 3d instantons (that is, monopoles) interact by a long range magnetic Coulomb $1/r$.
 The result is that the 3d theory is  confined by monopoles-instantons, while the 4d theory is not confined by its instantons.
 Moreover, it happens  even in the weak coupling regime, in which
 the instanton-monopole density is exponentially small.
 
 Specific mechanism for QCD on $R^3 \times S^1$ was discussed by Unsal
in a separate paper \cite{Unsal:2007jx} for periodic fermions,  it is condensation of 
magnetic charge 2 ``bions" -- pairs of certain dyons -- bound by fermion-induced
forces in spite of mutual magnetic repulsion. The binding is analogous
to instanton-antiinstanton molecule formation, the confinement at high
$T$ is like Polyakov mechanism in 3dim \cite{Polyakov:1976fu}.

  The  {\em third Shifman-Unsal deformation} 
  is done by an $addition$ to the  QCD  action of artificial potential for the $U$, e.g.
  \be V_{add}\sim\sum_n a_n |Tr U^n|
 \ee
with some coefficients $a_n$ chosen at will. The authors themselves
argued that if the coefficients are chosen in order to $oppose$ the $V_{eff}(U)$
generated by quantum fluctuations naturally, one should be able to
delay deconfinement (increase $T_{conf}$), to the extent that
it will occur in the weakly coupled domain and make it tractable in a (semiclassical)
controlled approximation.

Perhaps it is also interesting to go to another
direction as well, $decreasing$ $T_{conf}$,  reaching for the
regime in which electric theory is even stronger coupled than usual
but its dual --magnetic theory of monopoles -- will gets perturbative instead.
Both deformations can easily be done on the lattice: a possibility to
check continuity of the underlying physics of both deconfinement
and chiral restoration all the way from strong to weak coupling will
surely contribute a lot into our understand of both.

\subsection{Phenomenology of $V_{eff}(U)$ in QCD} 
The early history of perturbative derivation of the effective potential
for $U$ (or $A_0=const$) can be found in the classic review
\cite{Gross:1980br}.  Instead of repeating here
well known results, let me refer to more recent attempts to combine known
perturbative results with the lattice data include a paper by Pisarski  
\cite{Pisarski:2006hz} where one can find the details of the recommended
effective Lagrangian.

One more
motivation to study the QCD deformation via adding extra potential for
Polyakov loop on the lattice
comes from heavy ion phenomenology. Dumitru and
collaborators \cite{Bazavov:2008qh} have used this form of the effective Lagrangian
to study real-time evolution of $A_0$.
The main conclusion from their work is that
that $<A_0>$  belongs to the class of so called  {\em slow variables},
and its evolution in heavy ion collisions has to be treated
separately from the overall equilibration.  They have performed numerical solution
of the EOM for it, starting from ``suddenly quenched" value corresponding
to its vacuum form, moving toward its
 minimum at the deconfined phase at $T=2T_c$.
 The main finding of this work
is that  the relaxation of this variable is very slow, taking about 40 fm/c or so.
This time significantly
exceeds the QGP lifetime at RHIC  which is only about 5 fm/c or so,
which suggests that in real collisions we should treat $A_0$ as essentially
random variable frozen at some value and color direction
during hydro evolution. It means there is a chaotic out-of-equilibrium Higgsing,
slowly rolling down, like in cosmological inflationary models: thus
one would like to know as much as possible about phase transitions
and EoS for $all$ values of $<U>$.

Another active direction using effective potential $V_{eff}(U)$
is the so called PNJL model \cite{Fukushima:2003fw}, which combines
the Polyakov loop with well known Nambu-Iona-Lasinio model for chiral symmetry
breaking, see also \cite{Megias:2004hj}.
 Quite impressive results for QCD thermodynamics were
obtained along this path by the group of Weise \cite{Ratti:2006wg}.
Let me give their notations and the parameterization of the potential. 
A background color gauge field $\phi \equiv A_4 = iA_0$, 
where $A_0 = \delta_{\mu 0}\,g{\cal A}^\mu_a\,t^a$ with the $SU(3)_c$ gauge 
fields ${\cal A}^\mu_a$ and the generators $t^a = \lambda^a/2$. The matrix 
valued, constant field $\phi$ relates to the (traced) Polyakov loop as follows:
\be
\Phi=\frac{1}{N_c}\mathrm{Tr}\left[\mathcal{P}\exp\left(i\int_{0}^{\beta}
d\tau A_4\right)\right]=\frac{1}{ 3}\mathrm{Tr}\,e^{i\phi/T}~,
\ee
In the so-called Polyakov gauge one  chooses a diagonal
representation for the matrix $\phi$, $
\phi = \phi_3\,\lambda_3 +  \phi_8\,\lambda_8~,
$
which leaves only two independent variables, $\phi_3$ and $\phi_8$. The potential
involves the logarithm of $J(\Phi)$, the Jacobi determinant which results 
from integrating out six non-diagonal $SU(3)$ generators while keeping the two 
diagonal ones, $\phi_{3,8}$, to represent $\Phi$:
\be
\frac{{\cal U}(\Phi,
\Phi^*,T )}{T^4}=-\frac{1}{2}a(T)\,\Phi^*\Phi
+b(T)\,\ln\left[1-6\,\Phi^*\Phi+4\left({\Phi^*}^3+\Phi^3\right)
-3\left(\Phi^*\Phi\right)^2\right]
\label{u1}
\ee
with 
\be
a(T)=a_0+a_1\left(\frac{T_0}{T}\right)
+a_2\left(\frac{T_0}{T}\right)^2,~~~~~~b(T)=b_3\left(\frac{T_0}{T}
\right)^3.
\label{u2}
\ee
The logarithmic divergence of $\mathcal{U}(\Phi,\Phi^*,T)$ as 
$\Phi,~\Phi^*\rightarrow 1$ automatically limits the Polyakov loop $\Phi$ to be
always smaller than 1, reaching this value asymptotically only as 
$T\rightarrow\infty$. The parameters $a_i$ and $b_3$ are 
determined to reproduce lattice data for the
thermodynamics of pure gauge lattice QCD up to about 
twice the critical temperature\footnote{At much higher temperatures, where 
transverse gluons begin to dominate, the 
PNJL model is not supposed to be applicable.}. 
The values of these parameters are 
$
a_0 = 3.51~,~~a_1 = -2.47~,~~a_2 = 15.22~,~~b_3 = -1.75
$.
The critical temperature $T_0$ for deconfinement in the pure gauge sector is
fixed at 270 MeV in agreement with lattice results.

\begin{figure}
\centering
\includegraphics[height=12cm]{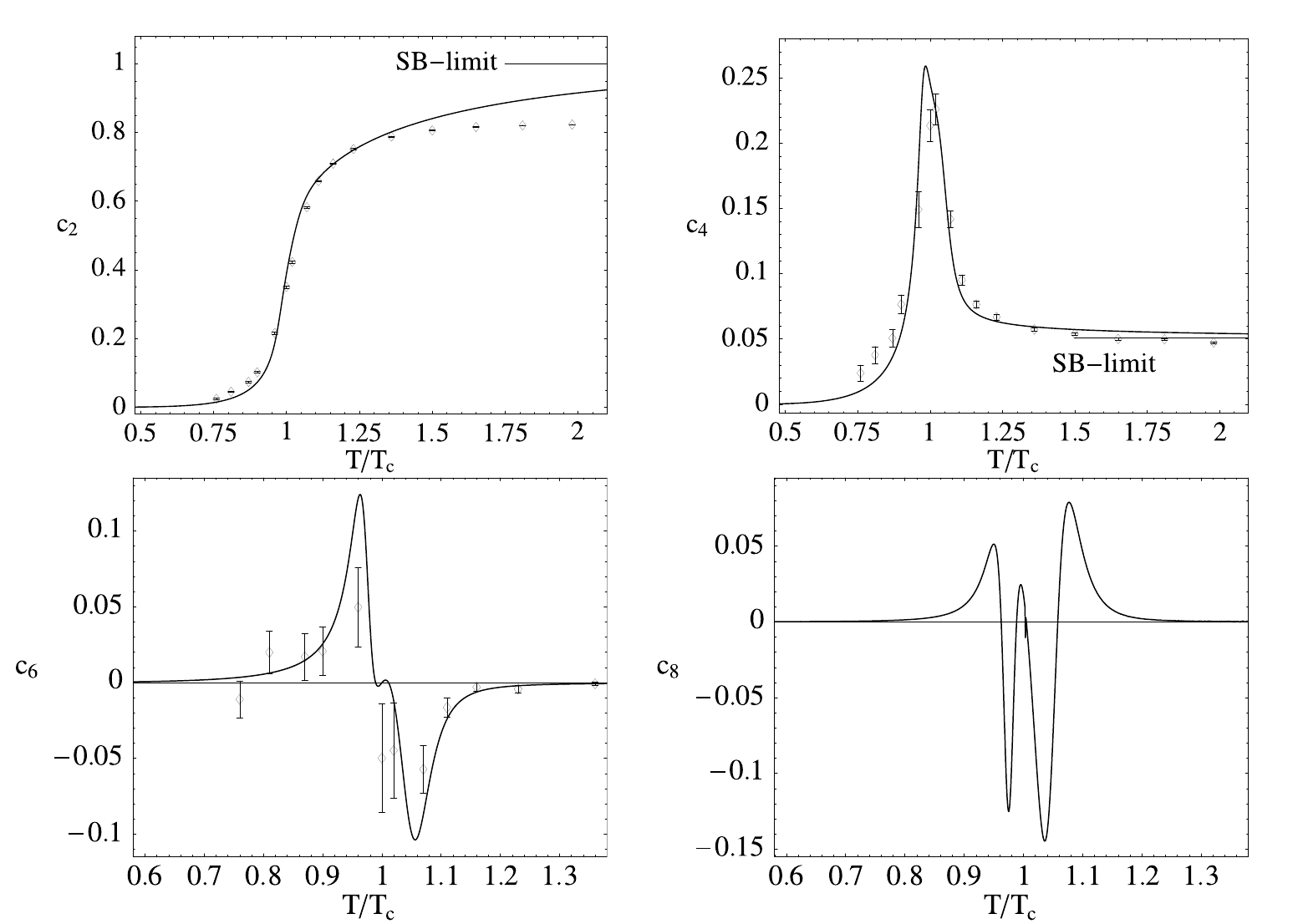}
\caption{
Comparison of PNJL model predictions 
\protect\cite{Ratti:2006wg}with susceptibilities of different
order, from 2 to 8 derivatives over $\mu$ at $\mu=0$ with
the lattice data \protect\cite{Allton:2005gk}.
\label{fig_pnjl}}
\end{figure}

After performing a bosonization of the PNJL action and introducing
scalar and pseudoscalar auxiliary fields, $\sigma$ and ${\vec\pi}$, the PNJL
thermodynamic potential becomes:
$$ \Omega\left(T,\mu_q,\sigma,\Phi,\Phi^*\right)\!=\!\,\mathcal{U}
\left(\Phi,\Phi^*,T
\right)+\frac{\sigma^{2}}{2G}
-2N_f\int\frac{\mathrm{d}^3p}{\left(2\pi\right)^3}
\left\{T\!\,\ln\!\left[1+3\,{\Phi}\,\mathrm{e}^{-\left(E_p-\mu_q\right)/T}\!\!+
3\,{\Phi^*}\,\mathrm{e}^{-{2}\left(E_p-\mu_q\right)/T}+
\!\!\,\mathrm{e}^{-3\left(E_p-\mu_q\right)/T}\right]\right.
$$
\be +\!\left.T\! \,\ln\!\left[1\!+\!\,3\,{\Phi^*}\,\mathrm{e}^{-\left(E_p+
\mu_q\right)/T}\!\!+\!\,
3\,{\Phi}\,\mathrm{e}^{-{2}\left(E_p+\mu_q\right)/T}\!\!+
\!\,\mathrm{e}^{-{3}\left(E_p+\mu_q\right)/T}\right]+3\,\Delta E_p\,\theta
\left(\Lambda^2-\vec{p}^{~2}\right)\right\}
\nonumber
\ee
where the quark quasiparticle energy is $E_p=\sqrt{\vec{p}^{\,\,2}+m^2}$ and the 
dynamical (constituent)
quark mass is the same as in the standard NJL model:
$m=m_0-\sigma=m_0-G\langle\bar{\psi}\psi\rangle$. The last term in the previous equation involves the difference $\Delta E_p$ between the quasiparticle energy $E_p$ and 
the energy of free fermions (quarks). 

In order to see how this model works, the reader can have a look at 
 Fig.\ref{fig_pnjl}: the agreement is very good.
The key feature of the expression above is the 
suppression by the Polyakov loop, which appears as $\Phi$
in the single quark (or antiquark) term in square brackets, as  $\Phi^*$ for diquarks
and finally there is no  $\Phi$ in the last term corresponding to 3 quarks in
colorless combination. The effect of all of that is that quark and diquark terms
are suppressed in the deconfined phase, while 3-quarks are $not$: this
drastically improves the prediction of the original NJL model. 
The reader should note, that unsuppressed colorless 3-quark term
is basically nothing else but a baryon -- thus the main lesson of this work
is actually quite similar to that of Liao and myself \cite{Liao:2005pa}, namely
that the QCD thermodynamics at finite $\mu$, both below and right above $T_c$, is
still dominated by $baryons$ rather than by individual quarks.


\section{ Electric-magnetic duality and finite-$T$ gauge theories} 
Electric/magnetic duality is repeatedly resurfacing during the history of physics,
serving as a source of inspiration for a while and then going dormant again.
I would put the first occurrence of that to famous Maxwell's guess about displacement
current, guided by a ``dual relation" containing time derivative of
magnetic flux discovered by Faraday. From then on, vacuum Maxwell eqns are of course
perfectly E-B dual. 

In the first years of quantum mechanics Dirac famously shown that
the wave function can only be consistently defined if the electric and magnetic
coupling constants are related by the
 celebrated  quantization condition
 \begin{equation} \label{Dirac_quantization}
\alpha(electric)\alpha(magnetic)=n \hspace{1cm} 
\end{equation}
where n is some integer. With advent of
  quantum field theory, renormalization and running couplings, this condition
elevates into a requirement that these two couplings must run in the $opposite$
directions:
\begin{equation} 
 \beta(electric)+\beta(magnetic)=0
\end{equation}
Thus
 when $\alpha(electric)=e^2/4\pi$ is small
(at high T),  $\alpha(magnetic)=g^2/4\pi$ should be strong.
We will discuss implications of that for QGP below.

In 1974 't Hooft and Polyakov have discovered monopole solutions
in gauge theories. The specific setting was Georgi-Glashow model,
which is a gauge theory supplemented by adjoint scalar providing
``Higgsing". Recall that when VEV is diagonalized, all $N_c-1$
diagonal color matrices commute with it, which leave those $U(1)$
``photons" massless\footnote{In SU(3) theory those are 2 -- $\lambda_3,\lambda_8$
-- massless and 6 massive gluons.}. (There is huge literature
about monopoles in this setting and in supersymmetric theories,
 for non-experts the book by Shnir \cite{Shnir:2005xx} can be recommended.) 
 
  G.'t Hooft and Mandelstamm \cite{'tHooft-Mandelstamm} 
famously suggested to explain 
confinement  by a ``dual superconductor'' made of Bose-condensed
 magnetically charged objects. 
 Seiberg and Witten \cite{hep-th/9407087}
have famously shown how it works in the \cal{N}=2 super Yang Mills
theory. They have shown spectacular example of ``triality" between
electric, magnetic and ``dyonic" descriptions near the corresponding
singularities on the moduli space.
Work continues on the supersymmetric front, populating
our ``topological zoo" by new species, for good pedagogical
review see e.g.\cite{Tong:2005un}.
 Let me just mention two recent additions: the non-Abelian
magnetic strings and monopoles which live $inside$ them as ``beads" 
\cite{Shifman:2008wv}
or duality between gauge theory on domain walls
and in the bulk \cite{Shifman:2006zz}.
All these developments are of course the pillars of our understanding
of magnetic objects: which however remains quite incomplete
for the very theory it is needed, the QCD.
One obvious way to find and study dynamical role of magnetic
 objects in various observables is the detailed analysis of
lattice configurations. Such approach was very successful 
for instantons (for  review see  e.g.\cite{Schafer:1996wv}). 
Let me comment on those works
related to monopoles, dyons and fermionic dyons, consequently. 

  Lattice observation of the
monopoles require fixing the gauge, as explained long ago e.g. in DeGrand 
\cite{DeGrand:1980eq},
and then calculating the total magnetic flux through all  elementary 3-d boxes.
Naively, since each link appears twice there should be always zero
flux: but since the phase is counted modulo $2\pi$ 
there are  nonzero fluxes in some boxes. It literally means that 
(in a particular gauge) we look where {\em singular Dirac strings $end$}. The
monopole current can then be defined: monopole boxes form closed loops.
Are those objects physical or just some UV noise in unphysical gauge
choice? The answer to this long standing question in general is still not
completely clear: however better posed specific questions recently produced
so reasonable and consistent answers -- apparently independent
of the particular lattice parameters -- that they should be physical 
(in my opinion). 

In particular, if one counts only the loops which are  winding around the lattice
 {\em at least once}, then
their density was shown to scale correctly with lattice parameters.
As we show in the next section, the correlation function between
those monopoles behave ``physically", in a way consistent with what one expects
from a running magnetic coupling. In vacuum,
they run around confining electric flux tubes, as ``magnetic coil"
should do. Above $T_c$ they show liquid-like correlation
 in quantitative agreement with
MD for classical electric/magnetic plasmas. All of this prompts me optimistically
assume that those objects are physical. (Yet we still don't quite
understand the objects themselves, don't know their masses and structure,
and we even cannot explain why in some other gauge one cannot see
them at all.)

  From the point of view of the definition, another kind of magnetic
  objects --
{\em the self-dual dyons} -- are much easier. Kraan-van Baal solution
\cite{Kraan:1998pm} has  shown that  already at the classical
level, in the presence of nonzero holonomy
or ``Higgsing",
 finite $T$ instantons are decomposed into $N_c$ self-dual dyons.
 
 Because these dyons are descendant of instantons,
they may be seen ``via the eyes of fermions" : and indeed such methods
were developed by Gattringer, Ilgenfritz and others, with stunningly
beautiful examples of such dyons in gauge configurations.
Diakonov  and collaborators have evaluated
quantum one-loop correction to Kraan-van Baal solution \cite{Diakonov:2005qa} 
and discussed the ensemble of dyons \cite{Diakonov:2007nv}.
Yet there are serious reasons to think that dyons are less important
than monopoles: .
(i) Their density is not peaking near Tc; (ii) They seem to be more massive
than monopoles (we will return to this issue in the summary of the chapter); and --
last but not least -- (iii) Bose-condensed dyons would not
create only electric confinement, as needed,
because they are selfdual  and  thus cannot tell $\vec E$ from $\vec B$ field.

Finally, massless fermions have zero modes
on topological objects, including the monopoles/dyons. Supersymmetric theories --
e.g. the celebrated Seiberg-Witten $\cal N$=2 theory --
must have it as a requirement, since all monopoles/dyons should have fermionic superpartners.
I would  think that in QCD quarks should also be able
to ``ride on a monopole": but their masses, density and
dynamical role 
is not yet investigated.

\subsection{A ``magnetic scenario" for near-$T_c$ region}

In the chapter related with applications of hydrodynamics
it was called traditionally ``the mixed phase", as the model EoS
used still had a phase order transition and Maxwell construction
at intermediate (entropy) density. This was done along the traditional
thinking, although we know for fact that none of the overheating-overcooling
 fluctuation phenomena (known  near any first order
 transition) are actually observed in heavy ion collisions. The so called
 event-by-event fluctuations there turn out to be remarkably small
 and simple. So, this is definitely $not$ a QGP and hadronic phase mixed
 together in some kind of emulsion.

The question is, what is the adequate picture for the near-$T_c$ region?
 Liao and myself \cite{Liao:2006ry} proposed a new and very
 simple view on this question --
 ``the magnetic scenario" -- which basically suggests it to be a magnetic
plasma of monopoles, in a liquid form.  Another line of work based on lattice 
data lead Chernodub and Zakharov  \cite{Chernodub:2006gu} to the same conclusion.

\begin{figure}[t]
  \includegraphics[width=0.48\textwidth]{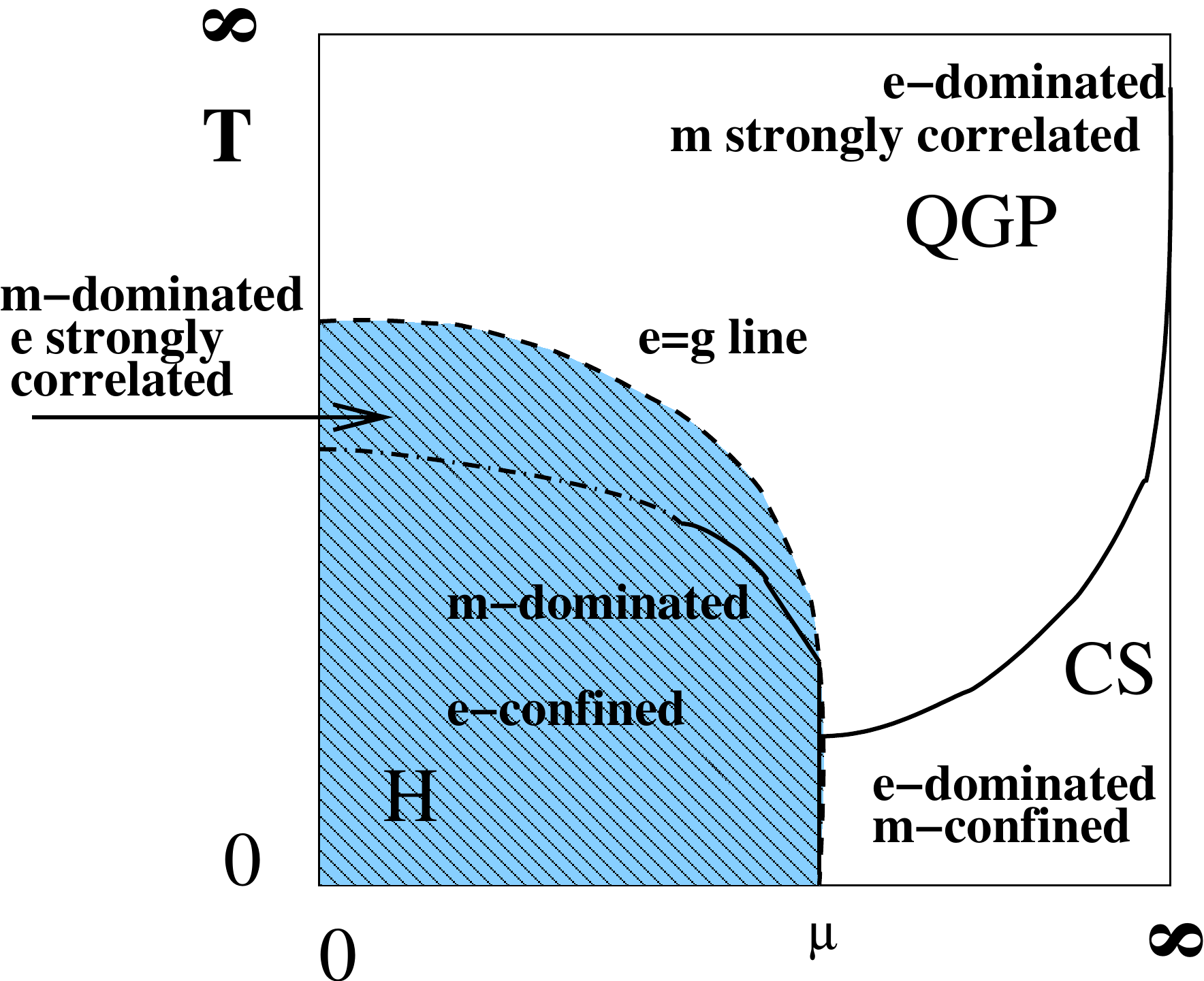}
 \includegraphics[width=0.4\textwidth]{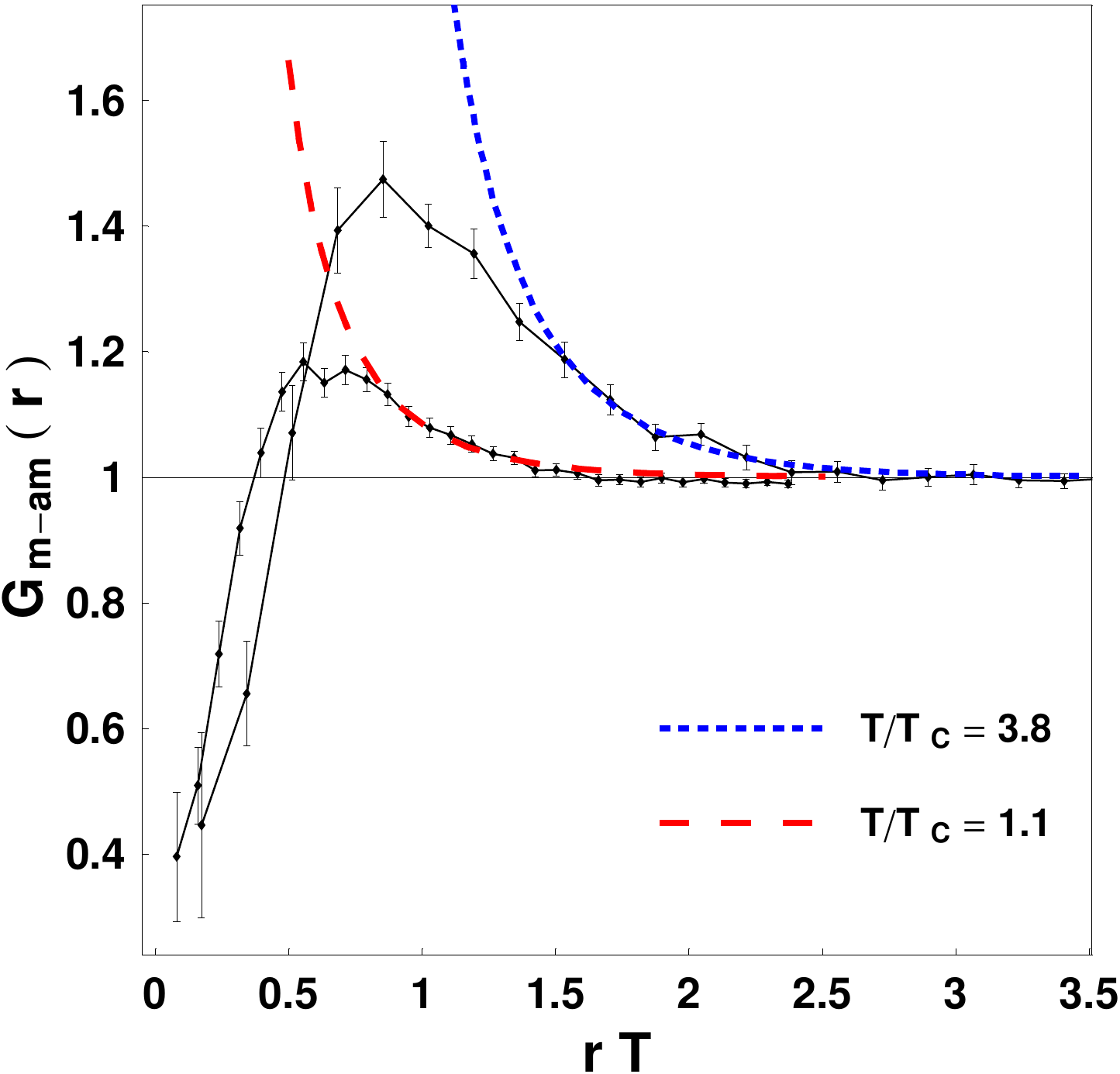}\\
  \includegraphics[width=0.48\textwidth]{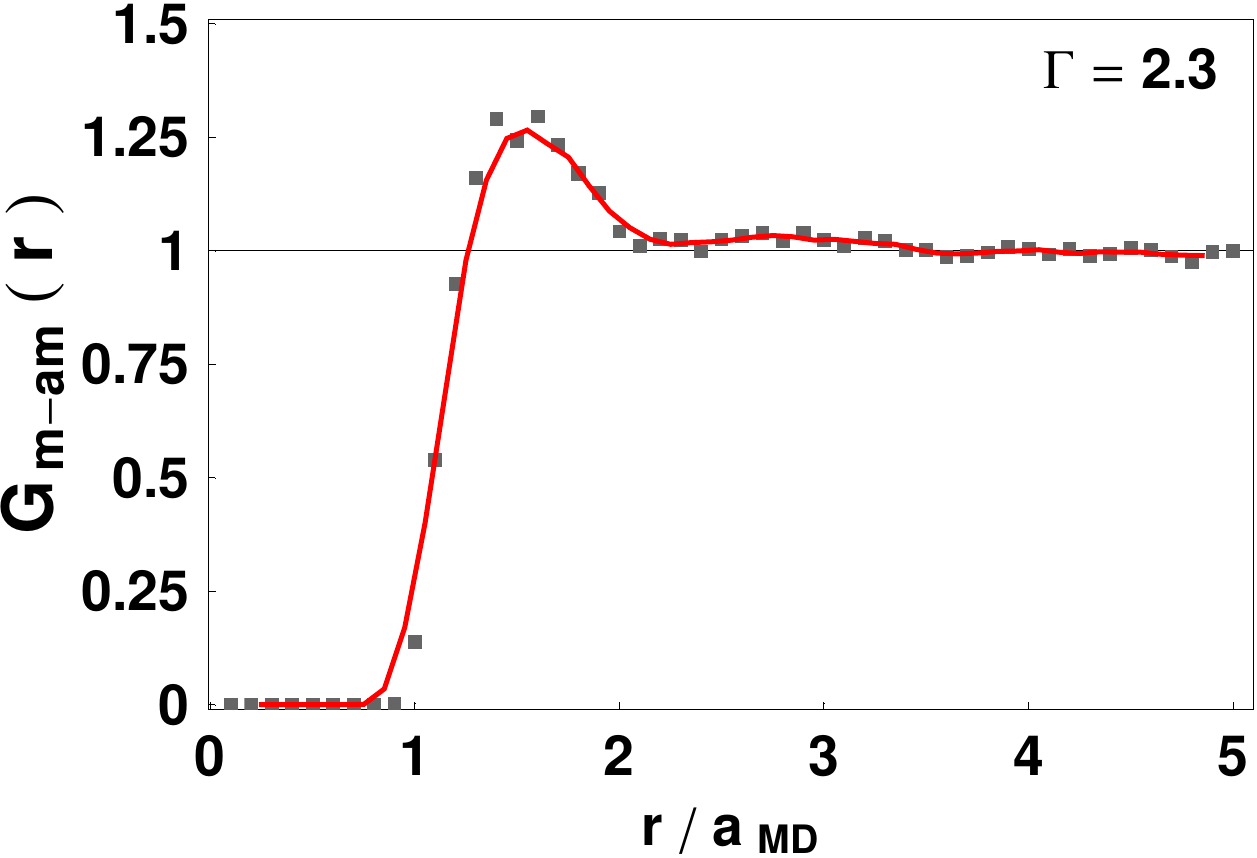}
 \includegraphics[width=0.45\textwidth]{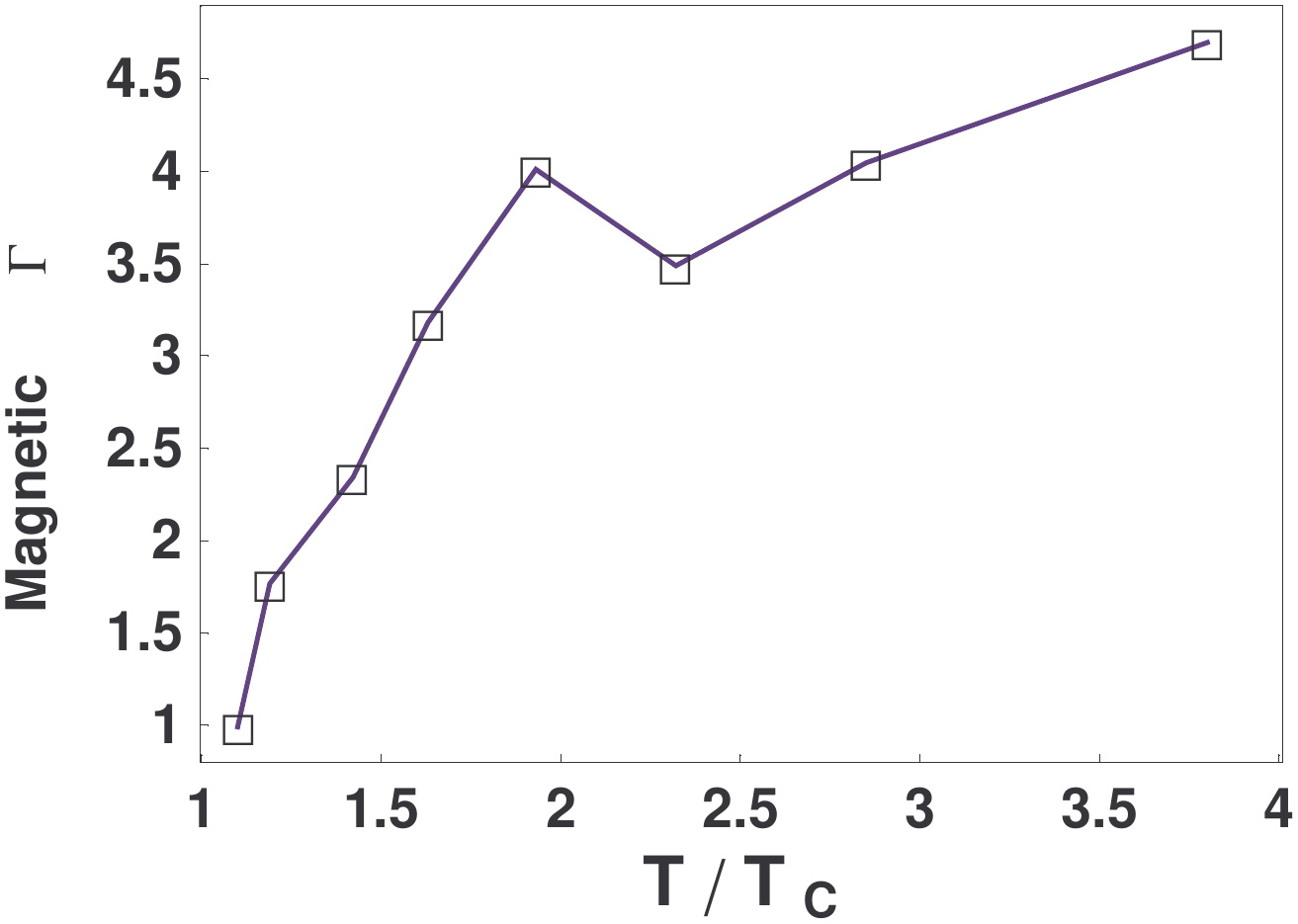}
 \vspace{0.1in}
\caption{\label{fig_em_phasediag} 
} 
 \end{figure}

Let me start with a qualitative discussion of the {\em running couplings}
on the phase diagram, which
is a crucial element of the argument.
As QGP produced is cooling down and 
$T$ decreases toward $T_c$, the electric coupling grows and 
the magnetic one
decreases, in accordance with asymptotic freedom and Dirac condition
(\ref{Dirac_quantization}). The electric objects -- quarks -- gets heavier
and more strongly coupled, while 
 the monopoles get lighter and less correlated: thus one should expect
 an equilibrium at some $T$, with dual language based
 on monopoles being simpler. A
 schematic phase diagram explaining these ideas
is shown in Fig.\ref{fig_em_phasediag}(top left). The main assumption
is that plasma becomes magnetic-dominated at $T$ $above$
the deconfinement line, interpreted as a BEC transition of magnetic
media.
 
In the beginning of this chapter we already defined the near-$T_c$ region
as $T/T_c=0.8-1.2$, based on the width of various ``specific heat" peaks
in static dipole energy.
Recent lattice data \cite{D'Alessandro:2007su} provided dramatic
 conformation of this scenario\footnote{
Even more recent lattice work by Meyer \cite{Meyer:2008dt}
have found liquid-like peaks in the correlators
of the energy as well. 
}.  Fig.\ref{fig_em_phasediag}(b)
shows two sets of these data, and the correlation 
(and thus magnetic coupling) is indeed stronger at $higher$ $T$.
Furthermore, the correlation function for 50-50 mix of
 electric/magnetic plasma obtained in our Molecular Dynamics (MD)
simulation  Fig.\ref{fig_em_phasediag}(c)
has the same shape and magnitude, provided one compare
at the same value of the magnetic plasma parameter
 $\Gamma \equiv \alpha(magnetic)  /  (\frac{3}{4\pi n})^{1/3}/{T}$:
its extracted values are shown in
 Fig.\ref{fig_em_phasediag}(d). It is very nice to find 
always $\Gamma>1$, which means that magnetic component of sQGP is also
liquid not gas, thus it does not spoil the ``perfect
liquid'' at RHIC. One may 
further think that viscosity has a minimum where both
electric quasiparticles (quarks) and magnetically ones
(monopoles) have similar difficulty propagating. We infer from
lattice data  that such {\em electric-magnetic equilibrium}
is at
$T\approx 1.5T_c$, right in middle of the RHIC domain.

We already reviewed the main properties of lattice potentials in the previous chapter.
A good starting point for development
of such a model is to think along the line of ``dual superconductor''
by t'Hooft-Mandelstamm \cite{'tHooft-Mandelstamm}.
If the
QCD vacuum at T=0 contains some magnetically charged
condensate, it will
expel electric flux between $\bar Q Q$ into a flux tube,
analogous to 
the Abrikosov-Nielsen-Olesen (ANO) vertex.

An explanation of the linear rise of
energy and entropy to very large values is due to persistence
of a flux tubes even in the plasma deconfined phase.
 This point of view was
 proposed in \cite{Liao:2006ry} and was further
developed in \cite{Liao:2007mj}, for  infinitely long flux tubes. 
In the latest paper \cite{Liao:2008vj} we extended the model to finite-distance
potentials, by
developing an analytic ``elliptic flux-bag'' model, with two static quarks at its
focal points. Omitting all technical details, we will only present two main
results for the tensions of ``slow" and ``fast" potentials. In the former case
we obtained
\begin{equation} \label{tension_slow}
\sqrt{\sigma_S}=1.69 \times \alpha_E^{1/4} \times {\cal P}_M^{1/4}
\end{equation}
and the saturated value of transverse radius to be
\begin{equation} \label{size_slow}
{\cal R}_S=0.82 \times \alpha_E^{1/4} \times {\cal P}_M^{-1/4}
\end{equation}
For ``fast" potential, related to ``normal" component of magnetic plasma,
the corresponding results are
\begin{equation} \label{tension_fast}
\sqrt{\sigma_F}=2.94 \times \alpha_E^{1/6} \times n_M^{1/3}
\end{equation}
and the saturated value of transverse radius to be
\begin{equation} \label{size_fast}
{\cal R}_F=0.57 \times \alpha_E^{1/3} \times n_M^{-1/3}
\end{equation}


\begin{figure}[h]
\begin{center}
\resizebox*{!}{6.cm}{\includegraphics{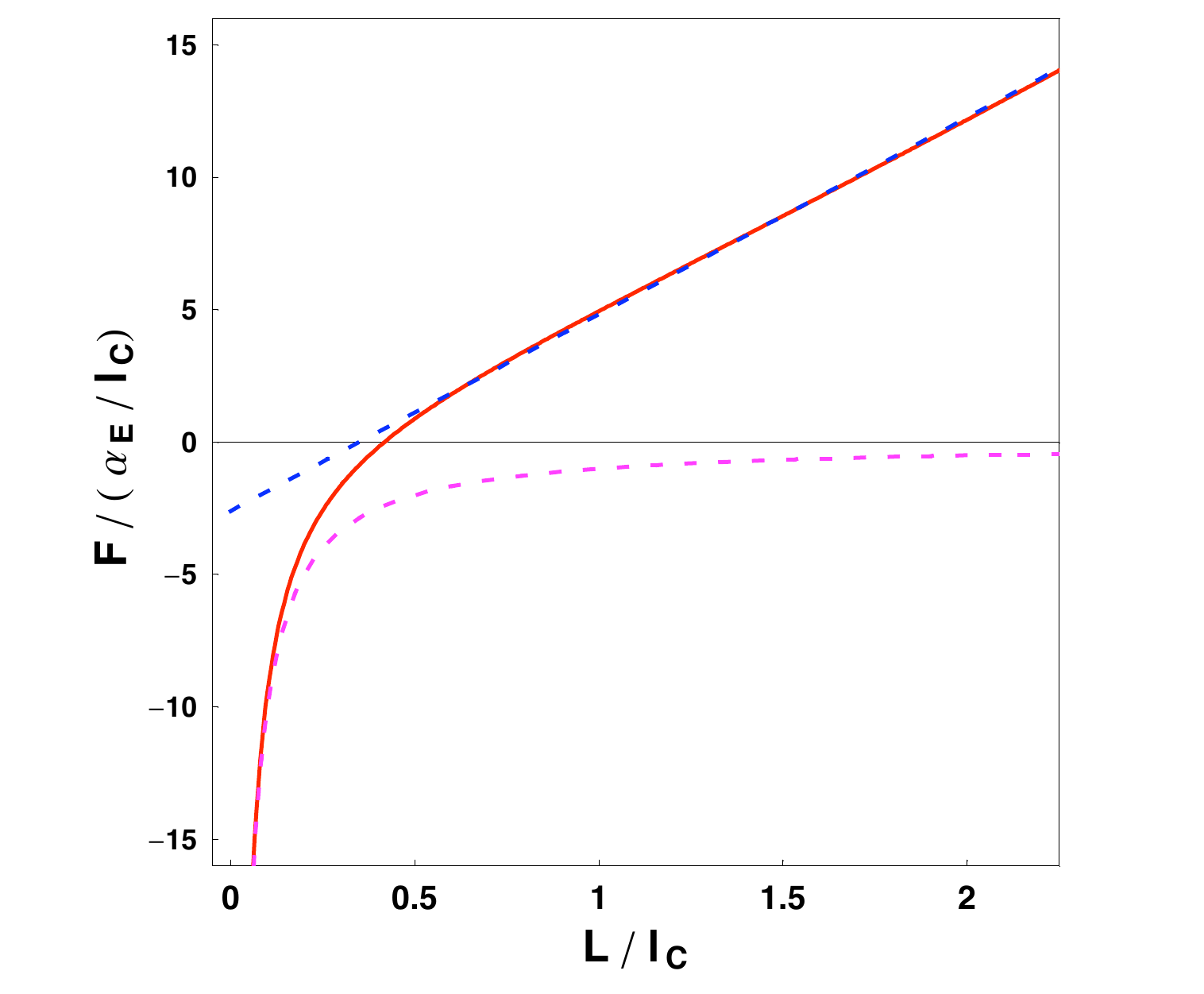}}\hglue 10mm
\resizebox*{!}{6.cm}{\includegraphics{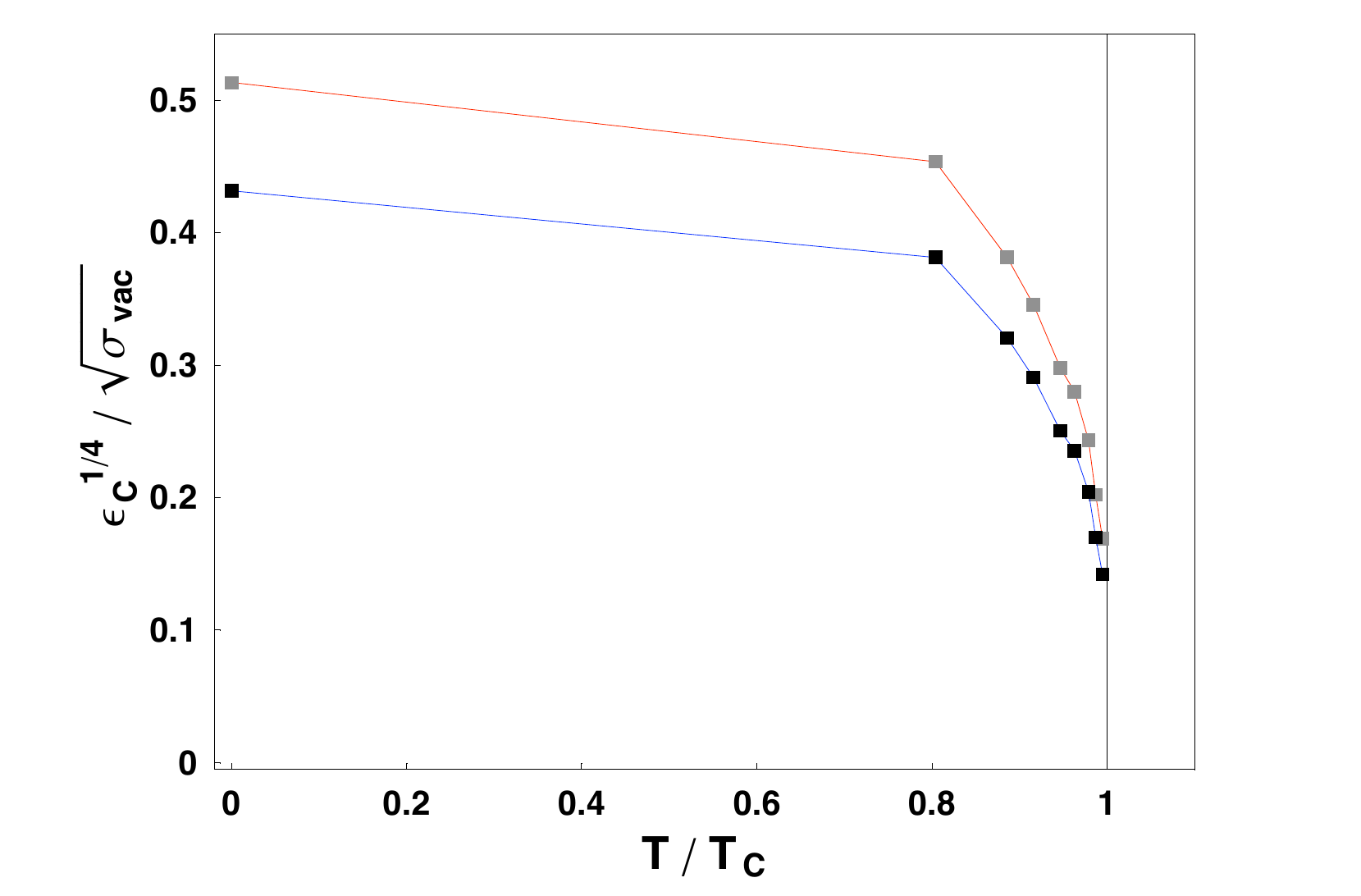}}\\
\resizebox*{!}{6.cm}{\includegraphics{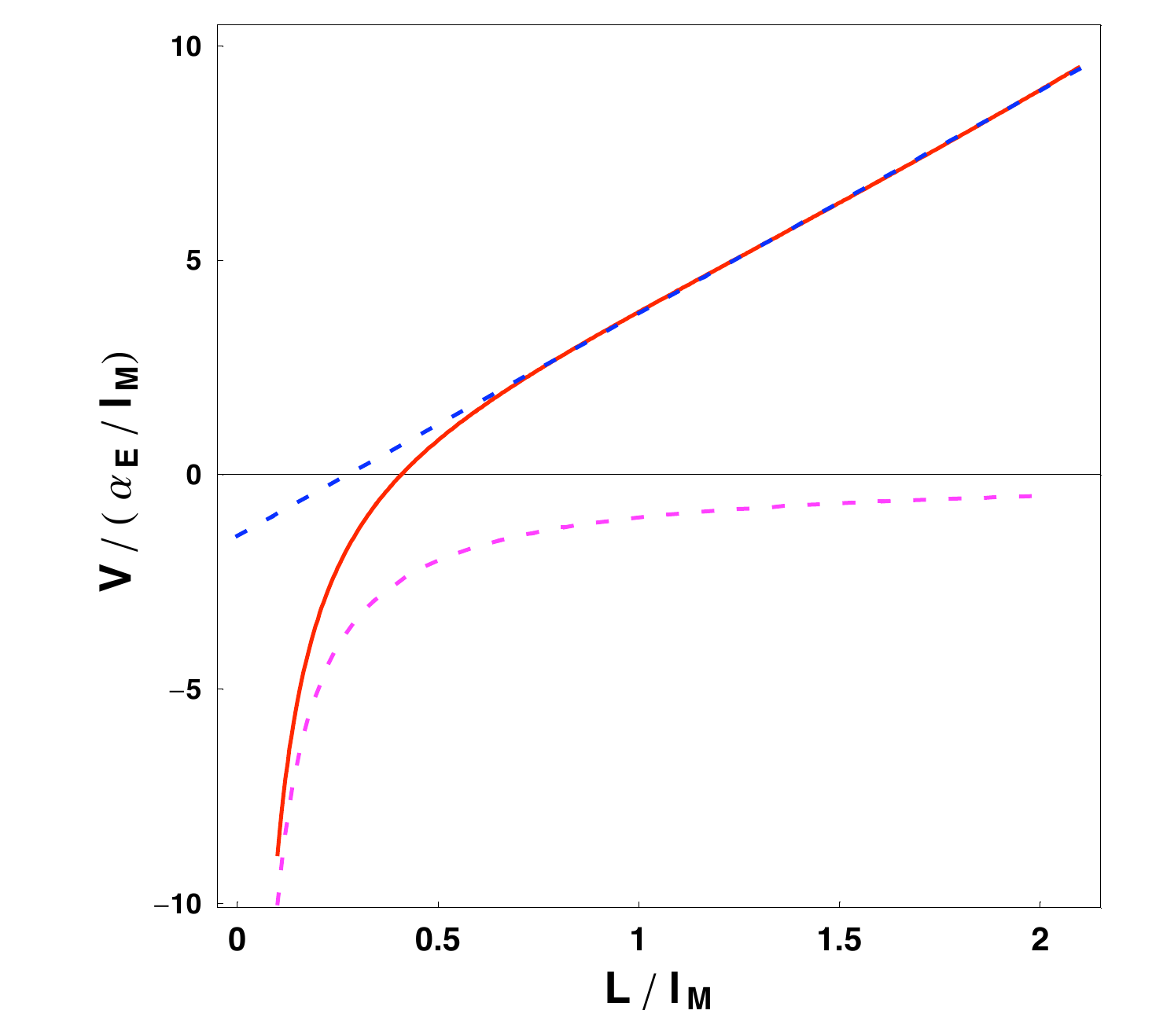}}\hglue 10mm
\resizebox*{!}{6.cm}{\includegraphics{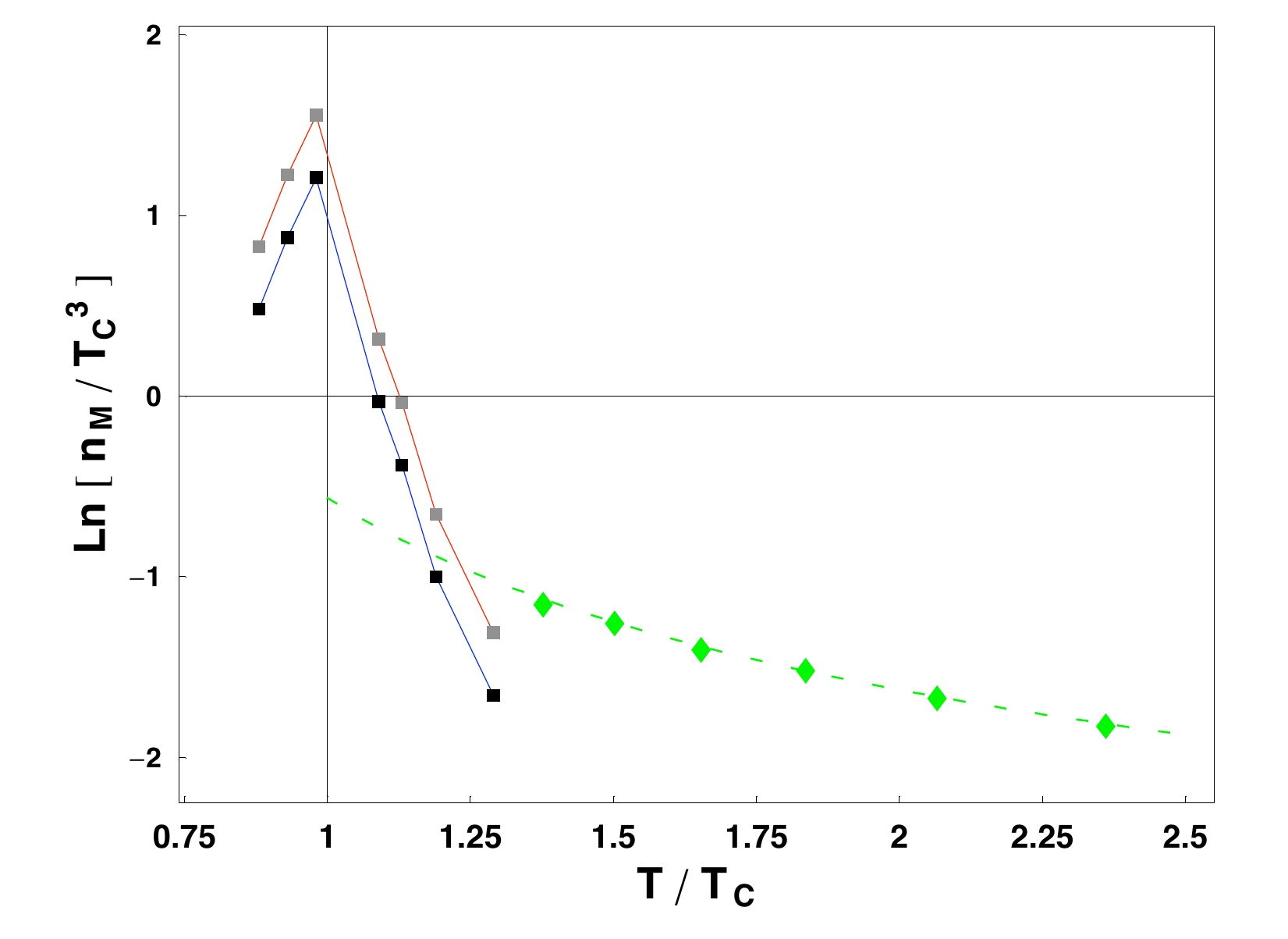}}
\end{center}
\caption{\label{fig_free_energy} (top left) free energy $F$ (in
unit of $\alpha_E/l_C$) versus separation $L/l_C$; (top right)
monopole condensate energy density ${\cal E}^{1/4}$ in unit of
$\sqrt{\sigma_{vac}}$, the two curves are for $\alpha_E$ being
0.5(upper) and 1(lower) respectively.\newline
(down left) The potential energy
$V$ (in unit of $\alpha_E/l_M$) versus separation $L/l_M$;
(down right) thermal monopole density $n_M/T^3$, the two curves
across boxes are for $\alpha_E$ being 0.5(upper) and 1(lower)
respectively, and green curve across diamonds shows data for
$T>1.3T_c$ from D'Elia and D'Alessandro.}
\end{figure}

Using these expressions and lattice potentials/tensions one can
 estimate the monopole density needed to
produce the energy string tension at $T_c$, $\sigma_E(T_c)\approx
5.15\, \sigma_{vac}$. From (\ref{tension_fast}) we obtain the
monopole density should be $n_M(T_c)\approx 4.65 fm^{-3}$ assuming
$\alpha_E= 1$, or $n_M(T_c)\approx 6.58 fm^{-3}$ assuming
$\alpha_E= 0.5$. This again is a reasonable number as compared
to direct lattice observations . The
transverse size is then ${\cal R}_F\approx 0.34 fm$ for $\alpha_E=
1$ and ${\cal R}_F\approx 0.24 fm$ for $\alpha_E= 0.5$. The results are
shown in Fig.\ref{fig_free_energy}, for Bose condensed component 
(upper) and ``normal" one (lower), the latter compared to 
lattice data on directly observed monopole paths (winding ones around
time axis). We will return to these numbers at the summary
of this chapter below.

\subsection{Molecular dynamics for magnetic/electric plasmas}
  
   Gelman, Zahed and myself \cite{Gelman:2006xw}
 proposed a classical model for the description of strongly interacting
colored quasiparticles  as a 
nonrelativistic Non-Abelian Coulomb gas. The sign and strength
of the inter-particle interactions are fixed by the scalar product
of their classical {\it color vectors} subject to Wong's equations.
Details should be looked at the papers: let us just explain here
its physical meaning. For SU(2) color group a color vector rotates
around the direction of the total color field induced
by all other particles at its position: same as magnetic moments
would do in a magnetic field. For arbitrary group precession 
on a group is determined by the Poisson brackets of color vectors:
$ \{ Q^a, Q^b\}= f^{abc}\,Q^c$
which are classical analogue of the SU(N$_c$) color commutators.
Thus for arbitrary gauge group one should  use its structure
constants  $ f^{abc}$ describing ``precession" of the color vector.
 
 For  the non-Abelian group SU(2) the  adjoint color vectors 
 resign on  a 3-d unit sphere: with one conserved quantity
($(Q^a)^2$) it makes $S^2$ or 2 degrees of freedom. For SU(3) the group has
8 dimensions: with two conserved combinations
$(Q^a)^2, d^{abc}Q^aQ^bQ^c$ it is 6 d.o.f., and so on.
Although color precession equations do not look like the usual
canonical EoM for pairs of coordinates and momenta, they actually
can be rewritten as pairs of conjugated variables, as can be shown via
 the so called Darboux parameterization. Thus one can even define
the phase space and use all pertinent theorems related to its classical
evolution, if needed.

\begin{figure}[t]
\begin{center}
    \includegraphics[width=7cm]{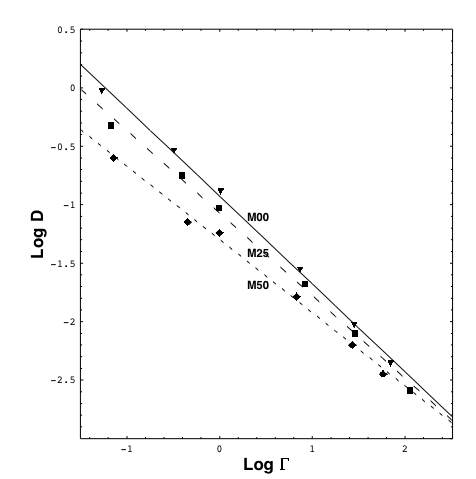}
\hspace{.2cm}
    \includegraphics[width=7cm]{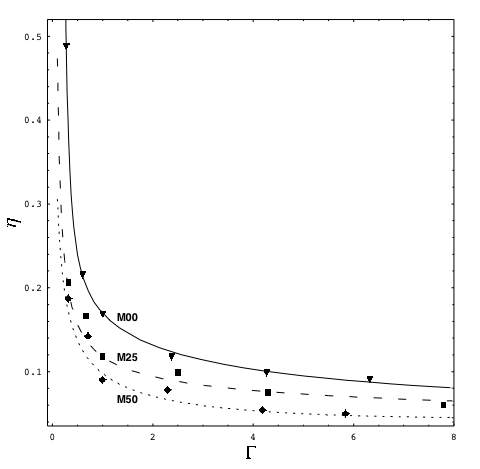}
\caption[a]{The diffusion constant (a) and shear viscosity (b)
  as a function of
the dimensionless coupling $\Gamma$.
Three sets of data are for only electric particles (M00), with the quoter (M25) and
the half (M50) particles being monopoles. Note the decrease in viscosity as the admixture of monopoles grows.}
\label{eos-shuryak-fig_diffusion}
\end{center}
\end{figure}

The model can be studied using Molecular Dynamics (MD),
which means solving numerically EoM for $10^2-10^3$ particles.
It  displays strong correlations and
various phases, as the Coulomb coupling is
increased ranging from a gas, to a liquid, to a crystal with
anti-ferromagnetic-like color ordering. There is no place for details
here: so we simply jump to results on transport properties.
In Fig.\ref{eos-shuryak-fig_diffusion} one can see the result for
 diffusion and viscosity vs coupling:
note how different and nontrivial they are. 
When extrapolated to the sQGP
suggest that the phase is liquid-like, with a diffusion constant 
$D\approx 0.1/T$ and a bulk viscosity to entropy density ratio 
$\eta/s\approx 1/3$.


 Transport properties for novel types of plasmas, including
electric and magnetic charges, have been calculated
by Liao and myself
 \cite{Liao:2006ry}: and $\eta$ is indeed minimal for most symmetric
 mixture  50-50\%. Before we turn to these results, let me 
qualitatively explain
 why in this case the diffusion/viscosity is maximally
reduced. Imagine
one of the particles - e.g. a quark. The Lorentz force makes it
rotate around a magnetic field line, which  brings it toward
one of the nearest monopoles. 
Bouncing from it, quark will go along the line to an antimonopole,
and then bounce back again: like electrons/ions do in the so called
``magnetic bottle''~\footnote{By the way, invented in 1950's by one of my teachers
G.Budker.}. Thus in 50-50 mixture all
 particles can be trapped between their dual neighbors, 
so that the medium can only
expand/flow collectively.

\begin{figure}[th]
 \hspace{4cm}   \includegraphics[width=8.cm]{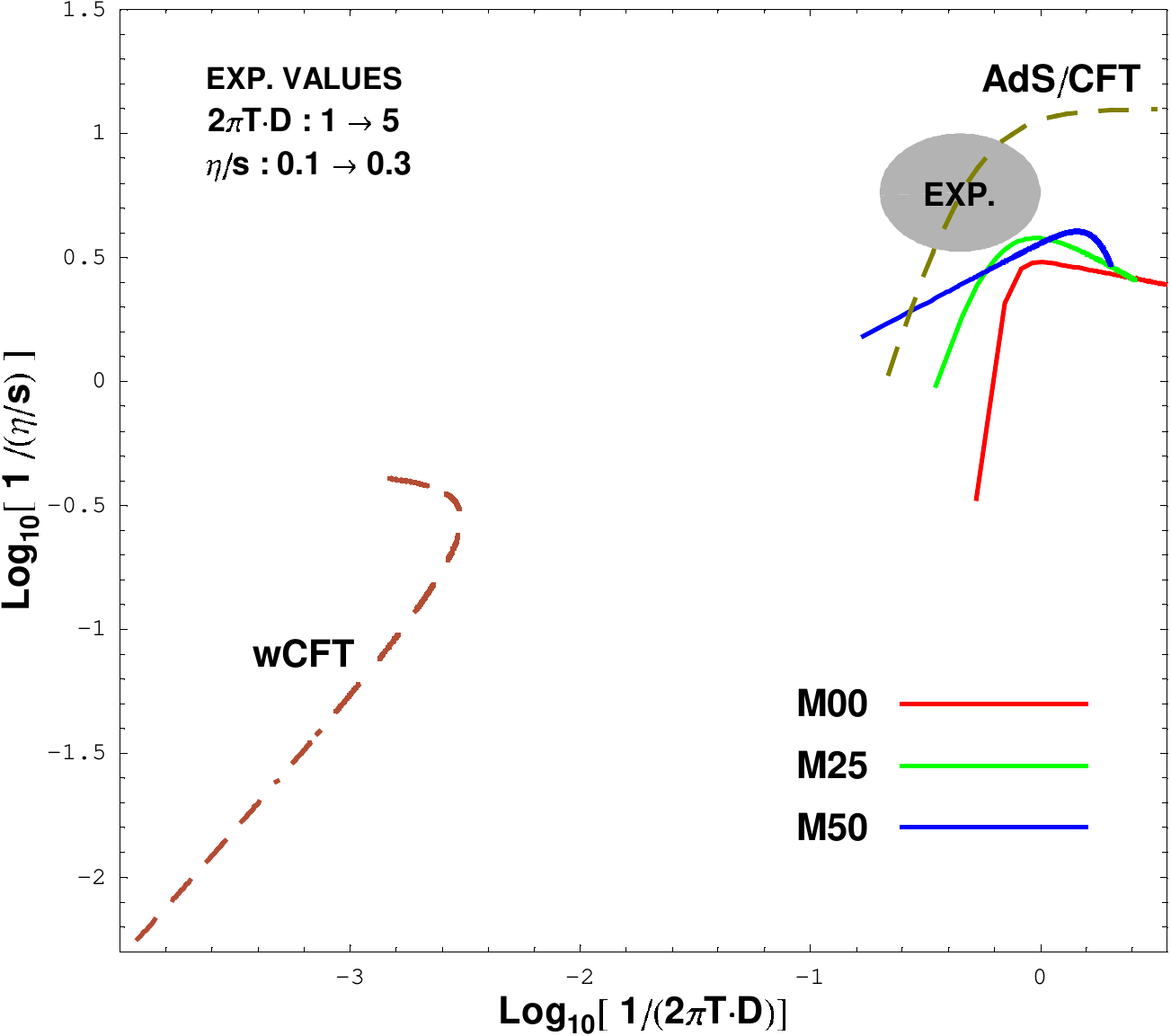}
 \caption{Transport summary from \protect\cite{Liao:2006ry}:
 $Log[1/(\eta /s)]$ v.s. $Log[1/(2\pi T D)]$ including
results from our MD simulations, the AdS/CFT calculations, the weakly coupled
CFT calculations, as compared with experimental values, see text. }
\label{fig_vis_d_mapping}
\end{figure}

Our MD results  are shown on viscosity-diffusion plane in
Fig.\ref{fig_vis_d_mapping} by three lines: they are
compared to those from the AdS/CFT correspondence
in weak and strong coupling as well as 
with empirical values
 from RHIC experiments ( gray oval).
The dashed curve in the left lower corner is for $\cal N$=4 SUSY
YM theory in weak coupling:
, where viscosity for this theory in weak coupling is from
\cite{Huot:2006ys} and diffusion constant
from\cite{Chesler:2006gr}. The curve has a slope of one on this
plot, as in weak coupling
 both quantities are proportional to the
same mean free path. These weak coupling results
are quite far from empirical data from RHIC
in the right upper corner. (Viscosity estimates follow from
deviations of the elliptic flow at large $p_t$ from hydro
predictions 
and diffusion constants are
estimated from $R_{AA}$ and elliptic flow of charm
.)
The  strong-coupling AdS/CFT results (viscosity according
to \cite{Policastro:2001yc} with $O(\lambda^{-3/2})$ correction, 
diffusion
constant from \cite{Casalderrey-Solana:2006rq})
are represented  by the upper dashed
line,  going right through the empirical
region. Our MD results -- three solid lines on the right --  
 are close to the
experiment as well, especially the 
version with the equal mixture of
electric and magnetic particles.

\subsection{Bose condensation: from liquid $He$ to monopole plasma }

Classical approximation discussed above can explain many properties
of liquids and solids, but it obviously ignores quantum
effects\footnote{Some of quantum effects can be put into effective potential, see
e.g. recent dedicated study related to charm quarks in sQGP \cite{Dusling:2007cn}.}.
It is well known that quantum effect at low temperatures  may lead to
 qualitative changes in the system's behavior, such
as superfluidity and superconductivity. A system of bosons may undergo
 Bose-Einstein condensation\footnote{
In fact ``Bose-Einstein" statistics is due  entirely to Bose while condensation
entirely to Einstein.}
 (BEC). While  for ideal gas BEC is a textbook material ,
 it is a difficult problem for interacting systems.
 Liquid $He^4$ remained for a long time the only example   and 
the relation between its superfluidity and BEC phenomenon
 was hotly debated and even denied in many classic works.
 The
 dilute atomic gases were finally cooled to BEC in 1990's, but 
 those are weakly coupled and we would not discuss them.
 
  Liquid $He^4$ problem became amenable to 
 direct numerical attack in 1980's, when simulation of the
 Feynman path integral via Monte Carlo algorithms
 was finally technically possible. Such first-principle
 approach  was a success: see e.g. Ceperley's review \cite{Ceperley}. After that, other systems were studied, filling the gap between weakly
 coupled gases and  liquid $He^4$. And yet  when we studied this literature 
 -- trying to understand conditions for BEC of the monopoles --it still
 looked that the net was not cast wide enough and many general questions
remained open. To name one particular example, such question of principle
 -- whether $solid$ $He^4$ is a $supersolid$ at $T=0$, and if not why and
whether a bit different atomic potentials may still produce it.

 One of the first applications of the path integral method 
  by Feynman himself were aimed at explanation of
 the  $He^4$ $\lambda$-point. His 
    classical papers \cite{fey1,fey2} introduced the idea
    of  {\em ``polygon clusters"}. Starting from some configuration
    of particles at Euclidean time $\tau=0$, Feynman identifies 
    such clusters as a group of atoms which exchange places
    during the Matsubara time $\tau=\beta=\hbar/T$.
 The polygons of course in principle may have any shape,
 but since a minimal additional
 action for ``jumping" atoms be required,  the most probable paths are those where particles interchange places 
 with their nearest-neighbors and thus the most probable jumps
 have length of one interparticle distance. And yet, the polygons themselves
 need not be small: as Feynman shown  BEC implies
 that there should be {\em infinitely large} cluster. Furthermore, the
 BEC transition temperature can be defined as  such $T$ at which
the series over infinitely long polygons start to be divergent. 
And, since the number of polygons
grow with its length in a certain geometrical way, one may argue
that there should be {\em universal action} per ``jumping" particle $\Delta S^*$
corresponding to all BEC transitions.

\begin{figure}[!ht]
\centering	\includegraphics[height=8cm]{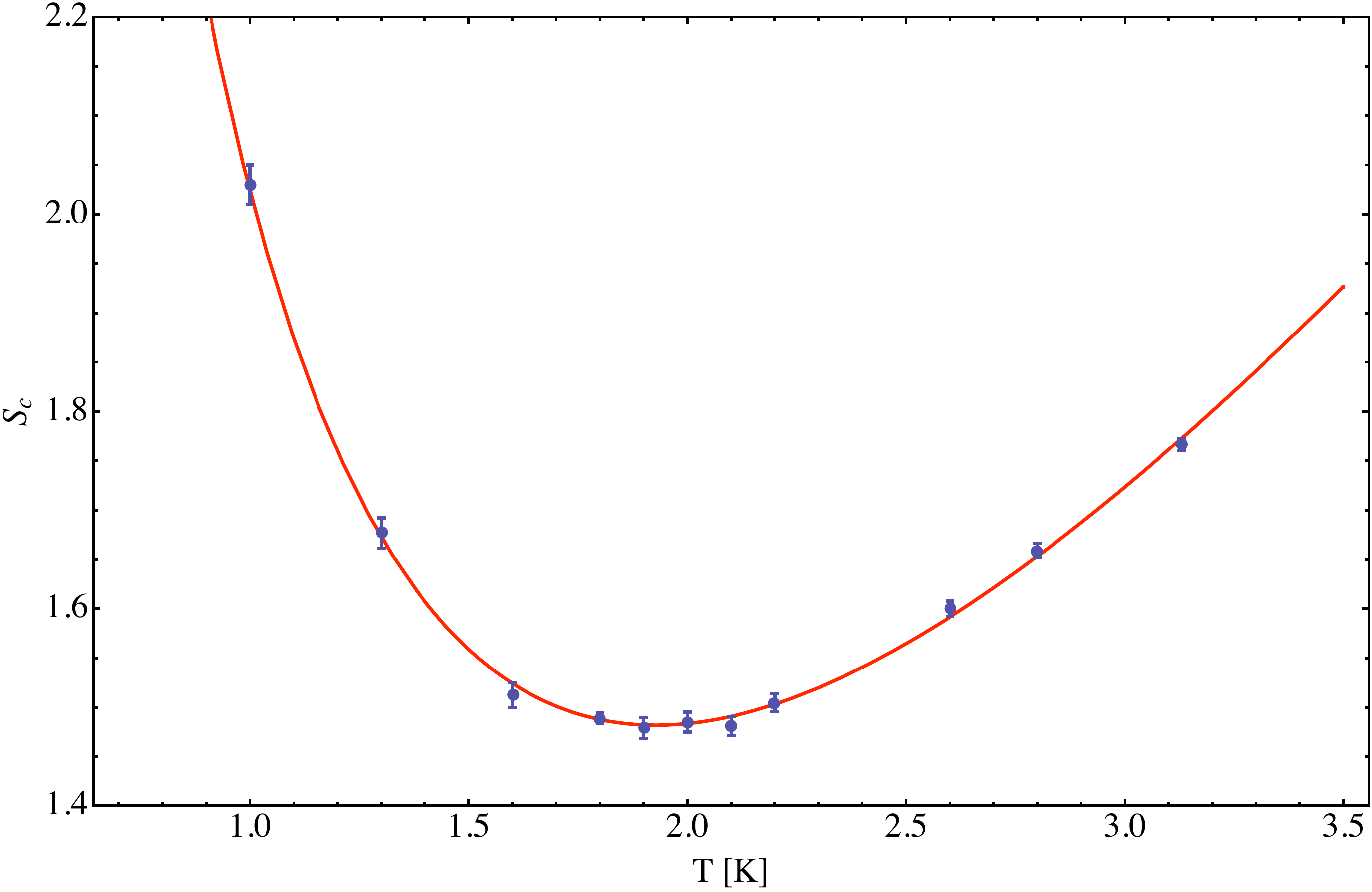}
\vspace{0.5cm}
\caption[h]{Total extra action for ``jumping" He4 atoms   as a function of the temperature. 
}\label{fig:Tc}
\end{figure}

Feynman's main approximation was  the estimate of $\Delta S^*$
using $kinetic$ energy  only, ignoring the potential one. He motivated
by ``rapid particle rearrangements" and wrote the thermal partition function in the form
\be
	Z=e^{-\frac{F}{T}}\\\nonumber
	=\frac{1}{N}\sum_P\int\Big(\frac{m^*T}{2\pi\hbar}\Big)^{3N/2}\exp\Big[-\frac{m^*T}{\hbar^2}\sum_i(\underline{R}_i-P\underline{R}_i)^2\Big]\\\nonumber
	\rho(\underline{R}_1...\underline{R}_N)\textrm{d}^3\underline{R}_1...\textrm{d}^3\underline{R}_N
\ee
where exponent comes from ``jumps" done in a straight lines and 
containing corresponding action related to kinetic energy. To correct
for potential energy somehow, Feynman introduced
an ``effective mass of a He atom" $m^*$. The sum is done over permutation $P$ of the particle coordinates, the function $\rho$ includes effects of the interparticle interactions. His main idea was that the function $\rho$ can be inferred from general properties of the liquid (or solid), its quasi-ordered local structure with peaked distribution over interparticle distances at some nearest-neighbor value $d$ (in the case of cubic lattice $d$ is the lattice spacing). 

The relative amplitude of a term with permutation of $n$ atoms is thus proportional to the $n$-th power of the "jump amplitude" 
\be\label{eq:yf}
	y_F=exp(-\Delta S)=\exp\Big[-\frac{m^*Td^2}{2\hbar^2}\Big]
\ee
which should at the transition point be exactly compensated the
divergence of the combinatorial prefactors, describing a number of corresponding to non-crossing polygons on the corresponding $3$d lattices. Feynman estimated that he expected the critical value  to be in the range $y_c=1/4$ -- $1/3$.
 Kikuchi et al.~\cite{kik} have argued that the critical action should be $S_c\approx 1.9$,  but later studies in the eighties
 such as by Elser~\cite{elser} determined numerically that the critical action is smaller  $S_c\approx 1.44$. 
 
 Let us see what critical temperature for He would follow from those actions,
 in Feynman approximation.
  We  consider the distance $d$ of a "jump" being fixed to the position of the nearest neighbor maximum in the static correlation function $g(r)$ for liquid He, which is $d\approx 0.35 \, nm$ . Kikuchi's critical action with the unmodified He mass
leads to  $T_c=3.57K$ while Elser's to   $T_c=2.72K$, still well above
 the correct position of the $\lambda$-point $T_c=2.17K$. 
 
  Cristoforetti and myself \cite{Cristoforetti_08} have tried to generalize Feynman criterion for interacting
gases by including the $potential$ energy effect 
``jumps" and their  actions. We have used two approaches,
(i) a
 $semiclassical$ one,  in which the path is found by minimizing classical
 equation of motion and the action given by this ``instanton"; and (ii)
 numerical one, simulating path integral for a ``jumping" particle in the
 potential created by ``non-jumping" particles in the background.  
 
 The setting for the calculation can be described as follows:
imagine Helium atoms filling the best-packing lattice and consider all atoms to
be non-jumping except  of  atoms in one raw of the lattice which
are all ``jumping" coherently by one step in the same direction.
This is a good approximation because one should consider  a very large polygon with infinite number of atoms. We found that many different systems,
from He atoms to Coulomb systems\footnote{Let me remind the reader that below
we hope to use this criterion of confinement for monopole plasma.}
have effective potentials for jump which can be quite accurately
be represented by universal periodic potential  with
only one harmonics $V\sim cos(\pi x/a)$.
It is simple to find
the instanton path for this potential and its action.
It improves substantially Feynman estimate, but 
since  the value of the critical action ($\sim 1.5$) is not really large,
semiclassical estimates cannot be very accurate here and have mostly pedagogical value.


In fact it is not difficult to perform 
 numerical evaluation of the ratio of two path integrals, for ``jumping" to ``periodic" paths
 can be performed numerically: in this way also quantum fluctuation around the classical trajectory are taken into account. These calculations are very cheap in terms of computational power because we have a system with only one particle.
We have done so for different values of the temperature, the result is presented in Fig.~\ref{fig:Tc}.  The value of the jump action at $T_c=2.17K$ is $S_c=1.48$,
which is very close to the value $S_c\approx1.44$ computed in~\cite{elser} using the combinatorial method\footnote{The figure clearly shown a minimum at  $T=1.94$ K
and then grows again
but presumably the Feynman argument cannot be used at $T<T_c$.}. 

Solid He is a bit denser than liquid, and thus its action never
reaches $S_c\approx1.44$: thus --
from the viewpoint of this criterion -- it seems  never become a ``supersolid".
(The experimental situation is complicated with conflicting experiments
about rotating solid He at very low $T$.)

Let us now return to QCD near-$T_c$ region.
 Before we return to our monopole, let me mentioned one school of thought 
(from \cite{DelDebbio:1996mh} to \cite{Gorsky:2007bi})
 who think differently about confinement, putting
 2d objects -- $vortices$
 --
  as a primary objects randomizing the vacuum fields and leading
 to the area law for the Wilson loop. If so, the vacuum should have
 a condensate of vortices, which are after all also bosonic objects. However
 to my knowledge neither the order parameter -- resembling Pisa one
for monopoles -- have been constructed, nor there is any understanding
how one can address the issue of BEC for such objects, whose worldvolumes are
2-dimensional.

The monopoles, on the other hand, are particles, and 
 the Feynman criterion
( $S=S^*$ at $T=T_c$)  discussed above should be applicable.
Let us use it  {\em a la} Feynman, using only kinetic energy first.
The motion of the monopole which tries to occupy the position
of another identical monopole at distance $a$ away 
during the Matsubara time $\beta=\hbar/T=1/(.27\, GeV)=.73
\,
fm$ is relativistic, with
the 
(Euclidean) velocity on the straight path 
 $v=a/\beta$. Since we speak about tunneling,
having imaginary action and velocity $>1$
is in fact quite appropriate; no negative roots should appear. The Euclidean
relativistic action is just the length of the path 
\be S_E=iS_M=im\int ds= m \beta \sqrt{1+a^2/\beta^2} \ee 
should be $S^*\approx 1.44$ at $T=T_c$. Since we know the monopole density
(and thus $a$) and $T_c$, we can use this criterion to estimate the 
effective monopole mass: this sets it in the range $m=(0.2-0.3) GeV$.
Putting Coulombic repulsion and monopoles into appropriate
cubic lattice, one can get sinusoidal
potential for jumps and evaluate semiclassically or numerically
the action for a ``jump" more accurately.

Since in principle lattice data on the whole path integrals 
of monopoles is known, in the future one should be able to learn 
the effective monopole masses and interactions in the near-$T_c$
region and verify the details of their Bose-Einstein condensation.

\subsection{Summary of the magnetic scenario in the near-$T_c$ QGP}
The main conclusion of this section is that lattice monopole-like
singularities --- located via
endpoints of the Dirac strings in certain gauges such as maximal abelian gauge -- are found
to behave as physical particles. Their density is independent of the
lattice details. The extracted magnetic coupling $grows$ with $T$, exactly as 
expected. Their spatial correlation functions agree with the idea that their ensemble
is just a
strongly coupled liquid driven essentially by the magnetic Coulomb
interaction: they  are quantitatively reproduced  by
simple classical MD. In addition, the transport properties of such a magnetic
plasma is in agreement with heavy ion phenomenology. 

But we don't yet know many important parameters, such as e.g.
the masses of all quasiparticles involved.  Current
understanding of those masses is vague, the only direct lattice measurements
have been done at $T=1.5T_c$ and $3T_c$
for quarks and gluons in \cite{Petreczky:2001yp},
both masses are about equal and $M \approx .6-.8 GeV$ at both $T$.
Let us assume that they are the same at $T_c$ as well.

We now proceed to self-dual dyons, a components of finite-T instantons.
The mass is derived from the instanton action
\be M_{dyons}=T({8\pi^2 \over g^2 N_c})\sim 4T_c\sim 1 \,GeV\ee
where we used a typical instanton action to be around 12.
Their density (for all $N_c$ types together) can also be estimated from lattice data on instantons
 \be n_{dyons}\sim {n_{inst}\over T N_c}\sim 2 \, fm^3\ee
So, although the dyon density is not negligible, it 
is not peaking near $T_c$ and is few times smaller than that of monopoles. 

The
monopole density numbers from Figures above does peak at $T_c$, indicating
that they do play significant role in confinement. It is numerically very high --
in absolute numbers it is about \be n_{mono}(T=T_c)\sim 10 fm^{-3} \ee 
This is larger than Bose gas of
non-interacting bosons even with zero mass, but for strongly correlated
liquid this is probably possible. Hopefully,
the
estimate of the monopole mass from the Feynman's
Bose condensation criterion $m_{mono}=(0.2-0.3) GeV$ 
will also work out. 
 At the moment, all we know about magnetic plasma
is its density and 
 magnetic coupling. Those have been combined into Gamma parameter
in the Fig.\ref{fig_em_phasediag} (down right) with a conclusion that it is always above 1,
so we do have a magnetic liquid. 

As explained above, we had classical MD simulations which lead to
reasonable diffusion and viscosity of such plasmas: and 
a 50-50 mixture of electric and magnetic particles (occurring somewhere
at $T\approx 1.5T_c$) is probably the
optimal point for the $lowest$ viscosity. How classical effects
will modify these numbers we do not know yet.

\section{ AdS/CFT duality}

AdS/CFT correspondence \cite{Maldacena:1997re} is a $duality$ between specific
gauge theory known as $\cal N$=4 super-Yang-Mills theory (SYM)
in 4 dimensions and the (10-dimensional) superstring theory in
a specific setting. Before we describe the results obtained,
here is a brief introduction, explaining its logic
and what all these words mean aiming at non-experts. Those who know
it may skip two next subsections.

\subsection{Black holes and AdS/CFT for pedestrians}
For applications we have in mind
the most important feature of string theories is the existence of massless modes of closed
strings, which include spin-2 ``gravitons" as well as lower spin members
of the multiplet, including vectors and scalars. String theory thus has aspiration
to be ``theory of everything" combining gauge and gravity interactions
known in nature. This aspect of it is not discussed in this paper at all:
all of these ``bulk" fields in 10 dimensions will only play a role to mimic
4-dimensional QCD-like gauge theories.

 One of several ``string theory revolutions" of 1990's was caused by a discovery of
 solitonic objects   of (nearly) any dimensions $p$ (ranging $p= -1 - 9$) called $D_p$
$branes$. One should think of them as some p-dimensional membranes, 
solitons made of strings and embedded in
9+1 dimensions of space+time. For our purposes the
most important example is the $D_3$ brane 
which has infinite extension in
 3 coordinates $x_1,x_2,x_3$, while it has no extension (and thus is
 just a point object) 
in the remaining 6 coordinates $x_4...x_9$.

Since branes are material
objects with certain masses and charges, they create gravity/Coulomb/scalar fields 
around them in the bulk, like planets and stars do. As they have no extension
in some coordinates (e.g. just mentioned $x_4...x_9$ coordinates for  $D_3$ brane),
their fields are that of a point objects: in general relativity this naturally implies
that they are black hole-like in such coordinates. As they are extended
in other coordinates, the proper name of such objects is {\em black branes}.
To find out their properties in classical (no loops or quantum fluctuations)
approximation, one has to solve
 coupled Einstein-Maxwell-Scalar eqns, looking for static spherically symmetric
 solutions.
This is no more difficult than to find the Schwartzschild solution for the usual
black hole, which we remind has a metric tensor in spherical coordinates
\be ds^2=g_{\mu\nu}dx^\mu dx^\nu= -(1-r_h/r)dt^2+{dr^2\over (1-r_h/r)}+r^2d\Omega^2\ee 
and the horizon radius (in fulll units) is $r_h=2G_NM/c^2$,
containing the mass $M$ and Newton constant $G_N$.  The horizon -- zero of $g_{00}$ -- is the ``event horizon": a distant observer
cannot see beyond it. (There $g_{00},g_{rr}$ change
sign.) Similar expressions, with appropriate powers of the distance
can be derived for p-branes.

Note that at large distance $r\ll r_h$ deviations from flat metrics is a small correction -- there the 
nonrelativistic potential description of Newtonian gravity is possible.
Similarly far from a brane  the fields are just given by Newton+Coulomb+scalar
formulae in corresponding dimensions, and it is not hard to figure
out what are their mutual interactions. Surprising (to anyone not involved
in monopoles/supersymmetry research) the answer is that all these forces
mutually {\em cancels out}, so there is no interactions
between branes  --they are so called BPS protected objects.
As a result, a ``brane engineer" may consider a number of parallel branes to be put
at some random points: and they will stay there.
Large ($N_c\rightarrow\infty$) number of branes put into 
the same point combine their mass and thus create strong gravity field, justifying
the use of classical  Einstein/Maxwell eqns. Open string states, which
keep their ends on some branes i and j lead to effective gauge theory
on the brane with the $U(N_c)$ group. 

When the black hole has a nonzero vector charge, the Gauss' law insists on constant
flux through sphere
of any radius,
thus there are nonzero fields at large $r$ and Schwartzschild solution gets modified
into a ``charged black hole". For asymptotically flat spaces there is the famous
statement, much emphasized by Wheeler:  there are ``no scalar hairs"
of black holes, as there is no Gauss theorem to protect them.
This happens to be not  true in general, and for spaces which are
AdS-like
thus scalars should $not$ be left over\footnote{In fact there is a whole recent direction
based on solution with a ``scalar atmosphere" around black branes, providing
gravity dual to superconductors on the boundary.}. 

 The solution for $D_3$ brane happens to be the so called
 $extremal$ (6-dimensional) charged black hole, which
has the lowest possible mass for a given charge. When the mass
decreases to the extreme value, the horizon shrinks to nothing,
and thus it is especially simple.
Spherical symmetry in 6d allows one to separate the 5 angles
(making the 5-dim sphere $S_5$) from the radius $r$ (in 6-dimensions
orthogonal to ``our world" space $x_1,x-2,x-3$)
and write the resulting 10-d metrics  as follows
 \begin{eqnarray} ds^2={-dt^2+dx_1^2+dx_2^2+dx_3^2 \over \sqrt{1+L^4/r^4}}+ 
\sqrt{1+L^4/r^4}(dr^2+r^2d\Omega_5^2)
 \end{eqnarray}
 Again, at large $r$ all corrections are small and we have
 the asymptotically flat space there. However
 at this point Maldecena told us to use instead further simplified version
 of this metric,
  in the ``near-horizon region"  at $r<<L$, when 1 in the roots can be ignored.
  If so,  in the last term two $r^2$ cancels out
and the 5-dimensional sphere element gets constant coefficient and thus
gets decoupled from 5 other coordinates. 
Thus quantum numbers or motion in $S^5$ becomes kind of internal
quantum numbers like flavor in QCD, and it will  be 
mostly ignored from now on. What is left
is very simple 5-dimensional metric known as  Anti-de-Sitter metric. 
Using 
a new coordinate $z=L^2/r$ we get it into the  ``standard $AdS_5$ form'' used below:
 \begin{eqnarray}  ds^2={-dt^2+dx_1^2+dx_2^2+dx_3^2+dz^2 \over z^2}  
\end{eqnarray}     
Note that $z$ counts the distance from ``the AdS boundary'' $z=0$.
This metric has no scale and is $not$ asymptotically flat even at the boundary.
Performing dilatation on these 5 coordinates we find that the metric remains
invariant: in fact one can do any conformal transformation.
It is this metric which
is AdS in the AdS/CFT correspondence, and string theory
in this background is ``holographically gravity dual" to some conformal gauge theory at 
the boundary.

So far  nothing
 unusual happened: all formulae came straight from string and general 
relativity textbooks. A truly remarkable
theoretical discovery is the so called
``holography'': the exact duality (one-to-one
correspondence) between the 5-dim ``bulk'' effective theory in $AdS_5$
to 4-dim ``boundary'' ($r\rightarrow\infty$)  gauge theory. 
There is a dictionary, relating any observable in the gauge
theory to another one in string theory: the duality implies that all
answers are the same in both formulations. We will see below how it
works ``by examples".

The last step, which makes it  useful, is the Maldacena relations between
the gauge coupling, the AdS radius $L$ and the string tension
$\alpha'$ (which comes from the total mass of the brane set):
 \begin{eqnarray} L^4=g^2 N_c (\alpha')^2=\lambda(\alpha')^2  
\end{eqnarray} 
It tells us that  large gauge 
coupling $\lambda>>1$ corresponds to large
AdS radius (in string units) and one can use
classical (rather than quantum) gravity. At the same time the string 
and gravity couplings $g_s\sim g^2$ may remain small:
 so one may do perturbative
calculations in the bulk! 

At this point many readers are probably very confused
 by new 5-th dimension of space. One possible approach is
to think of it as 
just a mathematical trick, somewhat analogous to more familiar
introduction of the complex variables. 

(Suppose an Experimentalist
 measured some complicated cross section which is approximately
a sum of Breit-Wigner resonances. His friend Phenomenologist
may be able to write the answer as an analytic function with certain
pole singularities in the complex energy plane,
which will help for fitting and for evaluating integrals.
 Even better, their other friend
Theorist  cleverly  developed a ``bulk theory'',
deriving the pole positions from some
 interaction laws on the complex plane.  ) 
 
 However, there is a perfectly 
  physical meaning of the
5-th coordinate.
One hint is provided by the fact that distance along it
$\int_a^b dl=\int_a^b dz/z=log(b/a)$ is the logarithm of the ratio.
Thus its meaning is the ``scale'',  the
argument of the renormalization group. If one takes a bulk
object and move it into larger $z$, its hologram at the boundary
(z=0) grows in size: this direction thus corresponds
to the infrared direction. 
The running coupling constant
would thus be a $z$-dependent field
called ``dilaton''.   Indeed, there are  theories with
gravity dual, in which this field (related to the coupling) 
does ``run" in $z$:
 unfortunately, known examples do not
(yet?) include QCD! In spite of that, there are efforts to built its gravity
dual ``bottom-up'', introducing weak coupling at ultraviolet
(small z) \cite{Shuryak:2007uq}
and confinement in infrared (large z) \cite{Karch:2006zz,Gursoy:2007cb}
 by certain modification of the
rules. These approaches -- known as AdS/QCD-- we would not discuss
in this review, except briefly in the subsection on bulk viscosity.

 Let me briefly remind few more facts
from the Black Hole toolbox. Studies of how
quantum field theory can sit in a  background of 
a classical black hole metrics have resulted in two major
discoveries: the {\em Hawking radiation}  and the {\em Bekenshtein entropy},
related to horizon radius and area, respectively. Hawking radiation
makes black holes in asymptotically flat space unstable: it 
and heats the Universe till the black hole disappears. Putting black hole
into a finite box also does not help: it is generically thermodynamically unstable
and gets smaller and hotter till it finally burns out. Only in appropriate curved spaces
black branes can be in thermal equilibrium with their ``Universe",
 filled with radiation at  some finite temperature $T$. As shown by Witten,
all one has to do to get its metric is to consider non-extreme (excited extreme)
black brane solution, which has a horizon.

 Let me now make a logical jump 
and  instead of just giving the metric for finite-T solution let me first
provide an ``intuitive picture'' for non-experts,
explaining
the finite-temperature Witten's settings
in which most\footnote{The exception is heavy quark diffusion
  constant  calculated by Casalderrey and Teaney\cite{Casalderrey-Solana:2006rq} which
 needs more complicated
settings, with a
 Kruskal metric connecting a World to an Anti-world through
the black hole.
}  pertinent calculations are done, shown in 
Fig.\ref{fig_relaxation}. The upper rectangle is the 
3-dimensional space boundary z=0 (only 2 dim shown),
 which is flat (Minkowskian) and corresponds
to ``our world'' where the gauge theory lives.
  Lower black  rectangles
(reduced in area because of curvature induced by $1/z^2$ in metric)  
is the corresponding (same in $x-1,x-2,x-3$) patch
of the horizon
(at $z=z_h$) of a black hole, whose center
is  located at $z=\infty$. Studies of finite-T 
conformal plasma by AdS/CFT famously started exactly by 
evaluation of the  Bekenstein entropy \cite{Gubser:1998nz}, 
$S=A/4$ via calculating the horizon  $area$  $A$. 

Now comes the promised ``intuitive picture'':
 this setting can be seen as a {\em swimming pool}, with
 the gauge theory (and us,  to be referred below as ``distant
 observers'')  living on its surface, at zero depth $z=0$, enjoying
the desired temperature $T$. In order to achieve that, the
 pool's bottom 
looks $infinitely$ hot for observers which are sitting at some fixed
$z$ close to its coordinate $z_h$: thus diving to such
depth is not recommended.
Strong gravity takes care and stabilizes
 this setting thermodynamically: recall that time units, as well as
those of energy and temperature are subject to ``warping''
with $g_{00}$ component of the metric, which vanishes at  $z_h$.

 When astronomers found evidences for black holes
 and accretion into those, the physics of black hole became a regular
 part of physics since lots of problem have to be solved.
  Here important step forward was the so called {\em membrane paradigm}
  developed by many people and best formulated
by Thorne and collaborators \cite{Thorne:1986iy}, known also under the name of
``stretched horizons". Its main idea is to imagine that there is a
physical membrane at some small distance $\epsilon$ away from
the horizon, and that it has properties  exactly such that all
the eqns (Maxwell's, Einstein's
etc) would have the same solutions outside it as $without$ a membrane but
with a continuation through horizon.
For example, a charged black hole would have a membrane
with a nonzero charge density, to terminate the electric field lines.
The fact that Poynting vector at the membrane must be pointed
inward means some (time-odd!) relation between $\vec E$ and $\vec B$:
this is achieved by giving the membrane finite $conductivity$, which in turn
leads to ohmic losses, heat and entropy generation in it. Furthermore,
as shown by Damour back in 1980's, displacements of the membrane
and relations for gravitational analog of the Poynting vector
leads to nonzero $viscosity$ of the membrane,
and its effective low frequency theory take the form
of Navier-Stokes hydrodynamics. 
For a bit more modern derivations of the effective action
and field theory point of view look at Parikh nd Wilczek \cite{Parikh:1997ma}.
As we will see below in this chapter,
all of those ideas have resurfaced now in AdS context, generating
new energy of young string theorists who now pushed the ``hydrodynamics
of the horizon memebrane" well beyond the
Navier-Stokes to a regular construction of systematic derivative expansions
to any needed order.

\subsection{CFT for pedestrians}
The $\cal N$=4 SYM theory is a cousin of QCD: it also has gauge
fields with $SU(N_c)$ color symmetry, but instead of quarks
it has  four ``flavors'' of fermions called gluinoes, as their
adjoint colors are the same as for gluons. There are also
6 adjoint scalars: with 2 polarizations of gluons it makes 8 bosonic
modes, same as 2*4 fermions. This makes supersymmetry possible
and leads to cancellations of power divergences. This theory
is the most symmetric theory possible in 4 dimensions: it has conformal symmetry and
its {\em coupling does not run}! 

How do we know this? Of course, one had calculated  first 
few perturbative coefficients
of the beta function, and indeed see that negative gauge
contribution is nicely canceled by fermions and scalars, 
order by order.
But there are infinitely many coefficients, and one has to check
them all! An elegant way to prove the case is based 
 on another outstanding feature of the 
 $\cal N$=4 SYM: this theory is $self-dual$ under electric-magnetic
duality. As we discussed above, 
the Dirac condition requires the product
of electric and magnetic couplings to be constant: and
so in QCD and other gauge theories they indeed run in the
opposite directions, electric becoming weak in ultraviolet
and magnetic weak in infrared. But
the multiplet of (lowest) magnetic
objects of  the $\cal N$=4 SYM theory include 
6 scalars (the monopoles), plus 4 fermions (monopoles plus one gluino
zero mode occupied), plus 2 spin-1 (monopoles with 2 gluinoes):
this turns out to be exactly the same set of states as the original 
electric degrees of freedom (gluons-gluinoes-Higgses).
That means that an
 effective magnetic theory has the $same$ Lagrangian
 as the original electric formulation: thus it must
have the same beta function. Since two couplings cannot run
in the opposite direction following the same
beta function, they cannot run at all!)

Recall that in QCD-like theories the scale  $\Lambda_{QCD}$ came about
because of running coupling.
If in the $\cal N$=4 SYM
theory  the coupling constant does not run, it means that there is no analog
of $\Lambda_{QCD}$ in this theory, and since all the fields are
massless and all the coupling dimensionless, thus the  $\cal N$=4 SYM
theory has no scale at all. It is
thus  conformal field theory --
 the CFT in the AdS/CFT correspondence.
One consequence is that the finite-T
version of this theory 
(we will be mostly interested in) is the same whether
T is large or small, since there is no other scale to compare with!
This is similar to QCD plasma
in the so called ``quasi-conformal regime'': at high enough $T>2T_c$
all dimensionless ratios like (energy density)/$T^4$ are practically
$T$-independent.  

Weakly coupled $\cal N$=4 SYM theory can be studied perturbatively,
like any other gauge theory. What makes it unique is that AdS/CFT
correspondence allows also to study it in the {\em strong coupling}
limit, defined by a large value of the so called 't Hooft coupling,
a combination of gauge coupling and number of colors which go together
\begin{eqnarray} \lambda=g^2N_c >>1
\end{eqnarray}

It is this combination which appears in physical effects, e.g. the
Coulomb law. Thus, before we embark on studies of strong coupling
regime $\lambda >>!$, the reader have all reason to say: wait a
moment,
is it not followed from Klein-Gordon (or Dirac) eqns that
for coupling larger than something  two charges will fall at each
other, as the square of the Coulomb potential $\sim \lambda^2/r^2$ 
will dominate the centrifugal term $l(l+1)/r^2$? Zahed and myself
\cite{Shuryak:2003ja} worried about this, and their semiclassical
approach to density of charge was even used later by
Klebanov, Maldacena and Thorn \cite{Klebanov:2006jj}. When the
spectrum
of heavy quarkonia was eventually found from AdS/CFT, 
all states including
the lowest s-wave ones were accounted for -- they have small but
positive masses, and string theorists
were not surprised.  But frankly
I still don't understand why no falling actually happens. 
This is one of many puzzles which shows that what can be derived
from gravity side may be very hard to understand in the gauge theory.

\subsection{The first example of AdS/CFT at work: new Coulomb law}
\begin{figure}[t]
\begin{minipage}[h]{7cm}
\vskip -2cm   \includegraphics[height=5.cm]{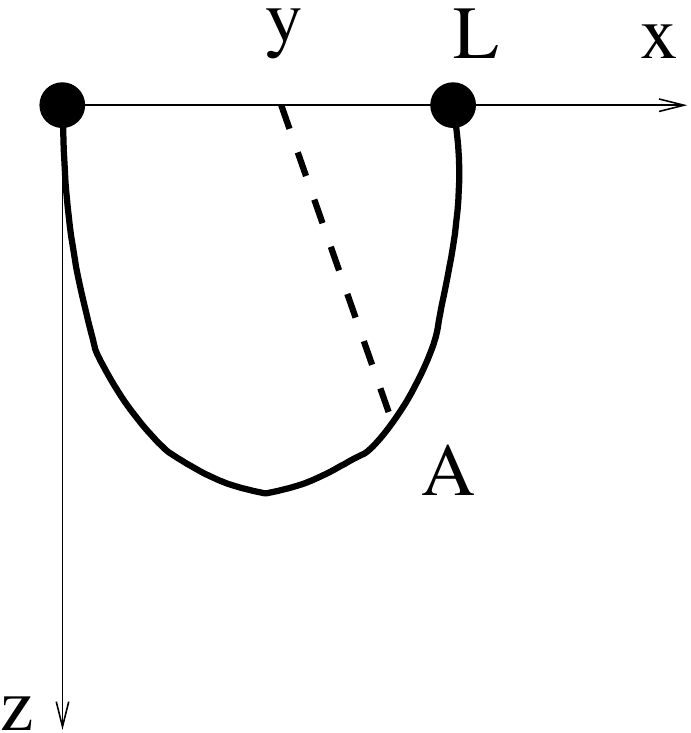} 
\end{minipage} \hspace{2cm}  
 \begin{minipage}[h]{10cm}   
\vskip -1cm \includegraphics[height=6.cm]{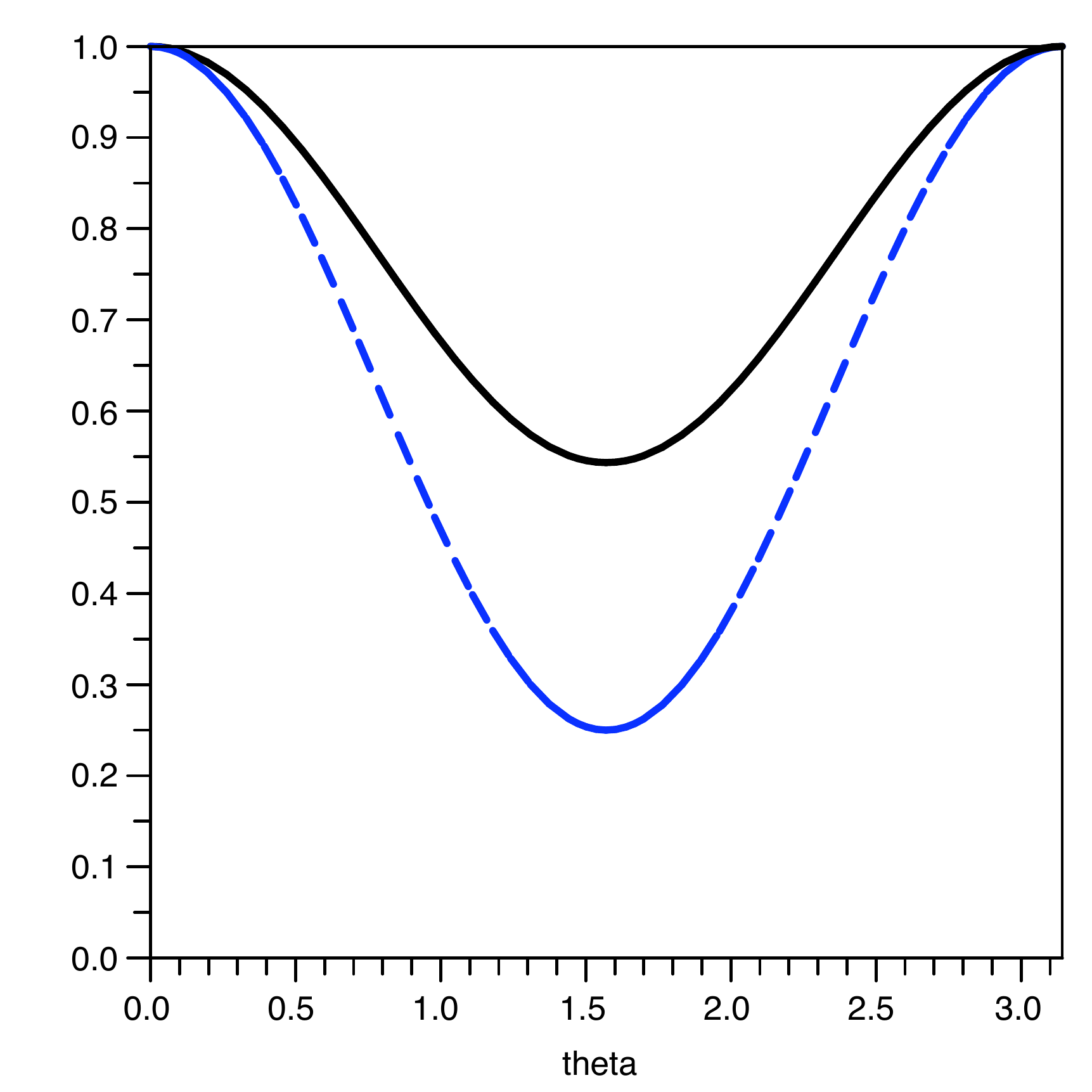}
 \end{minipage}  
 \vskip 1cm
\caption{ Setting of the Maldacena dipole: two charges
at the boundary (black dots) are connected by the string
 (shown by solid curve) pending under
gravity toward the AdS center $z\rightarrow\infty$. Classical graviton  propagator
(the dashed line) should be used
 to calculate the ``hologram'' -- the stress tensor at
the observation point $y$. The string is the gravity source;  
the point $A$ has to be integrated
over.
  (b) Angular distribution of the  far field energy versus the polar angle
 ($cos(\theta)=y_1/|y|)$.
 Solid black line is the AdS.CFT result,
  compared to the perturbative dipole energy $(3cos^2\theta+1)/4$ 
(the dashed blue line), both normalized at zero angle.
\label{eos-shuryak-fig_dipole}
}
\end{figure}

 Let me  start with 
 our first  example of the ``AdS/CFT at work'', related with
 the 
strong-coupling version of the Coulomb law calculated in
 \cite{MALDA2_REY}. 
The setting-- to be called ``the Maldecena dipole''
-- is shown in Fig.\ref{eos-shuryak-fig_dipole}(a), includes two static
charges (heavy fundamental quarks) separated by the distance $R$.

At weak coupling -- the usual QED -- we think of one
 charge creating the electric potential in which the other is
 placed, leading to the usual Coulomb law which in our notation is
 \be V(L)= -{g^2\over 4\pi L}  \ee
  is a sum of two Coulomb fields.
   \be E_m(y) =({g\over 4\pi})\left({y_m-(L/2)e_m \over
 |y_m-(L/2)e_m|^3}-{y_m+(L/2)e_m \over |y_m+(L/2)e_m|^3}\right)
 \label{eqn:dipole_weak} \ee
Because of a cancellation between two terms,
 at large distances from the dipole
the field decays as $E\sim L/y^3$. The corresponding  
 energy density (and other components of the
stress tensor) are thus of the order  $T_{00}\sim g^2 L^2/y^6$, with certain
``dipole" angular
distribution. 

In the $\cal N$=4 theory at $weak$ coupling the only difference is that one can
exchange massless scalars on top of gluons. It is always attractive, and for
two heavy quarks leads to cancellation of the force, with doubling for
quark-antiquark (we discuss now). The QED coupling $g^2$ changes to
't Hooft coupling $\lambda=g^2N_c$ proportional to   the number of colors $N_c$.

Now we turn to the  AdS/CFT for the $\cal N$=4 theory at $strong$ coupling 
\cite{hep-th/9803002,hep-th/9803001} .
The electric flux in the bulk forms a 
singular object -- the string (shown by the solid curve in Fig.\ref{eos-shuryak-fig_dipole}(a)) --
which pends from the boundary $z=0$ due to gravity force into the 5-th dimension,
like in the famous catenary (chain) problem\footnote{Another --
more  Einsteinian --way to explain it is to note that this is simply 
the shortest
string possible: it is not  straight because the space is
curved.
It is the same reason why  the shortest path from New York to London
does not look straight on the map.
}.  
The calculation thus follows from Nambu-Goto action for the string, whose general form
is
\be S={1\over 2\pi\alpha'} \int d\sigma d\tau \sqrt{det G_{MN} \partial_\alpha X^M \partial_\beta X^N  } 
\ee
where 2 coordinates $\sigma,\tau$ parameterize the string world 
line $X^M(\tau,\sigma)$, 
where
 $M,N$ are space-time indices in the whole space (10dim reduced to 5d in AdS/CFT). 
$G_{MN}$ is the space metric and $det$ stands for 2*2 matrix with all $\alpha,\beta$.
In the $AdS_5$ metric we need the components $-G_{00}=G_{11}=G_{55}=1/z^2$,
and we can think of $\sigma,\tau$ as our coordinates $x,t$: the string is then
described by only one function $z(t,x)$ and its action is reduced to
\be S\sim \int dt dx {1\over z^2}\sqrt{1+ (\partial z/\partial x)^2- (\partial z/\partial t)^2}\ee 
We will use this action for ``falling strings" below, and now proceed to further
simplifications for static string, for which there is no time derivative
and the function is $z(x)$. Maldacena uses  $u(x)=1/z(x)$ and thus the Lagrangian
becomes $L=\sqrt{(u_{,x})^2+u^4}$ with comma meaning the x- derivative.
 One more simplification comes from the fact
that $x$ does not appear in it: thus an ``energy" is conserved
\be H= p\dot q -L={\partial L \over u_{,x}} u_{,x}-L=E=const\ee 
which reduces the EOM from second order eqn to just $(u_{,x})^2=u^4(u^4/E^2-1)$ which 
can finally be directly integrated
 to \be x(u)=\int_{u_m}^u {du' \over (u')^2 \sqrt{(u')^4/E^2-1)} }\ee
 The minimum position of the string $u_m$ is related to $E$ by the relation
 following from this formula at
$x=L/2$. Plugging the solution back into action and removing
divergence (which is independent of $L$)
  gives finally the total string energy, which is
the celebrated {\em new Coulomb law at strong coupling}
\be \label{eqn_new_Coulomb}
V(L)= -{4\pi^2  \over \Gamma(1/4)^4 }{\sqrt{\lambda} \over  L}
\label{coulomb}
\ee
The power of distance $1/L$ is in fact the only one possible by dimension, as
the theory is conformal and has no scales of its own. What is 
remarkable is the (now famous) $\sqrt{\lambda}$ appearing instead of $\lambda$ in
the weak coupling. (The numerical coefficient in the first bracket is 0.228,
to be compared to the result from a diagrammatic
re-summation below.)

What is the reason for this modification? For pedagogical reasons
let me start with two ``naive but reasonable guesses", both to be
shown to be wrong later in this section:\\ (i) One idea is that
 strongly coupled vacuum acts like some
kind of a space-independent dielectric constant, $\epsilon \sim 1/\sqrt{\lambda}$
which is reducing the effect of the Coulomb  field, similarly at all points.\\
(ii) Perhaps such dielectric constant has nonlinear effects, and thus is
not the same at different points: but the fields created by static dipole
are still just the electric field $\vec E$.

 A good feature of the AdS/CFT is that one can get many further details about
 the problem and test
ideas like just expressed: but we cannot get $all$ features.
In order to understand the difference between dipole
fields in a weakly and strongly regimes it  be nice to calculate the electric and
the scalar
field distributions in both limits.
Unfortunately we cannot do that:  those fields do not belong to limited set
 of quantities one can calculate via AdS/CFT correspondence. 
In general AdS/CFT rules allows one to 
find  $holograms$  on the boundary of whatever happens
in the bulk. 
This can be done by solving bulk wave eqns
for massless bulk fields - which are scalars, vectors or gravitons. For example I
will give our results for boundary stress tensor obtained from
the (linearized) Einstein equation for the gravity perturbations,
shown by the dashed  line in Fig.\ref{eos-shuryak-fig_dipole}(a).
 
For example, in (relatively  recent) paper  Lin and myself
\cite{Lin:2007pv} have found the {\em stress tensor}
of matter  $<T_{\mu\nu}(y)>$ at any  point $y$ on the boundary,
induced by the Maldacena dipole. 
  The solution is too technical to be presented here and even the
  resulting stress tensor expression is too long\footnote{By he way,
   the stress tensor should always be traceless in conformal theory and
have zero divergence: those conditions are
used to  verify explicitly that no mistake in the calculation was made.}: let me just show
only the leading terms  far from the dipole
$y>>L$.

\ba\label{eq:ff}
T_{00}=\sqrt{\lambda} L^3
\left(\frac{C_1y_1^2+C_2 y^2}{|y|^9}\right)f(\theta)
\ea
where $C_1,C_2$ are numerical constants whose values can be looked up in the paper
and $f(\theta)$ is the angular distribution shown by solid line in  Fig.\ref{eos-shuryak-fig_dipole}(b), quite
 different that in  weak coupling (the dashed line). Note that in weak coupling
the energy density from the dipole is just electric
field (\ref{eqn:dipole_weak}) squared, leading to
a different power $T_{00}\sim \lambda L^2/y^6 $.


   As we get glimpse of some first results from AdS/CFT  we see that they are quite different from weak coupling and we would like
    to understand them.
    In fact we would like to do so both from the bulk (gravity) side
as well as  from the
   {\em gauge theory} side.
   It turns out the first is relatively easy. For example, both in the total energy
   and energy density we get $\sqrt{\lambda}$ because this factor
   is in front of the Nambu-Goto Lagrangian (in proper units). The reason
   the field decays as $y^7$ is extremely natural:
    in the $AdS_5$ space the
   function which inverts the Laplacean (analogously to Coulomb
   $1/r$ in flat 3d) has that very power of distance
   \be P_s={15\over{4\pi}}\frac{z^2}{\(z^2+r^2\)^{7\over2}}\ee
   with $z$ being the 5-th coordinate of the source and $r$ the 3-distance
   between it and the observation point\footnote{Please recall
   that in gravity there are no cancellations
   between different contributions: any energy source perturbs gravity with the same sign.}. 
   
    In order to understand the same results from the gauge side we will
    need a bit of pedagogical
   introduction: the resolution will be given by
   the idea  of {\em short color correlation time}
  by Zahed and myself \cite{Shuryak:2003ja}.
 In QCD, with its running coupling, 
 higher-order effects modify the zeroth order Coulomb field/potential. 
 Since we consider static problem, one can
rotate time into Euclidean time $\tau=it$, which 
not only makes possible lattice simulations but also 
simplifies perturbative diagrams.  
Let us do Feynman diagrams directly in the coordinate
representation. The lowest order energy, given by the diagram in which one
massless quantum (scalar or gauge component $A_0$) is emitted by one charge at time $\tau_1$ and absorbed by
another at time $\tau_2$  
is just 
\be V(L)(Time) \sim  -g^2\int_0^{Time} {d\tau_1 d\tau_2 \over L^2+
  (\tau_1-\tau_2)^2}\sim -{g^2 (Time)\over L} \ee
 where 'Time' is total integration time and
 the denominator -- the square of the 4-dim distance between
gluon emission and absorption  represents the Feynman propagator in Euclidean
space-time. The propagation time for a virtual quantum $(\tau_1-\tau_2)\sim L$,
thus the Coulomb $1/L$ in the potential.

 Higher order diagrams include self-coupling of gluons/scalars
and multiple interactions with the charges. A
famous simplification  proposed by 't Hooft is the large number of colors limit
in which only $planar$ diagrams should be considered. People suggested that
as $g$ grows those diagrams are becoming ``fishnets'' with smaller
and smaller holes, converging to a ``membrane'' or string worldline:
but although this idea
was fueling decades of studies trying to cast gauge theory into
stringy form it have not strictly speaking succeeded.
It may still be true: just nobody was smart enough to sum up 
all planar diagrams\footnote{Well, AdS/CFT is kind of a solution,
actually, but it is doing it indirectly.}.

  If one does not want to give up on re-summation idea, one may
consider a subset of those  -- the $ladders$ --
which can be  summed up.
Semenoff and Zarembo \cite{Semenoff:2002kk} have done that: let us look what have they found.
The first point is that in order that each rung of the ladder contributes a factor $N_c$,
 emission time ordering should be strictly
enforced, on each charge; let us call
these time moments $s_1>s_2>s_3...$ and $t_1>t_2>t_3...$.
Ladder diagrams must connect $s_1$ to $t_1$, etc, otherwise it is nonplanar
and subleading diagram.
 Thus the main difference from
the Abelian theory comes from the dynamics of the color vector.
The (re-summed) Bethe-Salpeter kernel $\Gamma(s,t)$, describing
the evolution from time zero to times $s,t$ at two lines,
satisfies the following integral equation
\be \label{eqn_BS}
\Gamma({\cal S},{\cal T})=1+{\lambda \over 4\pi^2}\int_0^{\cal S} ds
\int_0^{\cal T} dt{1\over
    (s-t)^2+L^2} \Gamma(s,t) \ee
If this eqn is solved, one gets  re-summation of all the
ladder diagram. The kernel obviously
satisfies the boundary condition $\Gamma ({\cal S},0) =\Gamma(0,{\cal T})=1$.
If the equation is solved, the ladder-generated potential is
\be
V_{\rm lad}(L) 
=-\lim_{T\to{+\infty}}{\frac 1{\cal T} \Gamma\, ({\cal T},{\cal T})}\,\,,
\label{0a}
\ee
In weak coupling $\Gamma\approx 1$  and the integral on the rhs is
easily taken, resulting in the usual Coulomb law. 
For solving it at any coupling, it is convenient
to switch to the differential equation

\be
\frac{\partial^2\Gamma}{\partial {\cal S}\,\partial {\cal T}} =
\,\frac{\lambda/4\pi^2}{({\cal S-T})^2+L^2}
\Gamma ({\cal S,T})\,\,\,.
\label{1a}
\ee
and change variables to
$x=({\cal S-T})/L$ and $y=({\cal S+T})/L$ through
\be
\Gamma (x,y) =\sum_{m}\,{\bf C}_m \gamma_m (x)\,e^{\omega_m y/2}
\label{2a}
\ee
with the corresponding boundary condition $\Gamma (x,|x|)=1$. The
dependence of the kernel $\Gamma$ on the relative times $x$ follows
from the differential equation

\be
\left(-\frac{d^2}{dx^2} -
\frac{\lambda/4\pi^2}{x^2+1} \right)
\,\gamma_m (x) = -\frac {\omega_m^2}{4}\,\gamma^m (x)
\label{3a}
\ee
For large $\lambda$ the dominant part of the potential in (\ref{3a})
is from {\it small} relative times $x$ resulting into a harmonic
equation~\cite{Semenoff:2002kk}

At large 
times ${\cal T}$,  the kernel is dominated by the lowest harmonic mode. For large times ${\cal S\approx T}$ that is small $x$ and large 
$y$ 

\be
\Gamma (x,y)\approx {\bf C}_0\,e^{-\sqrt{\lambda}\,x^2/4\pi}\,
e^{\sqrt{\lambda}\,y/2\pi}\,\,.
\label{5a}
\ee
From (\ref{0a}) it follows that
in the strong coupling limit the ladder generated potential
is \be V_{\rm lad}(L)= -\frac{\sqrt{\lambda}/\pi}L \ee which 
has {\em the same parametric form}  as the one derived from the
AdS/CFT correspondence (\ref{eqn_new_Coulomb}) except for the
overall coefficient. Note that the difference
is not so large,  since $1/\pi=0.318$ is larger than the exact value  
0.228 by about 1/3. 
Why did it happened that the potential is reduced relative to 
the Coulomb law by $1/\sqrt{\lambda}$? It is because the relative time
between gluon emissions is no longer $\sim L$, as in the Abelian case,
but reduced to parametrically small time of relative color coherence  $\tau_c\sim 1/L\lambda^{1/2}$. Thus  
we learned an important lesson: in the strong coupling regime even the
static charges communicate with each other via high frequency
gluons and scalars,  propagating
(in Euclidean formulation!) with a  super-luminal velocity
$v\approx \lambda^{1/2}\gg 1$. 

This idea --although it 
 seemed to be too bizarre to be true --
 as we will see below to explain some of the AdS/CFT results.
Klebanov, Maldacena and Thorn \cite{Klebanov:2006jj}
have pointed out that  the reason the stress tensor around the
dipole is  different from perturbative one by a factor
$\sim  (L/r)/\sqrt{\lambda}$ is actually explained by  
 limited   relative emission time by color coherence time.
 I am sure possible usage of this idea does not ends here.
 
Before we leave the subject of Maldacena dipole, one more interesting
question is what happen if one of the charges makes a small accelerated motion
near its original position. Will there be a radiation? In a somewhat different
setting than used above, the answer was provided by Mikhailov \cite{Mikhailov:2003er}. He studied perturbations of the string and found that 
the radiated energy is described by familiar classical Li\'enard formula
\begin{eqnarray}\label{Lienard}
\Delta E= A
\int_{-\infty}^{\infty} {\ddot{\vec{x}}^{\;2}-
\left.[\dot{\vec{x}}\times \ddot{\vec{x}}]\right.^2\over
(\,1-\dot{\vec{x}}^{\;2}\,)^{\scriptstyle 3\atop  }}dt
\end{eqnarray}
in which QED weak coupling constant $A={2\over 3}e^2$
is substituted by CFT strong coupling $A={\sqrt{\lambda}\over 2\pi}$.
This result is similar in its meaning to QCD synchrotron radiation we
already mentioned \cite{Shuryak:2002ai} and would provide
complete quenching of all jets beyond certain energy if the coherent fields
in  ``glasma" would produce such acceleration, see discussion in 
\cite{Kharzeev:2008qr}. 

\subsection{ Conformal plasma in  equilibrium and the idea of relaxation}
In brief, the finite-T setting is different from zero-T AdS space discussed
so far by existence of a horizon. Relaxation is due to the fact
that all objects in the bulk sink toward it and their energy and information
gets lost.

\begin{figure}[t]
\centerline{\includegraphics[width=12cm]{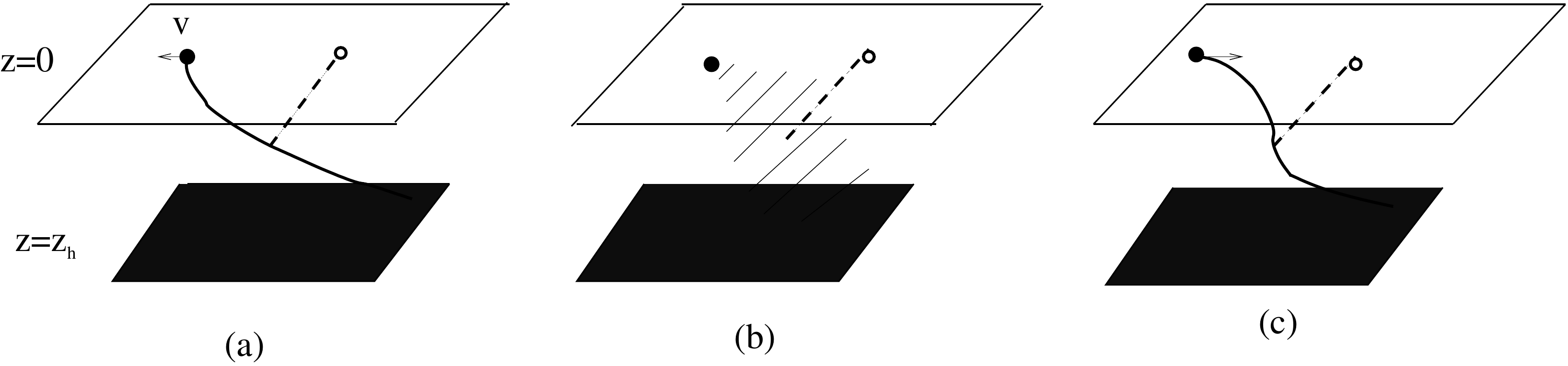}}
\caption{\small Schematic view of the relaxation settings,
a string (a), a wave (b) or a particle (c)
fall into the 5-th dimension toward the
black hole. 
} \label{fig_relaxation}
\end{figure}

After we have considered static heavy quarks and their strings,
it is natural to proceed to the next stationary situation with a heavy quark,
now in the finite-T setting. 
Fig.\ref{fig_relaxation}(a) shows a 
setting of heavy quark quenching \cite{hep-th/0407215,Herzog:2006gh,Gubser:2006bz,Buchel:2006bv,Sin:2006yz}:
a quark is  being dragged 
(at some hight $z_m$ related to the quark mass)
by an ``invisible hand'' (to the left): its electric flux goes into
 the 5-th dimension, into the so called
``trailing string''.  Its weight forces it to fall to the bottom
(horizon). (Think of 
a heavy quark  as a ship diligently laying
underwater cable to the pool's bottom.) The cost of that is the drag 
$$ {dP/dt}= -\pi T^2\sqrt{g^2 N_c} {v/2\over \sqrt{1-v^2}}
$$
connected to the diffusion constant  via
Einstein relation, a nontrivial  successful 
check on two very different calculations.

 Another form of relaxation is studied via propagating ``bulk
waves'' (b):  massless ones may have spin
 S=0 (dilaton/axion),1(vector) or 2 (gravitons).
Absorptive boundary condition at
the horizon (black bottom)leads to spectra of
``quasinormal\footnote{Quasinormal modes are those which
do not conserve the norm of the wave: it is like decaying
radioactive states in nuclear physics which are
distinct from scattering ones, with real energies. }
 modes''
with the imaginary part $Im(\omega_n) \sim \pi T n$,
setting the dissipation timescale of various fluctuations.

We will give the equations themselves in the next subsection:
let me just mentioned  near-zero
modes, corresponding to  two propagating ``surface waves'',
the longitudinal $sound$ and transverse $``diffuson''$.
Absorption at
the bottom (horizon) of both
 famously gives the viscosity $\eta/s=1/4\pi$\cite{Policastro:2001yc}.
The waves may have real (timelike) 4-momentum or virtual
(spacelike) one\footnote{
 This  case, named DIS in AdS,  is
 discussed  here   by E.Iancu.}. Rather 
complete spectra of quasinormal
 modes and spectral densities for S=0,1,2 correlators are  
available, unfortunately mostly extracted numerically.
 The case (c) -- a ``falling stone'' --  perhaps represent
colorless (no strings attached)
 ``mesons'', released to plasma and relaxing. 

 
The dashed lines in all Fig.\ref{fig_relaxation} corresponds to
the next-order diagram, describing back reaction
of the falling bulk objects onto the boundary,
the observation point denoted by a small open circle. This fields 
may also have
spins 0,1 or 2, providing 3 pictures -- the
(4-d) ``holograms'' -- of the boundary. Contrary to our intuition (developed from
our limited flat-world experience),  the hologram is $not$   
a reduced reflection of more complete 5-d dynamics in the bulk, but
in fact represents it fully. This phenomenon
 -- the AdS/CFT duality -- is a miracle occurring due to near-black-hole
setting.

\begin{figure}[t]
\includegraphics[height=6cm]{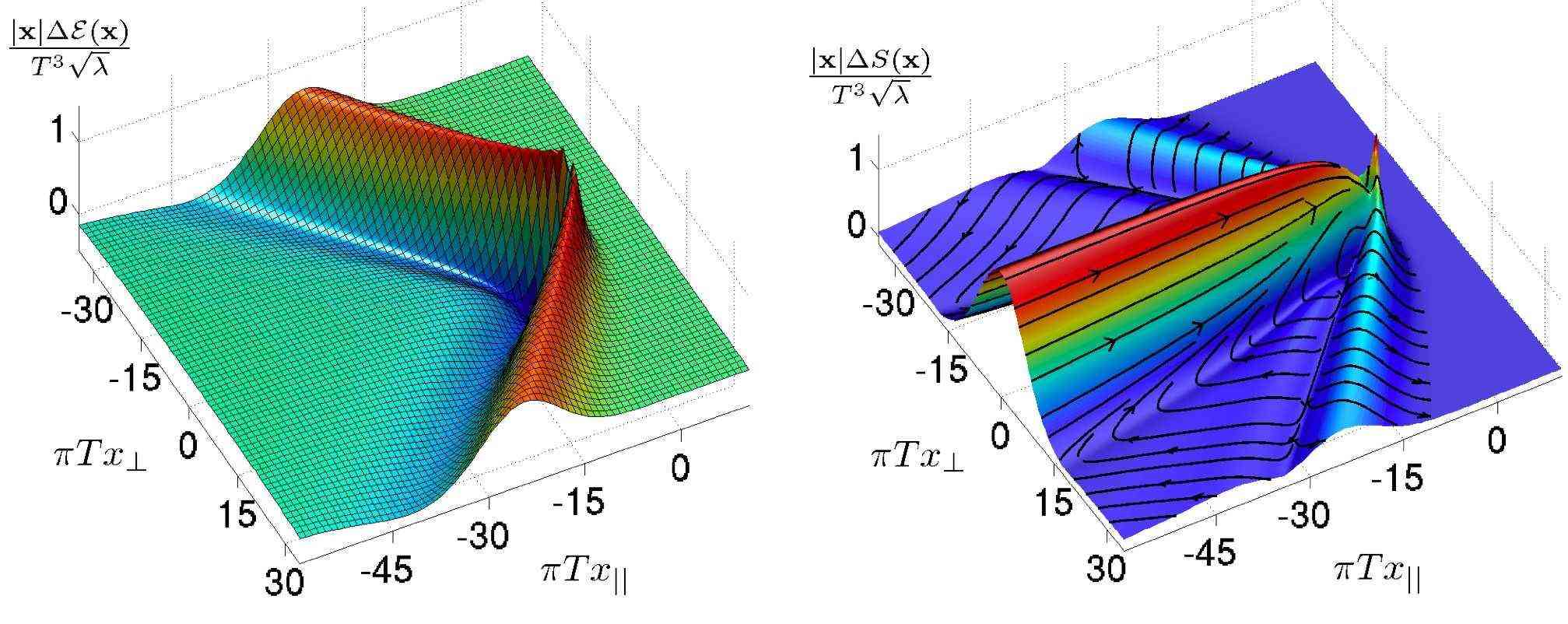}
\caption{\label{fig_holog}
 From \protect\cite{Chesler:2007sv}: 
hologram of the trailing string,
the normalized energy density 
 for one quark (supersonic jet) with $v = 3/4$ at nonzero $T$.}
\end{figure}

These holographic images are what the surface observer will
see.  Image of the trailing string was calculated in \cite{hep-th/0605292,Chesler:2007sv}:
 the recent example at nonzero $T$
is shown in Fig.\ref{fig_holog}(b,c): it accurately
displays hydro  conical flow. For a hologram of the  stone Fig.\ref{fig_relaxation}(c)
see recent paper \cite{Gubser:2008vz}: but to our knowledge
the holographic ``back reaction''
 of the falling waves remains to be done.

 How these predictions are related to experiment? 
Apart of those
shown in Fig.\ref{fig_vis_d_mapping},  important
 test is whether
the  drag force 
indeed depends only on the  $velocity$ (rather 
than momentum): can be done via single electrons from $c$ and $b$ 
decays. 
Another challenge is to test
if the effective
viscosity $\eta(k)$ is indeed $decreasing$  with increasing gradients
(or momentum $k$), 
as AdS/CFT nontrivially indicate. While in \cite{Lublinsky:2007mm}
only the issue of initial entropy production was discussed,
experimentally better control of this effect  can 
be inferred from studies of the viscous corrections to
elliptic flow at more peripheral collisions, when the gradient
in the direction of impact parameter $k\sim R_x$ 
can be made large.

Last but not least, the AdS/CFT correspondence predict a particular
ratio of the excitation amplitudes of the sound (conical flow) and diffuson
mode. So far we seem to see experimentally only
the conical flow, but not the second mode
(which would result in a peak in the original direction of the jet). Much more
detailed experimental and
theoretical studies are required to see if indeed there is a 
potential discrepancy here or not.

\section{ AdS/CFT and hydrodynamics}
\subsection{Linearized hydrodynamics} 
Son and collaborators \cite{Policastro:2002se}
started their program of studies in
linearized approximation: they wanted to see
if small amplitude  perturbations of the finite $T$
Witten's metric do in fact correspond to hydrodynamical modes:
the sound and diffuson. The answer was positive, for example
their famous value of viscosity turned out to be consistently
the same in all tests they have made. This paper and its sequels
are so clearly written that any reviewer can hardly hope to improve 
the presentation: so my advice is to look up these papers.

Small perturbations are described by linearized
Einstein eqns, and if time and 3-space derivatives are
substituted by $\omega,\vec k$ one has second order differential
eqns in the ``holographic coordinate''  which for now we will denote $u=z/z_h$. 
The three relevant eqns and their quasinormal spectra are to be found
in Kovtun and Starinets \cite{Kovtun:2005ev}, the WKB approach
to spectral densities for vector/stress tensor currents
 in Teaney \cite{Teaney:2006nc} and also in 
another paper by Kovtun and Starinets\cite{Kovtun:2006pf}.
The ``master equation" has the form
\begin{equation}
  \frac{d^2}{du^2} Z_a(u) +
  p_a(u) \frac{d}{du}Z_a(u) +
  q_a(u) Z_a(u) = 0 \,,
\label{eq:master-equation}
\end{equation}
where $a=1,2,3$ labels the three symmetry channels, called {\em shear,sound and scalar},
respectively.
Three sets of coefficients depend on the dimensionless frequency $\bf{w}\equiv\omega/2\pi T$
and momentum $\bf{q}\equiv k/2\pi T$ and are
\begin{equation}
  p_1(u){=}\frac{({\bf{w}}^2-{\bf{q}}^2 f)f + 2 u^2{\bf{w}}^2}{uf({\bf{q}}^2 f-{\bf{w}}^2)}\,, \, \, \,
  q_1(u){=}\frac{{\bf{w}}^2 - {\bf{q}}^2 f}{uf^2}\, .
\end{equation}
\begin{eqnarray}
  p_2(u) &=& -\frac{3{\bf{w}}^2 (1+u^2) + {\bf{q}}^2 ( 2u^2 - 3 u^4 -3)}
           {u f (3 {\bf{w}}^2 +{\bf{q}}^2 (u^2-3))}\ , \\
  q_2(u) &=&  \frac{3 {\bf{w}}^4 +{\bf{q}}^4 (3{-}4u^2{+}u^4) +
            {\bf{q}}^2 (4u^2{\bf{w}}^2{-}6{\bf{w}}^2{-}4u^3 f)}
            {u f^2 ( 3 {\bf{w}}^2 + {\bf{q}}^2 (u^2 -3))}\, . 
\end{eqnarray}
\begin{equation}
  p_3(u) = -\frac{1+u^2}{uf}\ , \ \ \ \ 
  q_3(u) = \frac{{\bf{w}}^2 - {\bf{q}}^2 f}{u f^2}\ ,
\end{equation}
where $f=1-u^2$. The procedure is to look for solution satisfying  
the incoming wave condition
at the horizon $u=1$. Those
can be written  as a
linear combination of two independent solutions with different behavior at $u=0$
\begin{equation}
  Z_a(u) = A_{a} Z_a^{I}(u) + B_{a} Z_a^{II}(u) 
\label{eq:Z-two-solutions}
\end{equation}
and standard general argument express the retarded Green function in terms
of the ratio of two coefficients, now written in the original units
\begin{equation}
   G_a(\omega,q) = -\pi^2 N_c^2 T^4 
    \frac{B_{a}(\omega,k)}{A_{a}(\omega,k)}\,.
\end{equation}

\begin{figure}[t]
   \includegraphics[width=8.cm]{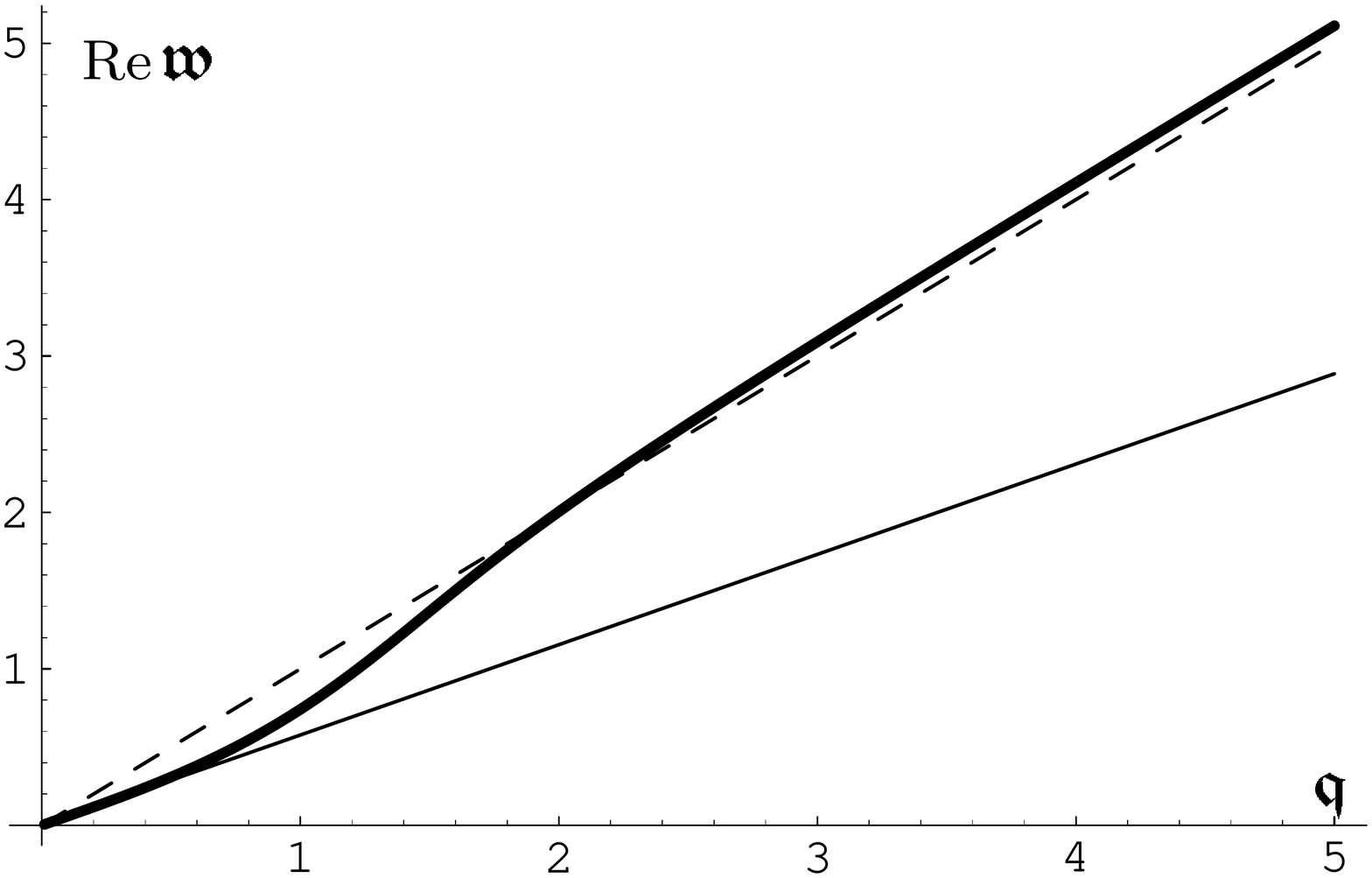} 
 \includegraphics[width=8.5cm]{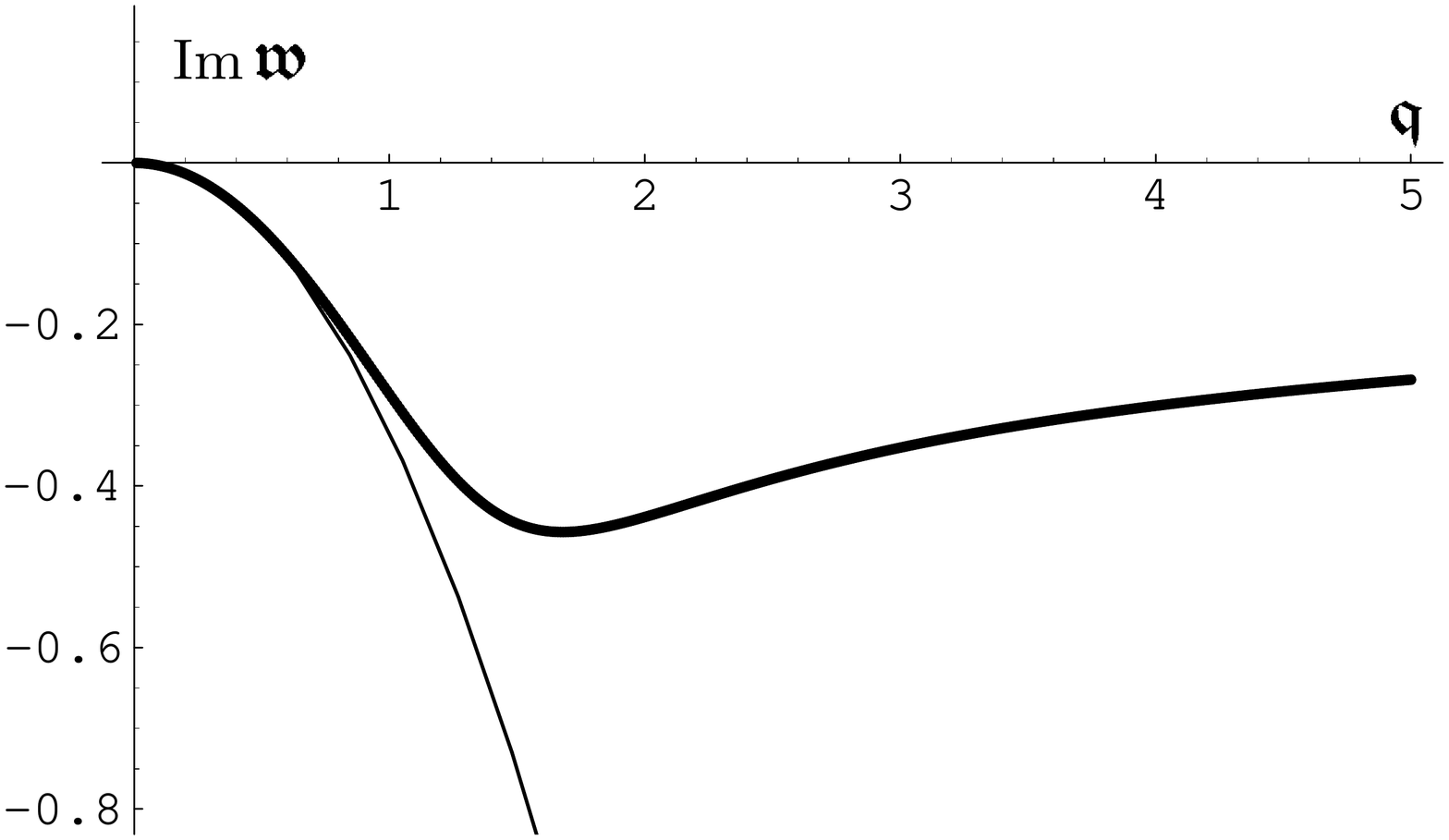}
\caption{ Sound dispersion (real and imaginary parts) obtained  from the analysis 
of quasinormal modes in the AdS black hole background, from  Ref.\protect\cite{Kovtun:2005ev}. Gothic omega and $q$ are frequency and momentum
in units of $2\pi T$.
\label{fig_sound1}}
\end{figure}
The quasinormal modes are poles of $G$ or zeros\footnote{So that only better
behaved solution $Z_a^{II}(u)$  is present at $u=0$.We remind that a
boundary condition at $u=1$ was satisfied already, and thus it is standard
quantization setting.} of $A$ in the lower
part of the frequency complex plane.

To make the equations more familiar they may be easily
put into the form of Schroedinger eqn.
Let me just comment on some issues which may be confusing
for non-experts. 
The issue I want to comment on is existence of the ``horizon''
in the thermal AdS metric at $z=z_h$, and from $z>z_h$ no wave
will return. The same situation also appears in quantum mechanics,
e.g. for radioactive alpha decay problem. There are two ways
to address the problem, which are orthogonal to each other:
(i) to consider scattering formalism with real $\omega,k$ and
multiple (but normalizable) scattering wave functions,
and (ii) use of discrete ``quasitationary states'' which have complex
  $\omega,k$ coming from ``black wall'' quantization condition. 
There is vast literature on how to use those: beware for example
that if the sign of $Im(\omega)$ corresponds to decay of the system,
$Im(k)$ would produce a wave exponentially growing at large distances.
So such states cannot be normalized, and need a lot of care to deal
with.
Quasinormal modes of black holes are also vast field, starting
from original setting  by Regge and Wheeler in 1950's for
Schwartzschild solution, to AdS black holes in various
dimensions:
recent review by Siopsis is recommended \cite{arXiv:0804.2713}.

Let me give one of the examples of the Kovtun-Starinets paper,
showing the real and imaginary part of the
lowest quasinormal mode in the sound channel, see Fig.\ref{fig_sound1}:
we will return to its discussion in section \ref{sec_beyond}.
Note that at small momenta we have a standard sound,
with velocity $c_s=1/\sqrt{3}$ and small dissipation described by
bulk viscosity (thin solid lines) but it departs from this behavior for large
$q>1$. 
Although I do not show spectral densities of the appropriate correlators,
they do contain a ``sound peak" whose position is given by
real part and width by the imaginary part of this quasinormal mode.
Thus (at least the low frequency part of) bulk gravitons
becomes the ``gravity dual" to sound, in a very direct way. 

\subsection{Bulk viscosity}
In the hydrodynamic discussions above we focused on shear viscosity $\eta$ and now
we will discuss recent progress related with
the bulk viscosity $\zeta$.
As their ratio $\zeta/\eta$ is known to be relatively small both
at small and large $T$, one may naturally expect it to be maximal
at or around $T_c$, when compressibility of matter is very small.  
The Kubo formula for $\zeta$ includes correlator of pressures.
Trace of  the stress tensor or the
dilatational charge can
also be used (with care),
and based on corresponding sum rules  Kharzeev et al
\cite{Kharzeev:2007wb} predicted integral of the spectral density. These authors
predicted a dramatic rise of the bulk viscosity
 near  $T_c$, see also Karsch et al \cite{Karsch:2007jc} who sugested
 that at $T_c$ the peak value be as large as $\zeta/s\sim .3$.  
 
 From the phenomenological point of view,one may expect limits
 on the $<\zeta>$ value (averaged over time in hydro evolution, in which
 the near-$T_c$ region makes about 1/3 of the time at RHIC) from
 the same sources as limits on shear viscosity $\eta$. One source
 of that is elliptic flow, which would be destroyed if  $<\zeta>$ be too large:
 in fact the limits obtained by Romatschke \cite{Romatschke:2007mq} and others
 is actually for a combination of $\eta$ and $\zeta$. While we expect $\eta/s$
 to be nearly constant and  $\zeta/s$ peaked near $T_c$, the latter
 gets a smaller weight in the time average: however since the QGP and ``near-$T_c$"
 eras have similar duration at RHIC this factor is about (1/2) or so.
 The same argument applies to conical flow, and perhaps this
 condition is even stronger because (as we argued in our discussion of he
 conical flow above) large angle implies that it is mostly formed
 exactly in the near-$T_c$ region. Unfortunately, to my knowledge
 such phenomenological limits are not yet worked out in the quantitative form:
 my initial guess is that Karsch et al value just mentioned is about as large as
 possible. 
 
 Violation of conformal symmetry in AdS/CFT setting would lead to
 (i) deviation of the thermodynamics, with nonzero $(c_s^2-1/3)$;
 and (ii) nonzero bulk viscosity $\zeta$. Both effects were studied in
  \cite{Buchel:2005cv,Parnachev:2005hh,Benincasa:2005iv} in a
  gauge theory deformed by a mass to $\cal N$=2 super-Yang-Mills.
  Recently Gubser and Nellore \cite{Gubser:2008yx} simplified
  the setting considerably, in the spirit of AdS/QCD\footnote{This large subject by itself is unfortunately outside the scope of this review: the reader
  is advised to look into \cite{Gursoy:2008bu} for a general introduction
and other references.}
  scalar field with some tunable potential which basically mimics
  the beta-function of the desired theory. The relevant action is
 \be \label{StartingAction}
  S = {1 \over 16\pi G_5} \int d^5 x \, \sqrt{-g} \left[
    R - {1 \over 2} (\partial\phi)^2 - V(\phi) \right] 
 \ee
and the corresponding equations of motion are
 \be \label{StartingEoms}
  \triangle \phi = V'(\phi) \qquad
  R_{\mu\nu} - {1 \over 2} R g_{\mu\nu} = \tau_{\mu\nu} 
\ee
where the triangle stands for covariant Dalambetian and the stress tensor for the scalar is
 \be \label{taumunuDef}
  \tau_{\mu\nu} = {1 \over 2} \partial_\mu \phi \partial_\nu \phi -
    {1 \over 4} g_{\mu\nu} (\partial\phi)^2 -
    {1 \over 2} g_{\mu\nu} V(\phi) 
 \ee
The backgrounds of interest have the form
 \be
  ds^2 = e^{2A(r)} \left[ -h(r) dt^2 + d\vec{x}^2 \right] +
    e^{2B(r)} {dr^2 \over h(r)} \qquad \Phi = \phi(r) 
\ee
The choice of radial variable $r$ is arbitrary: re-parameterizing it leads only to a different choice of $B$. 

\begin{figure}[h]
  \begin{centering}
 	\includegraphics[width=10cm]{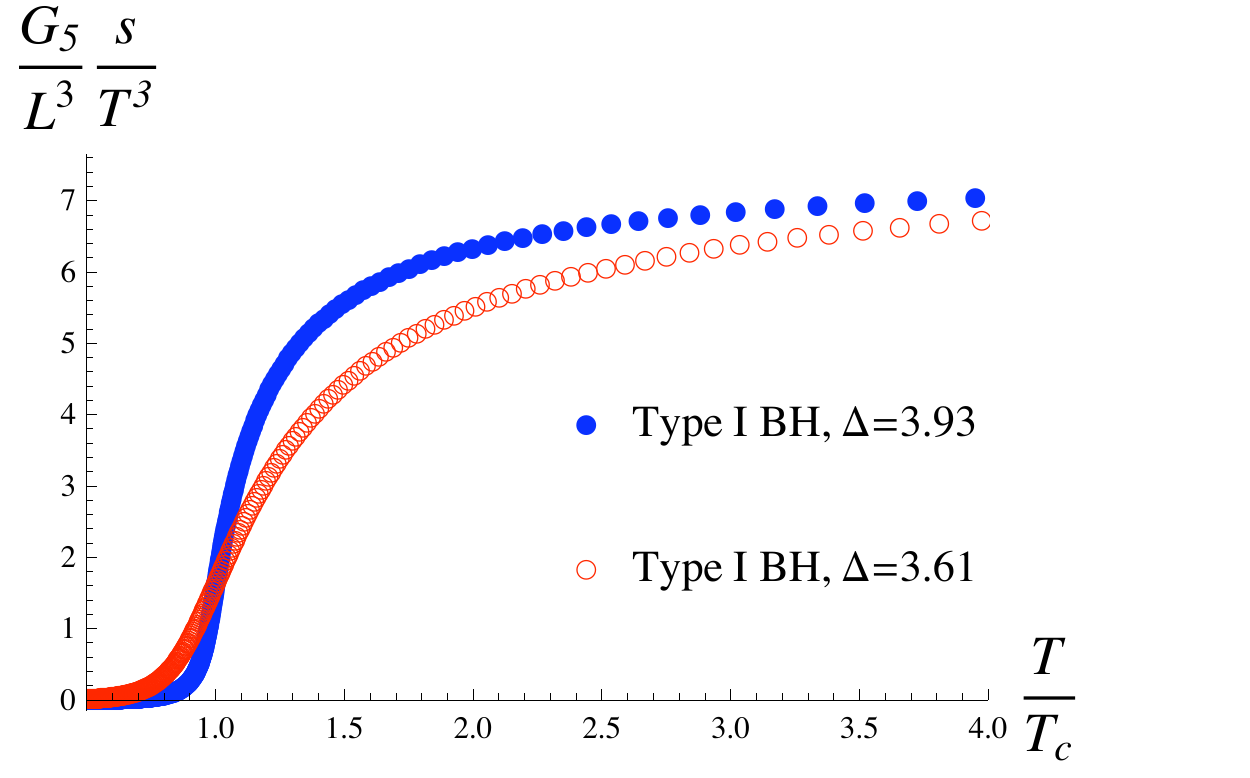}\\
	\includegraphics[width=15cm]{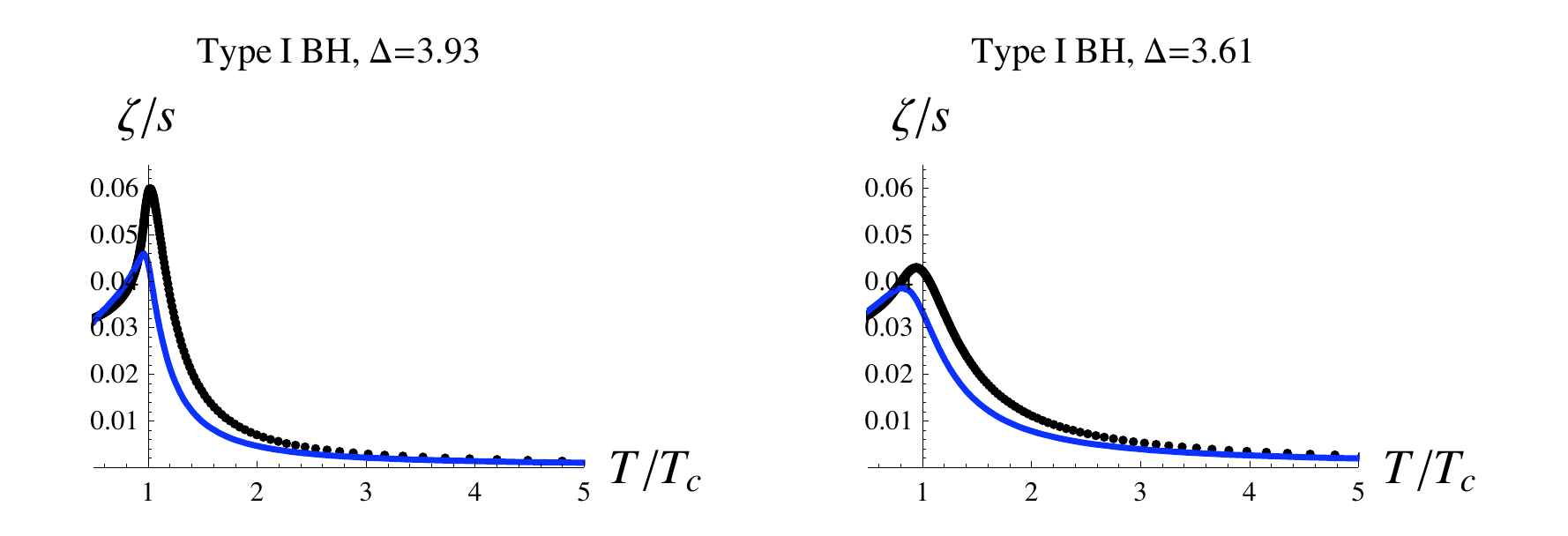}
\end{centering}
   \caption{
   (top)The normalized entropy density
$s/T^3$ as a function of $T/T_c$ for two potentials of the form
     (\ref{QCDLikePotential}) with parameters
	\mbox{$\{\gamma\approx 0.606,b\approx 2.06,\Delta\approx 3.93\}$}
	and $\{\gamma\approx 0.606,b \approx 1.503,\Delta \approx 3.61\}$.  
	(bottom) The ratio of bulk viscosity to entropy density
	$\zeta/s$  for those two potentials, respectively.
 The result of the calculation is
shown by the blue curve (the lower one, except at small T).
 The other (black) curve shows the
	 upper bound suggested by Buchel.
   } \label{ViscosityVsProability}
 \end{figure}
We will follow more recent paper by Gubser et al  \cite{Gubser:2008sz}, from which
Fig.\ref{ViscosityVsProability} is taken. The idea of the method is explained
at Fig.(a), which compares extraction of the shear (left) and bulk (right) viscosities:
in both cases it is given by a probability of a reflected waves. The QCD-like potential
is taken to be in th form   
\be \label{QCDLikePotential}
  V(\phi) = - {12\over L^2} cosh(\gamma \phi) + b \phi^2 \,.
\ee
The parameter $b$  can be adjusted so that the dimension $\Delta$ of the field theory operator dual to the bulk field $\phi$ matches that of $tr F_{\mu\nu}^2$ in QCD at a particular scale.  
$\gamma,\Delta$  can be tuned to get QCD-like behavior of the entropy shown
in the top figure:
 the corresponding bulk viscosity are shown by blue lines in the two
variants of the calculation shown in the bottom.
The peak in bulk viscosity is indeed there, but it is about 5 times smaller than
proposed by Karsch et al \cite{Karsch:2007jc}. If so,
 the bulk viscosity
is probably unimportant in comparison to the shear one at RHIC.
 
\subsection{Deriving gravity dual to (non-linear) hydrodynamics}

Janik and Peschanski \cite{Janik:2005zt}
have proposed an approach which I call ``top-down":
they proposed to take some well known hydro solution -- e.g. the rapidity-independent
Bjorken 1+1 dimensional expansion we discussed
above -- and look for its gravity dual by extrapolation. 
It
can be said to be a combination of a Big Bang expansion in one ``stretching" coordinate
and a black hole in all others. 

They have used  Fefferman-Graham
coordinates, which  is basically
a gauge in which $g_{55}=1/z^2,g_{5i}=0$. 
In this gauge the so called ``holographic renormalization''
 -- the Taylor expansion of the metric near the boundary
\be g_{AB}=g^{(0)}_{AB}+z^2g^{(2)}_{AB}+z^4g^{(4)}_{AB}\ee
 has the most direct physical meaning: the unperturbed
 term $g^{(0)}_{\mu\nu}$ is just flat Minkowski metric, $g^{(2)}_{\mu\nu}=0$ on general grounds 
and $g^{(4)}_{\mu\nu}\sim <T_{\mu\nu}>$ is related to the
induced stress tensor on the boundary.
If the latter is consider known, from some assumed solution to relativistic hydrodynamics,
 standard argument from the theory of differential equations ensures that
 for a second order differential equation
a knowledge of 2 subsequent Taylor coefficients is equivalent to knowledge
of all of them since the equation itself provides the recursive relation.  So
 one can in principle extrapolate from $z=0$ into finite $z$,
recovering the whole ``gravity dual" solution in the bulk.
 
Writing the metric as a late-time expansion of the form (schematically, with
different functions for different pertinent components)
\be g_{\mu\nu} \sim g_{\mu\nu}^{AdS}exp[\sum_n ({1 \over \tau^{2/3}})^n C_n(z^4/\tau^{4/3})]\ee
 with coefficients
depending only on a single scaling variable $z^4/\tau^{4/3}$, Janik and collaborators
have put this series into Einstein equations and got multiple coupled 
eqns for these coefficients.
The leading order solution indeed has   
 a departing $horizon$ located at $z_h\sim \tau^{1/3}$,
 in agreement with the entropy
conservation in ideal hydro.
 The next subleading terms $O(\tau^{-2/3})$ 
has been calculated by Sin and Nakamura  \cite{Nakamura:2006ih}  who 
 identified them with the viscosity effects. The
viscosity value was however only fixed by still further term calculated again by
 Janik, following the following principle:
  all physical observables -- the invariants made of curvature tensors --
 should be {\em nonsingular at the horizon}. However  still the next (the third) term
 calculated in \cite{Benincasa:2007tp} had bad logarithmic singularity
 which produced unphysical imaginary part in curvatures beyond the horizon.  
 This singularity indicated that  something
is serious wrong with the asymptotic solution expressed as series by Janik et al.
Although appearance of $some$ singularities in the bulk are to be expected
on general ground, one usually hope (as the next subsection would show
directly) that 
all singularities are hidden from the distant observers by  corresponding horizons,
at which locally nothing happens to curvature invariants. 
(I argued even before that  something is wrong already with Fefferman-Graham
coordinates, as they cannot show metric beyond the horizon.)

Very recently Heller et al \cite{Heller:2008mb} have
argued that the problem was somehow induced by Fefferman-Graham
coordinates, and in a
 different  (Eddington-Filkenstein) coordinates,
  a non-singular asymptotic solution for Bjorken flow
 is naturally obtained.
 Explicitly the metric found is up to second order
 in gradient expansion.
 The reason why it works in this settings is better explained from the opposite
approach to bulk ``gravity dual", which we will discuss now.
general setting.


 ``Down-up approach'' to derivation of hydro is based on two papers by 
 Bhattacharayya et al \cite{Bhatt:2008jc,Bhatt:2008xc},
see also \cite{Natsuume:2007ty}, who have
  worked out very intuitive and general derivation of hydrodynamics starting from
  (a parameterized) horizon singularity and solving Einstein eqns toward
  the boundary. I view it as the AdS/CFT incarnation
  of the old ``membrane paradigm": the details are different
  (e.g. conformal symmetry now forces the bulk viscosity and many similar
  coefficients to vanish). 
  
  Before we get technical, let me explain its quite intuitive idea for pedestrians.
  Recall our finite-T setting, which I compared to a pool heated from its
  floor. In equilibrium this floor was flat, located at the depth $z_h$ 
  which defined the temperature at the surface $z=0$. Now we generalize this
  problem into a situation which appears in case of $tsunami$ , when the ocean 
  bottom is moving and also not flat. It is intuitively clear that it is
  easier to get waves on the surface starting from the known position of the bottom,
  rather than recalculate what happened on the bottom from the
  observations on the top.

  Simplification of the situation, allowing for a systematic study, is obtained
  due to the assumption of small derivatives (in  3 physical coordinates plus time)
  \be  |{\partial z_h(t,x_i))\over \partial t}|\sim  |{\partial z_h(t,x_i))\over \partial x_i}|\sim \epsilon \ll 1
  \ee 
  In the zeroth order in $\epsilon$ one obviously have a simple ``tubewise"
  approximation, in which $T(t,x_i)$ on the boundary is given just by
  the bottom position at the same time and directly below the observation point.
  The authors worked out two next approximations, the orders $\epsilon, \epsilon^2$ ,
  corresponding to Navier-Stokes and the second-order hydrodynamics, respectively:
 apart of technical complexity of expressions, nothing seems to prevent
 calculation of further terms of this expansion.
   
   Mathematically speaking, 
  the Einstein tensor in 5d has 15 components (thus equations),
the metric after fixing the gauge has 10 components.
As we restricted zeroth order solution 
to an ansatz, the first order (in derivatives) part of metric
is unknown functions which is found from 10 eqns. The rest are
``constrains'' which should be valid without any freedom left:
  thus 4 functions in the original ansatz (the local temperature
and flow velocity $T(x_\mu),u_\mu(x_\mu)$, remember $  u_\mu u^\mu
  =1$)
cannot be arbitrary. Remarkably, the constraints happen to be
(zeroth order) hydrodynamical eqns!

 Let us return to ``equations'' 
for metric: as usual there are three versions of the $z$-eqns (for 
   4d spin-0,1,2 of the 5d gravity).
Keeping derivatives of zero order ansatz as sources,
all of them can be actually solved by
   straight integration  for arbitrary
   sources.    
   The starting point is the form of the metric
    the `boosted black branes'\footnote{The indices in the boundary  are  raised and lowered
with the Minkowski metric  $u_{\mu} = \eta_{\mu \nu} \, u^\nu$. }
\begin{equation}\label{boostedbrane}
ds^2 =-2\, u_{\mu}\, dx^{\mu} dr -r^2\, f(b\,r)\, u_{\mu}\, u_{\nu}\, dx^{\mu}dx^{\nu} + r^2\, P_{\mu\nu}\, dx^{\mu} dx^{\nu} \ , 
\end{equation}
with
\begin{equation} \label{defun} 
f(r)  =1- \frac{1}{ r^4}, \hspace{.3cm}
u^v=\frac{1}{\sqrt{1-\beta^2}}, \hspace{.3cm}
u^i=\frac{\beta_i}{\sqrt{1-\beta^2}} \ ,
\end{equation}
where the temperature  $T = {1\over \pi\, b}$ and velocities $\beta_i$ are all constants with $\beta^2 = \beta_j \, \beta^j$, and
\begin{equation}\label{defp}
P^{\mu \nu}= u^\mu u^\nu +\eta^{\mu \nu} 
\end{equation}
is the projector onto spatial directions.    The zero of $g_{00}$ is so-to-say proto-horizon: the true event horizon is to be discussed below, after the metric is determined.
   
   If all 4 parameters -- $T$ and velocity -- are promoted into {\em slowly varying}
   functions of space and time, the Einstein eqn are no longer automatically satisfied: yet
   it remains a good zeroth order approximation, with systematic corrections\footnote{In this respect the calculation is similar to a derivation of the effective chiral Lagrangian
   in its spirit.}
   organized in power of $\epsilon$.  At the end, a set of equations one get
   for the functions turned out to be the fluid dynamics eqns on the boundary.

Let me omit the solution itself: just saying that second order eqns in
holographic coordinate $r$ are as usual classified into 3 classes, with spin 0,1,2,
each solved for arbitrary ``source terms" coming from all those derivative
terms which Einstein eqns generate.   The general second order expressions are rather
lengthy: let me just give for illustration
the explicit global metric
to first order
in boundary derivatives about $y^\mu=0$
$$ \label{fullmettwo}
ds^2 =  2\, dv \, dr -r^2f(r) \, dv^2 + r^2 \, dx_i \, dx^i  
- 2\, x^{\mu}\, \partial_{\mu}\beta^{(0)}_i \, dr\, dx^i  -
2\, x^\mu \, \partial_\mu \, 
\beta^{(0)}_ir^2(1-f(r)) \, dv \, dx^i  $$ \be
  - 4 \frac{ x^\mu \partial_\mu \bz}{r^2} \, dv^2
 +2\, r^2\, F(r) \, \sigma^{(0)}_{ij} \, dx^i \, dx^j +{2\over 3}\, r \,\partial_i\beta^{(0)}_i \, dv^2 + 2\,  r \, \partial_v \beta^{(0)}_i \,dv \, dx^i .
\ee
where
\begin{equation}
\label{fdef}
F(r) = \int_r^{\infty}\, dx \,\frac{x^2+x+1}{x (x+1) \left(x^2+1\right)} ={1\over 4}\, \left[\ln\left(\frac{(1+r)^2(1+r^2)}{r^4}\right) - 2\,\arctan(r) +\pi\right] 
\end{equation}	
This metric solves  Einstein's equations to first order in the neighborhood of $x^\mu=0$ provided the functions $\bz$ and $\beta^{(0)}_i$ satisfy
\begin{equation}\label{eomsatisfy}
\partial_t b^{(0)} =  {\partial_i \beta^{(0)}_i \over 3} , \hspace{.3cm}
\partial_i b^{(0)} =\partial_t \beta^{(0)}_i .
\end{equation}
   Given $g^{(1)}$, it is in principle straightforward to use the AdS/CFT correspondence 
to recover the stress tensor. Transforming
the metric to Fefferman-Graham form and taking
   their boundary limit $z\sim 1/r\rightarrow 0$, one can finally
   read off the desired $T_{\mu\nu}$ 
\begin{equation}\label{tmn}
T^{\mu \nu} ={1 \over b^4}
\left( 4\,  u^\mu u^\nu +\eta^{\mu \nu}
\right)  - {2 \, \over b^3} \, \sigma^{\mu \nu} .
\end{equation}
where the last term contains the first order 
term which --as expected --has the  Navier-Stokes shear structure
 \be  \sigma^{\mu\nu}= P^{\mu \alpha} P^{\nu \beta} \, 
\, \partial_{(\alpha} u_{\beta)}
-\frac{1}{3} \, P^{\mu \nu} \, \partial_\alpha u^\alpha \ee
  with known viscosity-to-entropy ratio.

   The same story is repeated
  in the second order, with 1-st order hydro eqns needed for the bulk solution
  and the stress tensor on the boundary containing the second order terms
   \ba \label{fmst} 
T^{\mu\nu}=& (\pi \,T)^4
\left( \eta^{\mu \nu} +4\, u^\mu u^\nu \right) -2\, (\pi\, T )^3 \,\sigma^{\mu \nu} \\
& + (\pi T)^2 \,\left( \left(\ln 2\right) \, T_{2a}^{\mu \nu} +2\, T_{2b}^{\mu\nu} +
\left( 2- \ln2 \right)  \left[ \frac{1}{3} \, T_{2c}^{\mu \nu}
+ T_{2d}^{\mu\nu} + T_{2e}^{\mu\nu} \right] \right) \\
\ea
where
%
$$\sigma^{\mu\nu}= P^{\mu \alpha} P^{\nu \beta} \hspace{3mm} 
\, \partial_{(\alpha} u_{\beta)}
-\frac{1}{3} \, P^{\mu \nu} \, \partial_\alpha u^\alpha \hspace{3mm} 
T_{2a}^{\mu\nu}= \epsilon^{\alpha \beta \gamma (\mu} \, \sigma_{\;\;\gamma}^{\nu)} \, u_\alpha \, \l_\beta \hspace{3mm} 
T_{2b}^{\mu\nu}= \sigma^{\mu\alpha} \sigma_{\;\alpha}^{\nu} - \frac{1}{3}\,P^{\mu\nu}\, \sigma^{\alpha \beta } \sigma_{\alpha \beta} \\
$$
$$
T_{2c}^{\mu\nu}=\partial_\alpha u^\alpha\,\sigma^{\mu\nu} \hspace{3mm} 
T_{2d}^{\mu\nu}=  \CD u^\mu \,   \CD
 u^\nu  - \frac{1}{3}\, P^{\mu\nu}\, \CD u^\alpha  \, \CD u_\alpha  \hspace{3mm} 
  T_{2e}^{\mu \nu}= P^{\mu\alpha} \, P^{\nu\beta}\, \CD \left(\partial_{(\alpha} u_{\beta)}\right)
 - \frac{1} {3}  \,P^{\mu \nu}\,P^{\alpha \beta} \, \CD \left(\partial_\alpha u_\beta  \right) $$
$$
 \l_\mu =\epsilon_{\alpha \beta \gamma \mu} \, u^\alpha \partial^\beta u^\gamma .
 $$
%
 The second order coefficients
 agree with those   found previously by Baier et al \cite{Baier:2007ix}
 who worked out linearized approximation.    
Haack and Yarom\cite{Haack:2008cp}
generalized this analysis for conformal fluids in dimensions 3<D<7.
Thus, in a way we know
 a bit  more about the CFT plasma than about such usual liquids as water!

   The second paper  \cite{Bhatt:2008jc} of this (somewhat enlarged) group 
   of authors provides further important developments and clarifications. One of them
is the issue of finding the true dynamical horizon of the solution
-- which in dynamical situation is of course $not$ located at  zero of the 
$g_{00}$.    (This can be seen from the fact that true horizon must be a null surface.)
Its position can be found also via $\epsilon$ expansion, in a quasi-local
form. I will not here go into real calculation of it, but following the authors themselves explaining the issue with a simpler example. The so called Vaidya metric describes
massless matter accreting into the usual (3+1d) black hole 
\be  ds^2=-(1-{2m(t)\over r}) dt^2+2dtdr+r^2d\Omega^2 \ee
The horizon is at some position $r=r_h(t)$ which must define a null
surface: this condition means that it must satisfy the following eqn
\be r_h(t)=2m(t)+2r_h(t)\dot r_h(t)\ee
Without the last term we have static Schwartzschild black hole with
the usual $ r_h=2m$. With time-depending accretion,
 it is a differential eqn, so in general the horizon position is a global
 entity depending
on the whole history of accretion process $m(t)$. However this is drastically
simplified in the case of {\em slow accretion} with $\dot m=O(\epsilon)$,
to the extent that one can write down a {\em local ansatz}
\be  r_h(t)=2m(t) +am(t)\dot m(t)+bm(t)\dot m(t)^2+cm(t)^2\ddot m(t)\ee
with $a,b,c$ being just numbers fixed from the eqn to be 8,64,32, etc.
The second $O(\epsilon)$ term is positive: thus the event horizon
is $above$ the zero of the $g_{00}$ and calculation of its area is not
affected by it. In a general case, with a horizon slowly depending on all 4 coordinates,
the formulae are more involved but the principle is the same.

\begin{figure}[h!]
\includegraphics[width=5cm]{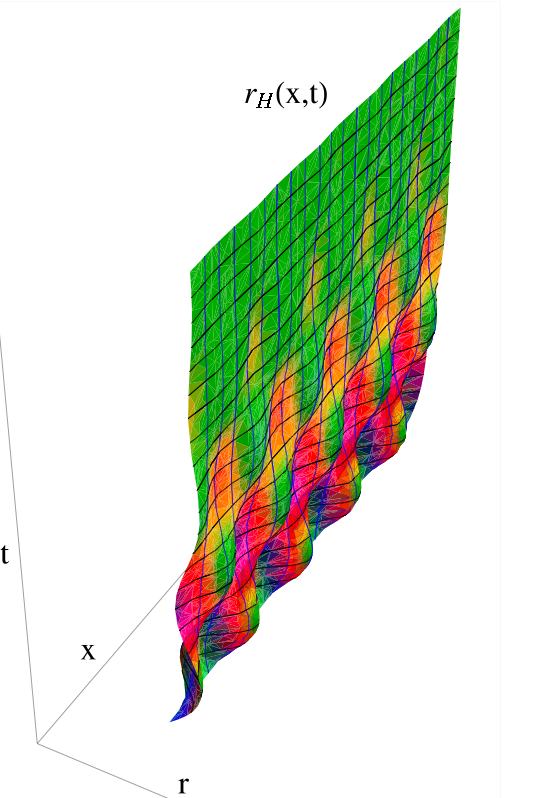} 
\caption{(from 
\protect\cite{Bhatt:2008jc}).The event horizon $r=r_H(x^\mu)$ sketched as a function of the time $t$ and one of the spatial coordinates $x$ (the other two spatial coordinates are suppressed).
} 
\label{wigglyhor}
\end{figure}

   A qualitative picture of evolution of the perturbed horizon is given in the Fig.
   \ref{wigglyhor}
   with time ripples on the horizon disappear 

The last issue they successfully addressed is that of {\em entropy production}.
Even in dynamical situation one expects that the Bekenstein entropy is still ultimately
is given by the horizon area, as it is a UV concept
and cannot depend on soft variation of the horizon shape.
 Since the horizon surface was located explicitly,
one may calculate this area. Furthermore, one may
check the $local$ form of the second law of thermodynamics, namely
that at any spatial point entropy's time derivative is strictly non-negative. 
    This is indeed demonstrated to be the case, to the $\epsilon^2$ order
    solution obtained.
    
    (That can be of course be 
    checked from the second order hydro eqns: this is the usual form
    of entropy production dissipative terms. What is shown in 
\cite{Bhatt:2008jc} is more than that: the  positivity of entropy production
     is true not only in our world, on the pool's surface, but all the way
    to the ``hell" with its  infinite temperature and even locally at any point!)
    
    We have pictorially argued in the preceding sections that
   gravity dual to dissipative relaxation processes means falling into
   the gravitational abyss.  Now, for pure gravity falling, 
   \cite{Bhatt:2008jc} found how to trace the
   entropy production locally in time (and $z$ coordinate), relating it to $r_z$ and its
   derivative terms -- the declined/time dependent bottom. Similarly one may
   ask how
the entropy is produced if other forms of bulk matter -- e.g. classical strings --
are falling as well.      
    
\subsection{Entropy production in hydro: beyond the derivative expansion}
\label{sec_beyond}
With derivative expansion being now under theoretical control, one
may  try to go beyond it and think about some kind of
re-summations of all dissipative effects.
This is not yet done: but let me provide two 
different reasons for doing 
that, one based on phenomenology and one on theory.  

Phenomenology of hydro, discussed at the beginning of the review,
hints that its applicability  starts
 very early, at a time of about 1/2 fm/c at RHIC.
At this time the longitudinal derivatives are not formally small enough
yet to make the dissipative (Navier-Stokes etc) terms
negligible even for the smallest $\eta/s=1/4\pi$; and yet
the ideal hydro seem to
do a good job even keeping the entropy.
Can it be that the higher order terms somehow compensate the first
 (viscosity), making its total effect smaller? 

The theory issue pointed out by  Lublinsky and
myself \cite{Lublinsky:2007mm}, comes from the behavior of the quasinormal
modes. Although there is nothing linearized about the early
stages of heavy ion collisions, let me still
use this example  to illustrate  the issue.

 Writing down
the Navier-Stokes eqn and solving it for sound dispersion 
produces the following  dispersion relation
\be \omega=-{i\Gamma_s k^2 \over 2} \pm \sqrt{c_s^2k^2-\Gamma_s^2 k^4/4} 
 \ee
where $\Gamma_s=\eta/(\epsilon+p)$ is the so called sound
absorption length. Reliable  hydro region is of course at small $k$,
so one should disregard the last higher order term under the square root
and get the textbook answer. Let me for pedagogical reason
keep this last term and see what it does: it curves the path of $\omega(k)$
on the complex plane so that
that left and right moving sound waves make half circle
 and collide at a finite $k$ when the square root vanishes: apparently there is no sound beyond that point, the Navier and Stokes would say.

  What we just did is of course not justified because higher order 
  gradient terms were not included. In the preceding section 
  we discussed the $second$ order gradients recently determined:
  it turns out those are  $O(k^3)$ real correction for the sound: 
they are increasing its speed a bit but keep
sound dissipation unchanged.
  For the diffusion (shear) mode there is a correction
  to the imaginary part \footnote{We remind that this expression is written in units of  $\pi T$ for $\omega,k$.} 
\be -Im\omega(k)=(1/4)k^2+{(2-ln2)\over 32} k^4+...\ee
which has the same sign as the first term, increasing the dissipation as $k$ (or gradients) grows.

Thus both examples seem to tell us that dissipation grows with
$k$ monotonously, even more so when we try to include corrections. 
Yet it is not clear how $Im\omega(k)$
depends on $k$ away from the origin, or  even if it is growing monotonously.
Condensed matter physics is full of such examples, with peaks in
the dissipation at some characteristic momenta/frequencies.

\begin{figure}[t]
\includegraphics[width=16.5cm]{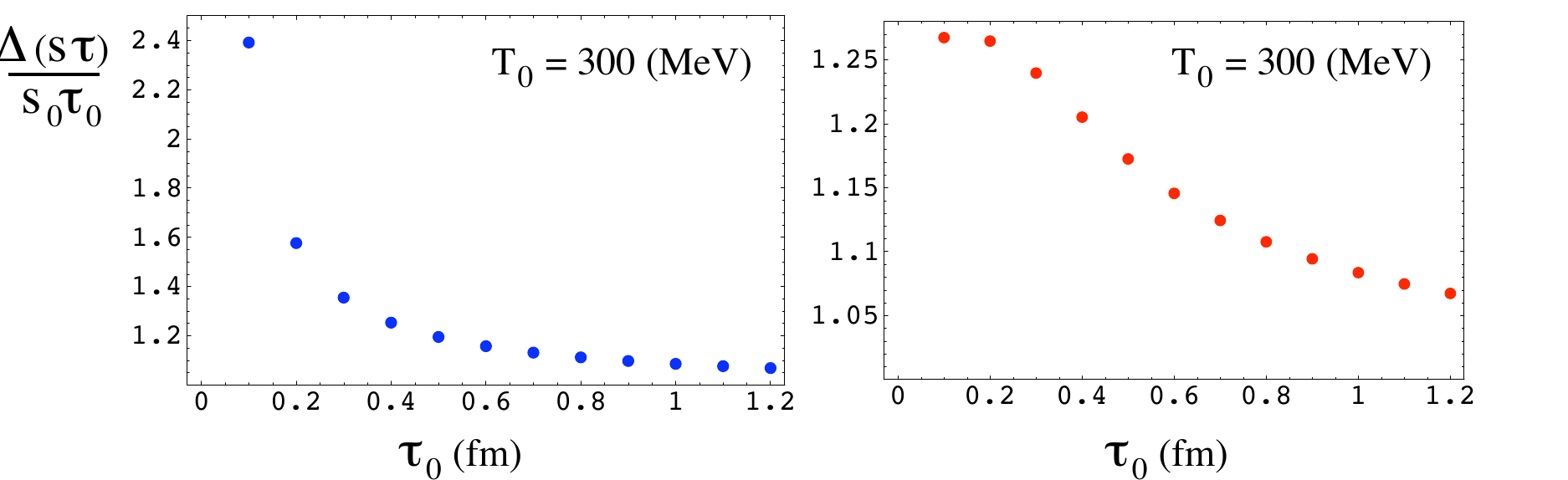}
  \vspace{0.1in}
\caption{(Fraction of entropy produced during the hydro phase as a function
of initial proper time. 
The initial temperature $T_0=300\,{\rm MeV}$. 
The left (blue) points
correspond to the first order
 (shear) viscosity approximation. The right (red) points are 
for the all order  re-summation.
\label{fig_sound}}
\end{figure}

In fact the lowest
quasinormal mode from AdS/CFT does turn around, see
Fig.\ref{fig_sound1}(b) and its imaginary part is reduced
after 
the
 maximum at $k/2\pi T\sim 1$.
  This means that the expansion in derivatives should  
have a radius of convergence $q/2\pi T\sim 1$.

The exercise Lublinsky and myself did was to
introduce effective momentum-dependent 
viscosity $\eta(k)$ and assume
that it is $decreasing$ with momentum k, reproducing
the path of the lowest quasinormal mode\footnote{For clarity:
we don't understand
its physics and do not claim it is the case:
we did it to see what possible effect this would have, if true.}.
 We found that changing constant viscosity to such effective
one turns out to be very 
 important at very early stages, see Fig.\ref{fig_sound}.
While the first order effects (left) more than double the initial entropy
of the flow, the modified viscosity (right) produce
much better controlled hydrodynamics and produces an order of
magnitude smaller amount of entropy.

\section{ AdS/CFT out of equilibrium}
\subsection{Black hole creation: qualitative ideas}
The most challenging
 frontier in the AdS/CFT-based studies
are those attempting to use it {\em  out of equilibrium}, addressing
 initial equilibration  and entropy production.
Non-equilibrium  state between the
collision moment and (``reasonably'') equilibrated QGP is now
called {\em ``glasma''} --  which  so far 
is modeled by
random classical glue by Venugopalan and collaborators, 
via classical Yang-Mills equations in $weak$ coupling. However 
the corresponding 
``saturation scale''  $Q_s$ at RHIC is only about 1-1.5 GeV -- not far
from parton momenta in sQGP, the perfect liquid as we now know --
 so one
may wander if a $strongly$ coupled regime should be
 tried instead.
 This is what I propose to call ``sglasma'' frontier.

If AdS/CFT is the tool to be used,  one has to start with
 high energy collision inside cold $T=0$ $AdS_5$ (the vacuum,
or a $bottomless$ pool) and then dynamically
solve two difficult problems: (i) explain how and with what accuracy
 ``the
collision debris'' act like 
a ``heater'' imitating black/hot patch of Fig.\ref{fig_relaxation};
(ii) find a consistent solution with  ``falling bottom'', $z_h(time)$,
and find its
hologram describing hydro explosion/cooling. 

Since the temperature/entropy only appear when
a horizon is dynamically created, 
leading to the information loss, one may say that 
``falling debris'' upgrade the extreme black hole into 
a non-extreme one.

A lot of work has been done on
gravitational collapse in asymptotically flat 4d space,
 for collapse into real black holes in our Universe.
With modified multidimensional gravity, 
people are thinking about their possible formation in  LHC
experiments, and this also created extensive literature.
 In the AdS/CFT language we are sure that
black hole production is not a rare event: in fact it must happen
in each and every
  RHIC heavy ion collision event, 
with an effective  gravity (imitating QCD) in the imaginary
(unreal) 5-th dimension. 

In heavy ion context, 
 Sin, Zahed and myself~\cite{Shuryak:2005ia}
first argued that exploding/cooling
fireball on the brane is dual to a
 $departing$ small black hole (separate from the AdS center),
  formed by the collision debris and then 
falling toward the AdS center. A specific solution 
 discussed in that paper 
was a brane departing from a static black hole, which generated
a ``spherical'' solution
(no dependence on all 3 spatial coordinates) with a time-dependent
$T$ 
(which however is more 
 appropriate for cosmology but not heavy ion
applications). We also discussed several idealized settings, 
with d-dimensional stretching, corresponding for d=1 to
a collision of two infinite  thin walls and subsequent Bjorken
rapidity-independent expansion, with 
 2d and 3d corresponding to cylindrical and spherical 
relativistic collapsing walls.

\begin{figure}[t]
\centerline{\includegraphics[width=12cm]{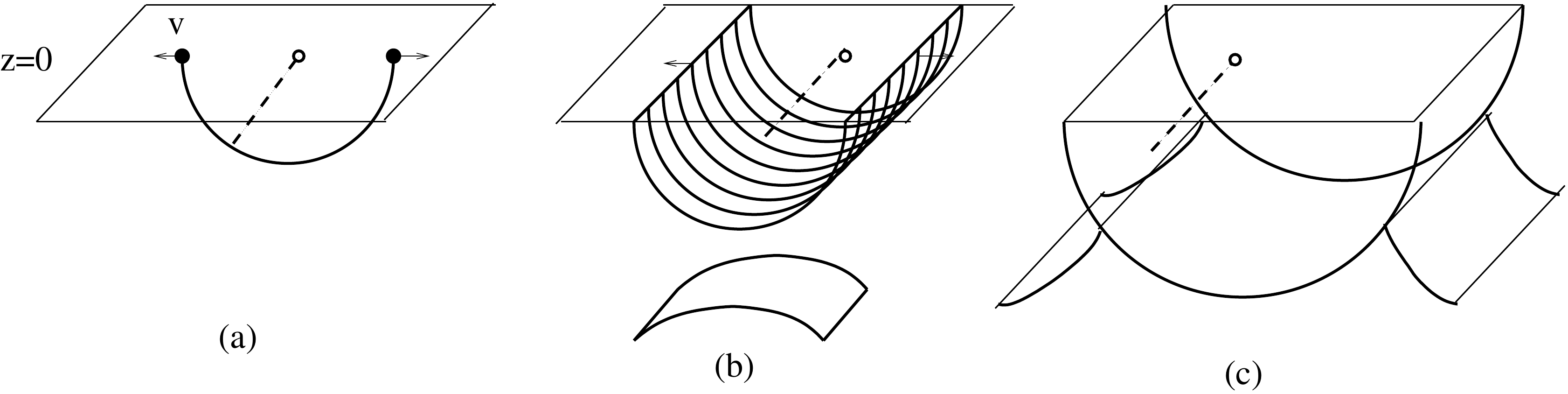}}
\caption{\small Schematic view of the collision setting.
Setting of the sGLASMA studies: (a) a single pair of heavy quark
jets,
moving with velocities $v$ and $-v$ and creating falling string.
Multiple strings create a 3-d falling membrane (2d shown),
which is (b) first far from trapped surface and then
very closed to it (c). 
} \label{fig_membranes}
\end{figure}

Before we embark on some details, let me explain 
what (I think) should be done to solve this problem, schematically
shown in
Fig.\ref{fig_membranes}. The first figure (a) shows the first
step -- a string which belongs to expanding dipole, with charges
moving away from each other. The string solution and the corresponding
hologram will be discussed in the next subsection: let us now just
note that in this case no horizon is produced and thus no temperature,
entropy or hydrodynamics are expected. The second figure (b)
display multiple strings with departing ends: those we think
can be a good approximation for heavy ion collisions, at least
if we think of heavy ions as being made of heavy quarks. If the 
combined mass of all the strings (or other collision debris)
is large enough, their gravity will induce a bubble of
trapped surface (a part of is shown in the lower part
of figure (b)). From the viewpoint of distant observer,
the two membranes -- the falling debris and rising horizon ones --
will be after some time glued and moved  together, as shown
in figure (c). This is what I call the {\em two membranes paradigm},
to which we turn in the last subsection.

Completing this introductory subsection, let
 me mention a separate direction by Kajantie
et al \cite{Kajantie:2007bn} addressing
the same issues in the 1+1 dimensional world. It is easier to work out
math in this case: and one can indeed see how a black hole
is produced. However,
 since the shear viscosity is absent in it
because of 1+1 dimensions  and
the bulk viscosity is prohibited by conformity, so it is a cute
toy case $without$ any dissipation.

\begin{figure}[h]
\includegraphics[width=7cm]{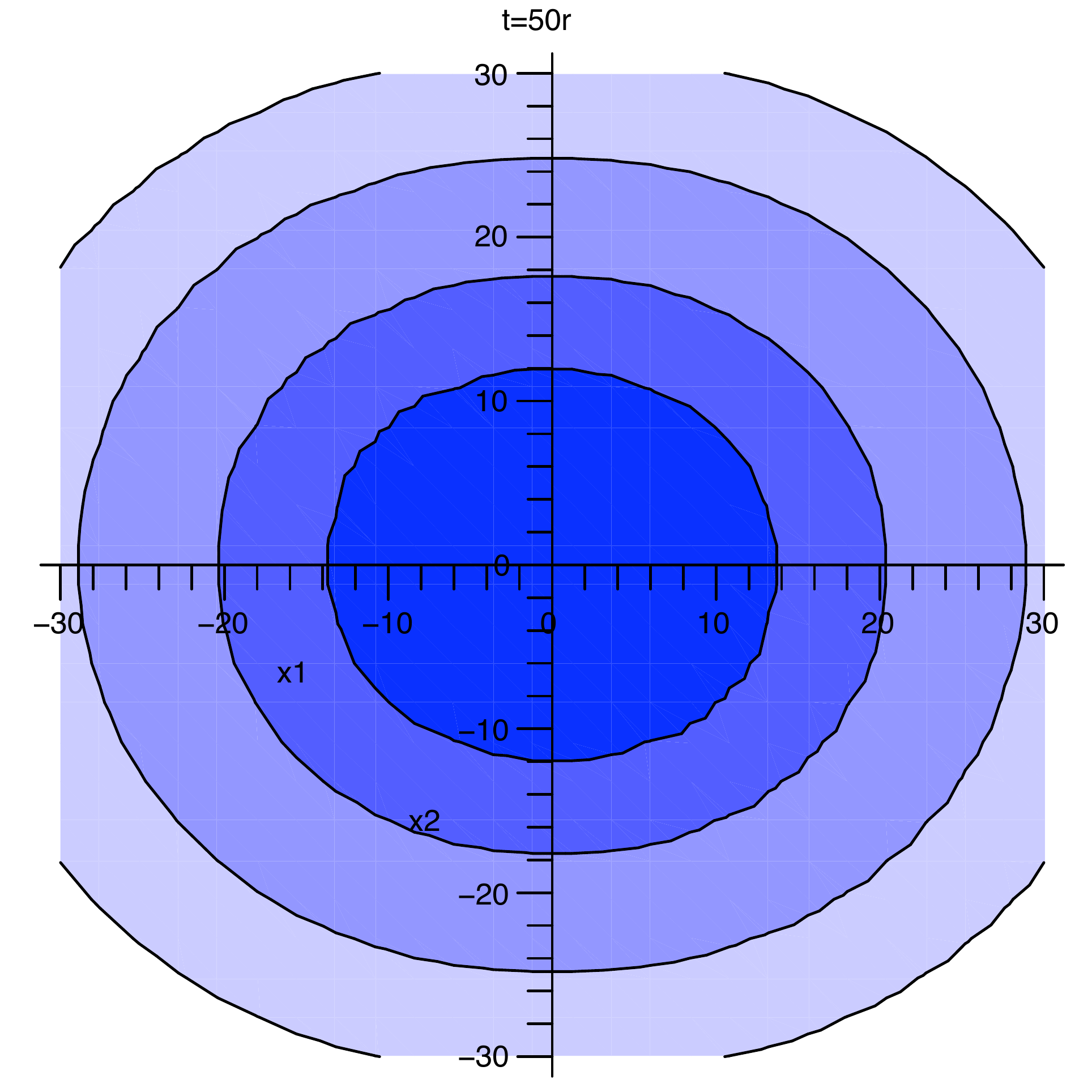}
\includegraphics[width=7cm]{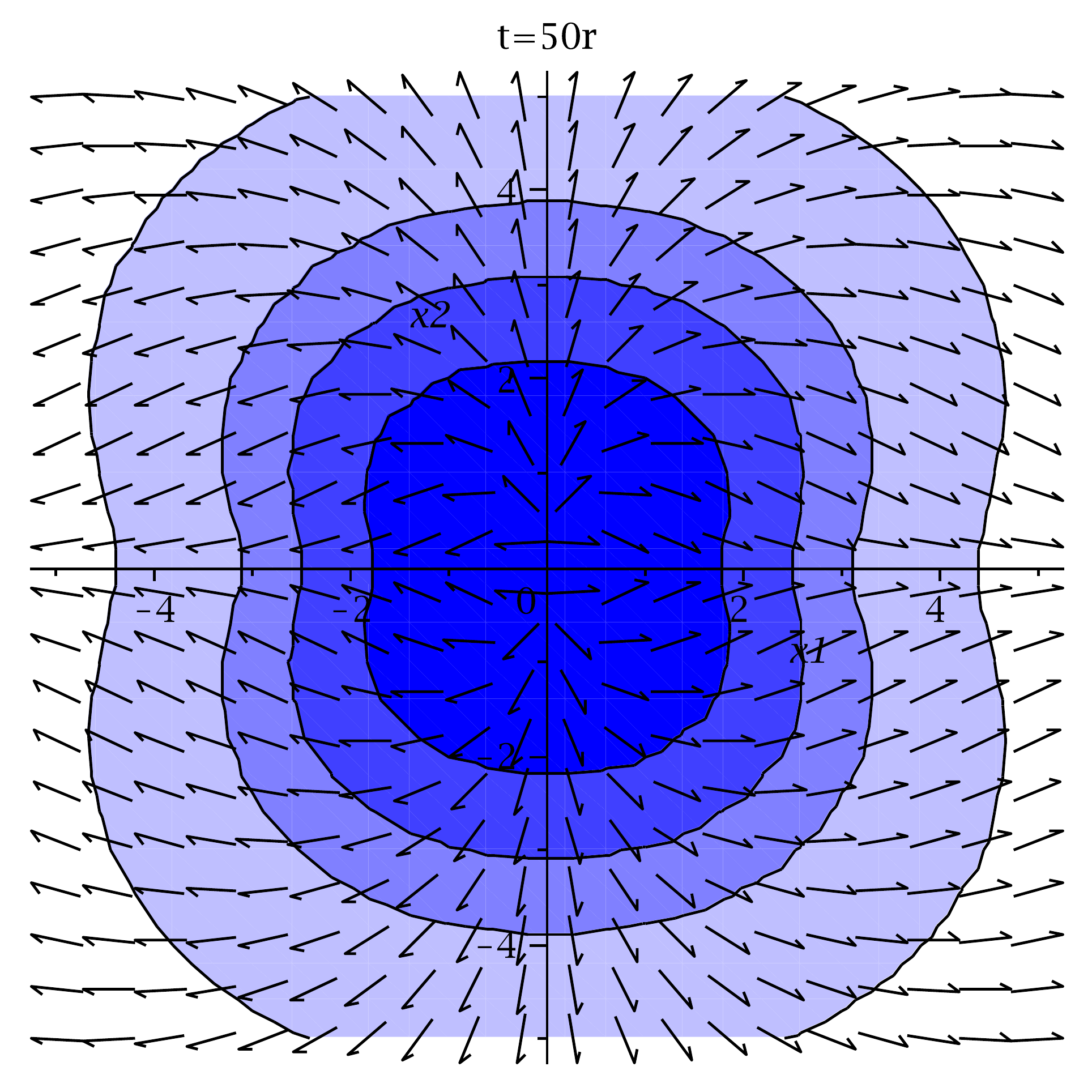}
 \caption{\label{fig_holo2} The hologram of a falling string.
The contours show the magnitude of the energy density (a) and the Poynting vector
$T^{0i}$ in the transverse plane (b) at some time.
The direction of the momentum flow is indicated by arrows.}
\end{figure}

\subsection{Falling strings and ``jets" at strong coupling }

Rather early in development of QCD, when the notion of confinement and
electric flux tubes -- known also as the QCD strings -- were invented in 1970's,
B.Andersen and collaborators \cite{Andersson:1983ia} developed what gets to be known as the Lund model of hadronic collisions. Its main idea is that during short time
of passage of one hadron through another, the strings can get reconnected,
and therefore with certain probability some strings become connected to
color charges in two different hadrons. Those strings get stretched longitudinally
and then break up into parts, making secondary 
mesons and (with smaller probability)
baryons. Many variants of string-based models were developed,
 and some descendants --e.g. PYTHIA -- remain
 widely ``event generators'' till today.
 
  AdS/CFT version of the Lund model has been developed in two
works by S.Lin and myself \cite{Lin:2006rf,Lin:2007pv}. Before we describe some
of its results, let me mention another possible usage of those results.
Some of the dreams high energy theorists have about possible discoveries
at LHC or beyond is existence of ``hidden" interactions which we don't see
because the so called ``mediators" -- particles which are charged both
under Standard Model and hidden theory -- are heavy. Perhaps the hidden sector
is strongly coupled: if so one may wander how production of a pair
of new charges would look like. Further discussion of this issue (and  more
references) can be found in a paper by
Hofman and Maldacena\cite{Hofman:2008ar}, who made further
steps toward understanding the  ``strongly coupled collider physics''.

 Let us now return to \cite{Lin:2006rf,Lin:2007pv} and consider
two heavy quarks $departing$ from each other with velocities $\pm v$  
and  connected by a flux tube (classical string): in AdS setting it
 is not  breaking\footnote{This is so for classical string
minimizing the action. However
 account for fluctuations of the string allows it to touch
the flavor brane and break: in fact Peeters, Sonnenschein and Zamaklar
\protect\cite{Peeters:2005fq} have calculated
the holographic decays of large-spin mesons this way.}
 but rather 
$falling$ into the 5-th $z$ direction, see Fig.\ref{fig_membranes}(a).
 Its action is the familiar Nambu-Goto action in $AdS_5$ background, and if one 
ignores two transverse coordinates $x_2,x_3$ and
uses as two internal coordinates 
 the    $t,x$ (time and longitudinal coordinate)
 the string is described 
by by one function of two variables $z(x,t)$. The corresponding string
action is t
\be S=-{R^2\over 2\pi\alpha'}\int dt\int {dx\over z^2}
 \sqrt{1+({\partial z\over \partial x})^2-({\partial z\over \partial t})^2}\ee
 Before solving the corresponding equation in full, we will
 first discuss ``scaling'' solutions
in
the separable form 
\be z(\tau,\eta)={\tau\over f(\eta)}\ee
where $\tau,\eta$ are the proper time and space-time rapidity
we discussed when we discussed Bjorken solution. This form,
suggested by conformal properties of the theory,
 were used
in literature \cite{Gross_Makeenko},
 in a different -- Euclidean -- context,
 for AdS/CFT calculation
of the anomalous dimensions of ``kinks'' on the Wilson lines
(of which our produced pair of charges is one).
The corresponding solution was obtained, but
we found that it can
 only exists for  sufficiently small\footnote{In particular, for
velocities going to zero one finds a strongly coupled version of Ampere's
law, the interaction of two nonrelativistic currents.} velocities.
Moreover, the analysis of the
 classical stability of such scaling solution
revealed that it gets unstable  for $Y>Y_m\approx 1/4$, where
quark rapidity is related to their velocity in the usual way $v=tanh(Y)$.
For larger velocity of the quarks the scaling solution has to be
substituted by a non-scaling one, depending on both variables in a
nontrivial way, which was analyzed numerically. 

The physical picture an observer at the boundary would see
is again given by a 
 (gravitational) hologram of the falling string calculated in the second paper
\cite{Lin:2007pv}. One feature of the result should not be
surprising for the readers who followed the description of the
hologram of the $static$ Maldacena string above: indeed, no trace
of a string is in fact visible. The results --
 shown in
 fig.\ref{fig_holog}(a) are near-spherical explosion.
 This means that {\em there are no jets at strong coupling}! Instead
  of giving rather lengthy formulae let me just comment that
  we found this explosion to be {\em non-thermal and
thus non-hydrodynamical}, in the sense that the stress tensor
found (although of course conserved and traceless)
cannot be parameterized by the energy density and isotropic
pressure. 

\subsection{Entropy production and the ``double membranes paradigm"}
As it was explained in the
previous section, ``top-down"
\cite{Janik:2005zt} and ``down-up"
significantly clarified how
   hydrodynamics can be derived in AdS/CFT. But we also would like
  in principle to understand how initial equilibration
  happened and the ``dynamical horizon" get formed: this means to solving the gravitational collapse problem following Einstein equations from some
initial  ``debris'' produced by  the collision all the way to black hole.

The initial question 
is what can an appropriate representation be
 of the colliding nuclei.
One straightforward approach is by Romatschke and 
Grumiller \cite{Grumiller:2008va} who literally
used the ``shock waves" which holographically leads to  the delta-funciton
like boundary $T_{\mu\nu} $. Their metric for one shock wave is
\be \ee
note that it
grows in into the  $z$ direction indefinitely, thus collision is that of ``icebergs"
which have their main weight deep down. Not surprisingly, the solution
is developing in time ``bottom-up", from infrared to the boundary, with
boundary $T_{\mu\nu} $ growing with the proper time. The solution is first
found analytically at small proper time, expanding in
its  powers. Perhaps, if the authors would be
 able to do the second part of their program -- solve numerically
 Einstein equations for finite proper time, they will be able to see
 approach to equilibrium and eventually transition to hydrodynamical
 expansion and cooling. Even then, the solution is not expected to 
 go to Bjorken flow because the setting is $not$ rapidity independent.

\begin{figure}[t]
\centerline{ \includegraphics[width=10cm]{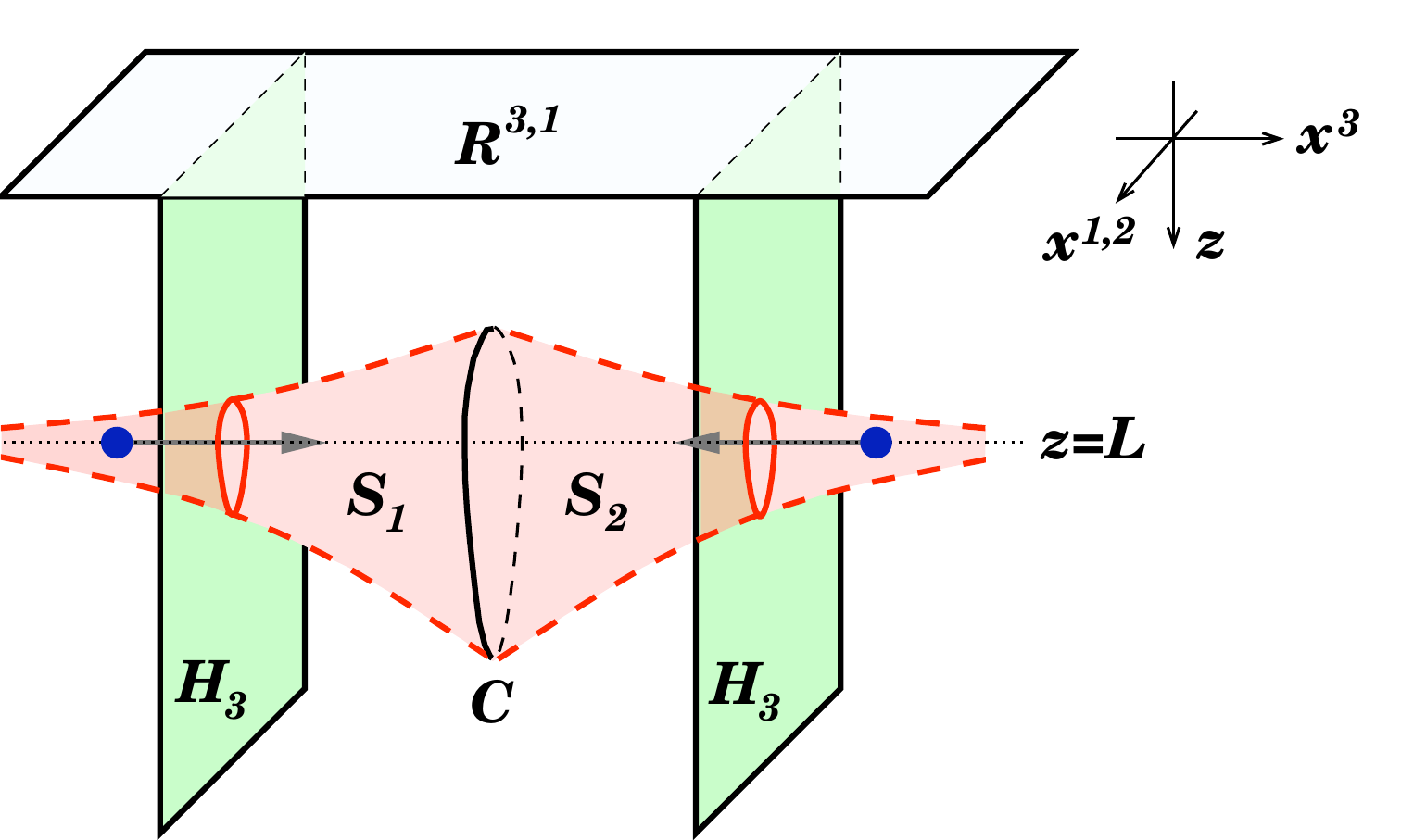}}
\caption{From \protect\cite{Gubser:2008pc}:A projection of the marginally trapped surface that we use onto a fixed time slice of the AdS geometry.  The size of the trapped surface is controlled by the energy of the massless particles that generate the shock waves.  These particles are shown as dark blue dots.
} 
\label{fig_trapped}
\end{figure}

A significant leap forward had been done recently by Gubser, Pufu and Yarom           
\cite{Gubser:2008pc}, who proposed to look at heavy ion collision as a
process of head-on collision of two point-like black holes, separated from the
boundary by some depth $L$ -- tuned to the nuclear size of Au to be about 4 fm,
see Fig.\ref{fig_trapped}.  By using global AdS coordinates,
these authors argued that (apart of obvious axial O(2)
symmetry) this case  has higher -- namely O(3)-- symmetry with the resulting
black hole at the collision moment at its center,
thus  in certain coordinate\be q={\vec x_\perp^2+(z-L)^2\over 4 z L } \ee
 the 3-d trapped surface C at the collision
moment should be just a 3-sphere, at constant $q=q_c$. 
(Here $x_\perp$ are two coordinates transverse to the collision axes.)
The picture of it is shown in Fig.\ref{wigglyhor}(b)

If so, one can find the radius at which it is the trapped null-surface 
and determine its energy and
Bekenstein entropy. For large $q_c$ these expressions are
\be E\approx {4 L^2 q_c^3 \over G_5}, \hspace{.2cm} S\approx {4 \pi L^3 q_c^2 \over G_5}, \ee
from which, eliminating $q_c$, the main result of the paper follows, namely that
the entropy grows with the collision energy as
\be S\sim E^{2/3} \label{eqn_Gubser_S}\ee
Note that this power very much depends on the 5-dimensional gravity and is different
from the 1950's prediction of Fermi/Landau, that this power 
should be 1/2. Simplistic comparison with data ignore the 
nonzero baryon number, important especially at lower collision energies,
which should be of course be removed from the $produced$ entropy\footnote{This 
point was clarified after this question was asked by E.Kiritsis at some meeting.}.

While the model is
far from being realistic, the math of this paper is extremely interesting. First of all, the ``depth" of the colliding objects should not
at all be related to the nuclear size (which can in principle
be taken to be infinite, for two colliding walls) , but rather to 
the typical ``saturation scale"
of parton's momenta. Second, this paper does not try to
answer the most difficult issue -- the dynamical formation of a  horizon.
Indeed, both colliding objects are already black holes, with nonzero
temperature and entropy: at the collision time those are just joint
together into a new shape. And, last but not least, it is not clear whether
the resulting black hole will retain the same ``spherical" shape of the horizon
as it falls into the AdS center: if not, further entropy is produced. All of that
would of course be subject of subsequent works: whether the result
(\ref{eqn_Gubser_S}) will survive or not remains to be seen.

Further work toward  a more realistic  ``gravity dual''
to heavy ion fireball is ongoing. A sketch
of our current thinking has already been presented
in Fig.\ref{fig_membranes}.
 If many strings are falling together their
 combined gravity is non-negligible -- they are partly falling
under their own weight. So
one should solve nonlinear Einstein eqns, which
tell us that (from the viewpoint of distant observer)
extra weight may actually slow down falling, eventually
leading to near-horizon
levitation. 
The trapped surface  is moving first upward (shown at the bottom
of   Fig.\ref{fig_membranes}(b))
 toward the falling
membrane, till two collide, get close and fall together, see 
Fig.\ref{fig_membranes}(c). After that distant observer finds
 a thermal hydrodynamical explosion as a hologram. This is the case
 at mid-rapidity
 but never
in the fragmentation regions.

So far we cannot solve it in realistic geometry, and used
a
simplifications instead. Lin and myself \cite{Lin:2008rw}
considered the case when falling shell (or membrane, made of collision
 debris)
is flat ($x_1,x_2,x_3$-independent). In this collapsing 
shell case one finds a
{\em quasiequilibrium} solution: the metric above the falling
membrane is static thermal AdS in spite of the fact that
the membrane is falling. We derived and solved equation of motion of it --
from the so called Israel junction condition-- and the metric
below the falling membrane, which is simple (vacuum) AdS.

The main question is by which experiments an observer on the
boundary can distinguish true thermal state from ``quasiequilibrium''. 
What we found is that a ``one-point observer" would simply see
equilibrium pressure and energy density, while more sophisticated 
``two-point observer" who can measure correlation functions will see
deviations from  the equilibrium ones in their spectral densities. Solving for various two-point functions
in the background with falling shell/membrane we
found such deviations: they
are oscillating in frequency or peak at certain ``echo" times
: see more on this interesting phenomenon in  \cite{Lin:2008rw}.

\section{Brief summary and two open questions}
As the reader have seen from this review, the field is actively
developing and is very much in flux. At such stage it would be
dangerous to make any firm conclusions: instead let me explain
in general terms where we are now.

The {\em near-$T_c$ temperature region} got a lot of new attention: from
old and experimentally discredited view of being a ``mixed phase" -- a mixture of dense QGP and
much more dilute hadronic phase -- it got a new name and picture, it is now viewed as
a {\em magnetic plasma} region. 
In it monopoles already dominate the bulk of plasma and expelled electric fields/objects,
but they are not yet Bose condensed, as in the confined phase.
 Its theory is rapidly developing, both
based on classical MD and quantum phenomena eventually  
leading to BEC of monopoles and confinement. 

The {\em ``quasi-conformal region"} $T>2T_c$ would perhaps be amenable to
AdS/CFT approach, on which we have many intriguing results
described above. In particular we have gained amazing
insights into the nature of hydrodynamics (deriving it directly from
Einstein eqn, order by order in derivatives) and dissipation in general,
relating those with classical information loss into the black holes.  

Here comes the main open $experimental$ question, originating from the following fact: {\em the RHIC experiments had never ventured into the ``quasi-conformal region"} .
Indeed one needs to run heavy ion collision at LHC to do that. Thus all of us are waiting for the first
heavy ion run and the first ALICE
data, which will tell us if hydrodynamics (elliptic flow especially)
will be as good there as it was at RHIC. The positive answer will put
AdS/CFT -- and our relations with string community -- to further heavy use.
The negative answer would mean that the ``perfect liquid" at RHIC is a near-$T_c$
 phenomenon, perhaps induced by an interplay of electric and magnetic quasiparticles
 we discussed above.
 
 Let me end with the central $theoretical$ question of the day: is there a direct relation between
 two explanations of ``perfect liquid", the one based on electric/magnetic duality/transition and that based on AdS/CFT duality? In essence, the former is based on
 the gauge coupling $g^2/4\pi$ passing from the value $<1$ to
 ``strong coupling" or magnetic domain where it is $>1$. AdS/CFT
duality operates with
the 't Hooft coupling $\lambda=g^2 N_c$. Since $N_c$ is thought of as
 infinitely large, $\lambda$ is large always and the value of $g$ itself
 is of no importance. Whether we are in electric or magnetic domain
 does not really matter! 
What this all means is that a progress in AdS/CFT at {\em finite $N_c$} is badly needed,
before we can unite the two main theory directions -- based on two famous dualities
-- into one consistent theory.

\section{Acknowledgements}
 The work was partially done during my participation
in Galileo Galilei Institute
spring 2008 program, and I am grateful to
the Institute for support and to my fellow co-organizers,
 especially to Adriano Di Giacomo and to Valentin Zakharov, who
made this interesting program happen. Many new things I learned
from its participants are embedded in this review.
 Many more people helped me, especially my
collaborators at Stony Brook 
Ismail Zahed, Derek Teaney, Jinfeng Liao and Shu Li. The
 work is partially
supported by the US-DOE grants DE-FG02-88ER40388 and
DE-FG03-97ER4014.

\end{document}